\documentclass[10pt,usenames,dvipsnames]{article}
\usepackage{enumerate}
\usepackage{amsmath,amssymb}
\usepackage[scr=boondoxo]{mathalfa}
\usepackage{natbib}
\usepackage{caption}
\usepackage{enumitem}
\usepackage{comment}
\usepackage[usenames]{color}
\usepackage{bm}
\usepackage{ying}
\usepackage{multirow}
\usepackage{rotating}
\usepackage{fullpage}
\usepackage{authblk}
\usepackage{mathtools}
\mathtoolsset{showonlyrefs}
\usepackage{geometry}
\usepackage{parskip}
\usepackage{float}
\usepackage{placeins}
\usepackage{afterpage}
\usepackage{etoolbox}
\usepackage{subcaption}
\usepackage{booktabs}
\usepackage{tikz,tabularx}
\usetikzlibrary{patterns}
\usetikzlibrary{arrows}
\usetikzlibrary{tikzmark, positioning, fit, shapes.misc}
\mathtoolsset{showonlyrefs}

\usetikzlibrary{decorations.pathreplacing, calc}
\tikzset{brace/.style={decorate, decoration={brace}},
  brace mirrored/.style={decorate, decoration={brace,mirror}},
}

\newcolumntype{g}{>{\columncolor{red}}c}

\usepackage{graphicx,amssymb}
\usepackage{xcolor}

\usepackage[colorlinks,
            linkcolor=red,
            anchorcolor=blue,
            citecolor=blue
            ]{hyperref}
\usepackage{algorithm}
\usepackage{algorithmic}

\let\emptyset\varnothing

\providecommand{\keywords}[1]
{{ 
  \fontsize{9}{12}\selectfont
  \textbf{\textit{Keywords:}} #1
}}

\theoremstyle{plain}

\allowdisplaybreaks

\usepackage{hyperref}
\def\@#1\@{\begin{align}#1\end{align}}
\def\$#1\${\begin{align*}#1\end{align*}}

\usepackage{color-edits}
\addauthor[Ying]{ying}{magenta}
\addauthor[Xuhe]{xuhe}{blue}

\usepackage{xcolor}
\title{Conformal Annotation with Randomized Expert Review: \\ Application to Online Cell-Type Labeling}
\author[1]{Xuhe Zhang}
\author[2]{Zhi Huang}
\author[1,3]{Ying Jin\thanks{Correspondence to \texttt{yjinstat@wharton.upenn.edu}}}
\affil[1]{Applied Mathematics and Computational Science, University of Pennsylvania} 
\affil[2]{Department of Pathology and Laboratory Medicine, University of Pennsylvania} 
\affil[3]{Department of Statistics and Data Science, University of Pennsylvania} 
\date{}

\begin{document}

\maketitle

\begin{abstract}
Artificial intelligence (AI) models can accelerate cell-type annotation in digital pathology, but safely scaling their use requires determining---with error control---which model-generated labels can be accepted without expert review. This is challenging in evolving workflows because cell morphology and type composition may change across tissue regions and slides, making one-time calibration unreliable. We propose \textit{Conformal Annotation with Randomized Expert review under Distribution shift} (CARED), a distribution-free framework  that leverages randomized expert audit to achieve long-run error control for AI-assisted annotation  under distribution shift. For each incoming batch, CARED uniformly selects a subset of cases for expert review and uses the acquired labels to calibrate later decisions. 
Among unreviewed cases, CARED determines a subset of them to safely skip expert labeling, while suggesting candidate labels for the others for downstream resolution. We prove that this workflow provides long-run control of cumulative error over both the full stream and the cells outside of the audit sample, without stationarity or exchangeability assumptions. We further enhance the framework by introducing class-conditional guarantees to ensure error control within important but rare cell types, together with
a utility module to target more automatic resolutions, smaller candidate sets, and greater biological coherence, without compromising validity. Experiments on pancreas, lung, and ovary annotation streams, together with a stream constructed from real expert annotations, show that CARED controls the cumulative errors while a substantial fraction of cells (up to $70\%$) bypass expert review.
\end{abstract}

\section{Introduction}
\label{sec:intro}

Expert labeling underlies a broad range of scientific, engineering, and medical tasks. Experts examine and assign labels   that guide downstream analyses or decisions on individual cases; the resulting labeled cases also form the datasets used to train and evaluate predictive models~\citep{deng2009imagenet,gulshan2016development}. 
Artificial intelligence (AI) models can accelerate this process by producing predictions at scale which, when sufficiently accurate, can serve as labels~\citep{goel2023llms,gilardi2023chatgpt}, thereby reducing or even bypassing expert review. However, accepting those ``labels'' without expert review leaves the prediction errors fundamentally unobserved~\citep{northcutt2021pervasive,bernhardt2022active}. This creates a tension between reducing expert effort and maintaining label quality~\citep{candes2025probably,huang2025model}.

We study this problem in image-based cell-type annotation, an important task in biomedical research and digital pathology~\citep{ergen2024consensus, hou2024assessing}. 
In this task, a digitalized pathology slide contains many cells, 
each of which must be assigned a cell type to support downstream analysis~\citep{angus2025ai,kim2026transparency,gaffney2025pathology}. 
Emerging AI systems now enable collaboration between pathologists and cell-level AI models: as pathologists proceed across a slide image, the model proposes labels for the cells, which can be reviewed by the pathologists to assign the labels. 
For example, the TissueLab platform~\citep{li2025co} developed by the author(s) allows pathologists to navigate the slides and annotate cells with the support of an in-house AI model~\citep{xu2025adaptive}. 
Such collaborative systems streamline expert annotation, but every model proposal must still be inspected and finalized by a pathologist. 
Here, our goal is to identify cells whose AI predictions can be accepted without expert review while still ensuring long-run error control.


Recent work has begun to provide statistical guarantees for deciding which model-generated labels can be accepted without expert review. 
Assuming access to i.i.d.~(or a uniform random subset of)  labeled data, PAC Labels~\citep{candes2025probably} calibrates a criterion for accepting AI labels in the unlabeled data while controlling the overall error rate, whereas Conformal Labeling~\citep{huang2025model} uses selective inference~\citep{jin2023selection} to select AI labels for unlabeled data while controlling the false discovery rate among the selected labels. 
Their success relies on the data generating process ensuring the AI labeling errors in the labeled data are representative of the unknown errors in the new samples. 

However, as a pathologist moves across regions of a whole-slide image (WSI) and then across slides, the cells may differ in both morphology and cell-type composition. Additional variation can arise from changes in specimen preparation, imaging devices, and data processing~\citep{tellez2019quantifying,stacke2020measuring,howard2021impact,aubreville2023mitosis}. 
Therefore, a rule calibrated on earlier data may fail to provide valid error control as annotation proceeds. 
Figure~\ref{fig:quantile_shift} visualizes such workflow shift using the  NuClass model of \cite{xu2025adaptive}. We partition a pancreas WSI containing more than $10^{5}$
cells into 206 approximately equal-sized spatial regions. 
We then order the regions to mimic a pathologist’s traversal across neighboring fields, with adaptive changes in viewing resolution; panel (a) shows this annotation order. For this diagnostic, we use the held-out label of each cell to compute the NuClass score assigned to its true class. Panel (b) reports the within-fraction empirical $\alpha$-quantiles of these scores, separately for each cell type. 
The dashed rose curve in each panel, read against the right vertical axis, gives the cell-type proportion within each fraction. The composition of cell types varies substantially, irregularly, and at times abruptly across batches.
Moreover, the in-class distributions also vary substantially: 
if the cells within the same type are identically distributed across batches, the quantile curves would remain approximately constant (up to sampling variation). Instead, the curves fluctuate markedly over the annotation order. These within-cell-type distributional changes, together with the composition shifts, make one-time calibration unreliable. 


\newlength{\quantileShiftPanelHeight}
\setlength{\quantileShiftPanelHeight}{0.3265\linewidth}

\begin{figure}[!htbp]
    \centering
    \begin{subfigure}[t]{0.287\linewidth}
        \centering
        \includegraphics[height=\quantileShiftPanelHeight]{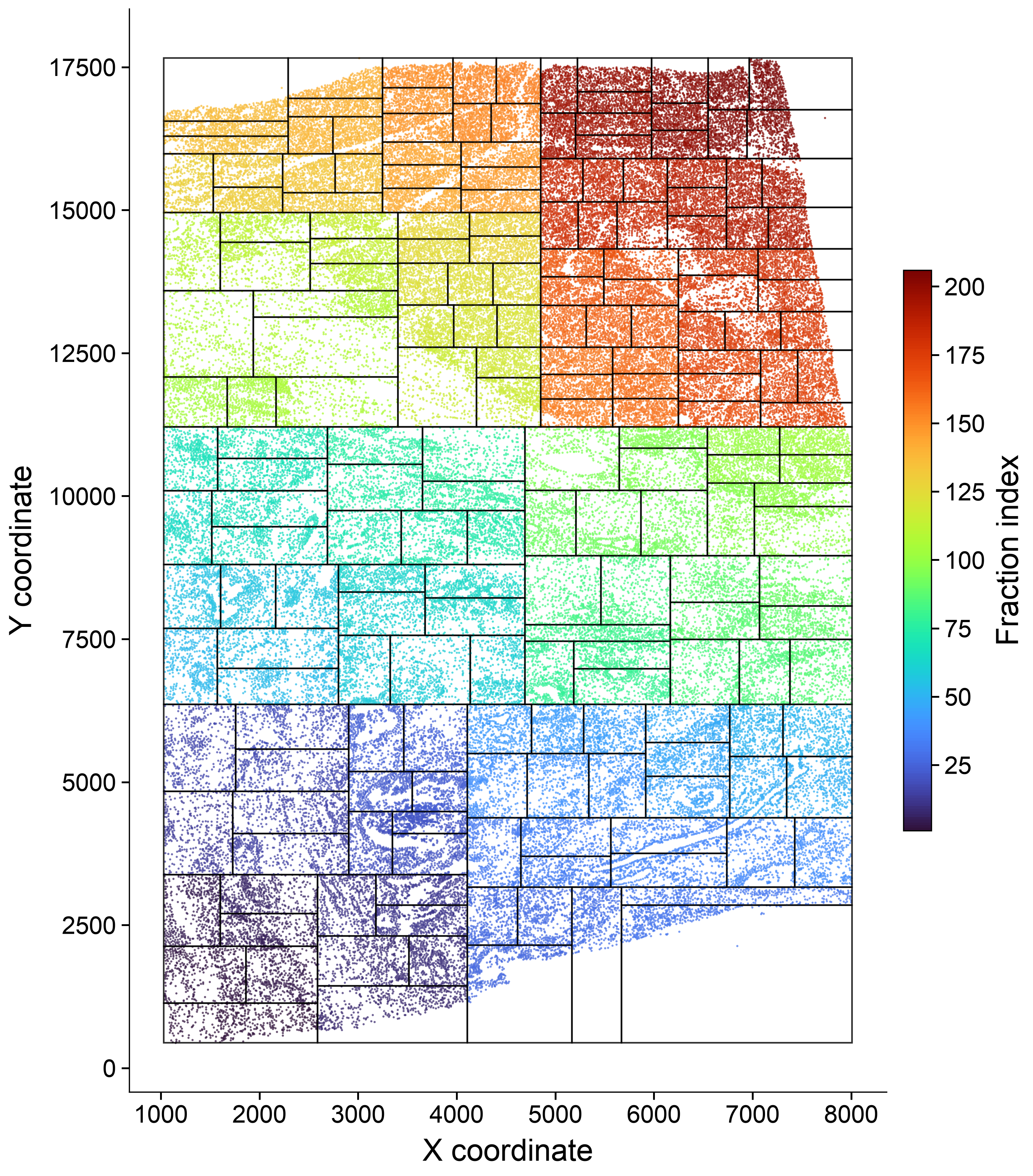}
        \caption{}
    \end{subfigure}
    \hfill
    \begin{subfigure}[t]{0.703\linewidth}
        \centering
        \includegraphics[height=\quantileShiftPanelHeight]{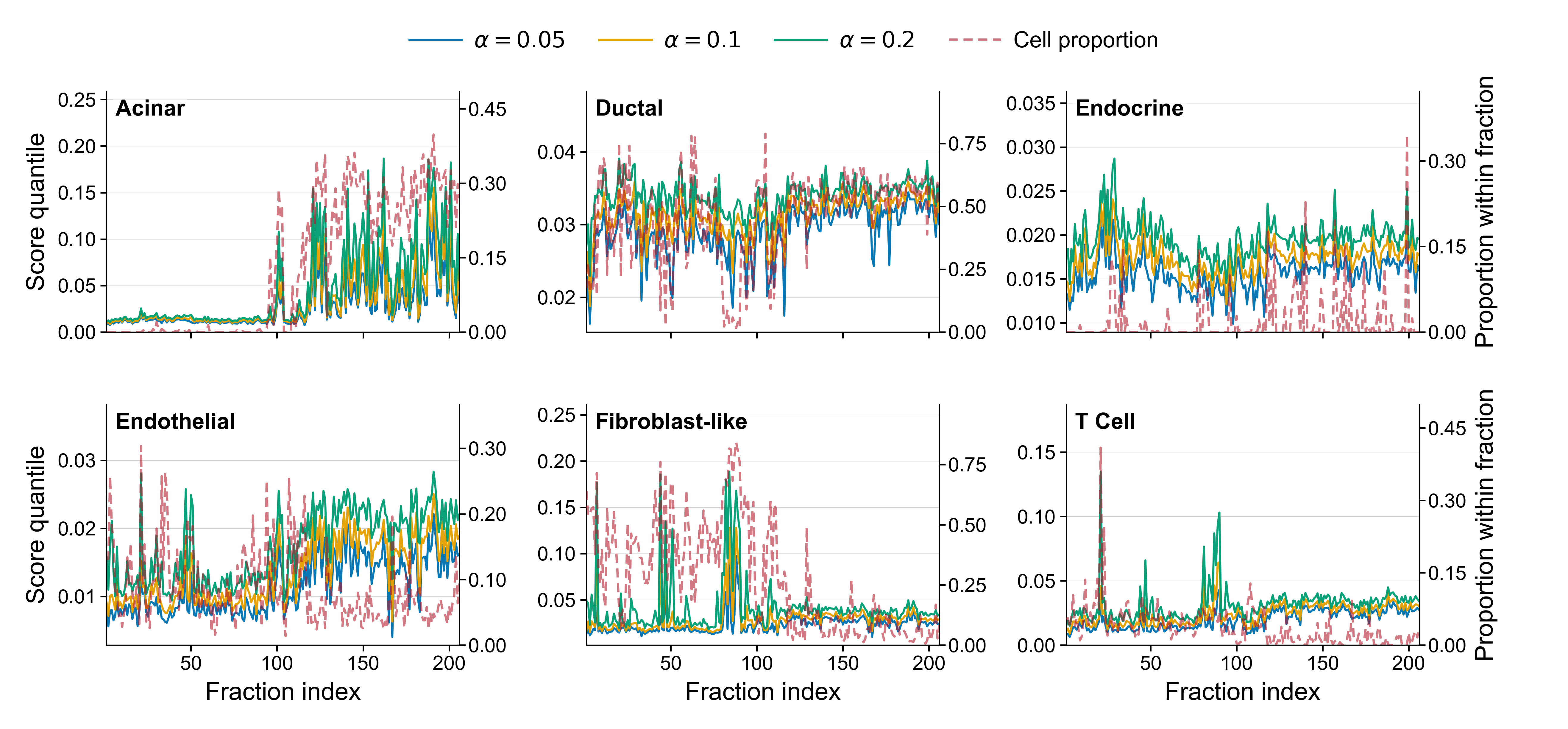}
        \caption{}
    \end{subfigure}
    \caption{\small \textbf{Evidence of workflow shift in image-based cell-type annotation.} A pancreas whole-slide image containing more than $10^5$ cells is partitioned into 206 approximately equal-sized spatial fractions. (a) Spatial locations of cells, with color indicating the order in which the fractions enter the annotation workflow. (b) Empirical $\alpha$-quantiles of the NuClass score assigned to the true class, shown separately by cell type for $\alpha\in \{0.05,0.10,0.20\}$. The dashed rose curve in each panel (right vertical axis) shows the corresponding cell-type proportion. Its irregular and, at times, abrupt changes reveal shifts in cell-type composition, while the variation in the solid quantile curves reveals within-cell-type score-distribution changes over the workflow.}
    \label{fig:quantile_shift}
\end{figure}

\subsection{Overview and contributions}

We introduce \textit{Conformal Annotation with Randomized Expert review under Distribution shift} (CARED), an online framework that uses conformal prediction sets and randomized expert audits to automate a subset of annotations in an evolving data stream.  CARED is designed for cell-type annotation, but  its core construction applies to other multiclass labeling workflows in which limited expert feedback is available.

The first ingredient of CARED is to derive long-run control of cumulative annotation error from coverage of prediction sets~\citep{vovk2005algorithmic}. At round $t$, let $D_t$ denote the incoming unlabeled batch. For each cell $x\in D_t$, the model scores are converted to a prediction set $\hat{C}_t(x)\subseteq \cY$ of plausible cell types, where $\cY$ denotes the label space.
CARED selects cells with $|\hat{C}_t(x)|\leq 1$ for automatic annotation; otherwise, the set is returned to the pathologist as a candidate list for later resolution. We refer to the fraction of cells that bypass expert review as the \emph{skipped ratio}. Let $I_t(x)$ denote whether an automated annotation is produced, let $y(x)$ denote the true cell type, and let $\hat{y}_t(x)$ denote the automated annotation when it exists. Then
\$
\ind\{\hat{y}_t(x)\neq y(x)\}\leq \ind\{y(x) \notin \hat{C}_t(x)\} \qquad \text{whenever } I_t(x)=1.
\$
Thus, the fraction of prediction sets failing to cover the true labels (miscoverage) over the unreviewed stream upper bounds the long-run fraction of unreviewed cells that receive erroneous automatic annotations.

A natural approach to maintain this guarantee as the annotation workflow changes is to recalibrate the prediction sets in an online fashion.
Adaptive conformal inference (ACI)~\citep{gibbs2021adaptive,gibbs2024conformal} provides a starting point: it updates prediction sets using newly observed labels and provides long-run control of cumulative miscoverage regardless of the distributional changes. The standard ACI, however, requires observing the true label after every prediction. Applied directly to annotation, it would request expert feedback for every incoming cell, leaving the burden of pathologist review unchanged.

\begin{figure}
    \centering
    \includegraphics[width=\linewidth]{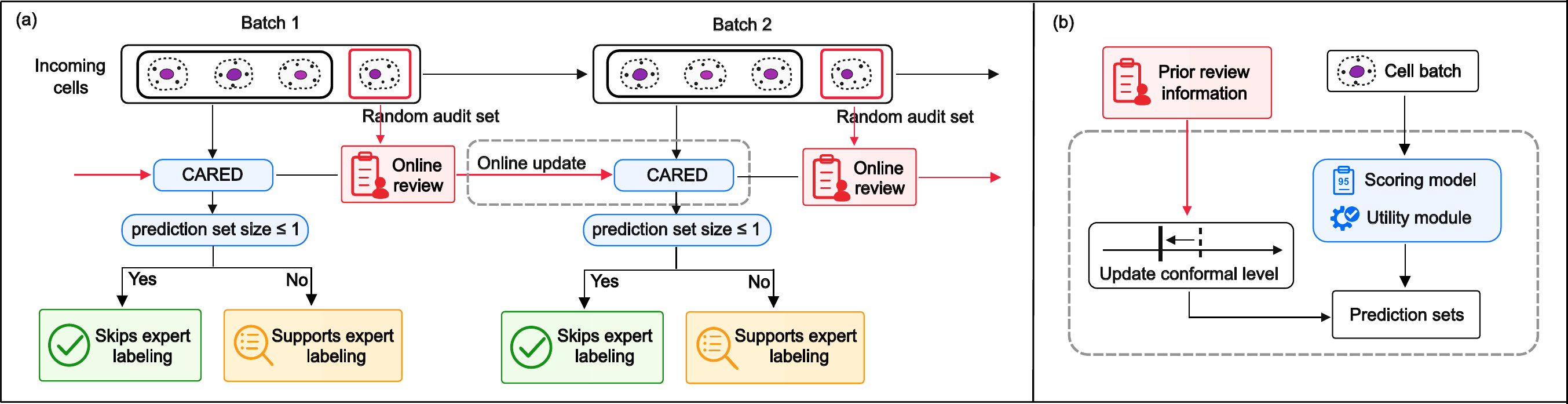}
    \caption{{\small Schematic of the CARED workflow. (a) Incoming cell batches are
    processed sequentially: CARED forms prediction sets, cells with set size at
    most one skip expert labeling, and a random audit set is reviewed to update
    the conformal level for subsequent batches. (b) Zoom-in of the prediction set generation process highlighted by the gray dashed box in panel (a), where prior review
    information updates the conformal level and the scoring model with the
    utility module produces prediction sets for the incoming batch.}}
    \label{fig:intro_workflow}
\end{figure}

CARED resolves this tension by proposing a novel random review mechanism. After prediction sets $\hat{C}_t(x)$ are constructed for $x\in D_t$, a subset $A_t\subseteq D_t$ with fraction $p\in (0,1)$ is sampled uniformly for expert review (audit). The reviewed cells provide unbiased feedback for the whole batch. 
We prove that updating with the audited labels is sufficient to provide long-run control of cumulative prediction-set miscoverage over both the full stream and the cells not selected for audit. Such a guarantee is ensured by the random review mechanism only, requiring no exchangeability or stationarity of the data stream, and allows the batches and prediction models to evolve arbitrarily over time. An overview of CARED is in Figure~\ref{fig:intro_workflow}.

For cell-type annotation, we further enhance the core framework in two ways. 
First, we use class-conditional calibration~\citep{vovk2003mondrian} to achieve the same long-run guarantee within each true cell type, ensuring accuracy in clinically important but rare and challenging cell types. 
Second, we develop a utility-aware module that updates the AI scores to favor more accepted annotations, smaller candidate sets, and biologically coherent outputs.

\begin{figure}
    \centering
    \begin{subfigure}[t]{0.4\linewidth}
        \centering
        \includegraphics[width=\linewidth]{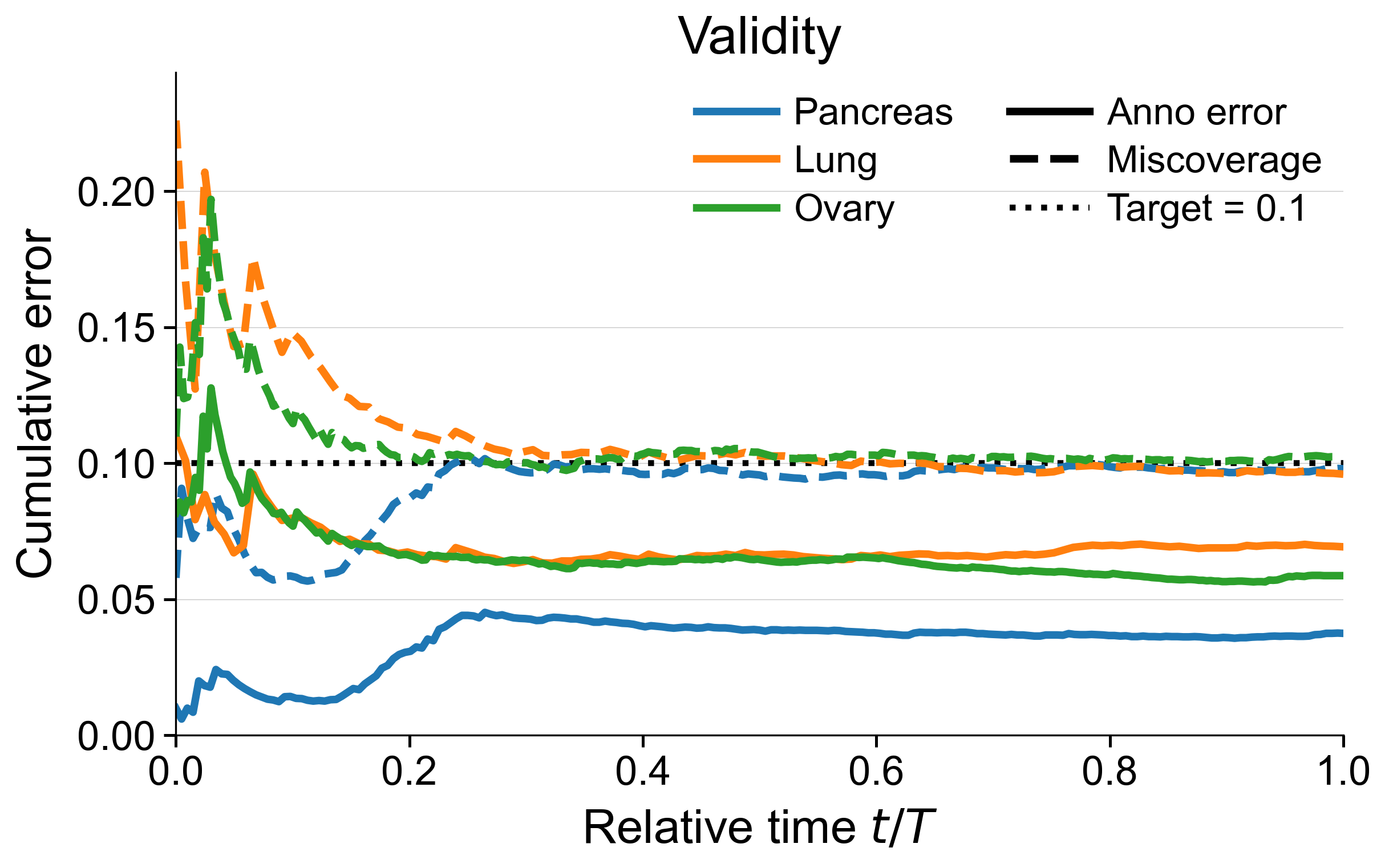}
    \end{subfigure}
    \hspace{0.1\linewidth}
    \begin{subfigure}[t]{0.4\linewidth}
        \centering
        \includegraphics[width=\linewidth]{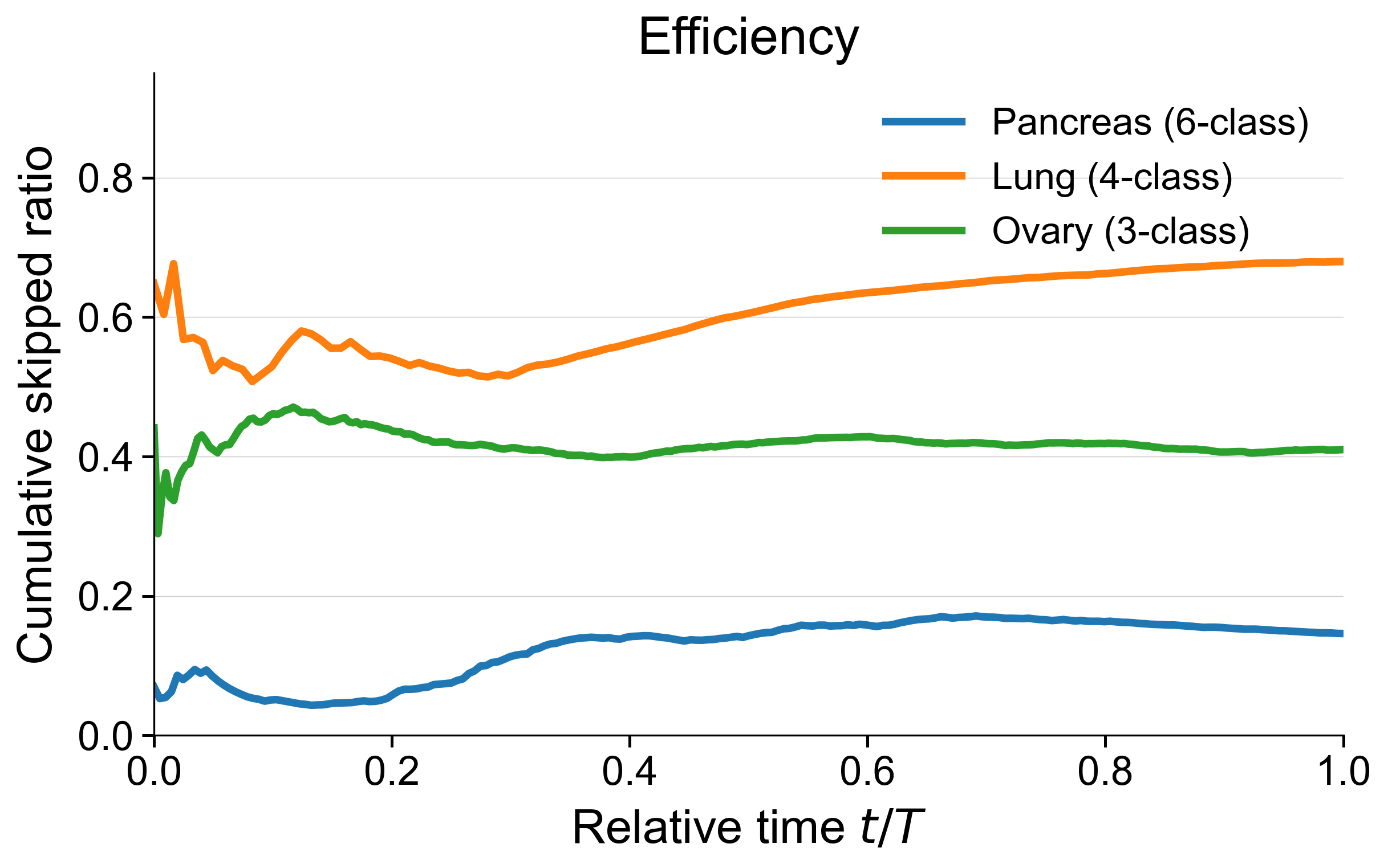}
    \end{subfigure}
    \caption{{\small Validity and efficiency of  CARED across three tissue-specific tasks using the global NuClass score source (Section~\ref{sec:experiments}). The $x$-axis indicates the fraction of processed batches in the stream.
    (a) Cumulative annotation error (solid lines) and miscoverage (dashed lines) for the three tasks; the dotted horizontal line denotes the
    target error level $\alpha=0.1$. (b) Fraction of cases bypassing expert review among all incoming ones for the three tasks at $\alpha=0.1$.}}
    \label{fig:intro_contribution_demo}
\end{figure}

We instantiate CARED with NuClass~\citep{xu2025adaptive} in image-based cell-type annotation and evaluate it on streams constructed from spatially partitioned WSIs across multiple tissue types, together with a stream based on real expert annotations from TissueLab~\citep{li2025co}. Our experiments emulate various workflow shifts in real annotation processes to examine the long-run error control, reduction in expert review, prediction-set utility, and sensitivity to the audit budget and target error level. Figure~\ref{fig:intro_contribution_demo} previews the main validity and efficiency results for CARED. Although only a fraction ($p=0.1$) of each incoming batch is selected for expert audit, the cumulative annotation error (solid lines) remains below the target level, whereas the cumulative miscoverage (dashed lines) moves toward and remains near the target level ($\alpha=0.1$) as the annotation streams progress. At the same time, panel (b) shows that the fraction of cells skipping expert review reaches as high as $70\%$, with its value varying across tissues and label spaces. Together, the two panels illustrate how randomized partial review can support online error control while allowing model-generated labels to replace expert annotation for part of the data stream.

\subsection{Related work}

\paragraph{Statistical guarantees for AI annotation.} Our work adds to the recent literature on how model predictions can replace a portion of expert annotations while retaining explicit guarantees on label quality. PAC Labels combines AI predictions with selectively acquired expert labels within a fixed target collection and controls the overall labeling error of the completed dataset with high probability~\citep{candes2025probably}. 
Conformal Labeling instead uses a labeled calibration sample to select AI-labeled cases from an i.i.d. test sample, controlling the false discovery rate among the selected labels~\citep{huang2025model}. 
These methods address settings where the labeled data for calibrating the annotation rule are representative of the unlabeled cases. CARED considers a different regime with potentially arbitrary workflow shift where expert labels are obtained through randomized audits. Its target is long-run error control over the full and non-audited streams, which in turn bounds the fraction of non-audited cases receiving erroneous automatic annotations. 

\vspace{-0.5em}
\paragraph{Human-in-the-loop annotation and partial-review workflows.}
Our workflow is closely related to human-in-the-loop annotation and partial-review diagnostic pipelines, where model outputs reduce expert effort. In computational pathology, prior work has studied semi-automated annotation interfaces and practical protocols for reducing expert-labeling time and labor \citep{cazzaniga2025improving,montezuma2023annotating,montezuma2025annotation} especially pathologist-supervised workflows~\citep{boden2021human,rau2024closing,gaffney2025pathology}. 
These works motivate the partial-review setting but focus mainly on workflow design and implementation, whereas our objective is to provide statistical guarantees through  partial review. A complementary literature on learning to defer jointly learns when a model should predict and when a case should be deferred to an expert~\citep{mozannar2020consistent,charusaie2022sample}. These methods optimize the prediction and deferral loss, whereas CARED aims for distribution-free long-run validity for the non-deferred cases. As CARED constructs prediction sets for deferred cases, the philosophy is close to~\cite{jin2026act}.

\vspace{-0.5em}
\paragraph{Conformal methods for cell-type annotation.}
Cell-type annotation is an important task in multiple areas. In single-cell genomics, representative procedures include reference mapping, consensus label transfer, atlas alignment, transfer learning, and, more recently, LLM-assisted annotation \citep{lotfollahi2024future,ergen2024consensus,korsunsky2019fast,lotfollahi2022mapping,hou2024assessing}. 
In parallel, image-based pathology has begun moving from generic tissue-level modeling toward direct cell-level annotation and classification from histopathology or multiplex tissue images \citep{li2025automated,xu2025adaptive}. More recently, conformal methods have also been developed for cell-type annotation~\citep{lopez2025conformal}. In particular, \citet{corbetta2026conformal} highlighted the distribution shift issue and modeled distribution shift as label shift (assuming the distribution of data within each cell type is invariant) or covariate shift (assuming the label distribution given features is invariant). However, these models do not resolve the adaptive, unpredictable changes prevalent in cell-type annotation settings (Figure~\ref{fig:quantile_shift}). In addition, we focus on annotation settings whereas these methods focus on the construction of prediction sets.


\vspace{-0.5em}
\paragraph{Adaptive conformal inference.}
Our random review mechanism extends the  adaptive conformal inference (ACI) framework~\citep{gibbs2021adaptive} whose representative extensions concern general update rules \citep{angelopoulos2024online,gibbs2024conformal} and time-series settings \citep{zaffran2022adaptive,xu2023sequential}.
Recent work starts to relax the full-feedback requirement, such as intermittently observed labels in imitation learning~\citep{zhao2025conformalized} and semi-bandit feedback in retrieval~\citep{ge2024stochastic}. 
In contrast, we present a random review mechanism to overcome the expert review bottleneck and yield error control guarantees on the entire stream and the unreviewed cases.

\vspace{-0.5em}
\paragraph{Label-conditional and utility-aware prediction sets.}
Apart from validity, a growing literature studies the utility of prediction sets. 
This includes studies of training or optimization schemes that seek smaller or better-shaped conformal outputs \citep{stutz2021learning,kiyani2024length}.
A related direction incorporates structure in the label space so that the candidate labels remain coherent with domain knowledge~\citep{goren2024hierarchical,mortier2025conformal,jin2026act}. 
Our utility-aware module is in spirit close to this literature: it incorporates an optimization module to improve the skipped ratio, reduce prediction set size, and encourage outputs that respect specified relations among cell types.


\section{Background and setup}\label{sec:problem_setup}

\subsection{Formal setup}\label{subsec:setup}

We now describe the formal setup of the AI annotation problem. 
For generality, we describe the setup for a generic annotation task, while referring to  cell-type annotation at times as examples. 
We consider a streaming workflow where the cases arrive sequentially in batches. At time step $t$, the AI model receives an unlabeled batch 
$D_t=\{x_{t,1},\dots,x_{t,B_t}\}$ 
with batch size $B_t:=|D_t|\geq 1$.  

The true labels $\{y_{t,i}\}_{i=1}^{B_t}$ belong to a finite label space $\cY=[K]=\{1,\dots,K\}$, where $K$ is the total number of labels (cell types). 
The classification model used at time $t$ produces for each case $x\in D_t$ a score vector 
$s_t(x)=\bigl(s_{1,t}(x),\dots,s_{K,t}(x)\bigr)\in[0,1]^K$, 
where $s_{k,t}(x)$ is a model-produced score for class $k$, with
larger values indicating stronger support for assigning $x$ to that class.
For each case $x_{t,i}$, based on the score $s_{t}(x_{t,i})$ and the information available at each time $t$, the task is to decide whether it can bypass the expert review that provides $y_{t,i}$ and, if so, to provide an automated annotation $\hat y_{t,i}$. This decision is denoted as $I_{t,i}\in \{0,1\}$, so $I_{t,i}=1$ means committing to the AI annotation. 

We define the cumulative \emph{annotation error} up to time $t$ as 
\$
\text{AnnoErr}_t:= \frac{\sum_{s=1}^t\sum_{i\in [B_s]:\,I_{s,i}=1} \ind\{\hat{y}_{s,i} \neq y_{s,i}\}}{\sum_{s=1}^t B_s}.
\$
Note that this quantity is fundamentally unobserved, since $y_{t,i}$ is not observed for the cases with $I_{t,i}=1$. 
Such an error rate is analogous to the notion considered in \cite{candes2025probably}. 

In addition, we assume that expert feedback is available throughout the process, so the procedure can actively request expert feedback, denoted by an indicator $E_{t,i}\in \{0,1\}$ where $E_{t,i}=1$ means the label $y_{t,i}$ is collected and observed before the next batch $D_{t+1}$. Together, cases in $D_{t}$ fall into three categories:
\vspace{0.25em}
\begin{itemize}
    \item $E_{t,i}=1$, whose true label $y_{t,i}$ is solicited from the expert and observed. 
    \item $E_{t,i}=0$ and $I_{t,i}=1$, whose automated annotation $\hat{y}_{t,i}$ is considered as trustworthy so $y_{t,i}$ is never collected from the expert. 
    \item $E_{t,i}=I_{t,i}=0$, whose annotation cannot be automatically resolved by CARED but whose true label $y_{t,i}$ need not be observed before time $t+1$; instead, it can be obtained later on using $\hat C_t(x_{t,i})$ as a candidate list, without interrupting our procedure. 
\end{itemize}
\vspace{0.25em}

We denote the audited set by $A_t:=\{x_{t,i}\colon E_{t,i}=1\}$. 
At each time point $t$, given the data stream $\{D_s\}_{s\leq t}$ and previously solicited feedback $\{y_{s,i}\colon E_{t,i}=1\}_{s\leq t-1}$, we aim to produce the decisions $I_{t,i}$ and $\hat{y}_{t,i}$ while controlling the unobserved error rate $\text{AnnoErr}_t$ below a user-specified level $\alpha\in(0,1)$ in the long run.

As previewed, we accompany each case with a prediction set $\hat{C}_t(x)\subseteq [K]$. This set is said to cover the label if $y_{t,i}\in \hat{C}_t(x_{t,i})$. Following the literature, we define the \emph{cumulative coverage error} up to time $t$ as
\$
\text{CovErr}_t:= \frac{\sum_{s=1}^t\sum_{i=1}^{B_s}  \ind\{y_{s,i} \notin \hat{C}_s(x_{s,i})\}}{\sum_{s=1}^t B_s}.
\$
As not all the $y_{t,i}$'s are observed, this coverage error is also unobserved for each time point $t$. Because every erroneous automated annotation necessarily corresponds to a prediction set that fails to contain the true label---either an empty set or a singleton set with the wrong label, the events counted by the numerator of $\text{AnnoErr}_t$ form a subset of those for $\text{CovErr}_t$, and therefore we have
$\operatorname{AnnoErr}_t\leq\operatorname{CovErr}_t.$

\paragraph{Additional notations for class-conditional errors.} In cell-type annotation, it is important to maintain error rates within each \emph{true} label class, because some classes can be rare but clinically important. We now introduce analogous notation for class-conditional errors. 
For each class $k\in[K]$, let 
$D_{k,t}\subseteq D_t$ 
denote the subset of batch-$t$ cells whose ground-truth class is $k$. Then $\{D_{k,t}\}_{k=1}^K$ forms a partition of $D_t$.

Correspondingly, let 
$A_{k,t}:=A_t\cap D_{k,t}$ 
be the reviewed cells from class $k$, and define the unreviewed subsets by 
$U_t:=D_t\setminus A_t$, and $U_{k,t}:=D_{k,t}\setminus A_{k,t}$. 
For a class-$k$ cell $x\in D_{k,t}$ and a corresponding prediction set $\widehat{C}_t(x)$, define the coverage error 
$\mathrm{err}_{k,t}(x):=\ind\{k\notin \hat C_t(x)\}$, 
which records whether the prediction set fails to cover the true label. For any horizon $T$, we write the group-wise annotation (within unreviewed cases) and coverage error as 
\[
\text{AnnoErr}_{k,T}^U =  
\frac{
\sum_{t=1}^T
\sum_{x_{t,i}\in U_{k,t}:\,I_{t,i}=1}
\ind\{\hat y_{t,i}\neq k\}}
{\sum_{t=1}^T|U_{k,t}|},
\quad 
  \text{CovErr}_{k,T}^{D}
=
\frac{\sum_{t=1}^T\sum_{x\in D_{k,t}}\mathrm{err}_{k,t}(x)}
{\sum_{t=1}^T|D_{k,t}|}.
\]
In addition, we define  the cumulative coverage error rates over reviewed and unreviewed cells as
\[  
\text{CovErr}_{k,T}^{A}
=
\frac{\sum_{t=1}^T\sum_{x\in A_{k,t}}\mathrm{err}_{k,t}(x)}
{\sum_{t=1}^T|A_{k,t}|},
\quad 
\text{CovErr}_{k,T}^{U}
=
\frac{\sum_{t=1}^T\sum_{x\in U_{k,t}}\mathrm{err}_{k,t}(x)}
{\sum_{t=1}^T|U_{k,t}|}.
\]

\subsection{Preliminaries on conformal prediction}\label{subsec:pre}

We next briefly review conformal prediction algorithms which will be useful in motivating our procedure. These algorithms may operate in different settings than ours, and we introduce them in their original setup. 

Split conformal prediction (SCP)
is a canonical variant of conformal prediction \citep{vovk2005algorithmic,lei2018distribution}. Assuming access to labeled data $\{(X_i,Y_i)\}_{i=1}^n$, the goal is to construct a prediction set $\hat{C}(X_{n+1})\subseteq \cY$ for the unknown label $Y_{n+1}$ of a newly observed test point $X_{n+1}\in \cX$. 
The SCP method leverages any conformity score $s:\mathcal X\times\mathcal Y\to\mathbb R$, which is a function whose training process is independent of $\{(X_i,Y_i)\}_{i=1}^n\cup\{X_{n+1}\}$. Oftentimes the score $s$ builds upon a pre-trained prediction model, and intuitively, the score $s(x,y)$ measures how well the value $y$ ``conforms'' to the predicted value based on $x$. 
The SCP prediction set takes the form 
\[
\widehat C^{\text{SCP}}_\alpha(X_{n+1})=\{y\in\mathcal Y:s(X_{n+1},y)\geq \widehat q_{\alpha}\},
\]
where $\widehat q_{\alpha}$ is the empirical $\alpha$-quantile of $\{S_i\}_{i=1}^n\cup\{-\infty\}$, and $S_i=s(X_i,Y_i)$ for $i=1,\dots,n$.  
Under the condition that $\{(X_i,Y_i)\}_{i=1}^{n+1}$ are exchangeable, without any other assumptions on the data distribution or the score $s$, the coverage guarantee of SCP holds in finite sample:  
$\mathbb P\bigl(Y_{n+1}\in \widehat C_\alpha^{\text{SCP}}(X_{n+1})\bigr)\ge 1-\alpha$. 

The Mondrian conformal prediction (MCP) method extends SCP to achieve class-conditional coverage~\citep{vovk2003mondrian}, and we state this method here in accordance with SCP (assuming a pre-trained score function). 
Assuming access to the same data as above, and writing
$\mathcal{Y} = \{1, \ldots, K\}$, we denote 
$\mathcal{I}_k = \{i \le n : Y_i = k\}$ as the calibration indices of class $k$. 
Then, 
let $\widehat q_{k,\alpha}$ denote the empirical $\alpha$-quantile of $\{s(X_i, k)\}_{i \in \mathcal{I}_k}\cup\{-\infty\}$. The MCP set for the new test point $X_{n+1}$ is given by 
\[
  \widehat C_{\alpha}^{\text{MCP}}(X_{n+1}) = \bigl\{k \in \mathcal{Y} : s(X_{n+1}, k) \ge \widehat q_{k,\alpha}\bigr\}.
\]
Assuming the \emph{within-class} exchangeability of $\{(X_i,Y_i)\colon Y_i=k\}_{i=1}^{n+1}$, these prediction sets satisfy 
\[
  \mathbb{P}\bigl(Y_{n+1} \in \widehat C_{\alpha}^{\text{MCP}}(X_{n+1}) \bigm| Y_{n+1} = k\bigr)
  \ge 1 - \alpha, \qquad \forall k \in [K].
\]
The essential advantage is that exchangeability
is no longer imposed on the full distribution of $(X, Y)$. In an online process, the label proportion
$\mathbb{P}(Y_t = k)$ may drift arbitrarily, and MCP maintains in-class coverage as long as the within-class distributions remain exchangeable.
Moreover, the class-conditional coverage guarantee it delivers is stronger than the marginal coverage guarantee and is particularly suitable for cell-type annotation in which certain classes can be consequential.

While SCP and MCP still rely on exchangeability structures within data, 
adaptive conformal inference (ACI) is an online procedure that maintains coverage even when the data distribution drifts arbitrarily~\citep{gibbs2021adaptive}. The key strategy is to adjust the sets adaptively based on newly observed labels. Consider an online setting where data points $(X_t)_{t\geq 1}$ arrive one at a time, and the true label $Y_t$ is revealed before time $t+1$. The algorithm maintains a running level $\alpha_t$,
and at step
$t$ issues the prediction set $\widehat C_{\alpha_t}(X_t)$ as in the SCP construction above. Once the true label $Y_t$ is
revealed, it updates the level by 
\[
  \alpha_{t+1} = \alpha_t + \gamma \bigl( \alpha - \mathrm{err}_t \bigr),
  \qquad \mathrm{err}_t := \ind\{ Y_t \notin \widehat C_{\alpha_t}(X_t) \},
\]
with a fixed step size $\gamma > 0$ and any initialization $\alpha_1 \in \mathbb{R}$. For any sequence
$(X_t, Y_t)_{t \ge 1}$, whether stationary, drifting, or
adversarially generated, this procedure delivers the long-run
guarantee  $\frac{1}{T} \sum_{t=1}^{T} \mathrm{err}_t \to \alpha$ as $T\to \infty$. However, the true label $Y_t$ must be revealed before
the next prediction can be issued.

\section{Method and theoretical guarantees}\label{sec:method}

In the following, we introduce the CARED framework in Section~\ref{subsec:online_conf}, the optional enhancement in Section~\ref{subsec:utility_module}, and the error control guarantees in Section~\ref{sec:theory}. 

\subsection{Conformal annotation with randomized expert review}
\label{subsec:online_conf}

At the beginning, we assume access to an initial labeled block $A_0$, among which the cells belonging to class $k \in [K]$ are denoted as $A_{k,0}$. This block provides the initial labeled pool used to construct class-wise prediction sets. Moreover, we initialize the class-wise levels by $\alpha_{k,1}=\alpha$ for every $k \in [K]$. 
We shall maintain the labeled sets $\{A_{k,t}\}_{k\in [K],t\geq 1}$ and the levels $\{\alpha_{k,t}\}_{k\in [K], t\geq 1}$  over time. 

The procedure proceeds over time $t=1,2,\dots,T$. At each time point $t$, for each class $k$, a calibration set $D_{k,\text{cal}}(t)$ is constructed to be the class-$k$ cells in the most recently reviewed $\min\{t, T_w\}$ batches, where $T_w\geq 1$ is a pre-specified parameter. That is, we define $D_{k,\text{cal}}(t) = \prod_{\tau=t-T_w}^{t-1}A_{k,\tau}$. This choice is computationally convenient and allows the calibration set to track local distributional changes. However, for theoretical validity, any subset of all already-reviewed class-\(k\) cells, i.e., any subset of \(\bigcup_{\tau=0}^{t-1} A_{k,\tau}\), are allowed. Also, a score $\tilde s_{k,t}(x) \in [0,1]$ is computed based on a prediction model (in this paper, the NuClass~\citep{xu2025adaptive} model), which might be updated over time, e.g., using the utility module described in Section~\ref{subsec:utility_module}. 
For any $u\in [0,1]$, we denote the $u$-th empirical quantile of the calibration set as 
\@\label{eq:quantile}
\hat{q}_{k,t}(u):=\inf\Big\{
q\in[0,1]: 
\textstyle{\frac{1}{|D_{k,\mathrm{cal}}(t)|}}
 \sum_{x\in D_{k,\mathrm{cal}}(t)}
\ind\{\tilde s_{k,t}(x)\le q\}  \ge u
\Big\}.
\@
For completeness, we define $\hat{q}_{k,t}(u)=-\infty$ for $u<0$ and  $\hat{q}_{k,t}(u)=+\infty$ for $u>1$.   
When $|D_{k,\mathrm{cal}}(t)|=0$, we set $\hat{q}_{k,t}(u)=-\infty$ for $u\leq 1$ and $\hat{q}_{k,t}(u)=+\infty$ for $u> 1$. 
Given the levels $\{\alpha_{k,t}\}_{k=1}^K$, we construct  
\begin{equation}\label{eq:set_cons}
\widehat C_t(x)
=
\left\{
k\in[K]:
\tilde s_{k,t}(x)
\ge
\hat q_{k,t}(\alpha_{k,t})
\right\},\quad x\in D_t. 
\end{equation}
Our updating rule below following ACI may lead to $\alpha_{k,t}\notin [0,1]$, yet the empirical quantiles are well-defined.   
To accommodate cases where $\widehat{C}_t(x)=\emptyset$, we define the enlarged set
\begin{equation}\label{eq:enlarged_set}
\widehat C_t^{+}(x)
=
\begin{cases}
\widehat C_t(x), & \widehat C_t(x)\neq\emptyset,\\
\{\operatorname*{arg\,max}_{k\in[K]}\tilde s_{k,t}(x)\},
& \widehat C_t(x)=\emptyset,
\end{cases}
\end{equation}
where ties in the second case are broken randomly. Thus,
$\widehat C_t^{+}(x)$ is a singleton whenever
$|\widehat C_t(x)|\leq1$.  
Since $\widehat C_t(x)\subseteq\widehat C_t^{+}(x)$, we know $\ind\{k\notin\widehat C_t^{+}(x)\}
\leq
\ind\{k\notin\widehat C_t(x)\}$.  
Our theory in Section~\ref{sec:theory} 
establishes the long-run coverage of $\widehat C_t(x)$, which translates to the long-run coverage of $\widehat C_t^+(x)$ and the annotation error bound. 


The levels $\{\alpha_{k,t}\}$ are updated through an \emph{unbiased randomized rounding scheme}. 
Fix a target reviewed fraction $p\in(0,1)$. At time $t$, CARED chooses a realizable fraction $p_t$ so that  $p_t B_t$ is an integer and the cumulative reviewed fraction $(\sum_{t=1}^T p_t B_t)/(\sum_{t=1}^T B_t)\to p$ as $T \to \infty$; see Appendix~\ref{subsec:rounding_scheme} for details. 
Then we sample 
uniformly without replacement from $D_t$ the audit set of size $|A_t|=p_t B_t$.
The annotation decision is
\begin{equation}\label{eq:automated-annotation-decision}
I_{t,i}
:=
(1-E_{t,i})
\ind\big\{|\widehat C_t(x_{t,i})|\leq1 \big\}.
\end{equation}
Whenever $I_{t,i}=1$, we define the automated annotation $\hat y_{t,i}$ as
the unique element of the prediction set, 
$\widehat C_t^{+}(x_{t,i})=\{\hat y_{t,i}\}$. 
No automated annotation $\hat y_{t,i}$ is produced when $I_{t,i}=0$. The
audited labels are then used to update the online levels for subsequent steps similar to ACI:
\begin{equation}\label{eq:alpha-update}
    \alpha_{k,t+1}
=
\alpha_{k,t}
+
\gamma_k
\sum_{x\in A_{k,t}}
\bigl(\alpha-\operatorname{err}_{k,t}(x)\bigr),
\qquad k\in[K],
\end{equation}
where $\gamma_k$ is a pre-specified step size for class $k$, and $\text{err}_{k,t}(x) = \ind\{k\notin\widehat C_t(x)\}$ is the miscoverage indicator. In our experiments, we fix the class-wise step sizes according to cohort-specific class frequencies; the specific choices are given in Appendix~\ref{subsec:appendix_experimental_details}. Appendix~\ref{sec:discussions} further discusses the performance of fixed step sizes compared with time-varying ones. The updated online levels are then used to compute the next batch's class-wise quantile cutoffs through \eqref{eq:set_cons}.  
We summarize the procedure of CARED in Algorithm~\ref{alg:cared}.

\begin{algorithm}[htbp]
\caption{CARED}\label{alg:cared}
\begin{algorithmic}[1]
\REQUIRE Initial labeled block $A_0$ with labels in $[K]$; incoming batches $\{D_t\}_{t=1}^T$; target error level $\alpha\in(0,1)$; target reviewed fraction $p\in(0,1)$; calibration window $T_w\geq 1$; fixed step sizes $\gamma_k>0$, $k\in[K]$.
\STATE Set $A_{k,0}\leftarrow\{x\in A_0:y(x)=k\}$ and $\alpha_{k,1}\leftarrow\alpha$ for every $k\in[K]$.
\FOR{$t=1,\ldots,T$}
  \STATE Receive $D_t=\{x_{t,i}\}_{i=1}^{B_t}$ and score functions $\{\tilde s_{k,t}:\mathcal X\to[0,1]\}_{k=1}^K$.
  \FOR{$k=1,\ldots,K$}
    \STATE Set $D_{k,\mathrm{cal}}(t)\leftarrow\bigcup_{\tau=\max\{0,t-T_w\}}^{t-1}A_{k,\tau}$.
    \STATE Compute $\hat q_{k,t}(\alpha_{k,t})$ according to~\eqref{eq:quantile}.
  \ENDFOR
  \STATE Compute $\widehat C_t(x_{t,i})$ and   $\widehat C_t^+(x_{t,i})$ according to~\eqref{eq:set_cons} and~\eqref{eq:enlarged_set} for every $i\in[B_t]$.
  \STATE Draw $\xi_t\sim\operatorname{Bernoulli}(pB_t-\lfloor pB_t\rfloor)$ and set $p_t\leftarrow(\lfloor pB_t\rfloor+\xi_t)/B_t$. \COMMENT{See Appendix~\ref{subsec:rounding_scheme}}
  \STATE Sample $A_t\subseteq D_t$ uniformly without replacement with $|A_t|=p_tB_t$.
  \STATE For every $i\in[B_t]$, set $E_{t,i}\leftarrow\ind\{x_{t,i}\in A_t\}$ and $I_{t,i}\leftarrow(1-E_{t,i})\ind\{ |\widehat C_t(x_{t,i})|\leq1\}$.
  \STATE Compute the unique $\widehat y_{t,i}\in\widehat C_t^+(x_{t,i})$ for each $i\in [B_t]$ with $I_{t,i}=1$.
  \STATE Obtain expert labels $y_{t,i}$ for all $E_{t,i}=1$, and set $A_{k,t}\leftarrow\{x_{t,i}\in A_t:y_{t,i}=k\}$ for every $k\in[K]$.
  \STATE For every $k\in[K]$, update $\alpha_{k,t+1}\leftarrow\alpha_{k,t}+\gamma_k\sum_{x\in A_{k,t}}\bigl(\alpha-\ind\{k\notin\widehat C_t(x)\}\bigr)$.
\ENDFOR
\ENSURE Prediction sets $\widehat C_t(x_{t,i})$, annotation decisions $I_{t,i}$ and annotation labels $\hat y_{t,i}$ for $I_{t,i}=1$.
\end{algorithmic}
\end{algorithm}

\subsection{Optional model training with utility improvement in cell-type annotation}\label{subsec:utility_module}

The procedure introduced in Section~\ref{subsec:online_conf} works for any score functions $\tilde{s}_{k,t}(\cdot)$, but the choice of the score function determines the form of the prediction sets. In this part, we introduce an optional module that adapts the score functions over time. Specifically, we parameterize the transformed score as 
$\tilde s_t(x)=f_{\Theta_t}\bigl(s_t(x)\bigr)$, 
where $\Theta_t$ is updated over time and $s_t(\cdot)$ is the original score given by the NuClass model. The construction of the score transformation $f_{\Theta_t}$ is provided in Algorithm~\ref{alg:score_update} in Appendix~\ref{subsec:utility_score_transform}.

\paragraph{Utility metrics and training objective.}
Recall that the $t$-th batch is $D_t=\{x_{t,1},\ldots,x_{t,B_t}\}$, where $B_t=|D_t|$.
For any parameter $\Theta$, let $\widehat C_{t,i}(\Theta)$ denote the prediction
set $\widehat C_t(x_{t,i})$ obtained using the transformed scores
$f_\Theta\{s_t(x_{t,i})\}$. The construction follows
\eqref{eq:quantile}--\eqref{eq:set_cons}.
Before updating $\Theta_t$ and deciding $\tilde s_t(\cdot)$, we first define what properties are good for prediction sets.
Below we quantify the utility of prediction sets for a batch $D_t$ based on
three criteria: (1) skipped ratio surrogate, (2) set size, and (3) structural coherence. 

First, we encourage the score function to lead to a large fraction of automated annotations.  
This fraction (among cases not selected for expert audit) can be estimated using the entire batch: 
  $\mathcal J_{\mathrm{skip}}(\Theta)
  =
  \frac{1}{B_t}
  \sum_{i=1}^{B_t}
  \ind \{
  |\widehat C_{t,i}(\Theta)|\leq 1
   \}$ 
In addition, since non-singleton sets will be used as references for experts to resolve the ambiguity, it is sensible to encourage small set sizes. Smaller sets leave fewer candidate labels for the subsequent narrowing-down
step and are therefore more useful in human-in-the-loop decision making
\citep{zhang2024evaluating,dhillon2024expected}. We thus consider the second metric 
  $\mathcal J_{\mathrm{size}}(\Theta)
  =
  \frac{1}{B_t}
  \sum_{i=1}^{B_t}
  |\widehat C_{t,i}(\Theta)|$. 
Finally, a non-singleton prediction set is more useful when the candidate labels respect the biological structures among the cell types~\citep{kaymaz2021hierfit,theunissen2024uncertainty,corbetta2026conformal}. We design a metric $\mathcal J_{\mathrm{struct}}(\Theta)$ to reward structurally coherent prediction sets based on the subtype tree-structures in cell types; the detailed construction
is given in Appendix~\ref{subsec:utility_structure_surrogate}.  
The three components are combined into  
\begin{equation}\label{eq:hard_obj}
  \mathcal J(\Theta)
  =
  \lambda_{\mathrm{size}}\mathcal J_{\mathrm{size}}(\Theta)
  -
  \lambda_{\mathrm{struct}}\mathcal J_{\mathrm{struct}}(\Theta)
  -
  \lambda_{\mathrm{skip}}\mathcal J_{\mathrm{skip}}(\Theta),
\end{equation}
where the weights are nonnegative; see Appendix~\ref{subsec:appendix_experimental_details} for the choice in our applications. Thus, minimizing
\eqref{eq:hard_obj} penalizes large sets while rewarding structural coherence
and a larger skipped-ratio. 

\paragraph{Online score training.}
To train the score function, we first build a smooth surrogate for the objective $\cJ(\Theta)$. Specifically, since several terms involve indicators that are difficult for gradient descent updates, we replace these quantities  via standard differentiable approximations. 
At each time $t$, we initialize $\Theta=0$ and perform $N$ gradient steps on
the surrogate objective evaluated on the current unlabeled batch $D_t$. Before
each update, we compute the class-wise conformal thresholds from the labeled
calibration sets $\{D_{k,\mathrm{cal}}(t)\}_{k=1}^K$ using the current
transformation $f_\Theta$. After each update, we recompute these thresholds and
evaluate the original empirical objective $\mathcal{J}(\Theta)$ on $D_t$.
Among the initialization and all gradient iterates, we select the parameter
value with the lowest objective value as $\Theta_t$.
Notably, CARED allows \(\Theta_t\) to depend on the current batch \(D_t\), provided that \(\Theta_t\) is constructed before the audit set \(A_t\) is sampled. The online training procedure is summarized in Algorithm~\ref{alg:online_selection} and detailed in Appendix~\ref{subsec:utility_online_selection}.


\subsection{Theoretical guarantees}\label{sec:theory}

In this part, we present the main theoretical guarantee, which establishes long-run class-wise
validity in CARED based on the random audit alone, regardless of the distribution drift and the score models. 
At a high level, the batch-wise ACI-style recursion controls the
error within the reviewed cells, and the uniform random audit transfers this control to
the full stream, including the cells that are not reviewed. These theoretical results help contextualize the quality guarantee for the automated annotations from CARED. 

Recall the notations from Section~\ref{subsec:setup}: $\text{CovErr}_{k,T}^{A}$, $\text{CovErr}_{k,T}^{D}$, and
$\text{CovErr}_{k,T}^{U}$ denote the cumulative conformal errors over
the audited cells, all cells, and unaudited cells that belong to class $k$, respectively, up to time
$T$. The true class membership is unknown unless the case is reviewed by experts (i.e., $E_{t,i}\in\{0,1\}$), and is used only for stating the guarantees.  
Let $\{\mathcal G_t\}_{t\ge0}$ be the
pre-review filtration, where $\mathcal G_t$ is the sigma-field generated by
the information after the prediction sets for $D_t$ are formed 
but before the audit set $A_t$ is sampled. In particular, the class
counts $|D_{k,t}|$ and the coverage indicators
$\mathrm{err}_{k,t}(x)$ for any $x \in D_{k,t}$ are
$\mathcal G_t$-measurable. 
Recall that $\{p_t\}_{t=1}^T$ are the fraction of cases sent for expert review in the batches.

We impose mild conditions on the data and the random audit, without imposing exchangeability or
stationarity of the incoming batches or any parametric model on the data distribution.

\begin{assumption}\label{ass:main_guarantee}
Fixing a target reviewed fraction $p\in(0,1)$, we assume that:
\begin{enumerate}[label=(\roman*),leftmargin=0.08\linewidth]
\item The batch size $B_t=|D_t|$ is uniformly controlled, obeying  
$B^-\le B_t\le B^+$ 
for some finite, positive integers  $0<B^-\leq B^+ <\infty.$
\item For every $k\in[K]$, the cumulative number of class-$k$ cells diverges, i.e., 
$\sum_{t=1}^{\infty}|D_{k,t}|=\infty$ as $T\to\infty$.
\item Conditional on $\mathcal{G}_t$, the audit set $A_t\subseteq D_t$ is drawn
uniformly without replacement with $|A_t|=p_tB_t$.
\item The score function $\tilde{s}_t(\cdot)$ is adapted to $\cG_t$. 
\end{enumerate}
\end{assumption}

Assumption~\ref{ass:main_guarantee}(i) and (iii) can be ensured through appropriate batch construction and audit sampling. Moreover, (ii) is a mild condition that every cell type continues to occur and can be updated throughout the workflow so the within-class asymptotics apply. Finally, (iv) requires the score function to be determined before drawing the random audit, but can depend on the batch $D_t$. 

We begin with the guarantee on the expert-reviewed cells. The proof of Proposition~\ref{prop:reviewed_error} is in Appendix~\ref{proof:reviewed_error}. 

\begin{prop}[Long-run coverage error control for reviewed cells]\label{prop:reviewed_error}
Fix any class $k \in [K]$, suppose the accumulated class-$k$ frequency is bounded away
from zero, i.e., 
$\liminf_{T\to\infty} 
\frac{\sum_{t=1}^T |D_{k,t}|}{\sum_{t=1}^T B_t}
\geq q_k $ for some constant $q_k>0$.  
Under assumption \ref{ass:main_guarantee}, almost surely there exists some $T_0<\infty$ such that for all $T\geq T_0$, 
\[
\left|\textnormal{CovErr}_{k,T}^{A}-\alpha\right|
\le
\frac{\max\{\alpha,1-\alpha\}}
{\gamma_k\sum_{t=1}^T |A_{k,t}|}
+
\frac{C}{T},
\] 
where $C = \frac{2B^+}{B^- q_k}(1+\frac{1}{pB^-})$.
Consequently,  
$\textnormal{CovErr}_{k,T}^{A} \to \alpha$ almost surely as $T\to\infty$. 
\end{prop}

Intuitively, the coverage guarantee on the reviewed cases inherit the guarantees from ACI. Our main result is that such guarantee translates to that on the entire stream, due to the unbiased random design. The proof of Theorem~\ref{thm:long_run_validity} is in Appendix~\ref{proof:long_run_validity}. 

\begin{theorem}[Long-run coverage error control for all cells]\label{thm:long_run_validity}
Under Assumption~\ref{ass:main_guarantee}, for every class $k$, as $T\to \infty$, the cumulative miscoverage over 
all cells and the cells not reviewed by the experts satisfies
\[
\left|\textnormal{CovErr}_{k,T}^{D}-\alpha\right|\to 0,\quad \left|\textnormal{CovErr}_{k,T}^{U}-\alpha\right|\to 0
\quad\textnormal{a.s.}
\]
\end{theorem}

These within-class coverage guarantees, together with the growing conditions on the group sizes, imply the overall coverage guarantee, i.e., $\text{CovErr}_T\to \alpha$ almost surely, and the coverage guarantee for all cases not reviewed by the experts by the lower-bounded audit probability, whose proofs are omitted for brevity. In a nutshell, it means that even though CARED only requests expert labels for a fraction of $p\in (0,1)$ of cases (in our experiments, the fraction can be as low as $10\%$ or even lower), the prediction sets for all the incoming cases are still asymptotically well-calibrated and adapted to the arbitrarily drifting workflow. 

More importantly for deployment,
because $\widehat C_t(x_{t,i})\subseteq\widehat C_t^+(x_{t,i})$, the annotation error is upper bounded by miscoverage. We present this result in Corollary~\ref{cor:annotation_error_control}. 

\begin{corollary}[Long-run automated-annotation error control]
\label{cor:annotation_error_control}
Recall the overall cumulative annotation error $\textnormal{AnnoErr}_T$ and class-wise annotation error $\textnormal{AnnoErr}_{k,T}$ defined in
Section~\ref{subsec:setup}.  
Under Assumption~\ref{ass:main_guarantee},
\[
\limsup_{T\to\infty}\textnormal{AnnoErr}_{k,T}^{U}
\le\alpha
\quad\textnormal{a.s.},\quad k\in [K].
\]
\end{corollary}



Like Theorem~\ref{thm:long_run_validity}, these guarantees are
distribution-free, requiring no exchangeability or stationarity conditions on the data stream. 
They also allow the AI models and scores to evolve arbitrarily over time. 
In addition, the per-class calibration ensures that the annotation error is rigorously controlled within every class, even though the true class memberships are unknown. This avoids neglecting clinically consequential but difficult-to-classify cell types. In practice, if the AI prediction quality is poor for one particular class, then CARED will accept fewer cases for automated annotation to maintain the error control. 

For completeness, we establish an explicit
convergence rate for the cumulative miscoverage error. Additional theory characterizing cumulative batch-wise deviations from the target level is in Appendix~\ref{subsec:sum_guarantee}.

\begin{prop}[Long-run error convergence rate]\label{prop:convergence_rates}
Under the CARED procedure, suppose that the class-frequency condition in
Proposition~\ref{prop:reviewed_error} holds for every class. Then, for every
class $k$,
\[
\left|\textnormal{CovErr}_{k,T}^{D}-\alpha\right| \vee 
\left|\textnormal{CovErr}_{k,T}^{U}-\alpha\right|
=
\mathcal{O}_p\Bigl(\bigl(\sum_{t=1}^T |D_{k,t}|\bigr)^{-1/2}\Bigr)
=
\mathcal{O}_p(T^{-1/2}).
\]
\end{prop}


\section{Case study in cell-type annotation workflows}\label{sec:experiments}

\subsection{Data and annotation workflow}\label{subsec:exp_set}

This case study evaluates CARED in the motivating cell-type annotation task where pathologists annotate cell types on whole-slide images (WSIs) in TissueLab~\citep{li2025co}.
We describe the AI model, the tissue cohorts and their labels, and the construction of
annotation streams.  

We use the NuClass~\citep{xu2025adaptive} model as the AI scores; it is a multi-scale histopathology model that detects nuclei and outputs a
class-wise score vector for each detected cell. It provides three score
sources, each used as the raw score $s_t(x)$: the
\emph{global} path, which incorporates the surrounding tissue context; the
\emph{local} path, which relies on nucleus-centered morphology; and the
\emph{gated} output, which adaptively fuses the two at the cell level.
We use the three sources to show how CARED behaves under scores of different
quality and sensitivity to tissue context. All cohorts below are
fully held out from NuClass training and tuning.

The main experiments use three held-out cohorts from pancreas, lung, and ovary
tissue, with label spaces of six, four, and three cell types, respectively
(Table~\ref{tab:data}). The cohorts consist of one H\&E-stained WSI per tissue, taken from three publicly available 10x Genomics Xenium datasets: FFPE pancreatic adenocarcinoma, non-small cell lung cancer, and ovarian serous carcinoma \citep{tenx_pdac_2024,tenx_lung_2024,tenx_ovary_2024}; see Appendix~\ref{sec:experimental_details} for details on the H\&E processing of the slide. 
We exclude cells on background or too close to the slide edge. 
Each cell was assigned a reference label from its spatial gene-expression profile: the 10x Genomics cell type with the highest normalized marker-gene score, which is viewed as the ground-truth cell type. 
Because these labels come from molecular measurements, they provide a reference independent of the H\&E morphology. 
For audited cells, these reference labels are revealed to CARED to emulate expert review. 
For unaudited ones, they are used to compute the error rates only, and are never available to the procedure.


The label spaces reflect the cell populations. These cell types are critical because pathologists routinely distinguish these tissues based on cell's spatial arrangements. Furthermore, they define tissue function and the tumor microenvironment: endocrine cells form the islets of Langerhans in the pancreas, while T cells, macrophages, and tumor-associated fibroblasts shape tumor progression and response to immunotherapy in lung and ovarian cancer. In this dataset,
several clinically relevant cell types are especially rare: in the pancreas cohort, endocrine
cells and T cells account for only 1.95\% and 3.77\% of all cells.
Due to the class imbalance, a marginal
error guarantee could be met while misannotating most of these rare cells, so we primarily target the class-conditional error control throughout. 

\begin{table}[t]
\centering
\caption{Data used in the case study. Class proportions are computed over
all cells in each cohort.}
\label{tab:data}
\small
\setlength{\tabcolsep}{6pt}
\renewcommand{\arraystretch}{1.18}
\begin{tabular}{@{}lll>{\raggedright\arraybackslash}p{0.65\linewidth}@{}}
\toprule
Tissue cohort & \# Cells & \# Batches & Cell type (fractions \%) \\
\midrule
Pancreas & 103{,}166 & 206
  & Acinar (12.8), Ductal (45.2), Endocrine (2.0),
    Endothelial (8.6), Fibroblast-like (27.6), T cell (3.8) \\
\addlinespace[4pt]
Lung & 61{,}490 & 123
  & Endothelial (27.6), Epithelial malignant (61.8),
    T cell (5.9), Tumor-associated fibroblast (4.7) \\
\addlinespace[4pt]
Ovary & 150{,}748 & 301
  & Endothelial (5.0), Epithelial malignant (85.3),
    Macrophage (9.7) \\
\bottomrule
\end{tabular}
\end{table}

\vspace{-0.5em}
\paragraph{Annotation data streams.}
To emulate how cells reach an annotator, we partition each cohort into
spatially contiguous regions containing between $B^-=490$ and $B^+=510$ cells.
Each region plays the role of a field of view and forms one batch. One region,
chosen at random, serves as the initial labeled block $A_0$. The order in which
the remaining regions are released is determined by annotation practice
rather than by the tissue, so we consider three release schemes, each
corresponding to a different practice:
\begin{itemize}
  \item \emph{Adjacent}: the next region is chosen among the unreleased
  neighbors of the current one, mimicking a pathologist panning across
  neighboring fields (Figure~\ref{fig:quantile_shift}(a)).
  \item \emph{Random}: regions are released in random order, mimicking the
  sampling of tiles or regions of interest when building annotated datasets
  \citep{qiu2023adaptive,hagele2020resolving}.
  \item \emph{Error-guided}: the next region is selected by identifying the
  class whose cumulative error on audited cells deviates most from $\alpha$
  and releasing the region with the most cells predicted as that class. This mimics tools that steer annotators
  toward regions where the model currently struggles. Because each release
  depends on past audit outcomes, this scheme serves as an adversarial test with strong dependence; our guarantees still hold as long as the audit set is  uniformly sampled given the past
  (Assumption~\ref{ass:main_guarantee}(iii)).
\end{itemize}
The three schemes use the same cohorts, scoring model, and procedure, and
differ only in the release order which determines the distribution drifts. 
Under the adjacent scheme, spatial autocorrelation yields gradual drift with abrupt changes at tissue-compartment boundaries (Figure~\ref{fig:quantile_shift}); under the error-guided scheme, the drift is driven by the procedure's  past outcomes. Both of them lead to strong non-exchangeability. In the random scheme, due to random ordering, the batches are exchangeable. This exchangeability, however, holds at the level of regions only, since cells within a region share morphology and composition. 
With a limited number of regions (Table~\ref{tab:data}), such exchangeability does not suffice to justify error control on the cell level in each realized stream as the guarantees in Section~\ref{sec:theory}.

\begin{figure}[!t]
    \centering
    \includegraphics[width=0.8\linewidth]{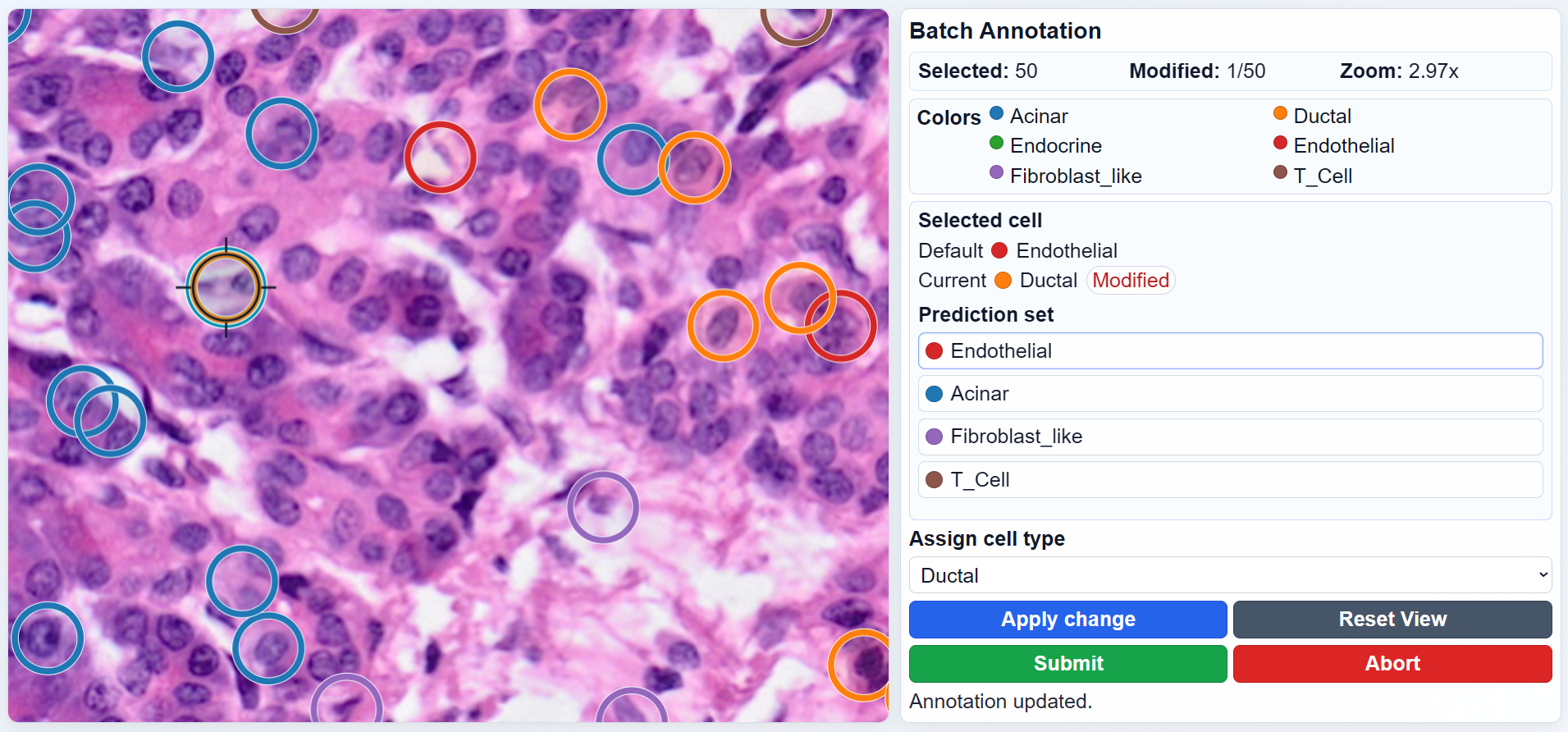}
    \caption{Interface for one pancreas batch in the partial-review
    workflow. The left panel shows a local histology field with reviewed cells
    highlighted by class-specific colored outlines. After clicking on a reviewed cell, the
    right panel displays its annotation information, including the default
    label, current label, and conformal prediction set. This example 
    illustrates a reviewed cell whose label has been modified.}
    \label{fig:batch_review_demo}
\end{figure}

\vspace{-0.5em}
\paragraph{Expert review protocol and interface.} 
Figure~\ref{fig:batch_review_demo} presents a demo interface for the annotation protocol. When a subsequent batch is processed, the interface displays all cells in the
batch to preserve the spatial and morphological context needed for expert judgment, but only the uniformly sampled audit set $A_t$ is flagged for review. 
Each flagged cell is outlined in the
color of its highest-scoring class. The annotator inspects the flagged cells and corrects the displayed label whenever it is wrong. 
The resulting labels are then used in CARED to update the class-wise levels $\{\alpha_{k,t}\}$ before the next batch is released. 
For unreviewed cells, CARED determines which of them to annotate automatically; otherwise, the cell is deferred, and its
prediction set is shown as a candidate list. In the three held-out cohorts,
the annotator's response is emulated by revealing the reference label of each
flagged cell.
Additional implementation details are reported in
Appendix~\ref{sec:experimental_details}.

\vspace{-0.5em}
\paragraph{Real expert annotation stream.}
To evaluate CARED on an annotation order produced by actual practice rather
than by a release scheme, we also use expert annotations collected in
TissueLab on a colon WSI containing 220{,}501 cells.
One annotator (a research scientist trained in histopathology annotation under pathologist supervision) annotated 1{,}374 cells over 37 rounds, in the order they chose while
navigating the slide. The annotated cells are grouped into three classes:
tumor (20.5\%), muscle (45.0\%), and others (34.5\%).\footnote{This category pools the original labels negative control,
epithelial cells, lymphocytes, eosinophils, and plasma cells.} Both the stream
order and the
labels are determined by the expert behavior; results are reported in
Section~\ref{subsec:real_stream}.

\subsection{CARED controls cumulative annotation  error}\label{subsec:cumulative_coverage_results}

We present the cumulative coverage results using the global NuClass score;
additional results for other score sources are reported in
Appendix~\ref{sec:experimental_details}. 
Unless otherwise stated, the experiments below use target error level
$\alpha=0.1$, target reviewed fraction $p=0.1$, and a fixed random seed $0$. 

\begin{figure}[!tp]
    \centering
    \captionsetup[subfigure]{font=scriptsize,justification=centering}
    \begin{subfigure}[t]{0.22\linewidth}
        \centering
        \includegraphics[width=\linewidth]{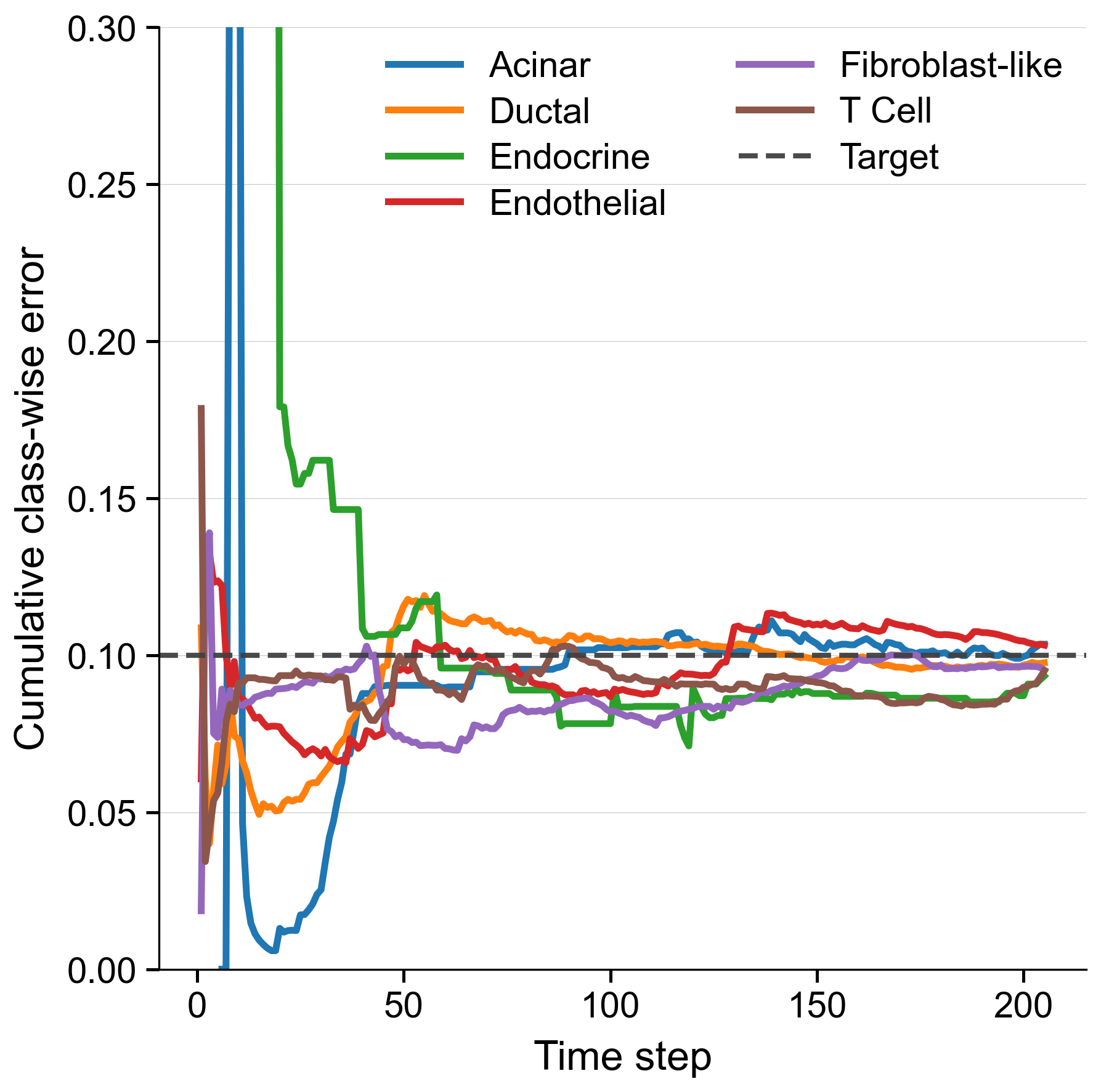}
        \caption{Error-guided, pancreas}
    \end{subfigure}
    \hfill
    \begin{subfigure}[t]{0.22\linewidth}
        \centering
        \includegraphics[width=\linewidth]{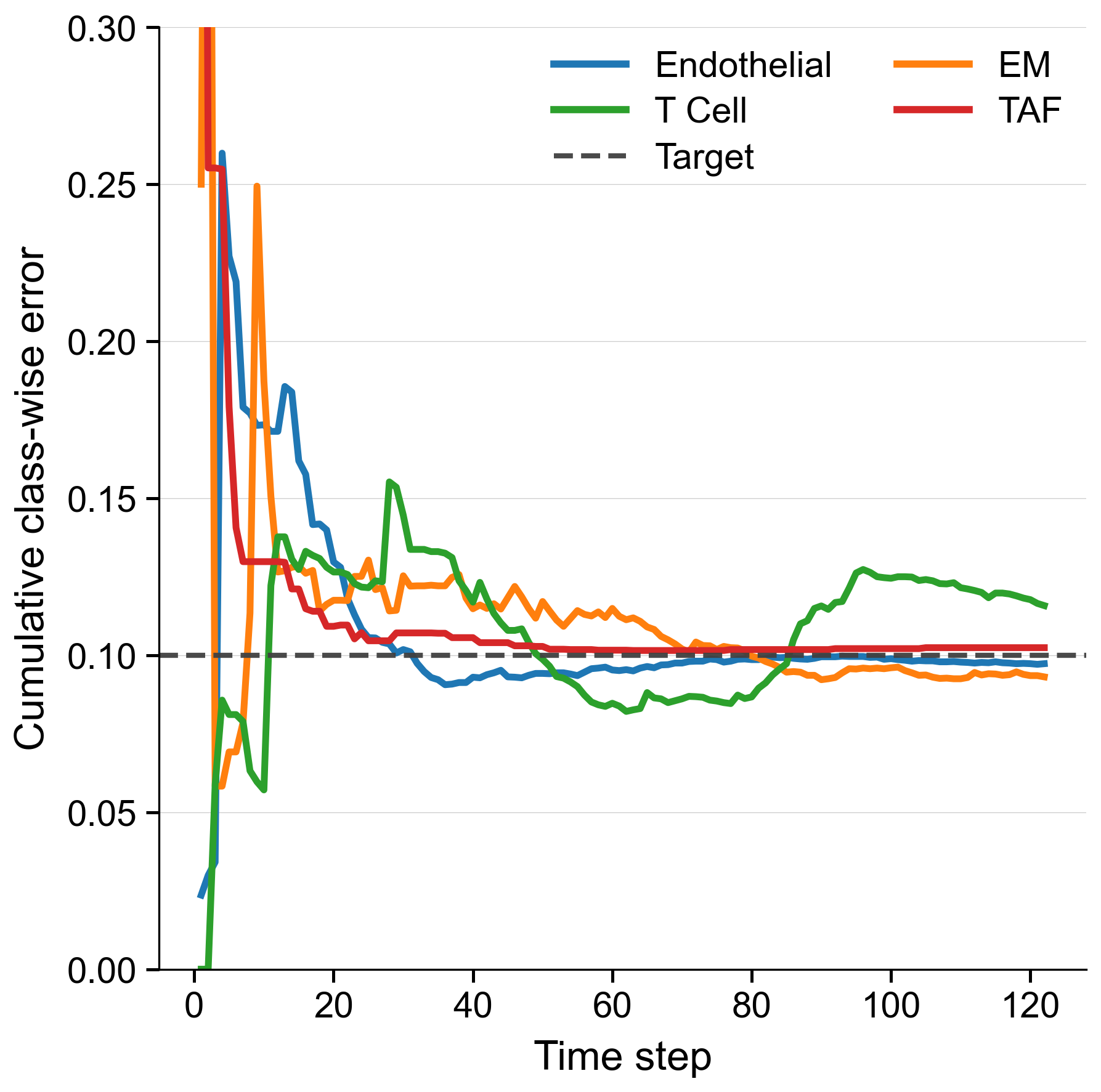}
        \caption{Error-guided, lung}
    \end{subfigure}
    \hfill
    \begin{subfigure}[t]{0.22\linewidth}
        \centering
        \includegraphics[width=\linewidth]{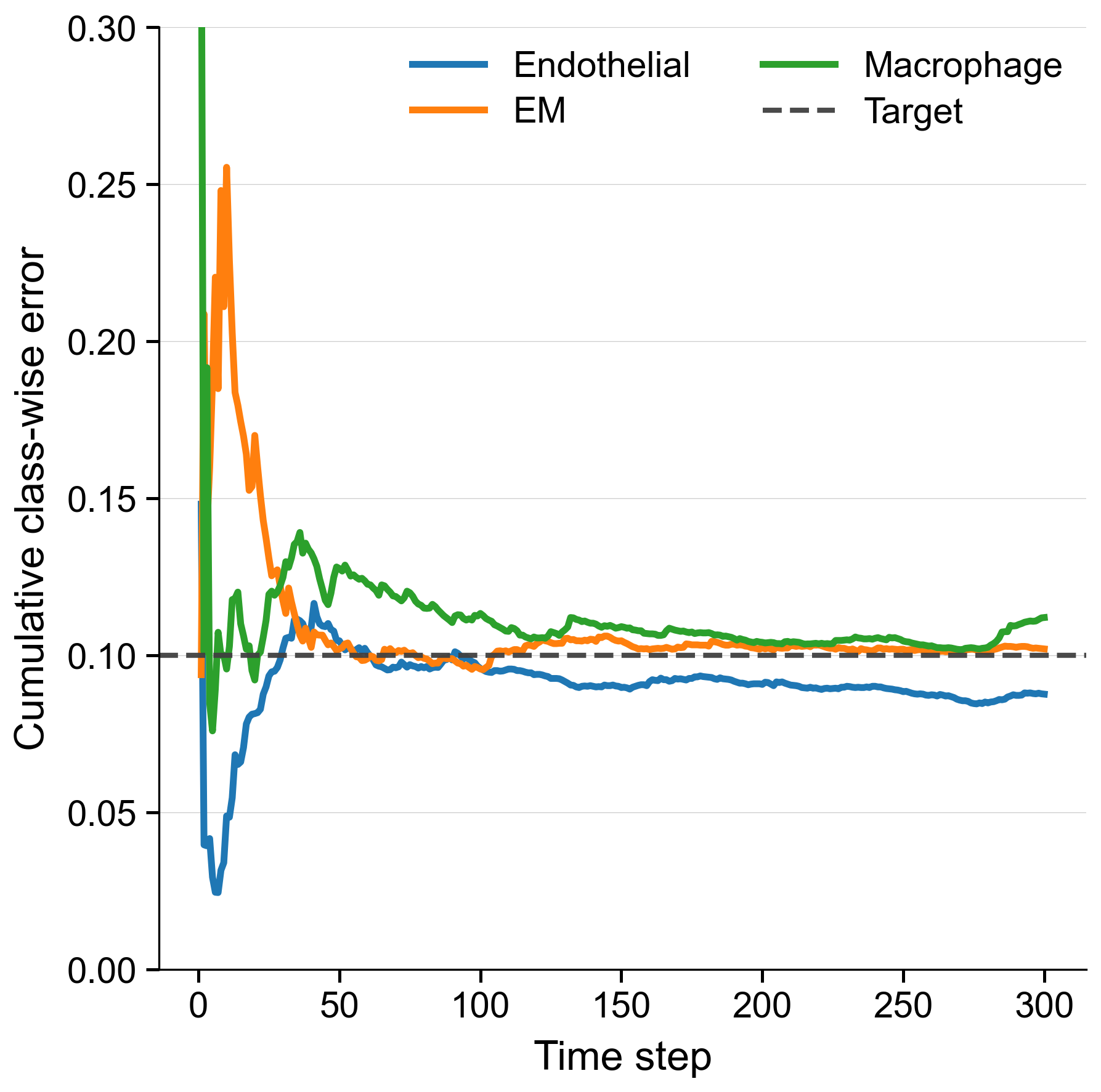}
        \caption{Error-guided, ovary}
    \end{subfigure}
    \hfill
    \begin{subfigure}[t]{0.22\linewidth}
        \centering
        \includegraphics[width=\linewidth]{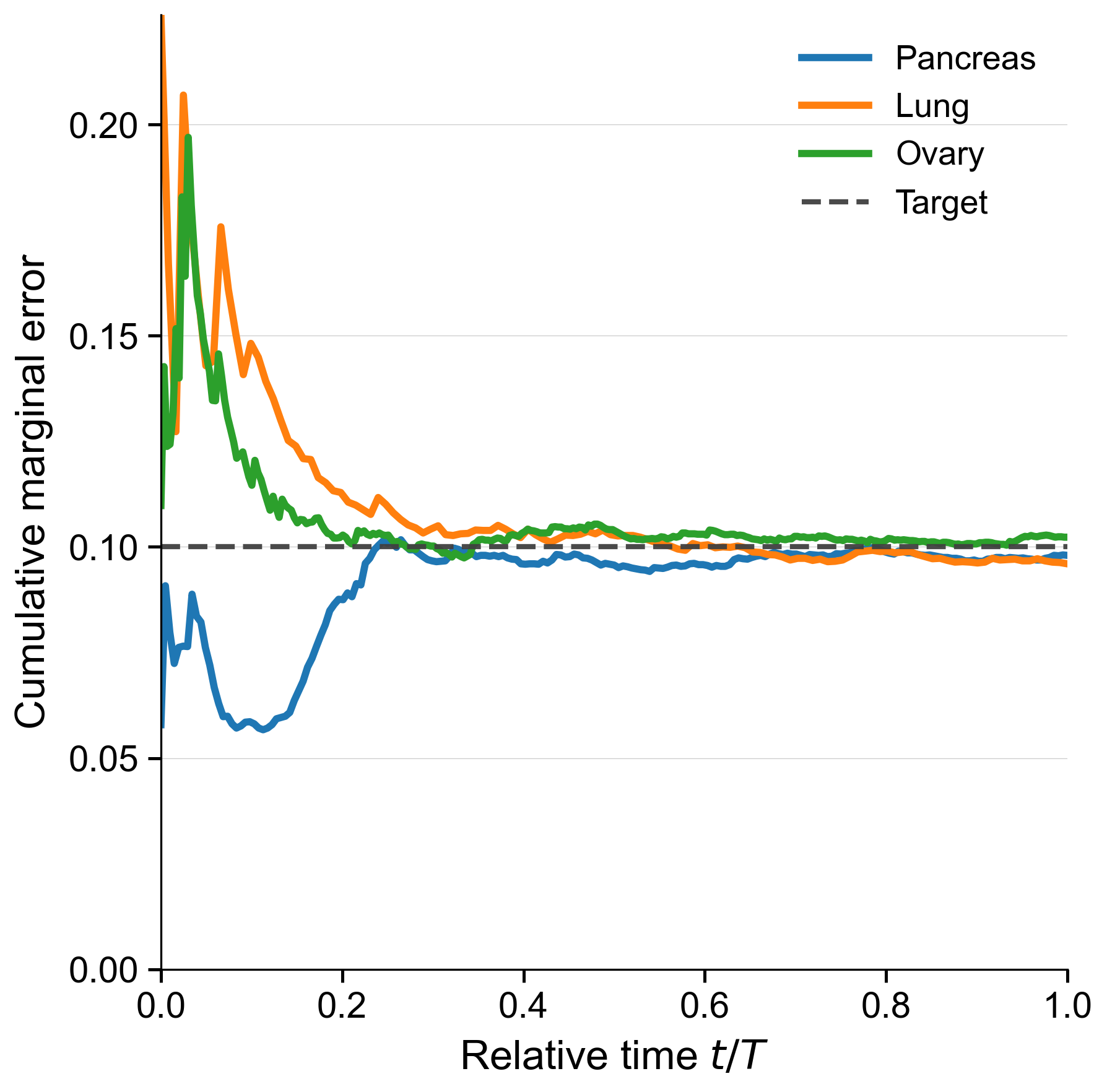}
        \caption{Error-guided, marginal}
    \end{subfigure}

    \smallskip

    \begin{subfigure}[t]{0.22\linewidth}
        \centering
        \includegraphics[width=\linewidth]{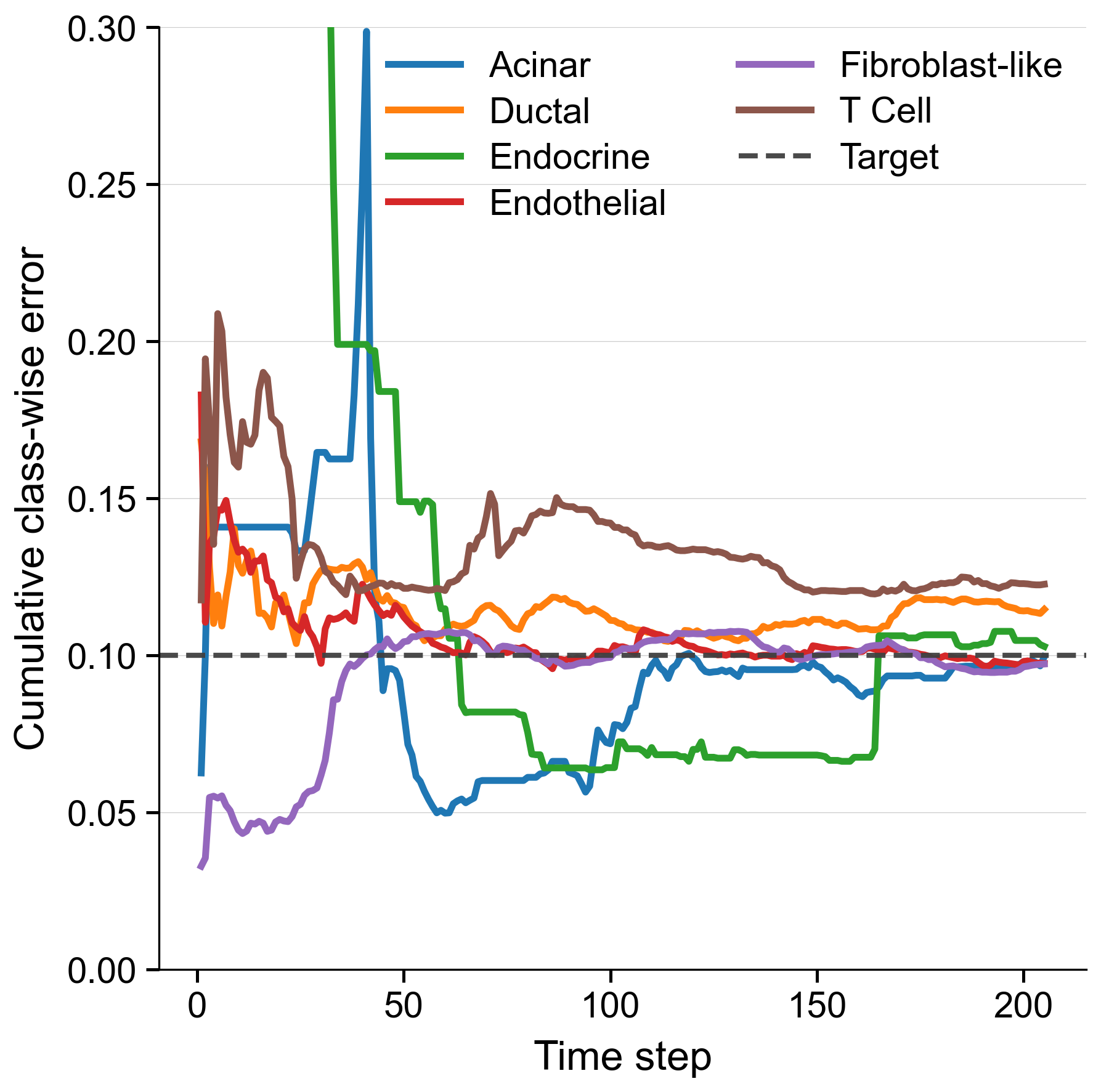}
        \caption{Adjacent, pancreas}
    \end{subfigure}
    \hfill
    \begin{subfigure}[t]{0.22\linewidth}
        \centering
        \includegraphics[width=\linewidth]{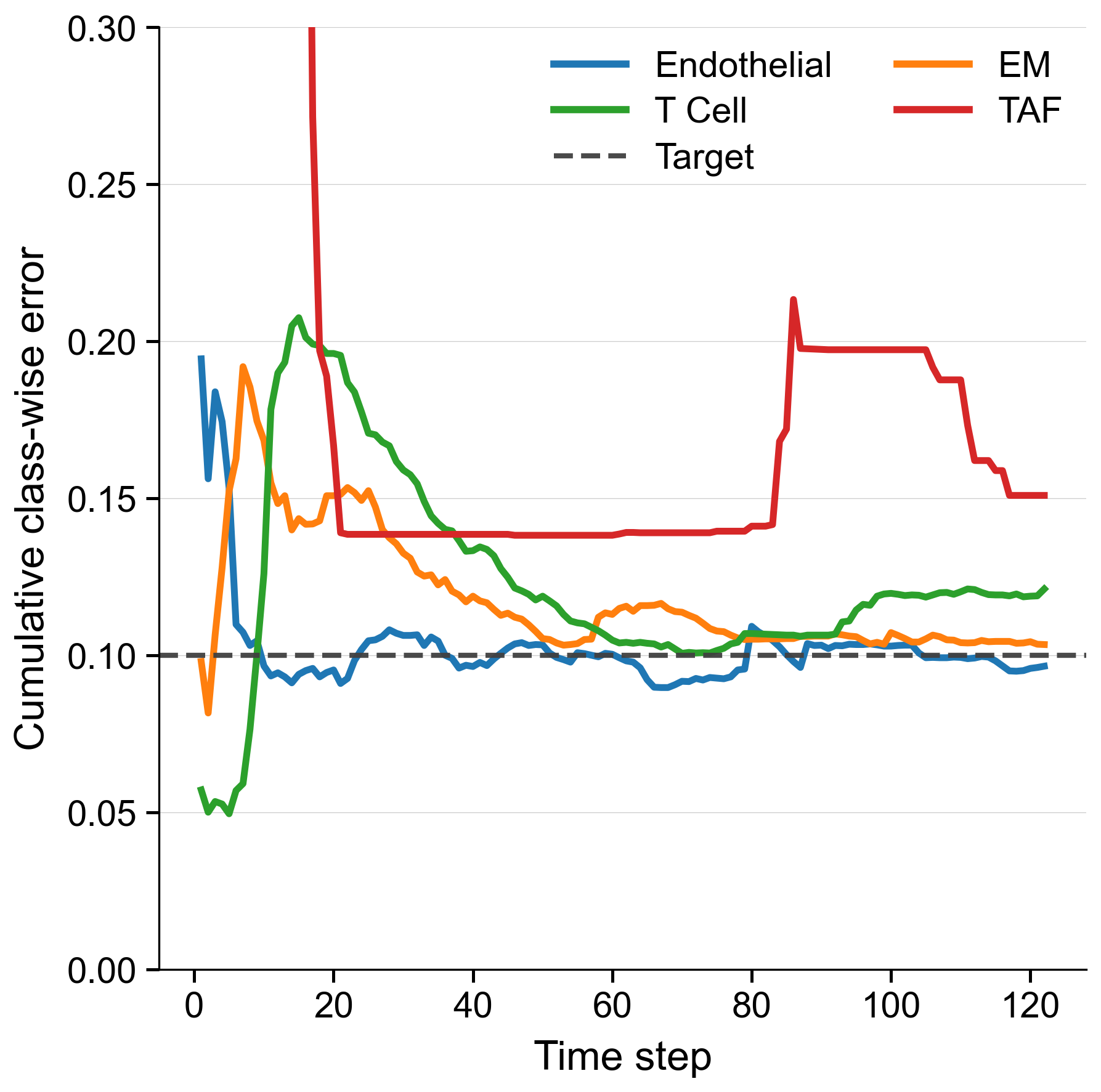}
        \caption{Adjacent, lung}
    \end{subfigure}
    \hfill
    \begin{subfigure}[t]{0.22\linewidth}
        \centering
        \includegraphics[width=\linewidth]{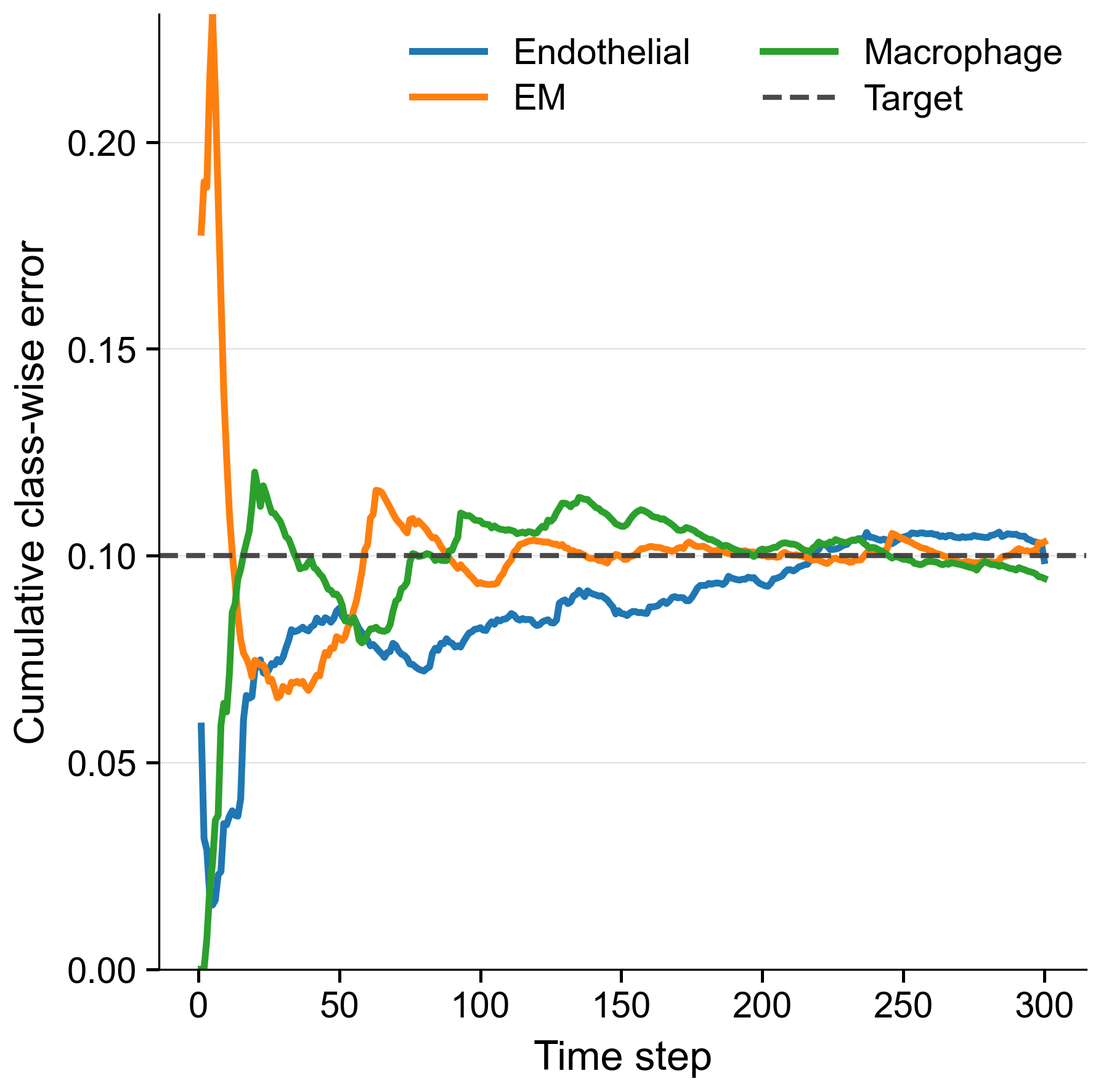}
        \caption{Adjacent, ovary}
    \end{subfigure}
    \hfill
    \begin{subfigure}[t]{0.22\linewidth}
        \centering
        \includegraphics[width=\linewidth]{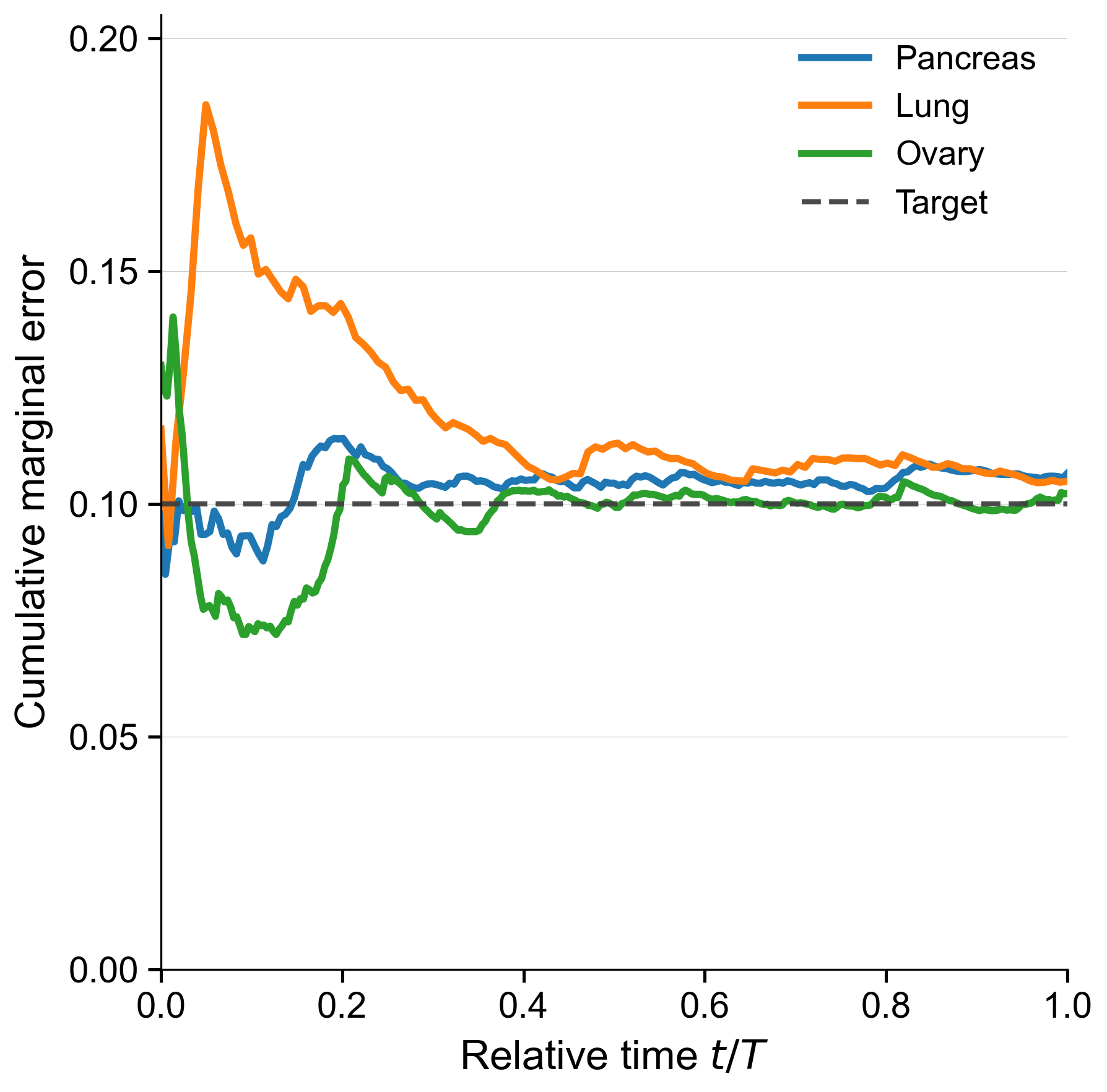}
        \caption{Adjacent, marginal}
    \end{subfigure}

    \smallskip

    \begin{subfigure}[t]{0.22\linewidth}
        \centering
        \includegraphics[width=\linewidth]{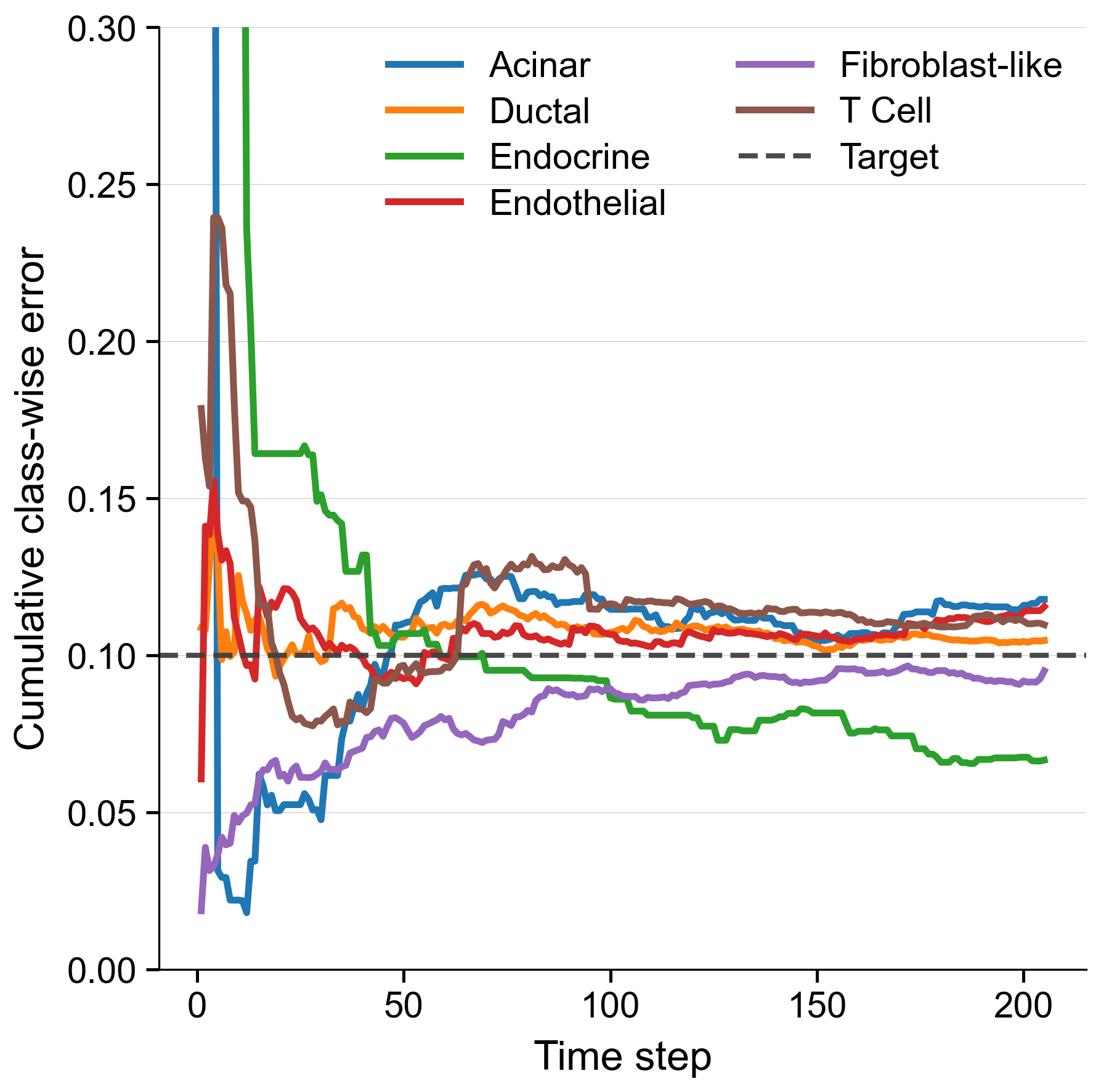}
        \caption{Random, pancreas}
    \end{subfigure}
    \hfill
    \begin{subfigure}[t]{0.22\linewidth}
        \centering
        \includegraphics[width=\linewidth]{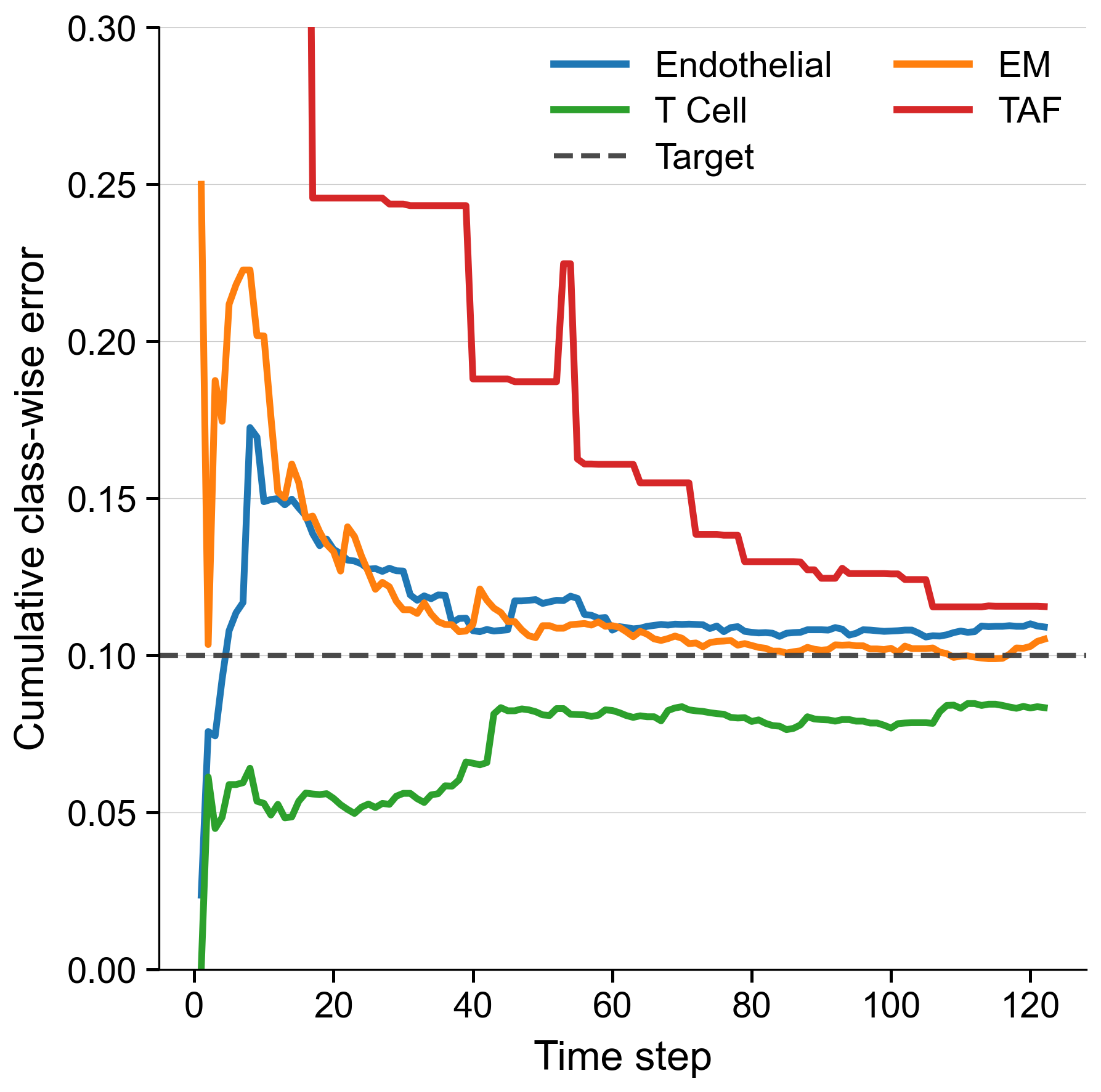}
        \caption{Random, lung}
    \end{subfigure}
    \hfill
    \begin{subfigure}[t]{0.22\linewidth}
        \centering
        \includegraphics[width=\linewidth]{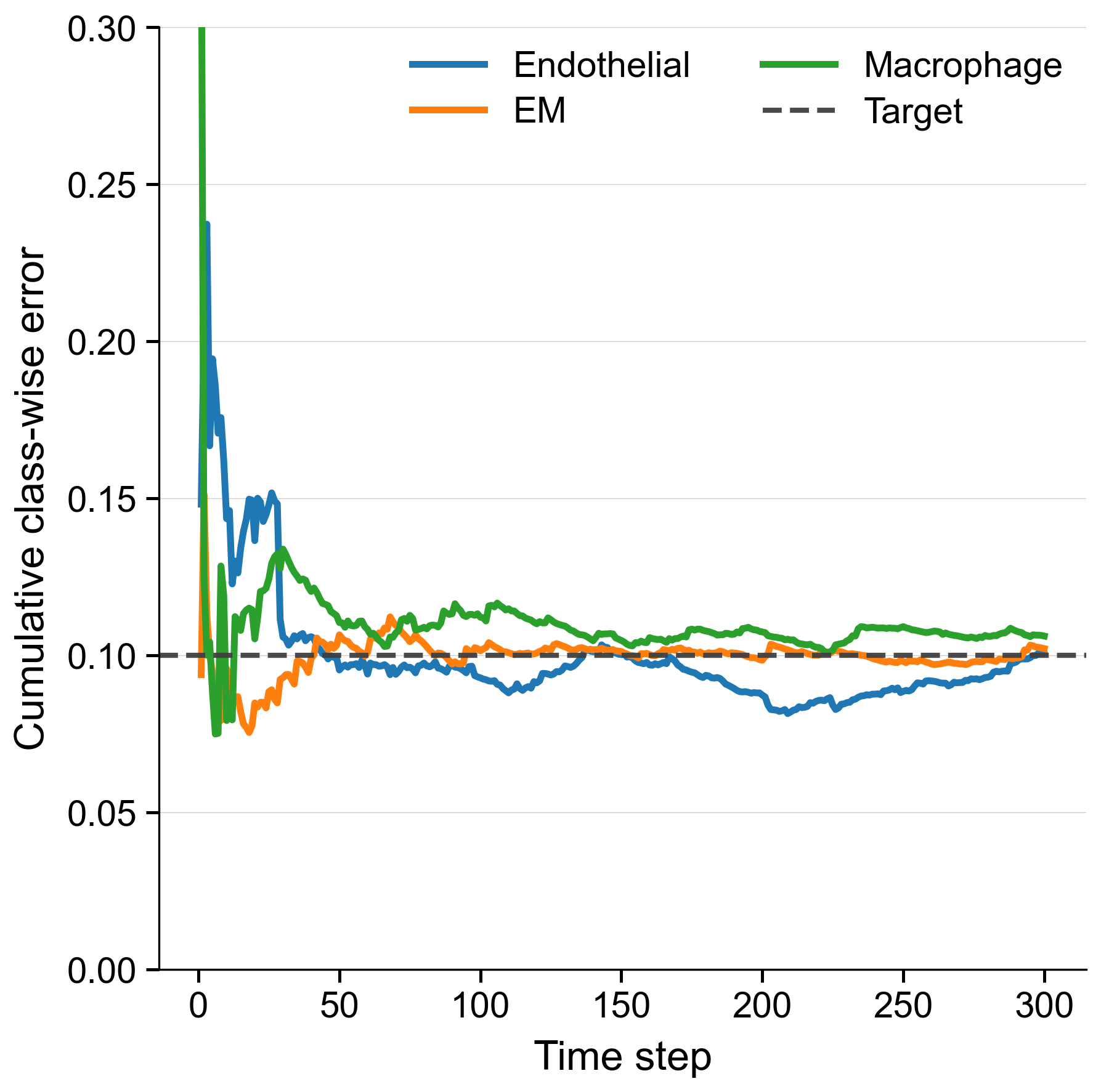}
        \caption{Random, ovary}
    \end{subfigure}
    \hfill
    \begin{subfigure}[t]{0.22\linewidth}
        \centering
        \includegraphics[width=\linewidth]{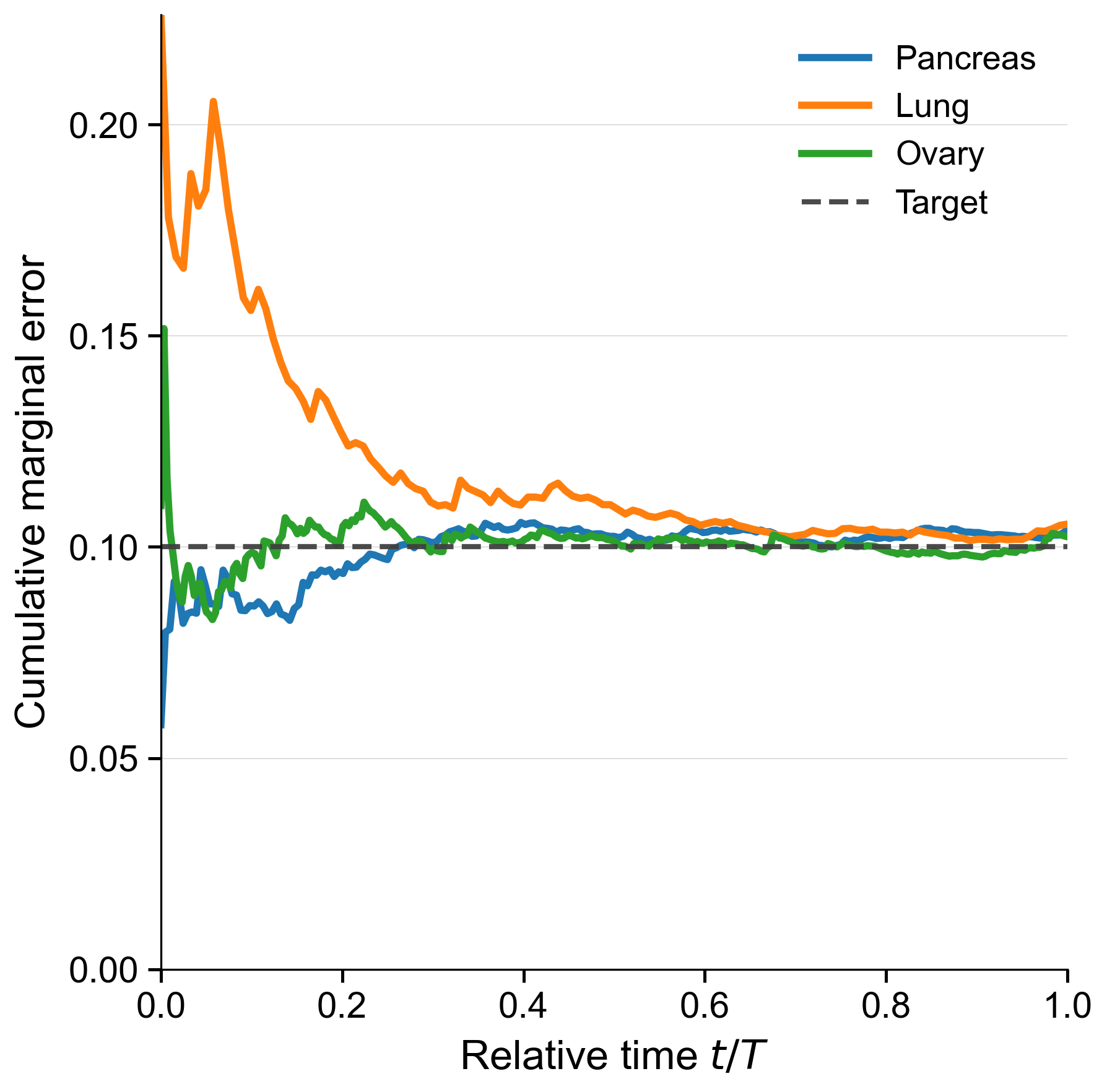}
        \caption{Random, marginal}
    \end{subfigure}
    \caption{Cumulative error trajectories under the global NuClass score for
    three batch-release schemes. Rows correspond to the error-guided,
    adjacent, and random schemes, while columns correspond to pancreas, lung,
    ovary, and marginal results. The tissue-specific panels report class-wise
    cumulative error over time; the marginal panels report cumulative  
    error by tissue against the relative time $t/T$. All experiments use target
    error level $\alpha=0.1$ and reviewed fraction $p=0.1$. Dashed horizontal
    lines mark the target error level.}
    \label{fig:cumulative_coverage_pancreas_global}
\end{figure}

\begin{figure}[!tp]
    \centering
    \begin{subfigure}[t]{0.49\linewidth}
        \centering
        \includegraphics[width=\linewidth]{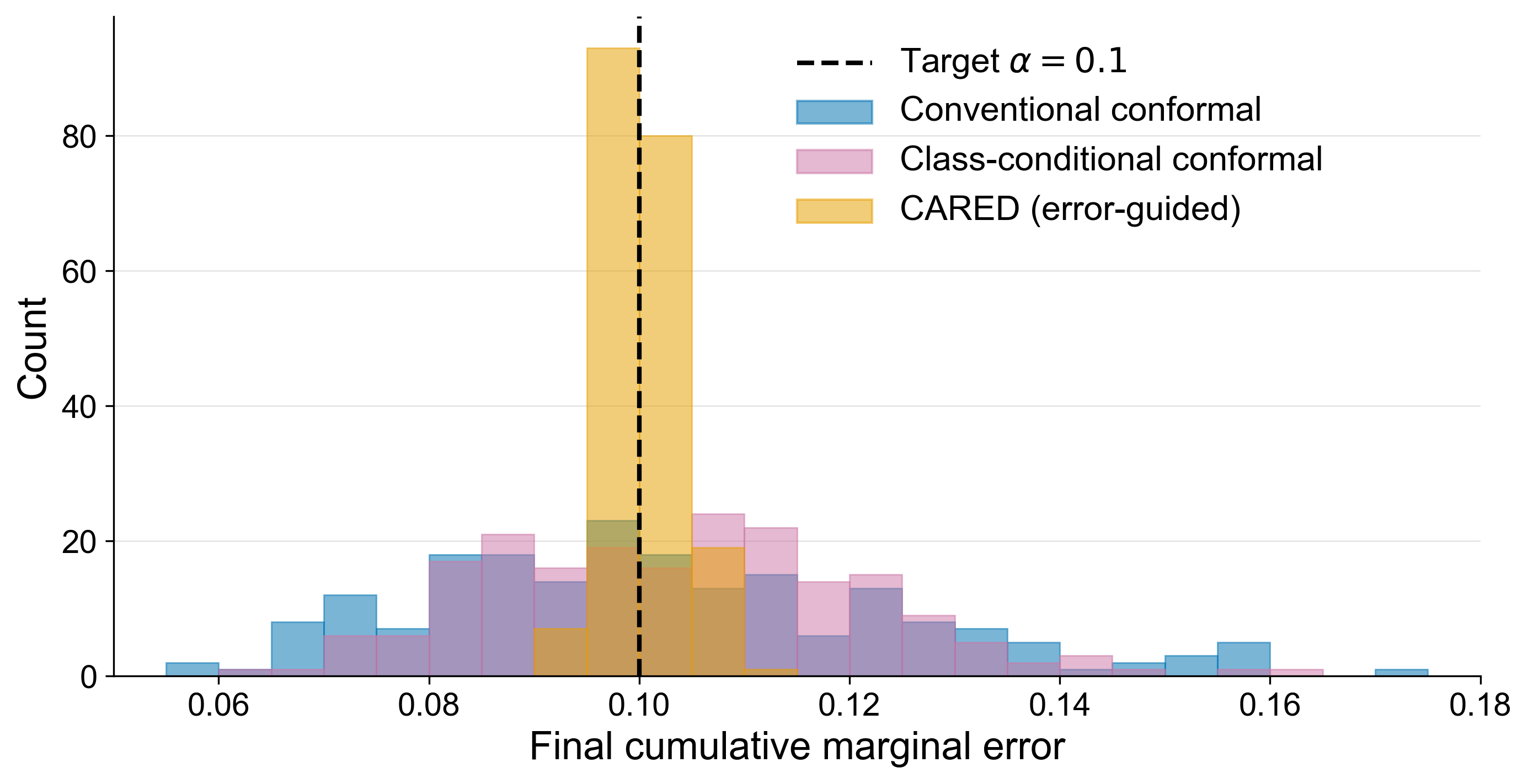}
        \caption{}
    \end{subfigure}
    \hfill
    \begin{subfigure}[t]{0.49\linewidth}
        \centering
        \includegraphics[width=\linewidth]{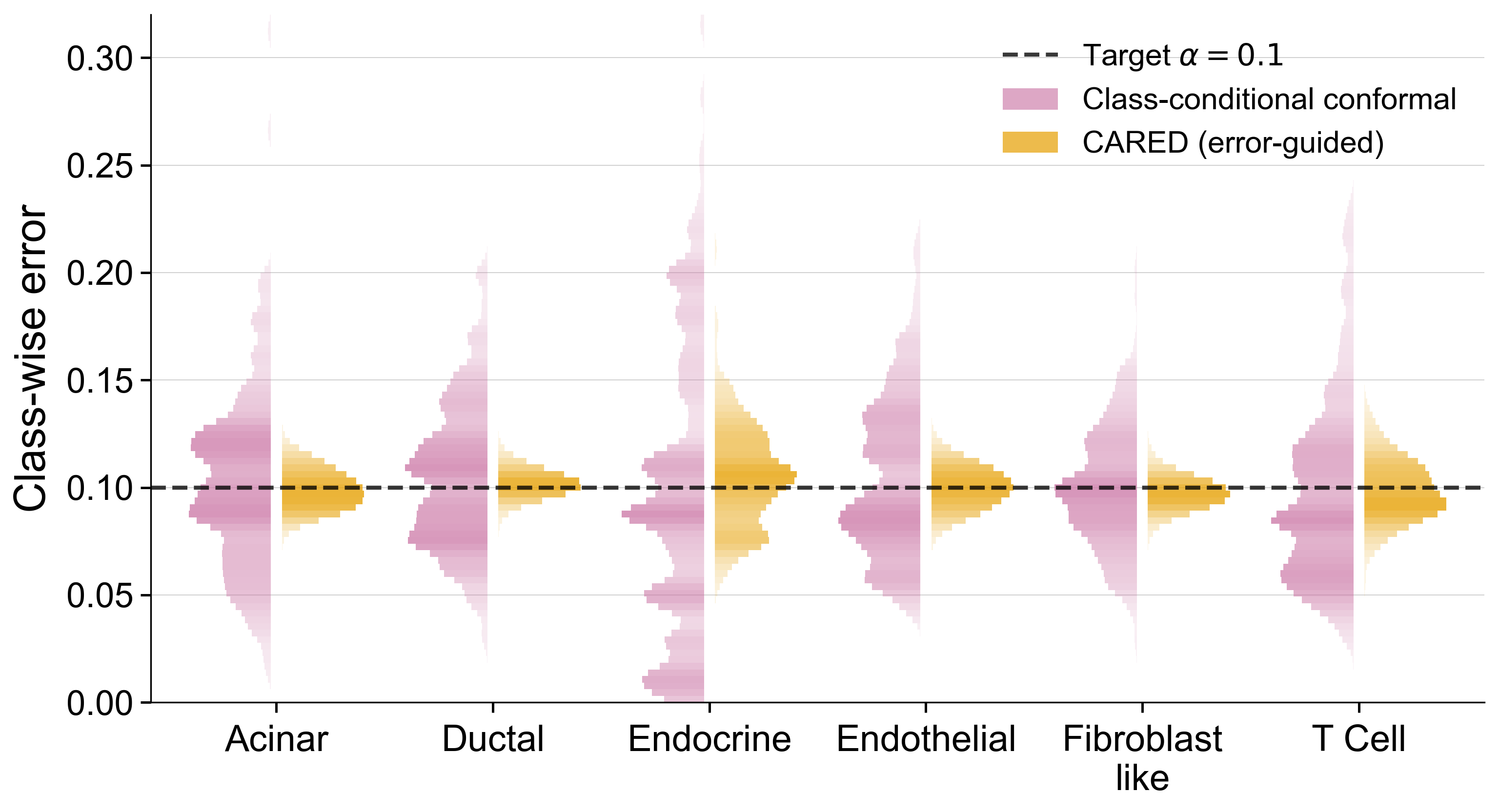}
        \caption{}
    \end{subfigure}
    \caption{Comparison of final cumulative error on pancreas tissue using the
    global score source. Each experiment randomly selects 10 batches as
    calibration data for the baselines and as the initial
    batches for CARED. (a)
    Distribution of final cumulative marginal error over 200 repeated random
    experiments, where the dashed vertical line marks the target error level
    $\alpha=0.1$. (b) Empirical distributions of final class-wise error over
    the same 200 experiments for class-conditional conformal prediction and
    our method, shown separately by cell type. The dashed horizontal
    line is the target level $\alpha=0.1$.}
    \label{fig:localized_calibration_comparison}
\end{figure}

Figure~\ref{fig:cumulative_coverage_pancreas_global} reports the whole-stream 
cumulative coverage error trajectories (which upper bounds the annotation error) under the
three batch-release schemes. Within each true cell type, miscoverage over the entire stream approaches $\alpha=0.1$ in most settings, even though true types are observed only for the $10\%$ audited cells. 
The largest remaining gaps occur for rare types, such as tumor-associated fibroblasts in the adjacent lung stream, consistent with slower convergence for rare classes (Proposition~\ref{prop:convergence_rates}). 
We also summarize the marginal errors per tissue, which also approaches the target level in each batch-release scheme.  

\vspace{-0.5em}
\paragraph{Comparison with static calibration.} A common alternative is to calibrate once on a few labeled regions, as when annotating selected ROIs or tiles~\citep{qiu2023adaptive,hagele2020resolving}. In this case, both covariate shift
and label shift can be amplified as we show in the introduction. 
We emulate this setting on the pancreas cohort: in each of 200 repetitions, we randomly choose $10$ batches as the calibration data in standard split conformal prediction and class-conditional conformal
prediction in Section~\ref{subsec:pre}, which are then applied to all remaining cells. CARED starts from the same batches and runs under the error-guided scheme with $p=0.1$, without the utility module.  All of them target the error control of $\alpha=0.1$. 
Across repetitions, the baselines' final marginal errors spread widely around $\alpha$, whereas CARED's concentrate at $\alpha$ (Figure~\ref{fig:localized_calibration_comparison}(a)). The gap widens within cell types (Figure~\ref{fig:localized_calibration_comparison}(b)): class-wise calibration adjusts for class composition but not for morphological differences across regions, so its errors vary most for the rare endocrine cells. Because exchangeability holds only at the region level, such one-time calibration is valid on average over randomly sampled regions but not in a given run. 

\subsection{Skipped ratio and prediction set utility}\label{subsec:additional_experiment_results}

\begin{figure}[!htbp]
    \centering
    \setlength{\parskip}{0pt}
    \captionsetup[subfigure]{font=scriptsize,justification=centering}
    \begin{subfigure}[t]{0.32\linewidth}
        \centering
        \includegraphics[width=\linewidth]{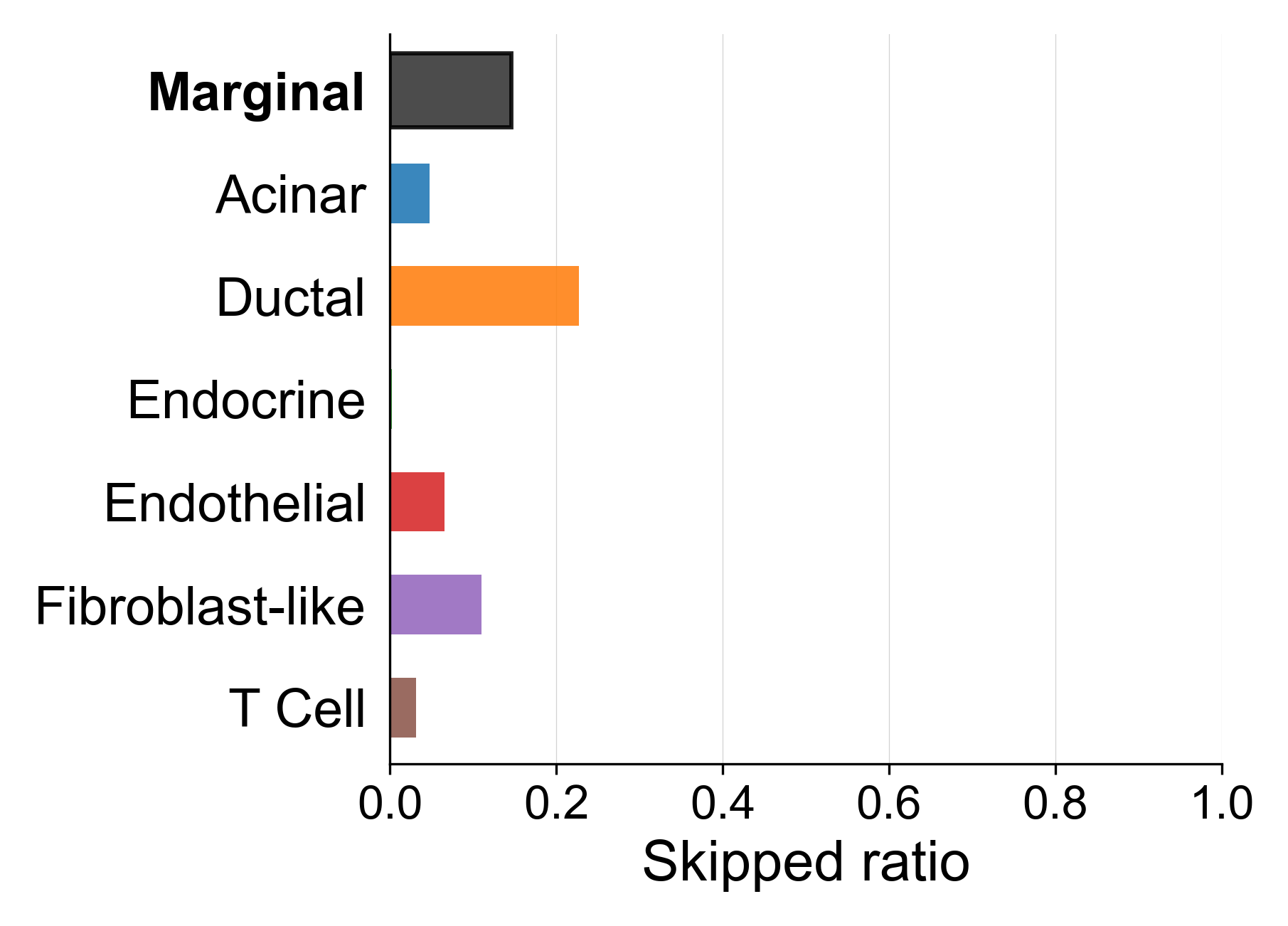}
        \caption{Error-guided, pancreas}
    \end{subfigure}
    \hfill
    \begin{subfigure}[t]{0.32\linewidth}
        \centering
        \includegraphics[width=\linewidth]{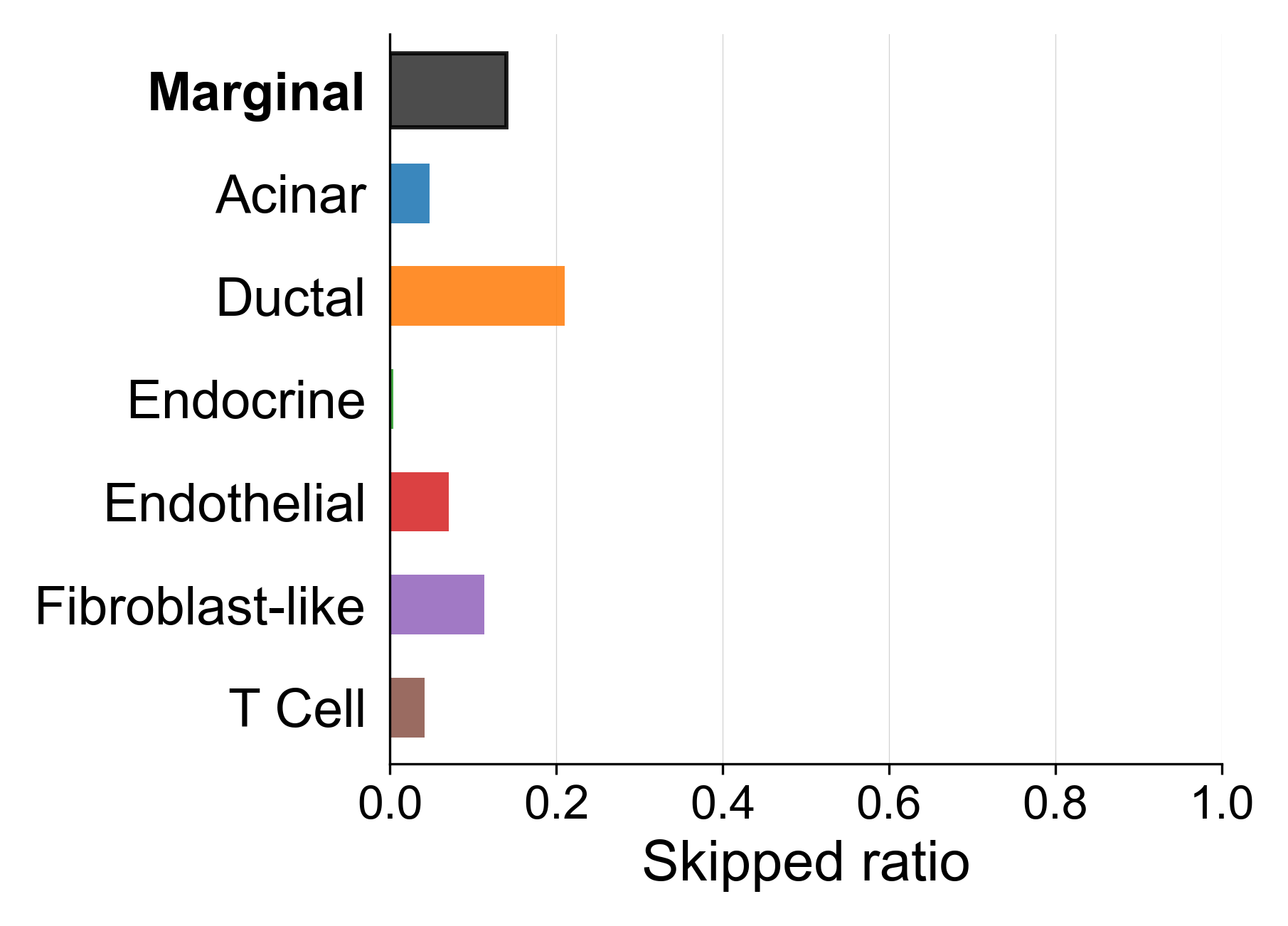}
        \caption{Adjacent, pancreas}
    \end{subfigure}
    \hfill
    \begin{subfigure}[t]{0.32\linewidth}
        \centering
        \includegraphics[width=\linewidth]{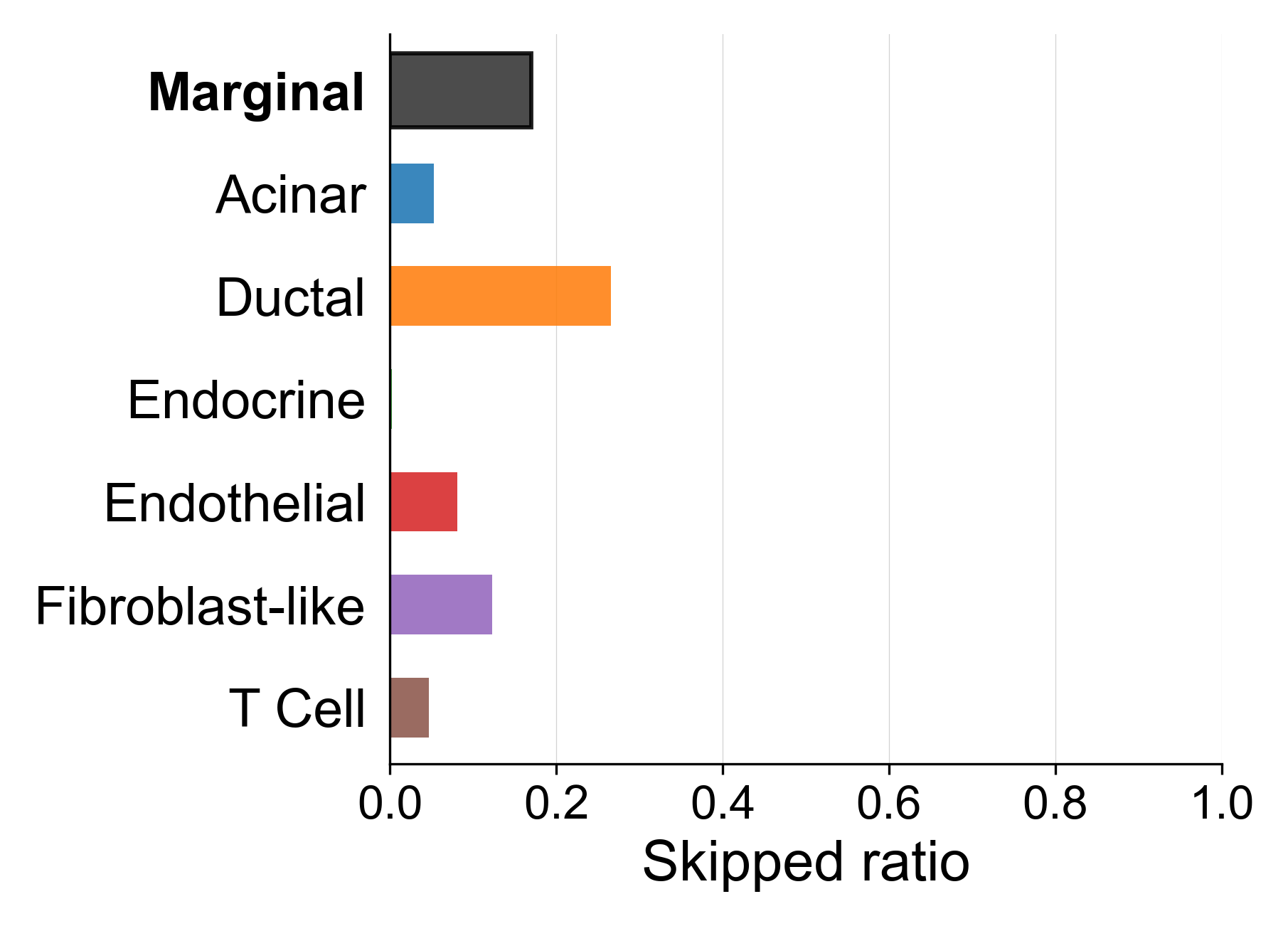}
        \caption{Random, pancreas}
    \end{subfigure}

    \vspace{1pt}

    \begin{subfigure}[t]{0.32\linewidth}
        \centering
        \includegraphics[width=\linewidth]{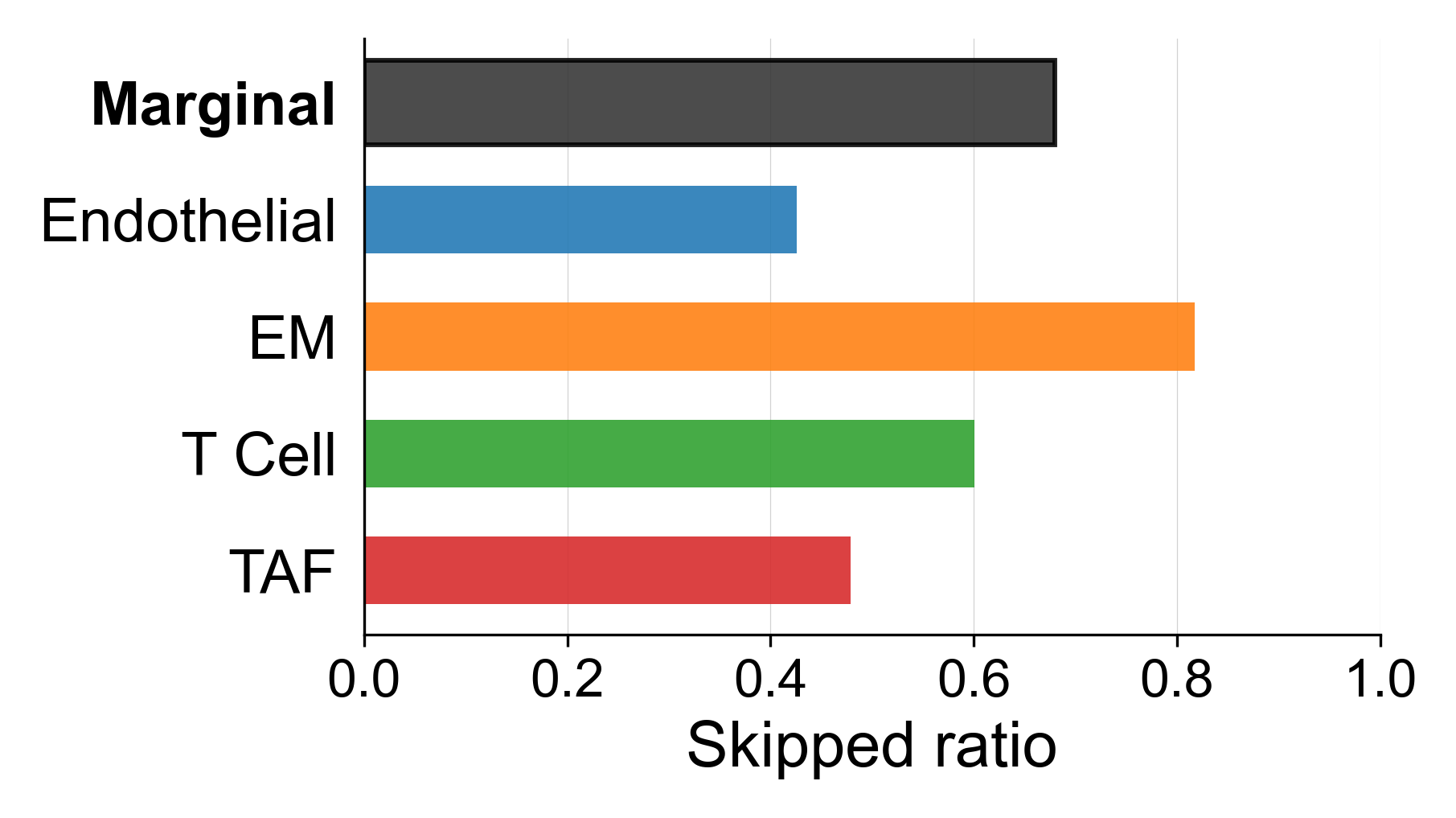}
        \caption{Error-guided, lung}
    \end{subfigure}
    \hfill
    \begin{subfigure}[t]{0.32\linewidth}
        \centering
        \includegraphics[width=\linewidth]{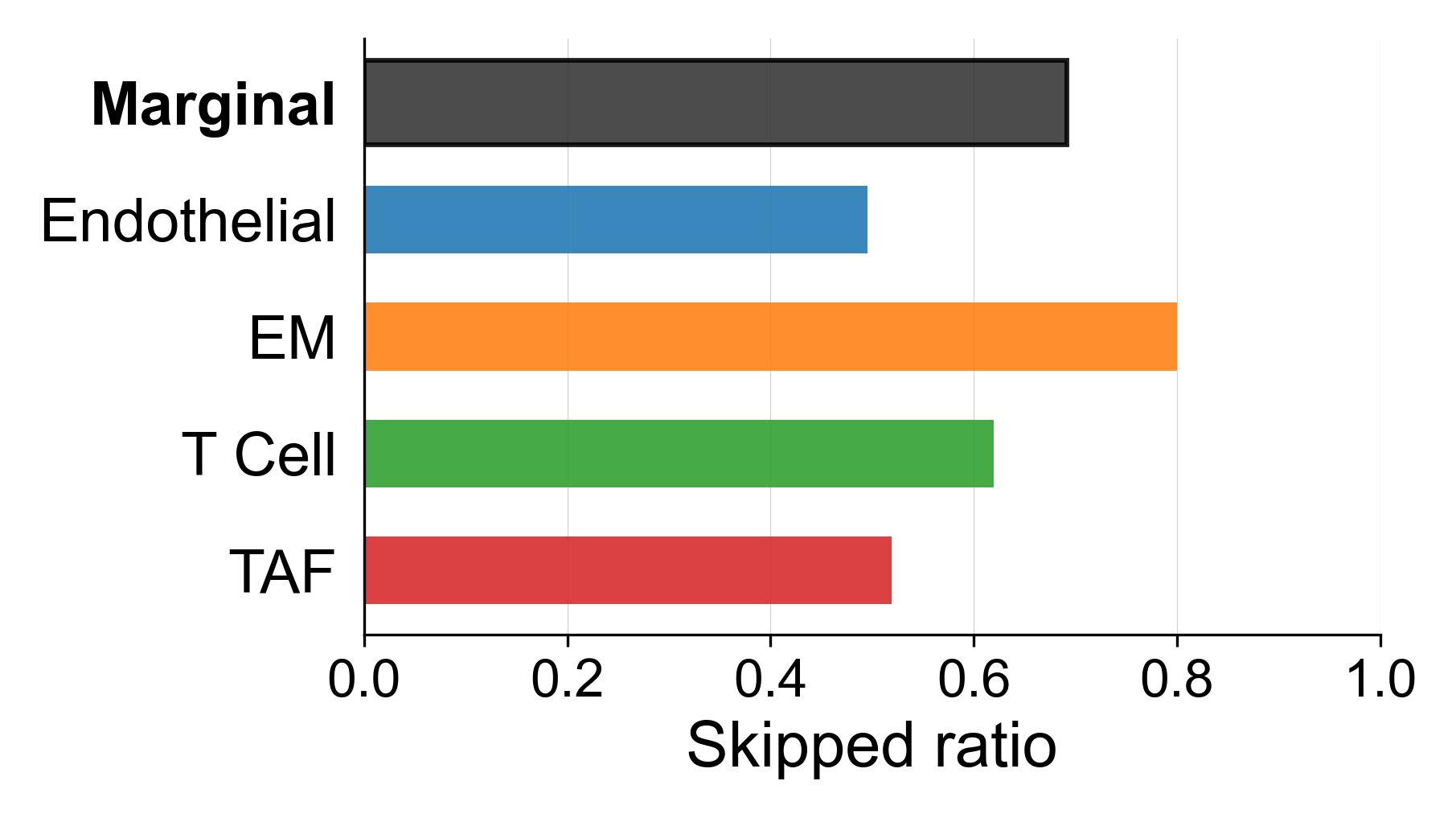}
        \caption{Adjacent, lung}
    \end{subfigure}
    \hfill
    \begin{subfigure}[t]{0.32\linewidth}
        \centering
        \includegraphics[width=\linewidth]{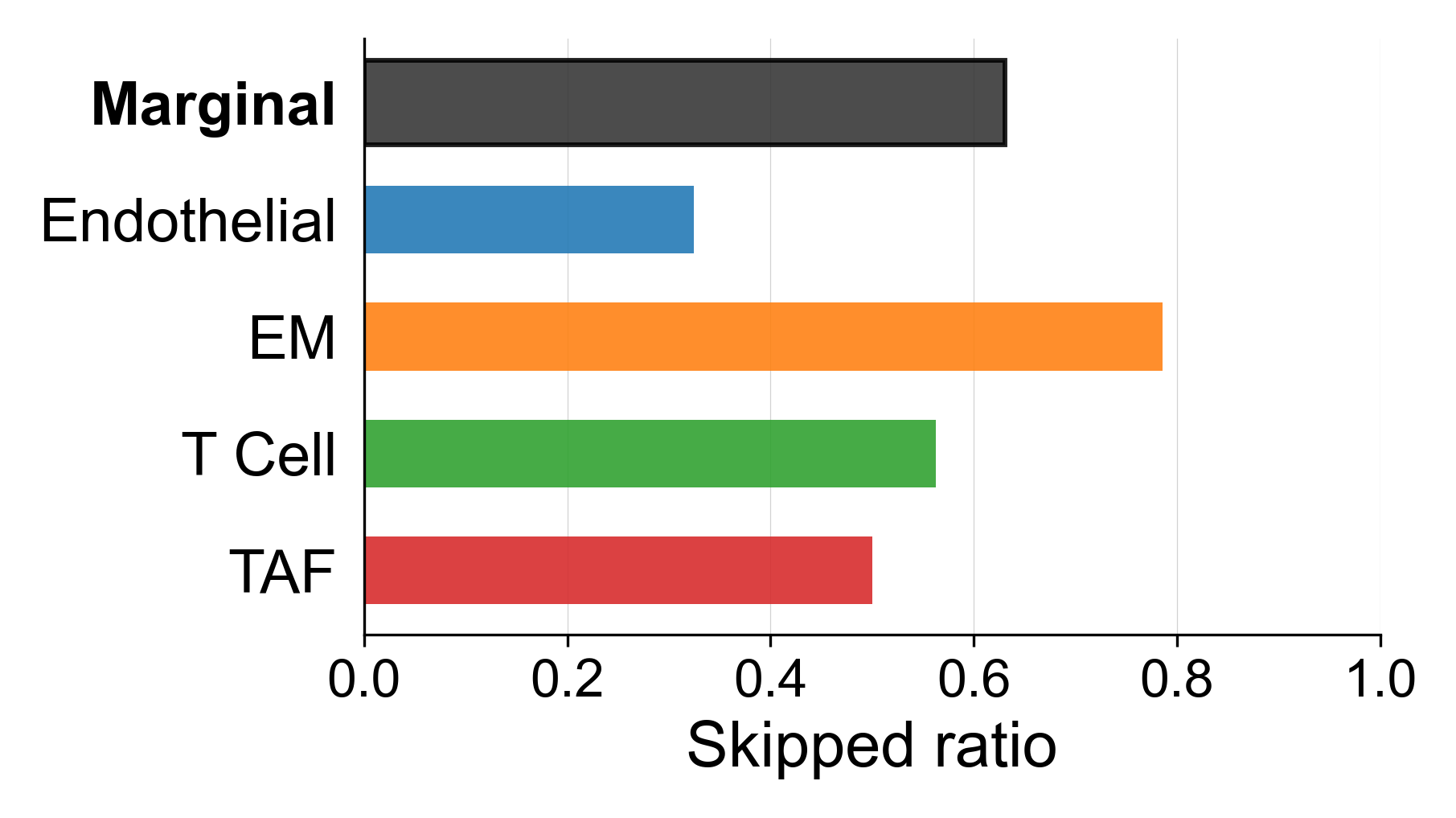}
        \caption{Random, lung}
    \end{subfigure}

    \vspace{1pt}

    \begin{subfigure}[t]{0.32\linewidth}
        \centering
        \includegraphics[width=\linewidth]{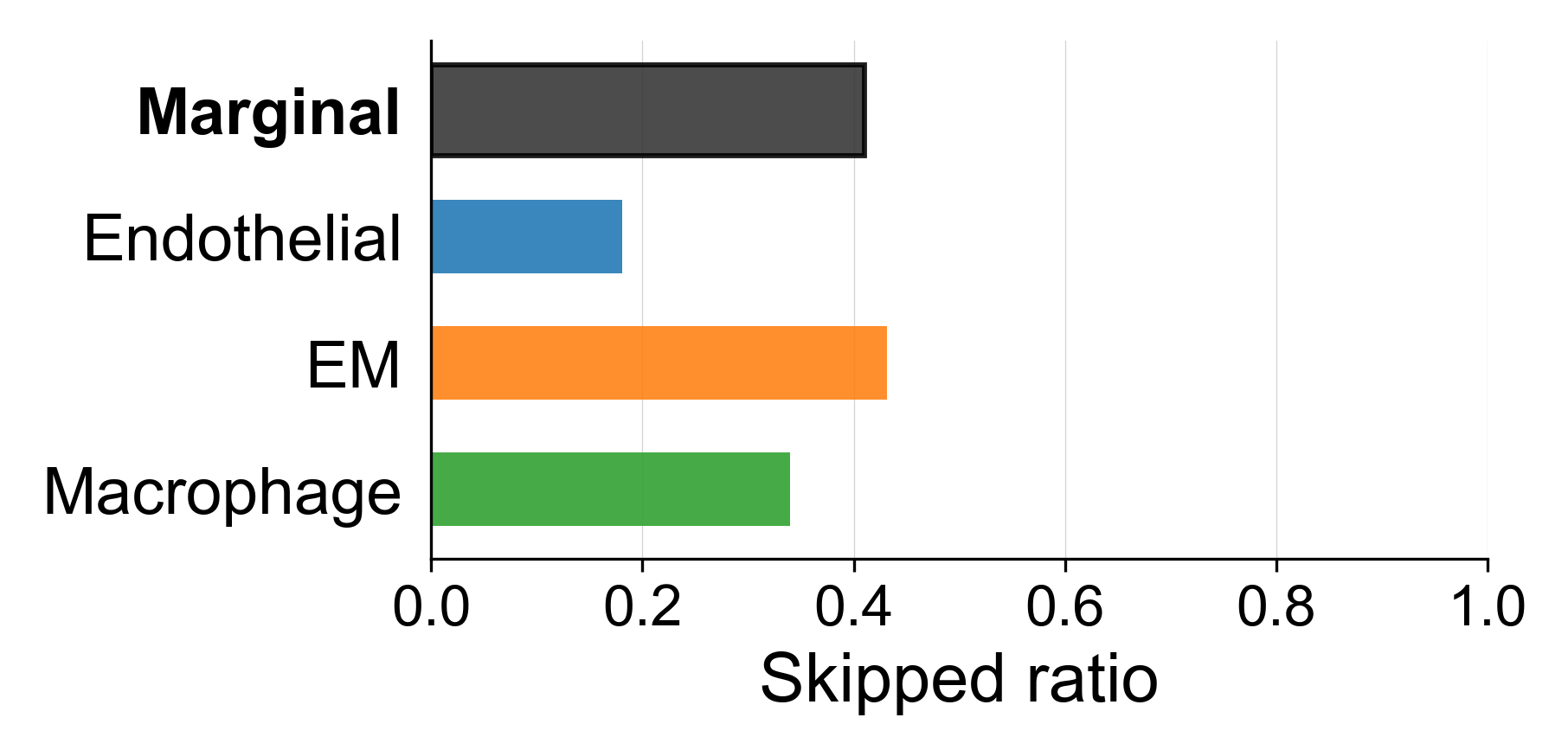}
        \caption{Error-guided, ovary}
    \end{subfigure}
    \hfill
    \begin{subfigure}[t]{0.32\linewidth}
        \centering
        \includegraphics[width=\linewidth]{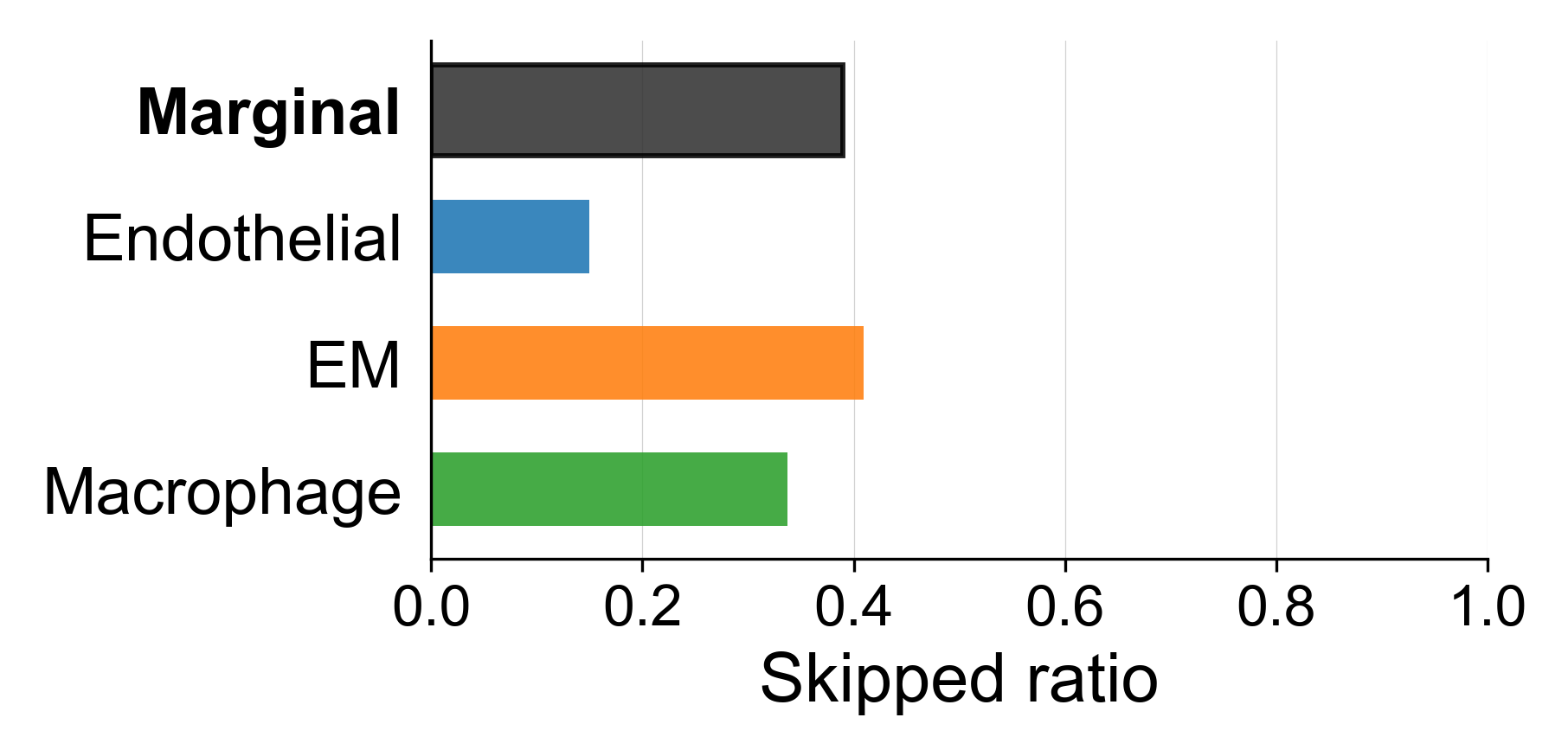}
        \caption{Adjacent, ovary}
    \end{subfigure}
    \hfill
    \begin{subfigure}[t]{0.32\linewidth}
        \centering
        \includegraphics[width=\linewidth]{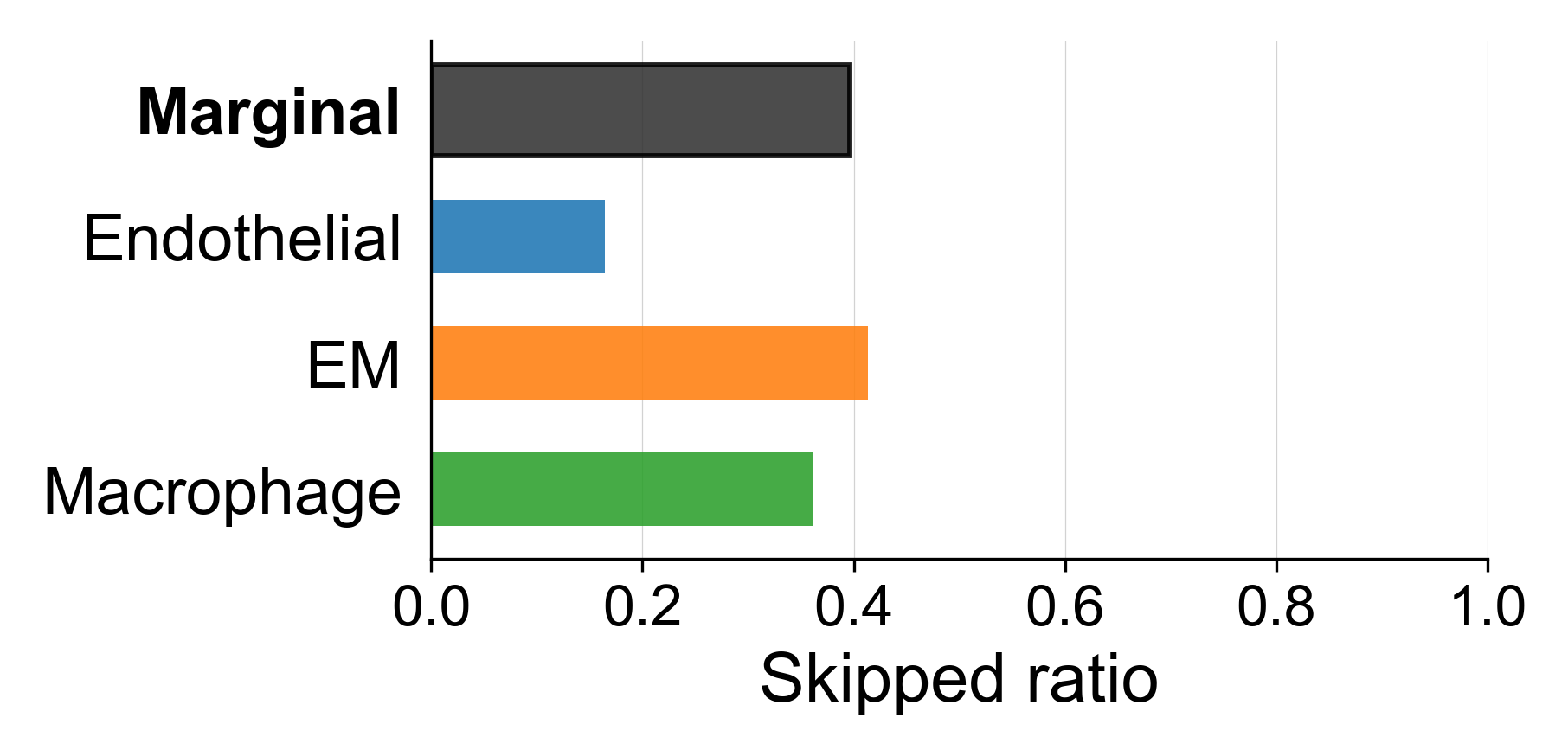}
        \caption{Random, ovary}
    \end{subfigure}
    \caption{Final skipped ratios using the global NuClass score. Rows correspond to tissues, while columns
    correspond to batch-release schemes. Each panel reports
    the marginal skipped ratio (bold) together
    with ratios stratified by true label. Here, EM denotes Epithelial Malignant
    and TAF denotes Tumor Associated Fibroblast. All experiments use target
    level $\alpha=0.1$ and reviewed fraction $p=0.1$.}
    \label{fig:final_skipped_ratio_placeholder}
\end{figure}

We measure the expert effort saved by the skipped ratio, the fraction of all cells that are not audited and receive an automated annotation, i.e., whose (restored) prediction set is a singleton~\eqref{eq:enlarged_set}. Figure~\ref{fig:final_skipped_ratio_placeholder} reports the 
skipped ratio at the final time $T$ across three schemes and three tissue cohorts, both over all cells and stratified by the true cell types. The ratio computes the number of automatically annotated cells among all, including reviewed, cells. CARED allows a substantial fraction of cells to bypass expert labeling across the
evaluated settings. 
In the four-class
lung annotation task at $p=0.1$ and $\alpha=0.1$, CARED allows skipping as much as 70\% of cells on average and over 80\% of epithelial malignant cells. The skipped ratio also varies across cell types within a tissue: for instance, almost no endocrine cells in pancreas are annotated automatically; this is because the AI predictions are less reliable for this class so CARED defers more cells of this class. 

\begin{figure}[!t]
    \centering
    \setlength{\parskip}{0pt}
    \captionsetup[subfigure]{font=scriptsize,justification=centering}
    \includegraphics[width=0.44\linewidth]{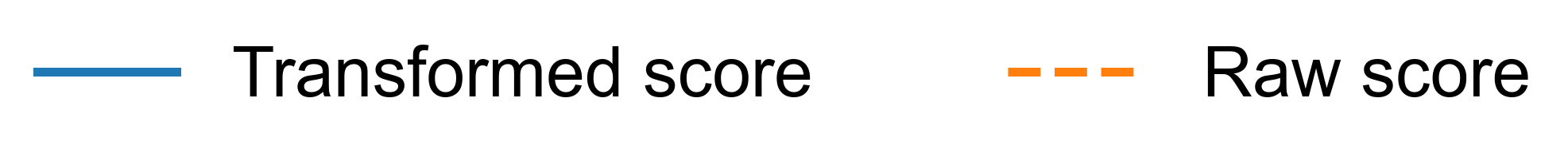}

    \vspace{1pt}

    \begin{subfigure}[t]{0.29\linewidth}
        \centering
        \includegraphics[width=\linewidth]{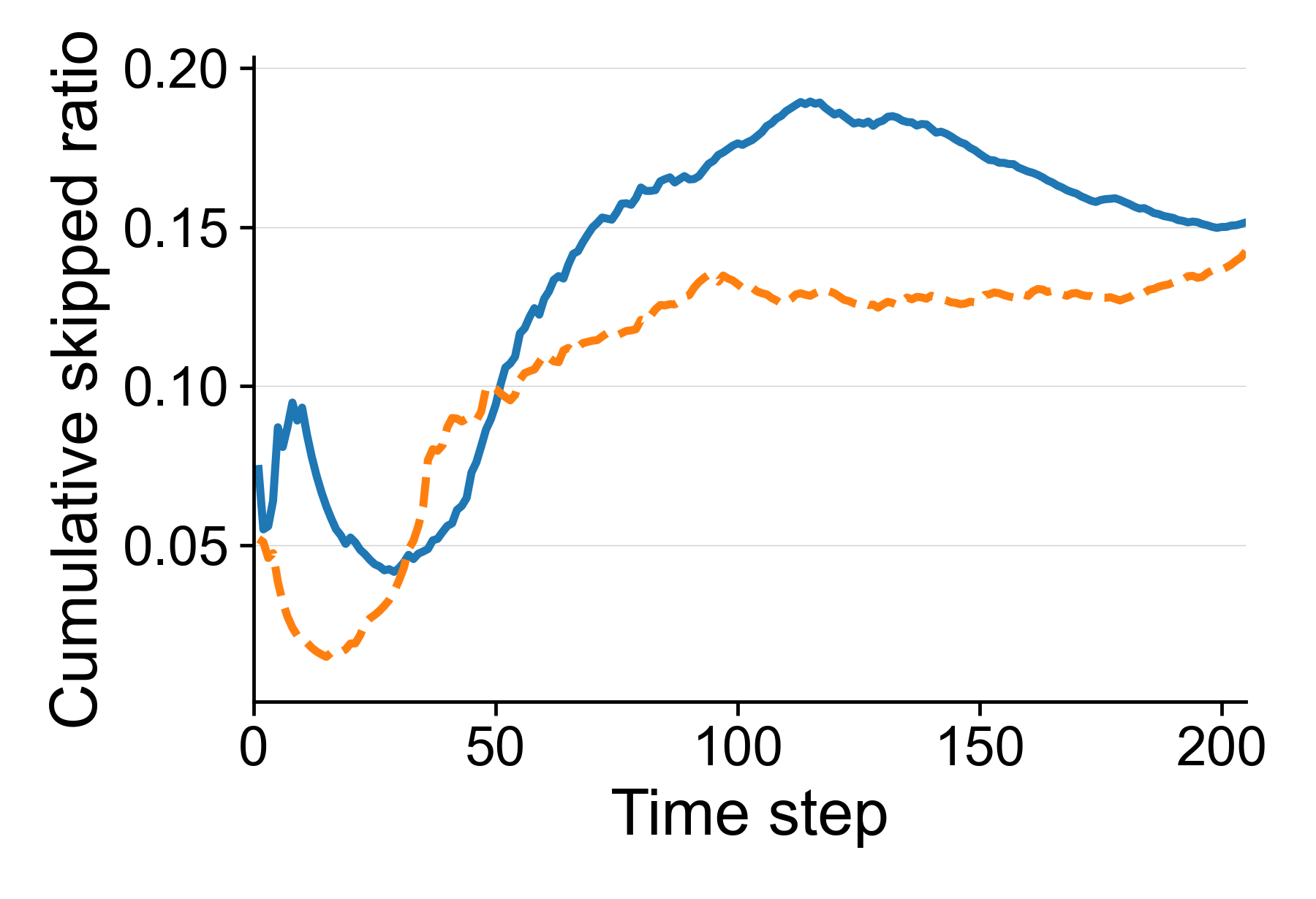}
        \caption{Error-guided, skipped ratio ($\uparrow$)}
    \end{subfigure}
    \hfill
    \begin{subfigure}[t]{0.29\linewidth}
        \centering
        \includegraphics[width=\linewidth]{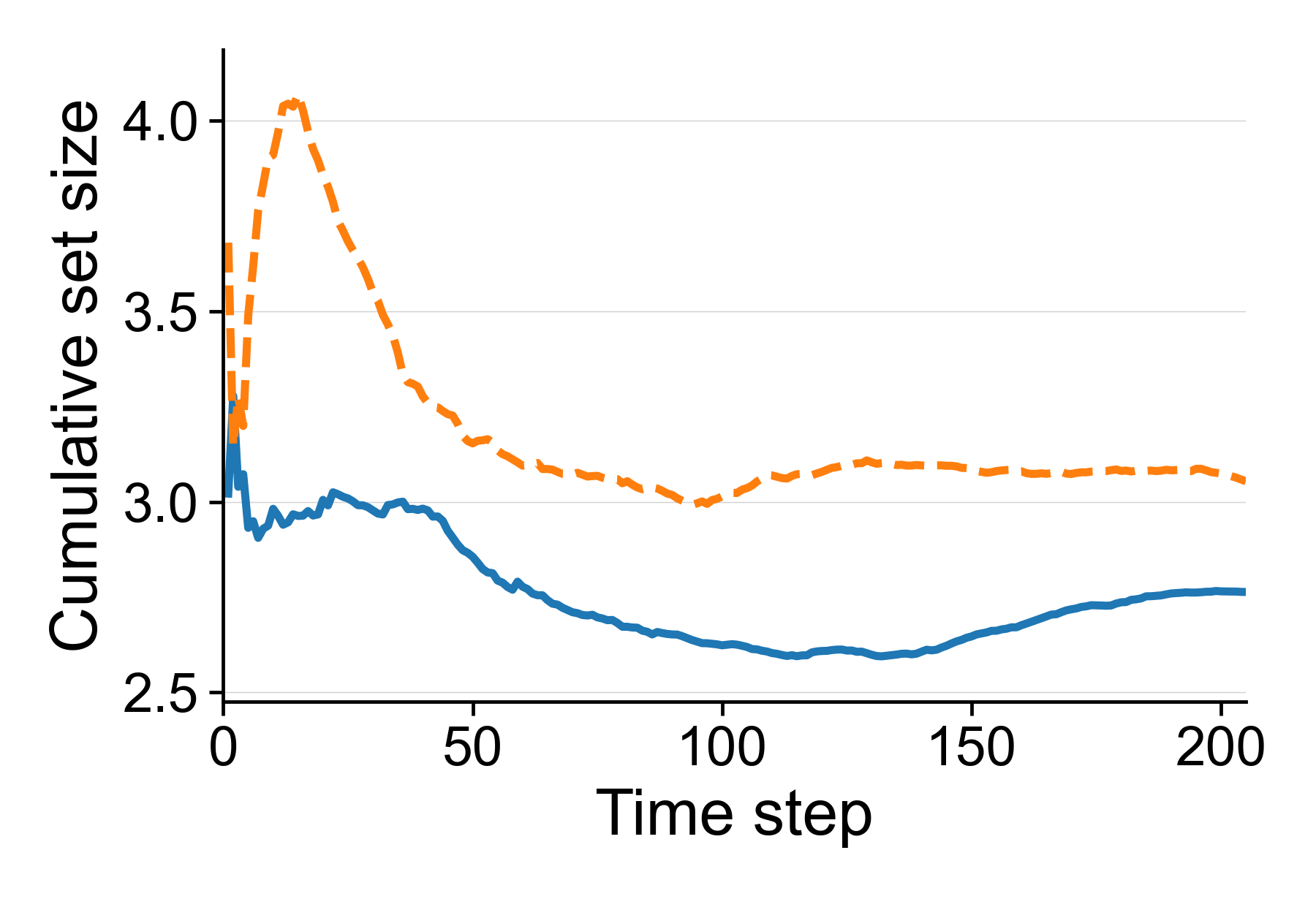}
        \caption{Error-guided, set size ($\downarrow$)}
    \end{subfigure}
    \hfill
    \begin{subfigure}[t]{0.31\linewidth}
        \centering
        \includegraphics[width=0.93\linewidth]{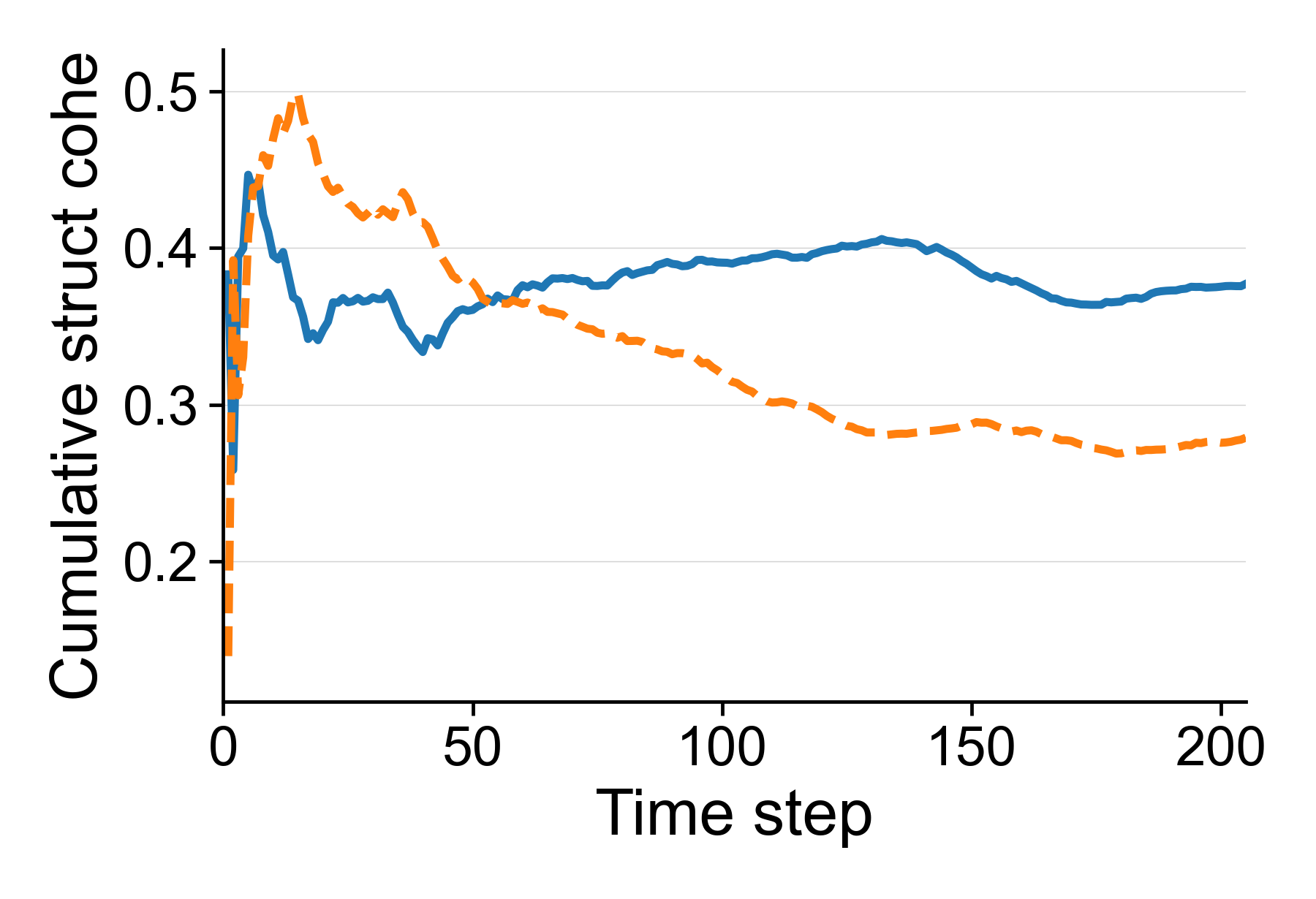}
        \caption{Error-guided, structural coherence ($\uparrow$)}
    \end{subfigure}

    \vspace{1pt}

    \begin{subfigure}[t]{0.29\linewidth}
        \centering
        \includegraphics[width=\linewidth]{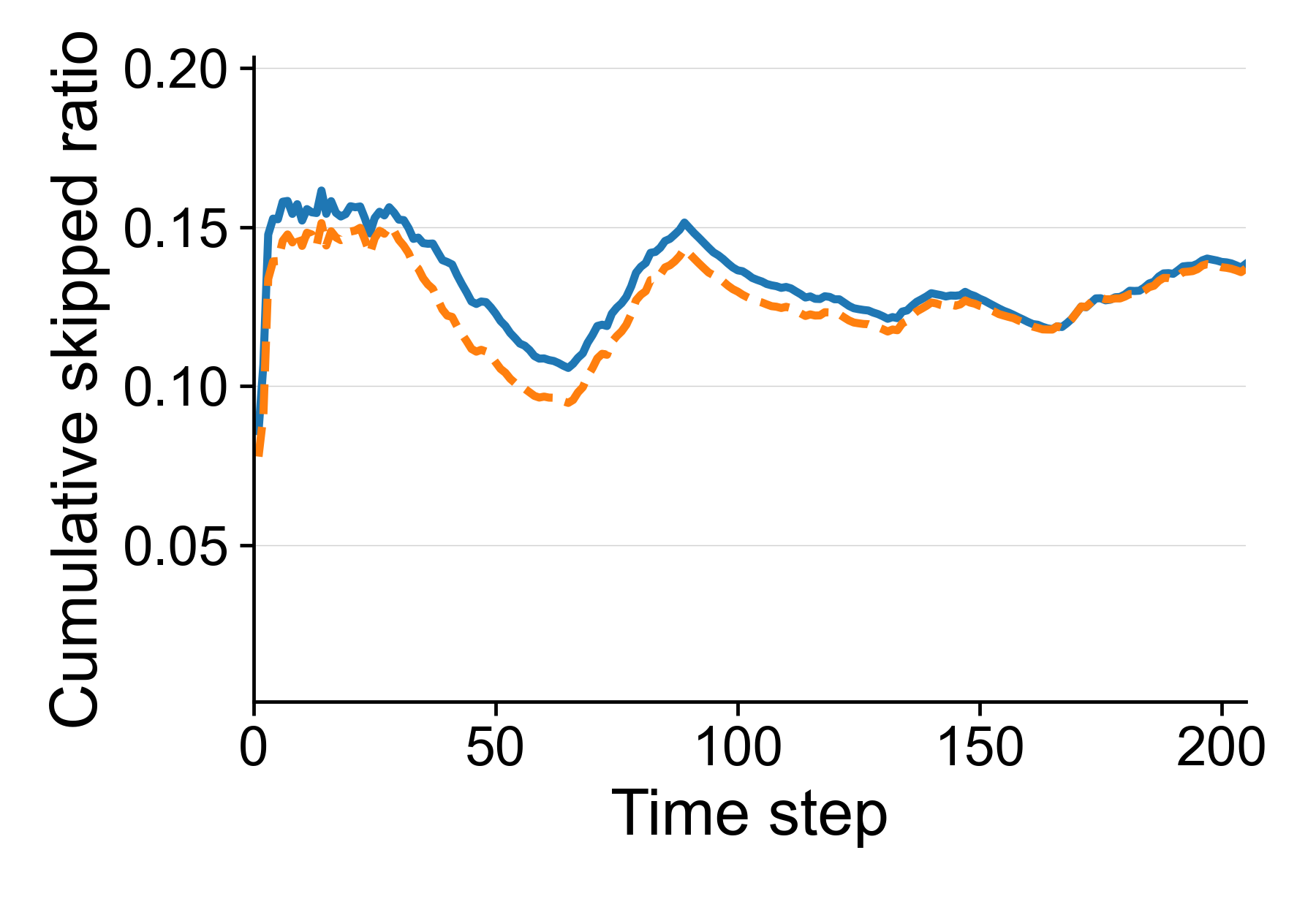}
        \caption{Adjacent, skipped ratio ($\uparrow$)}
    \end{subfigure}
    \hfill
    \begin{subfigure}[t]{0.29\linewidth}
        \centering
        \includegraphics[width=\linewidth]{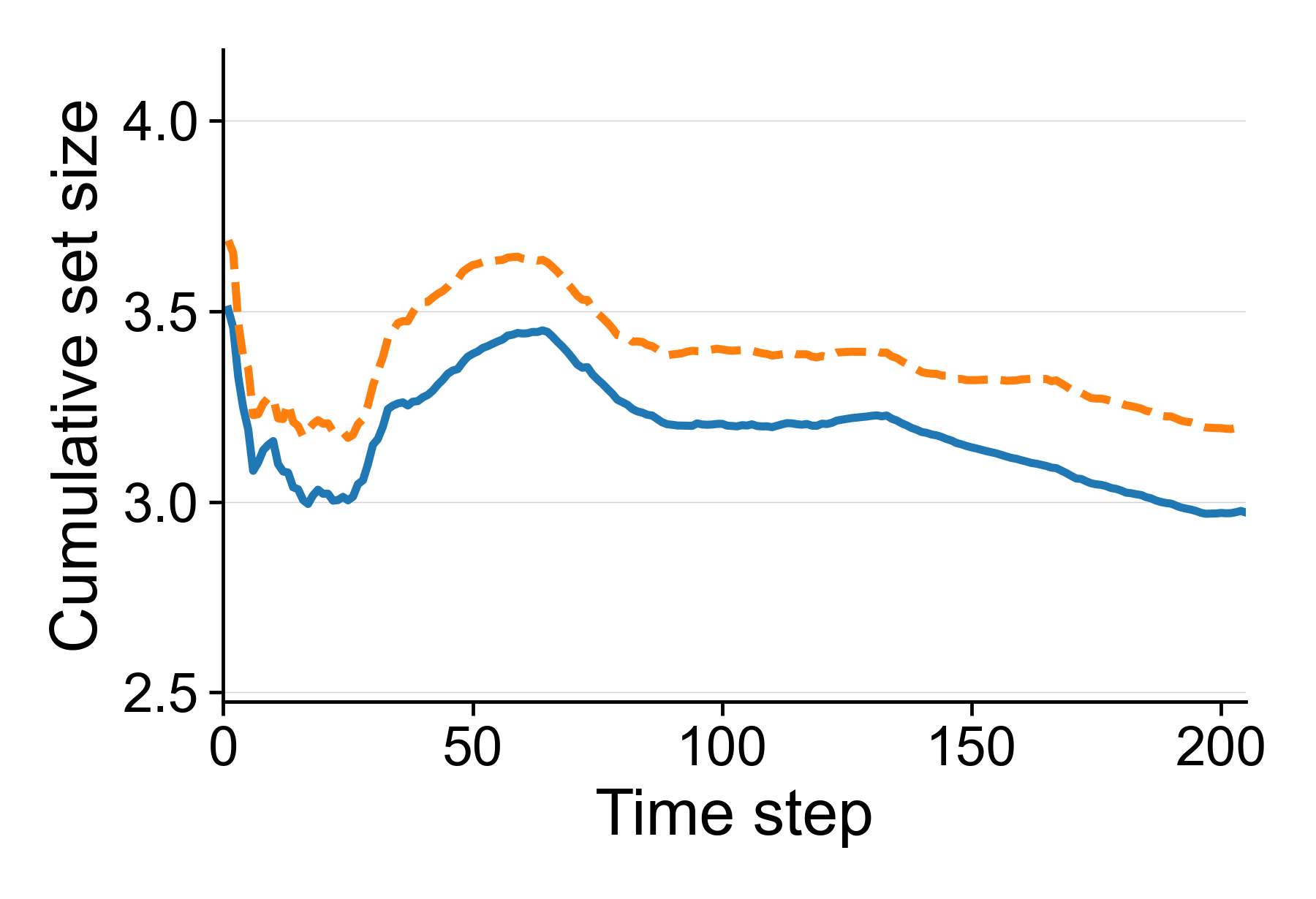}
        \caption{Adjacent, set size ($\downarrow$)}
    \end{subfigure}
    \hfill
    \begin{subfigure}[t]{0.31\linewidth}
        \centering
        \includegraphics[width=0.93\linewidth]{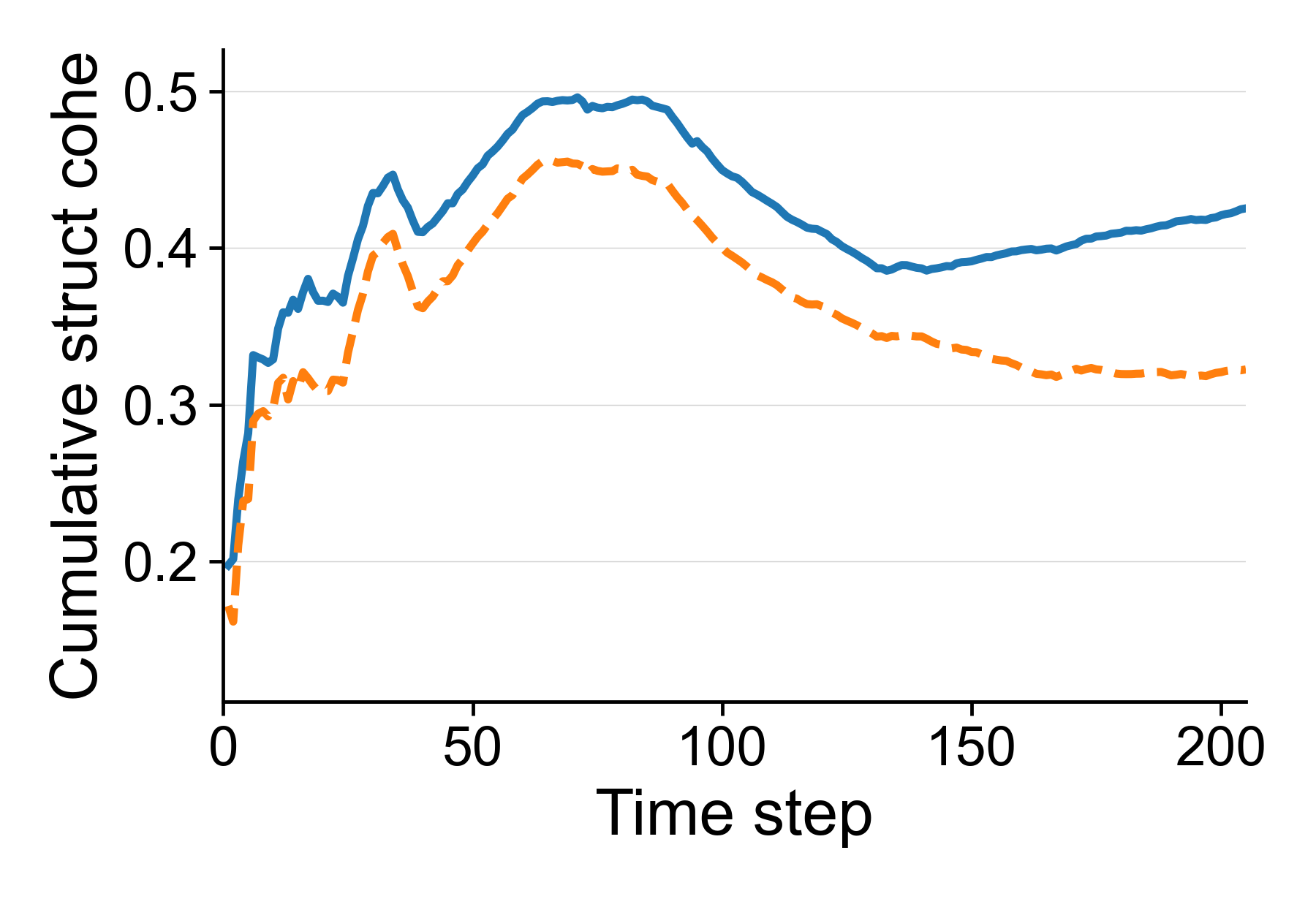}
        \caption{Adjacent, structural coherence ($\uparrow$)}
    \end{subfigure}

    \vspace{1pt}

    \begin{subfigure}[t]{0.29\linewidth}
        \centering
        \includegraphics[width=\linewidth]{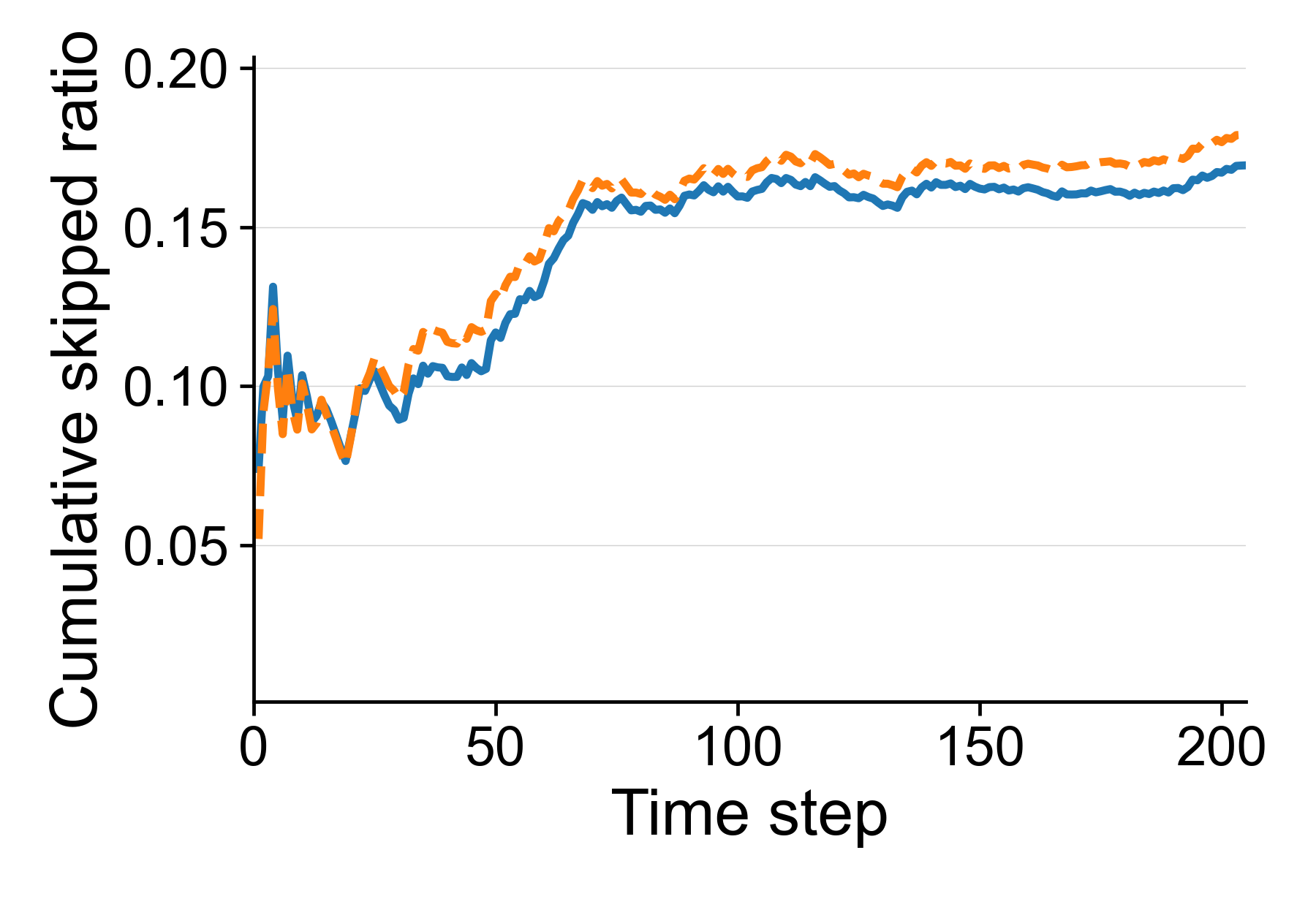}
        \caption{Random, skipped ratio ($\uparrow$)}
    \end{subfigure}
    \hfill
    \begin{subfigure}[t]{0.29\linewidth}
        \centering
        \includegraphics[width=\linewidth]{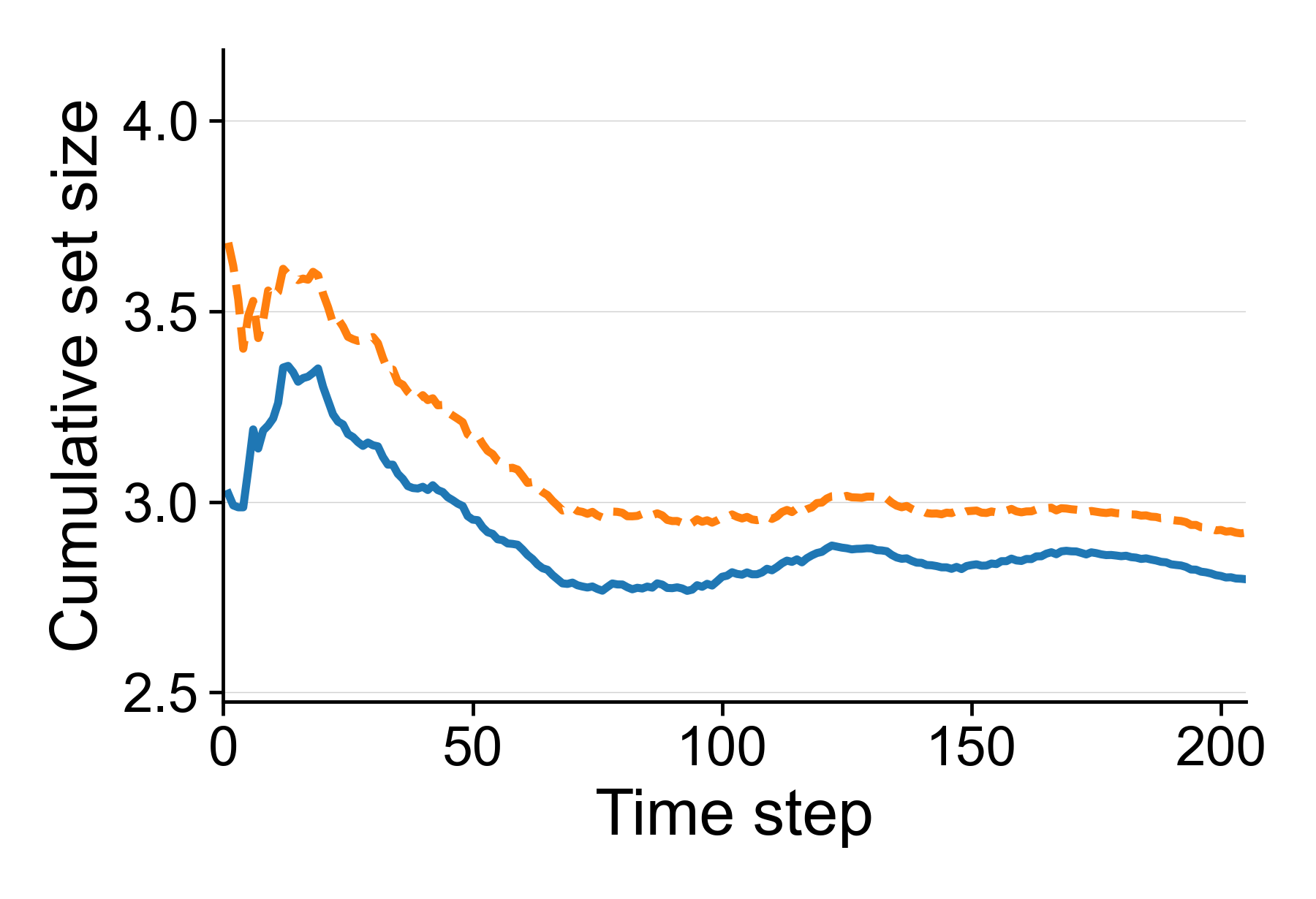}
        \caption{Random, set size ($\downarrow$)}
    \end{subfigure}
    \hfill
    \begin{subfigure}[t]{0.31\linewidth}
        \centering
        \includegraphics[width=0.93\linewidth]{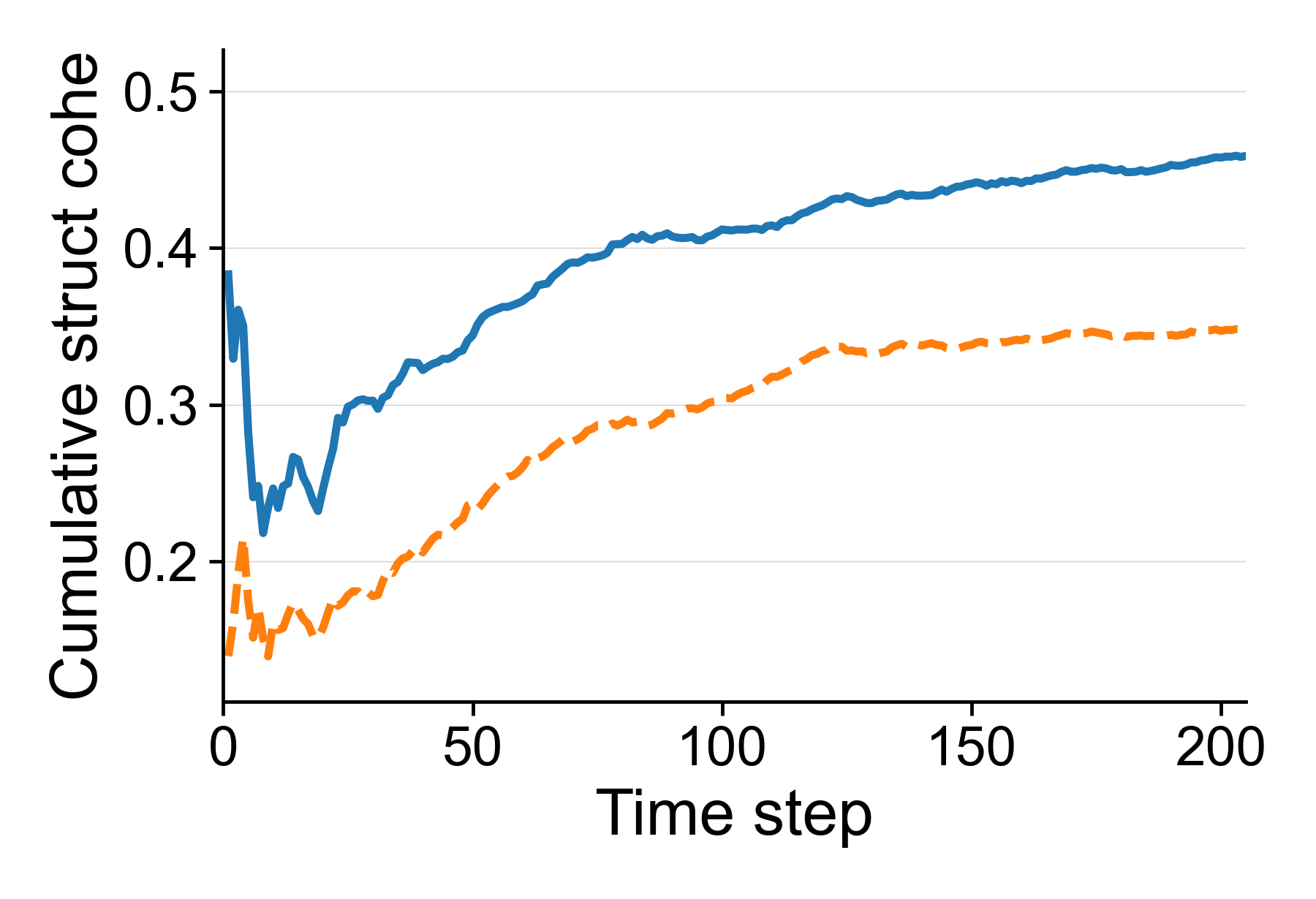}
        \caption{Random, structural coherence ($\uparrow$)}
    \end{subfigure}
    \caption{Utility of the raw and transformed scores over time on pancreas tissue. Rows correspond to the batch-release schemes, while columns correspond to the metrics. Higher skipped ratio and structural
    coherence and smaller prediction sets indicate better utility (indicated by the arrows in panel captions).}
    \label{fig:utility_pancreas_global}
\end{figure}

Cells that are not skipped receive a prediction set of candidate labels, whose value depends on how easily it can be resolved later. We evaluate the utility-aware score transformation introduced in
Section~\ref{subsec:utility_module} on the pancreas cohort where non-singleton candidate lists are common.  Figure~\ref{fig:utility_pancreas_global} compares the raw and
CARED-transformed scores in terms of skipped ratio, average prediction set size, and
structural coherence. Across all three release schemes, the utility-aware transformation consistently yields smaller candidate lists and improves structural coherence. It
also increases the skipped ratio under error-guided release while yielding
comparable skipped ratios under adjacent and random release. These results show
that the utility module reduces human effort while producing more coherent
candidate sets across different batch-release orders. Results for lung and ovary are reported in Figure~\ref{fig:appendix_lung_ovary_utility} in Appendix~\ref{subsec:more_experimental_results}: the utility-aware transformation yields pronounced improvements for ovary, while its benefits are limited in lung, where the skipped ratio is already high and most candidate lists are already coherent. Sensitivity of the utility metrics to the utility learning rate, random seed, and target error level is further discussed in Figure~\ref{fig:appendix_sensitivity_utility_metrics} in Appendix~\ref{subsubsec:sensitivity_utility_metrics}.

\subsection{Results with real expert annotations}
\label{subsec:real_stream}
We examine CARED on the colon data described in Section~\ref{subsec:exp_set}, where both the order of the 37 annotation rounds and the $1{,}374$ labels come from an expert navigating the slide. The distribution drift is generated by real practice but the dataset is also small: with $p=0.1$, CARED observes about $137$ audited labels in total, about $4$ per round. 
To avoid extremely scarce cell types which make error evaluation challenging, we group the cells are grouped into three classes: tumor, muscle, and others,
with \emph{others} combining several rare cell types. We run CARED with target error level
$\alpha=0.1$ using the global, local, and gated NuClass scores.

\begin{figure}[!htbp]
    \centering
    \begin{subfigure}[t]{0.48\linewidth}
        \centering
        \includegraphics[width=\linewidth]{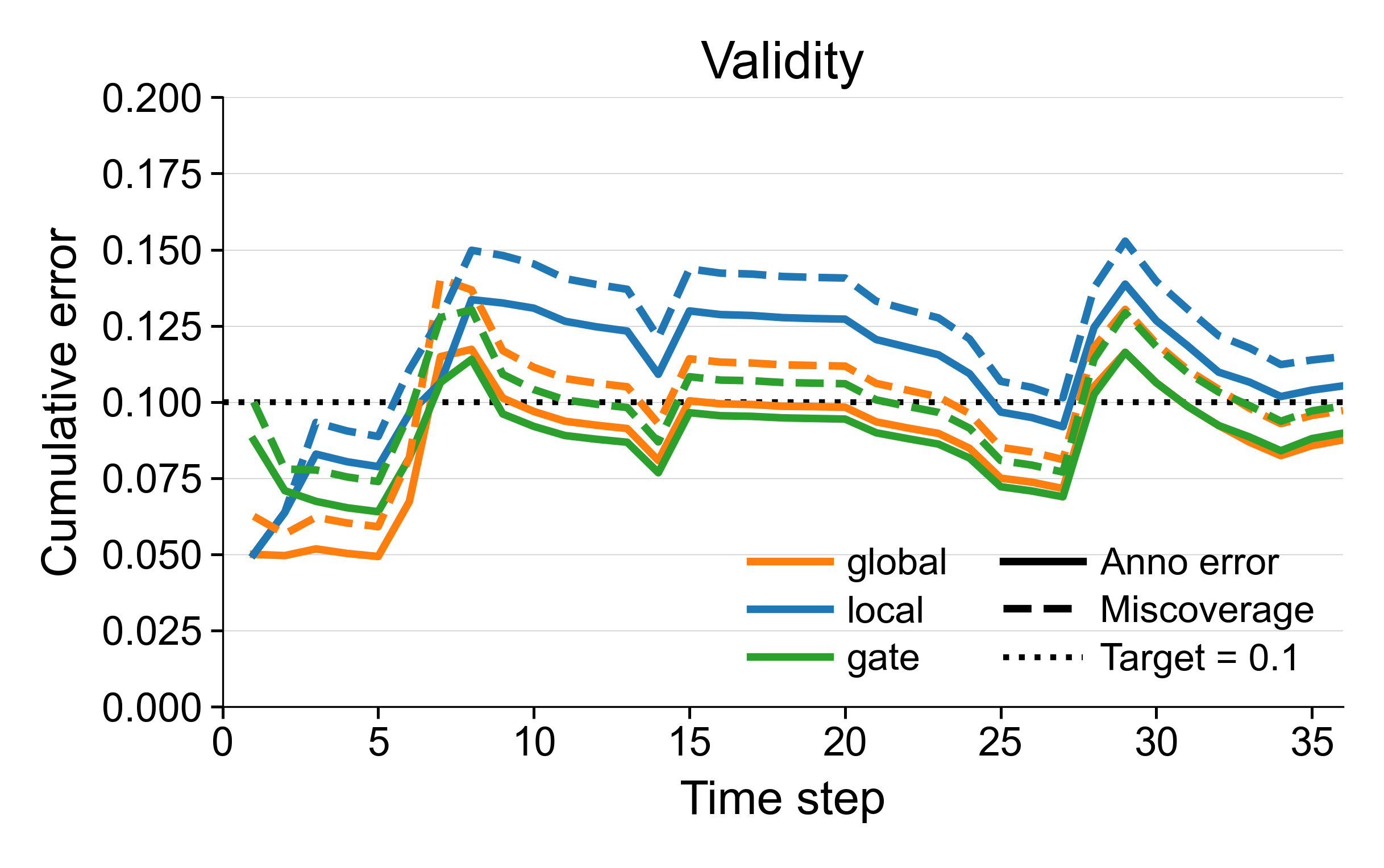}
        \caption{Cumulative annotation error and miscoverage}
        \label{fig:real_annotation_marginal_error}
    \end{subfigure}
    \hfill
    \begin{subfigure}[t]{0.48\linewidth}
        \centering
        \includegraphics[width=\linewidth]{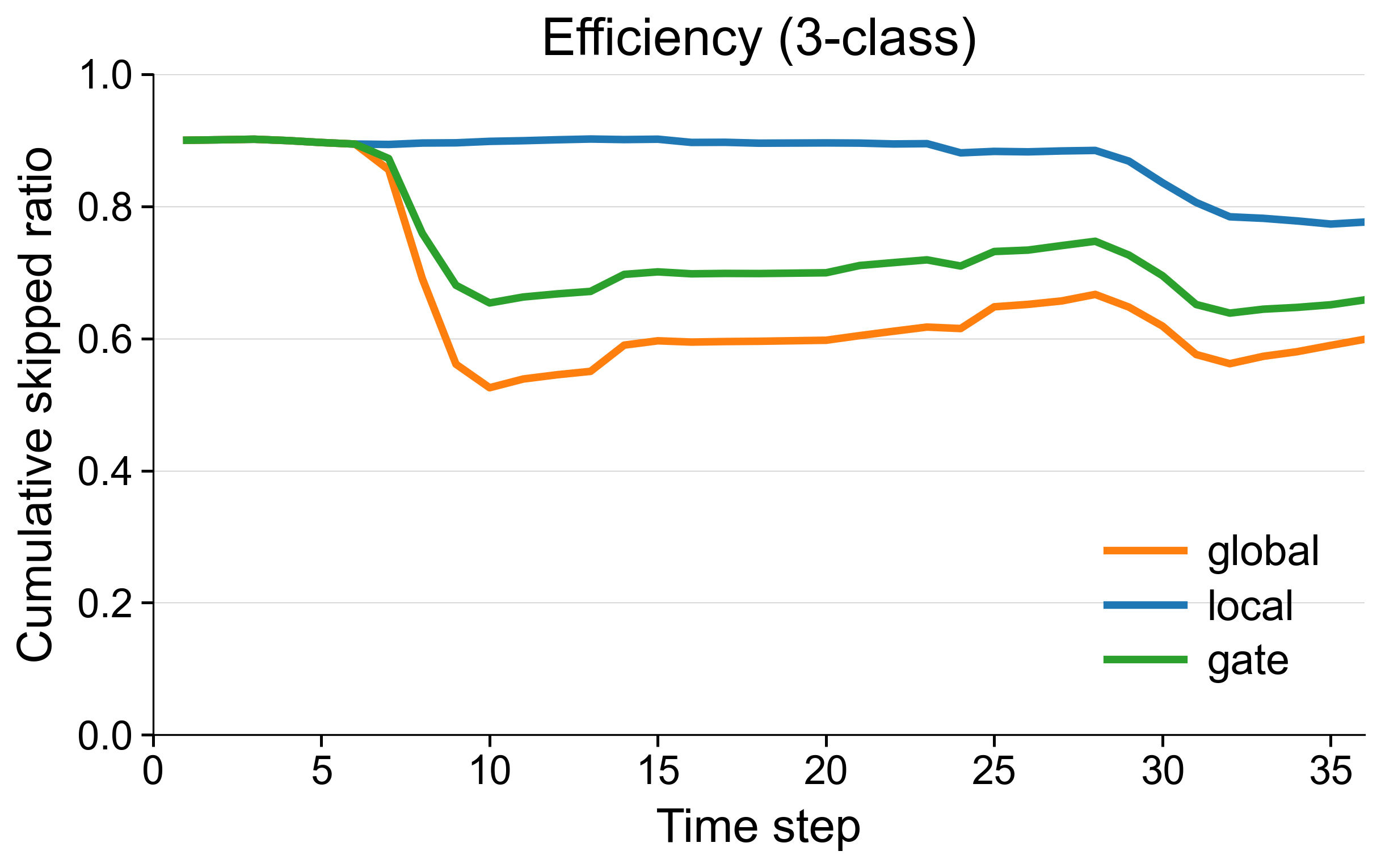}
        \caption{Cumulative saved annotation cost}
        \label{fig:real_annotation_efficiency}
    \end{subfigure}
    \caption{Results with real expert annotations on the colon dataset.
    (a) Cumulative annotation error (solid curves) and cumulative miscoverage
    (dashed curves) under the global, local, and gated NuClass score sources,
    with reviewed fraction $p=0.1$ and target error level $\alpha=0.1$; the
    black dotted horizontal line marks this target. (b) Cumulative skipped
    ratio under the same setting.}
    \label{fig:real_annotation_results}
\end{figure}

Figure~\ref{fig:real_annotation_results}(a) shows the cumulative
annotation-error (solid) and miscoverage (dashed) trajectories compared with the expert annotations used in evaluation. Across the three NuClass scores, the cumulative annotation error and miscoverage remain close to the target level, showing the robust performance of CARED in real data drifts.
Figure~\ref{fig:real_annotation_results}(b) reports the 
cumulative saved annotation cost, indicating that the proposed framework remains effective under distribution drifts created by real expert annotations.

\subsection{Sensitivity analysis on cumulative error}\label{subsec:sensitivity_cumulative_error}

\begin{figure}[!htbp]
    \centering
    \setlength{\parskip}{0pt}
    \captionsetup[subfigure]{font=scriptsize,justification=centering}
    \begin{subfigure}[t]{0.29\linewidth}
        \centering
        \includegraphics[width=\linewidth]{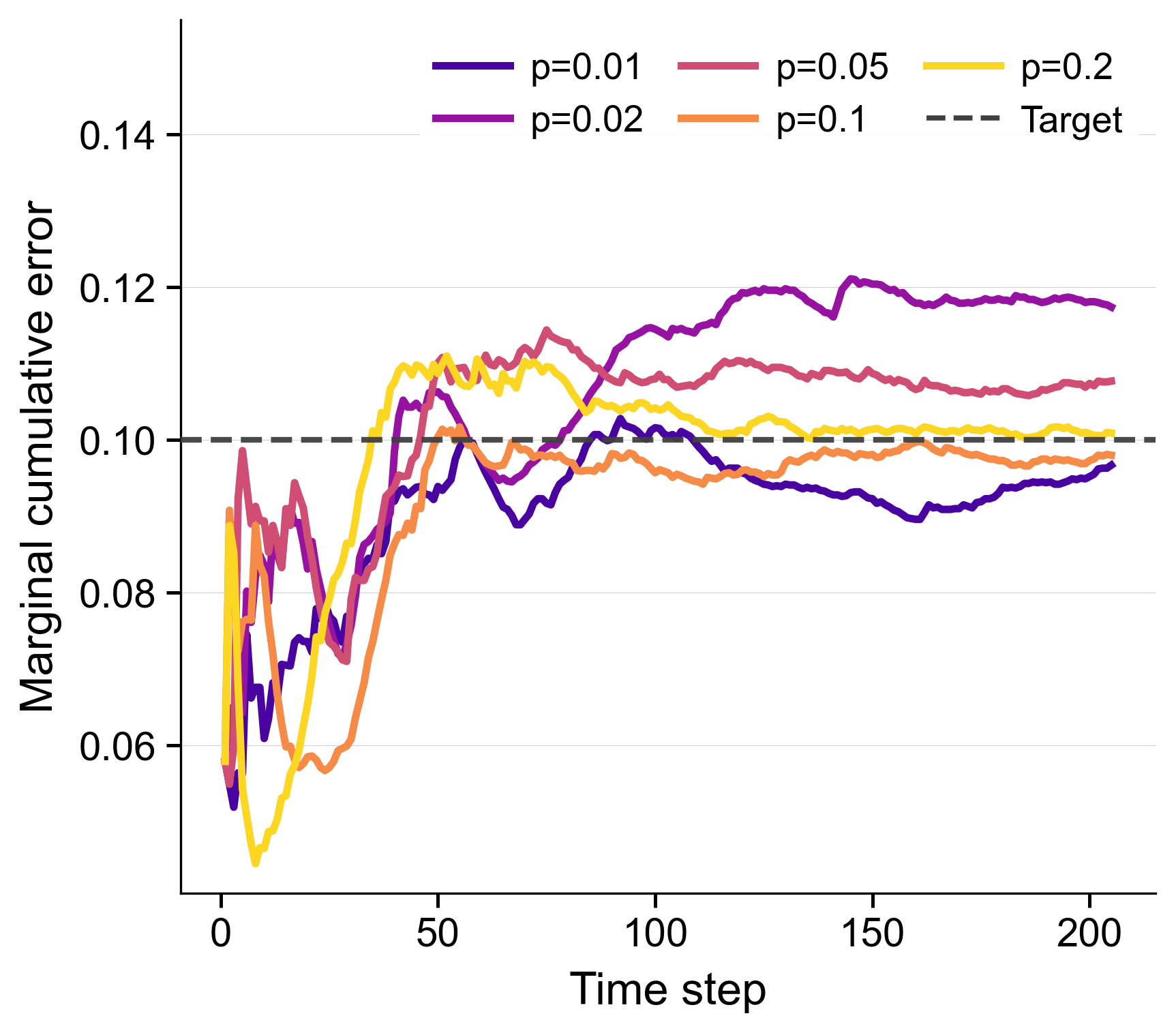}
        \caption{Error-guided, across reviewed fractions}
        \label{fig:sensitivity_cumulative_error_errorg_p}
    \end{subfigure}
    \hfill
    \begin{subfigure}[t]{0.29\linewidth}
        \centering
        \includegraphics[width=\linewidth]{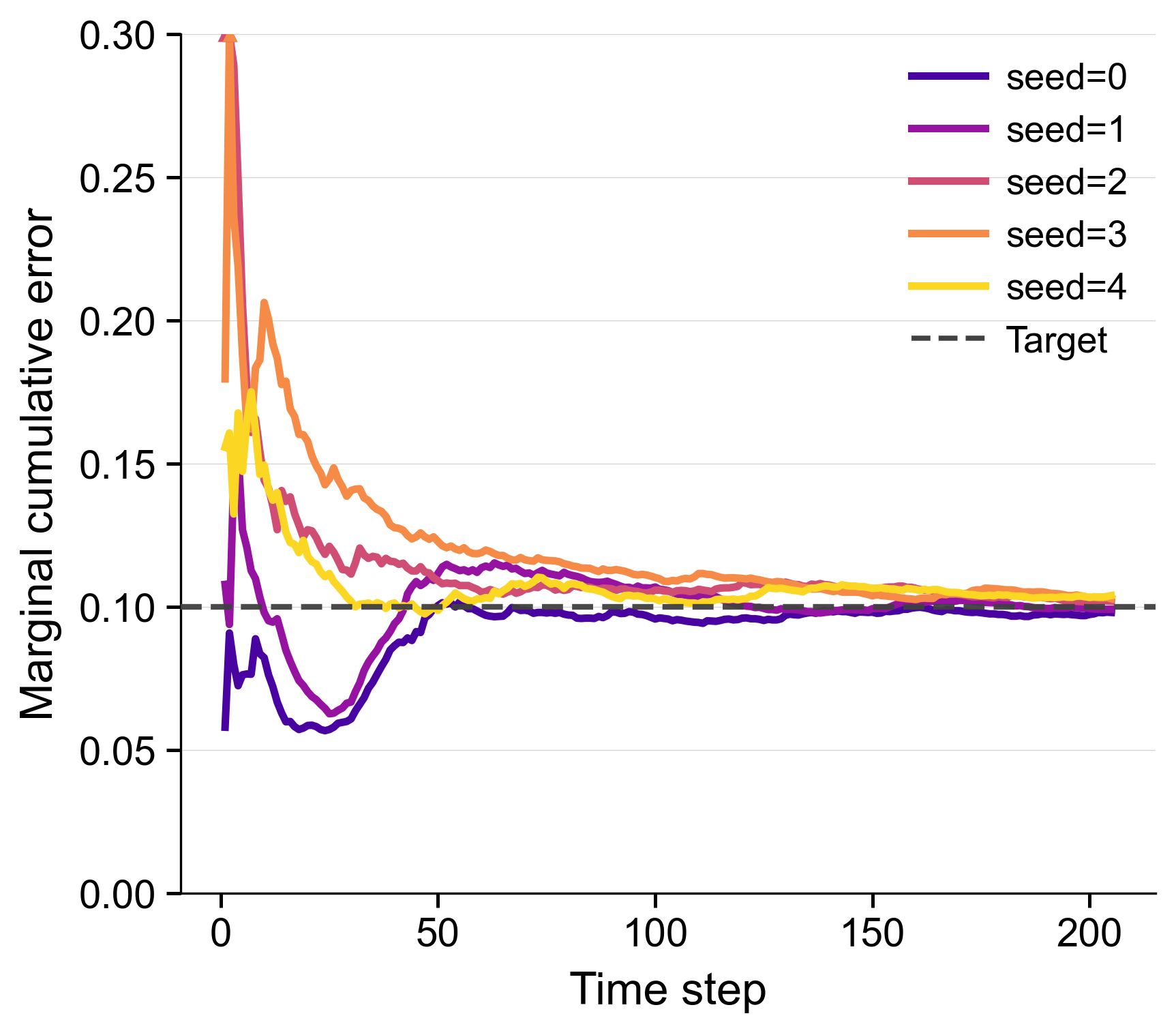}
        \caption{Error-guided, across seeds}
        \label{fig:sensitivity_cumulative_error_errorg_seed}
    \end{subfigure}
    \hfill
    \begin{subfigure}[t]{0.29\linewidth}
        \centering
        \includegraphics[width=\linewidth]{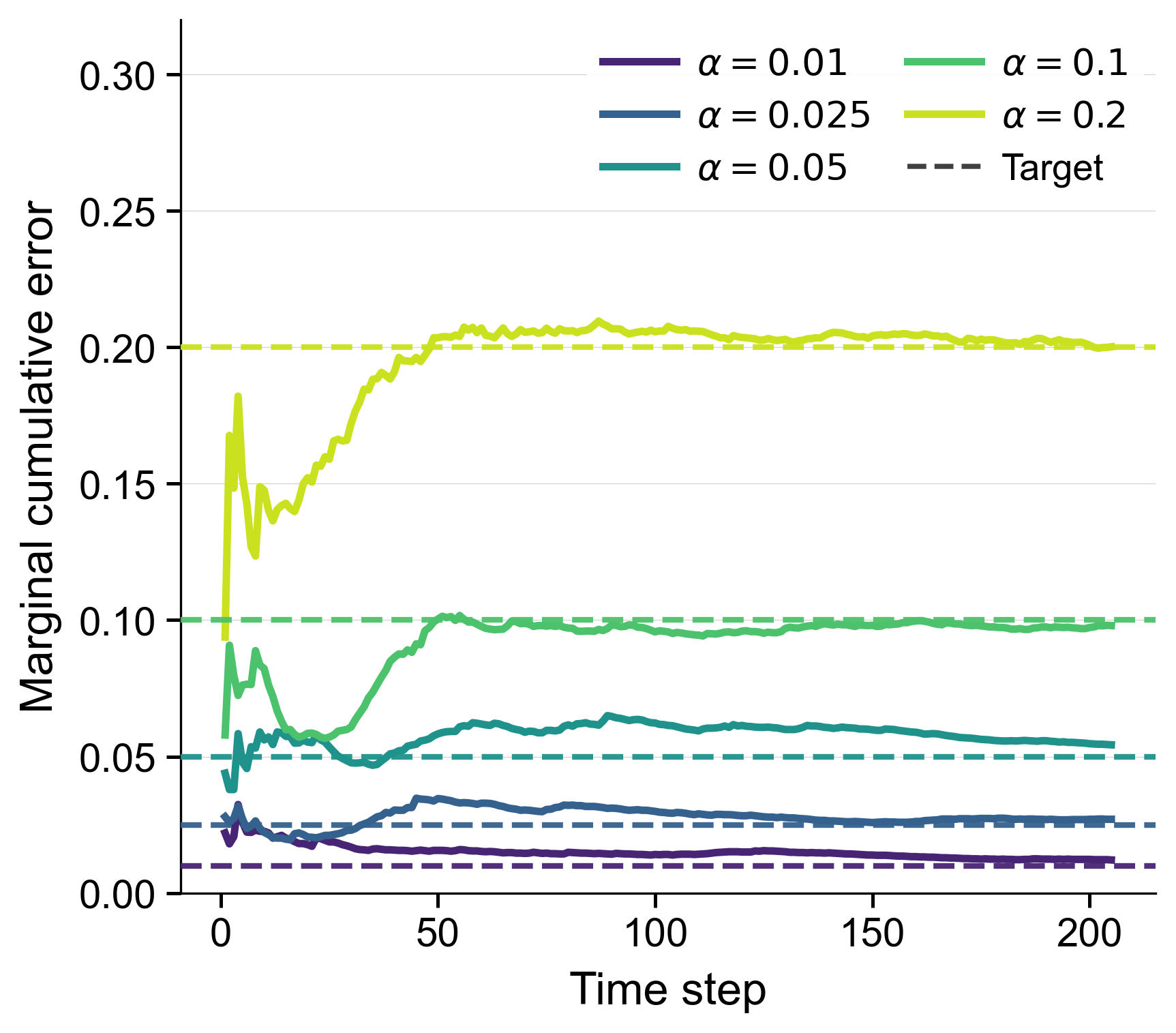}
        \caption{Error-guided, across target levels}
        \label{fig:sensitivity_cumulative_error_errorg_alpha}
    \end{subfigure}
    
    \vspace{1pt}
    
    \begin{subfigure}[t]{0.29\linewidth}
        \centering
        \includegraphics[width=\linewidth]{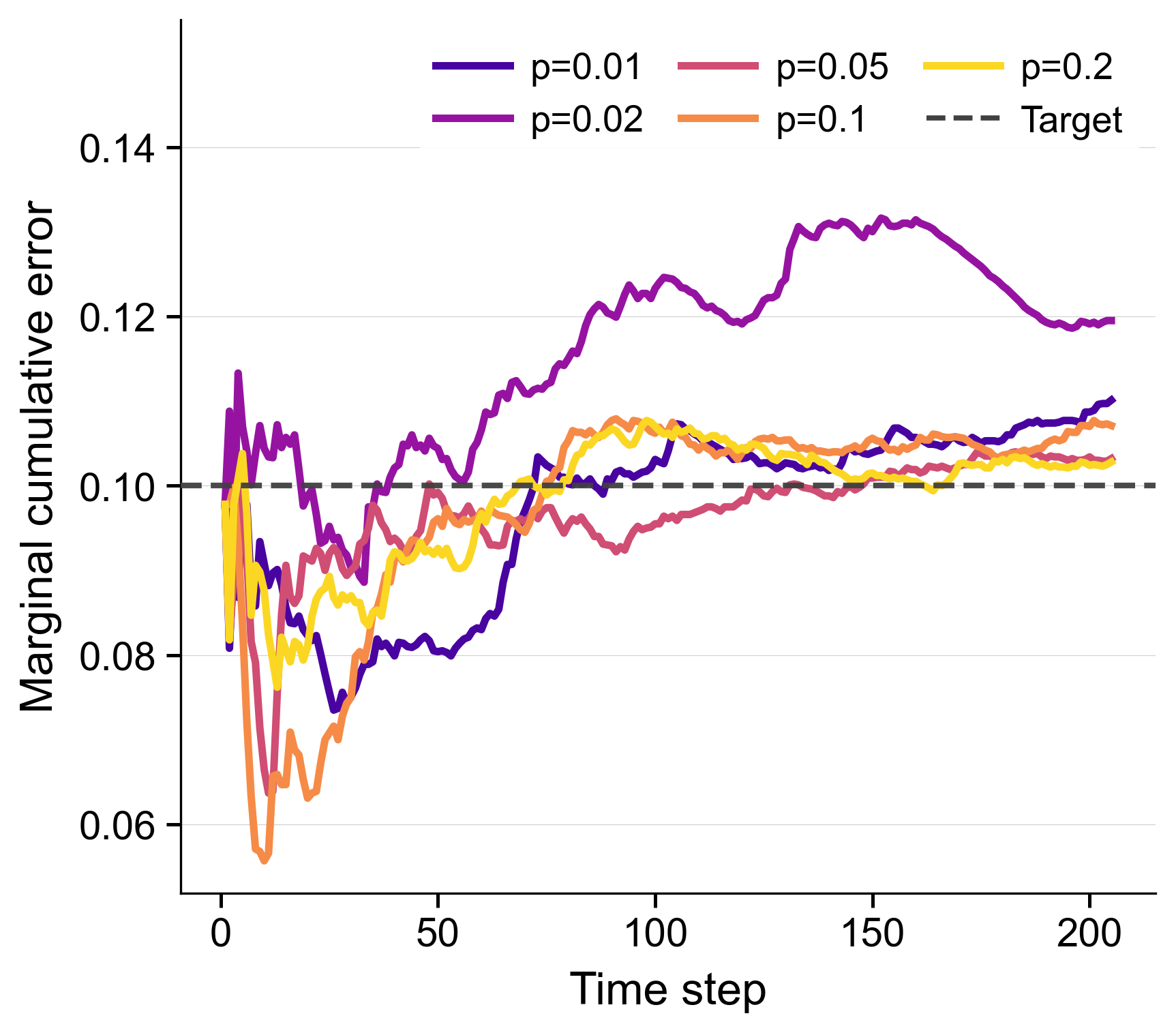}
        \caption{Adjacent, across reviewed fractions}
        \label{fig:sensitivity_cumulative_error_adjacent_p}
    \end{subfigure}
    \hfill
    \begin{subfigure}[t]{0.29\linewidth}
        \centering
        \includegraphics[width=\linewidth]{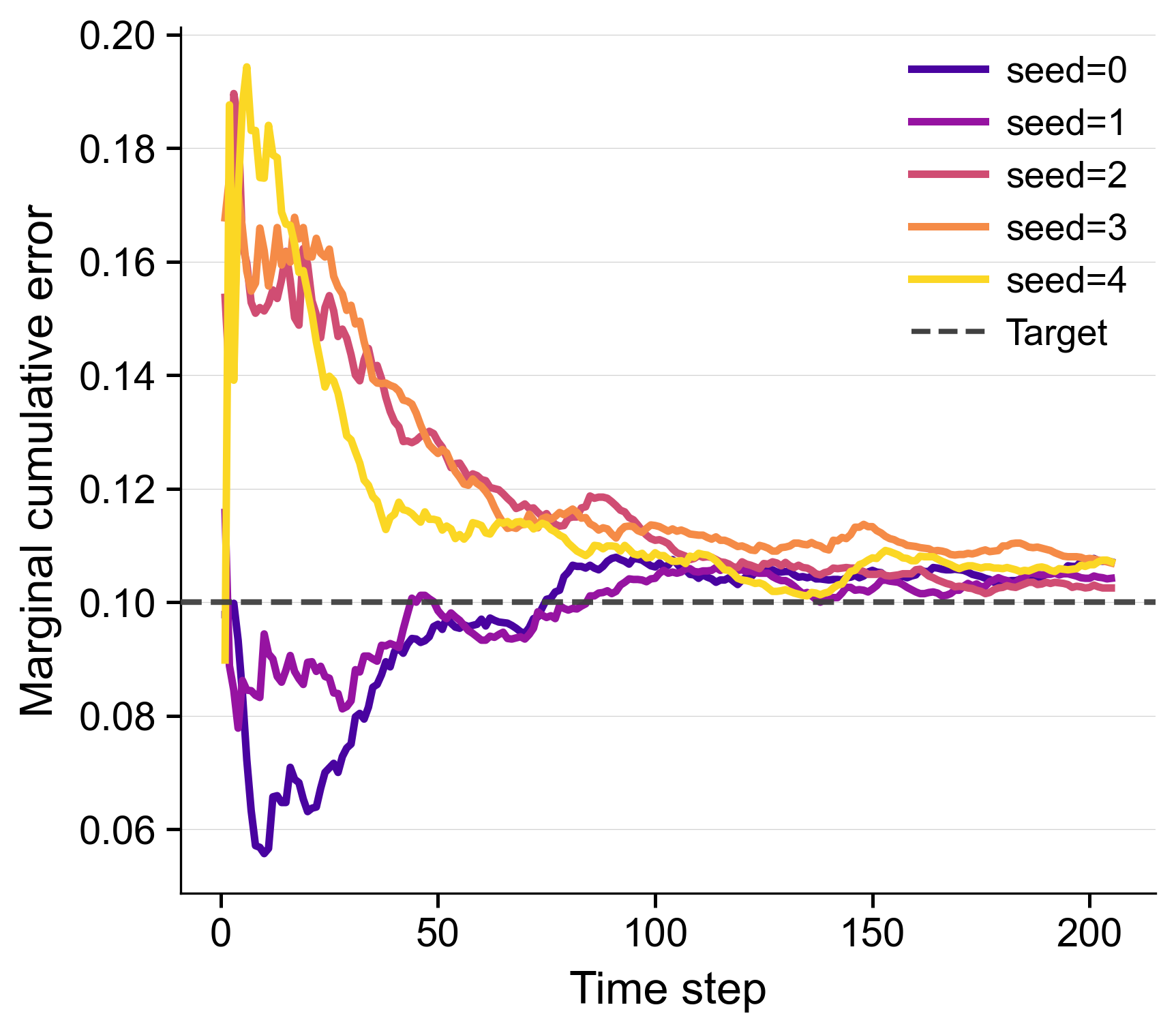}
        \caption{Adjacent, across seeds}
        \label{fig:sensitivity_cumulative_error_adjacent_seed}
    \end{subfigure}
    \hfill
    \begin{subfigure}[t]{0.29\linewidth}
        \centering
        \includegraphics[width=\linewidth]{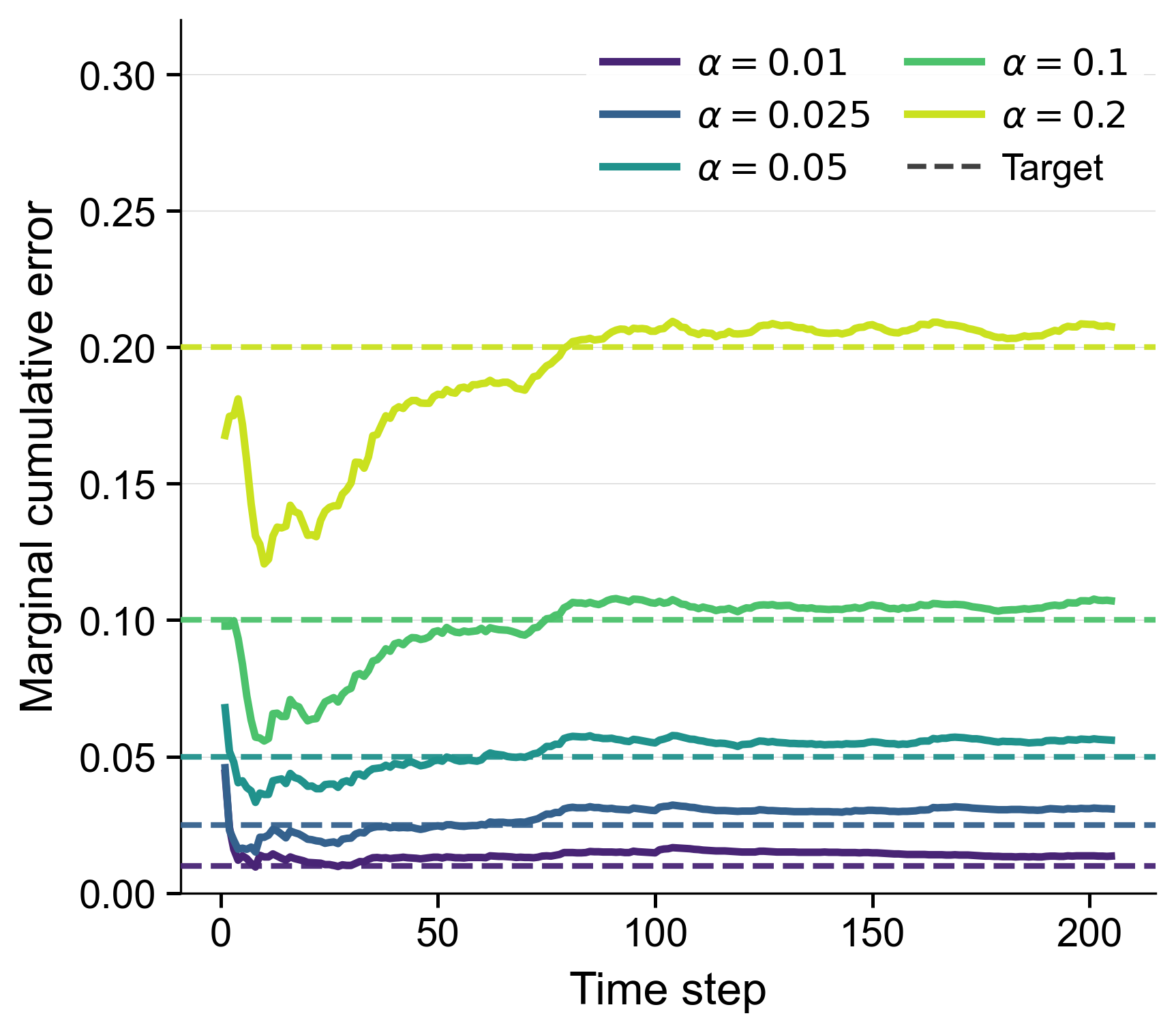}
        \caption{Adjacent, across target levels}
        \label{fig:sensitivity_cumulative_error_adjacent_alpha}
    \end{subfigure}
    
    \vspace{1pt}
    
    \begin{subfigure}[t]{0.29\linewidth}
        \centering
        \includegraphics[width=\linewidth]{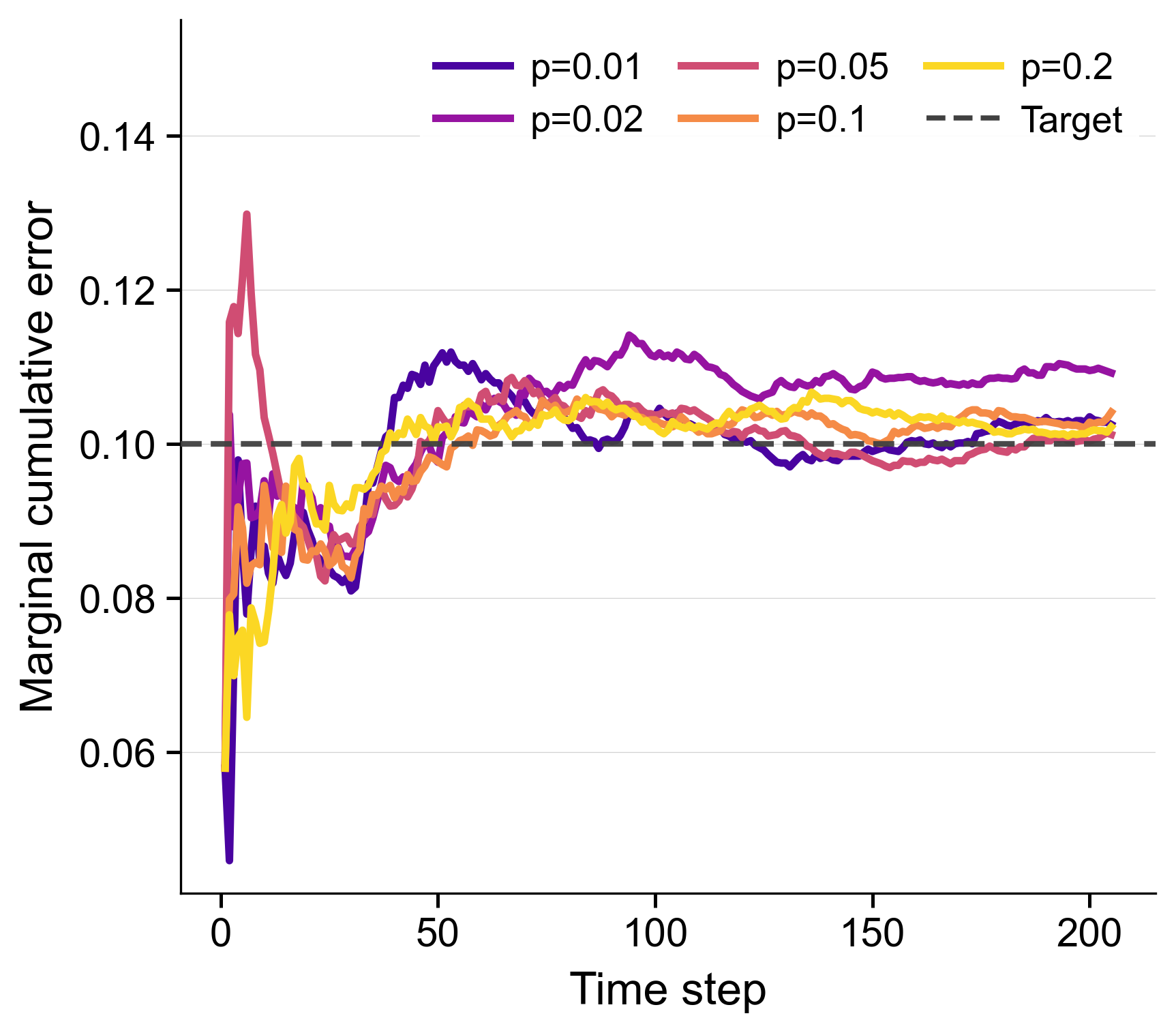}
        \caption{Random, across reviewed fractions}
        \label{fig:sensitivity_cumulative_error_p}
    \end{subfigure}
    \hfill
    \begin{subfigure}[t]{0.29\linewidth}
        \centering
        \includegraphics[width=\linewidth]{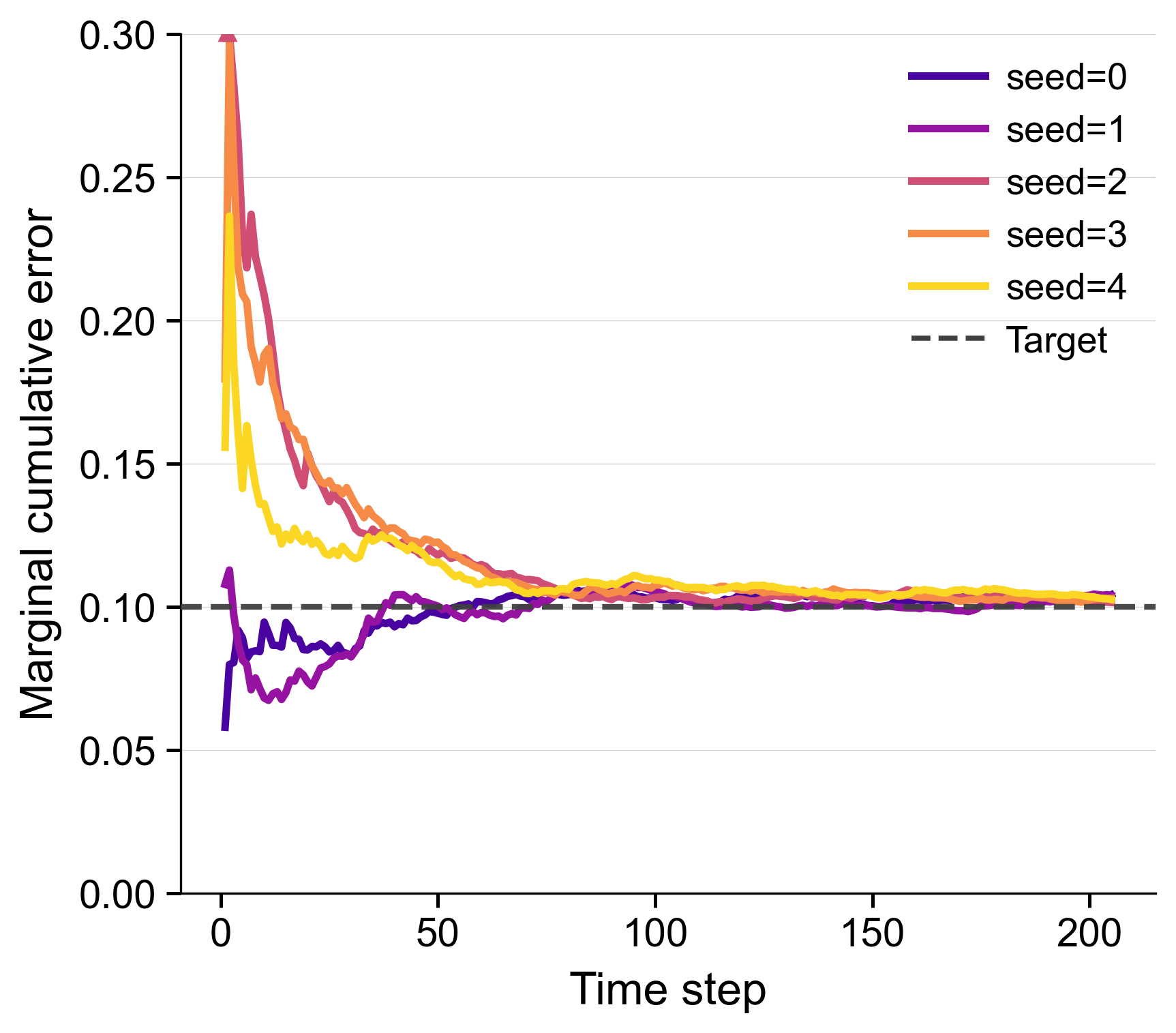}
        \caption{Random, across seeds}
        \label{fig:sensitivity_cumulative_error_seed}
    \end{subfigure}
    \hfill
    \begin{subfigure}[t]{0.29\linewidth}
        \centering
        \includegraphics[width=\linewidth]{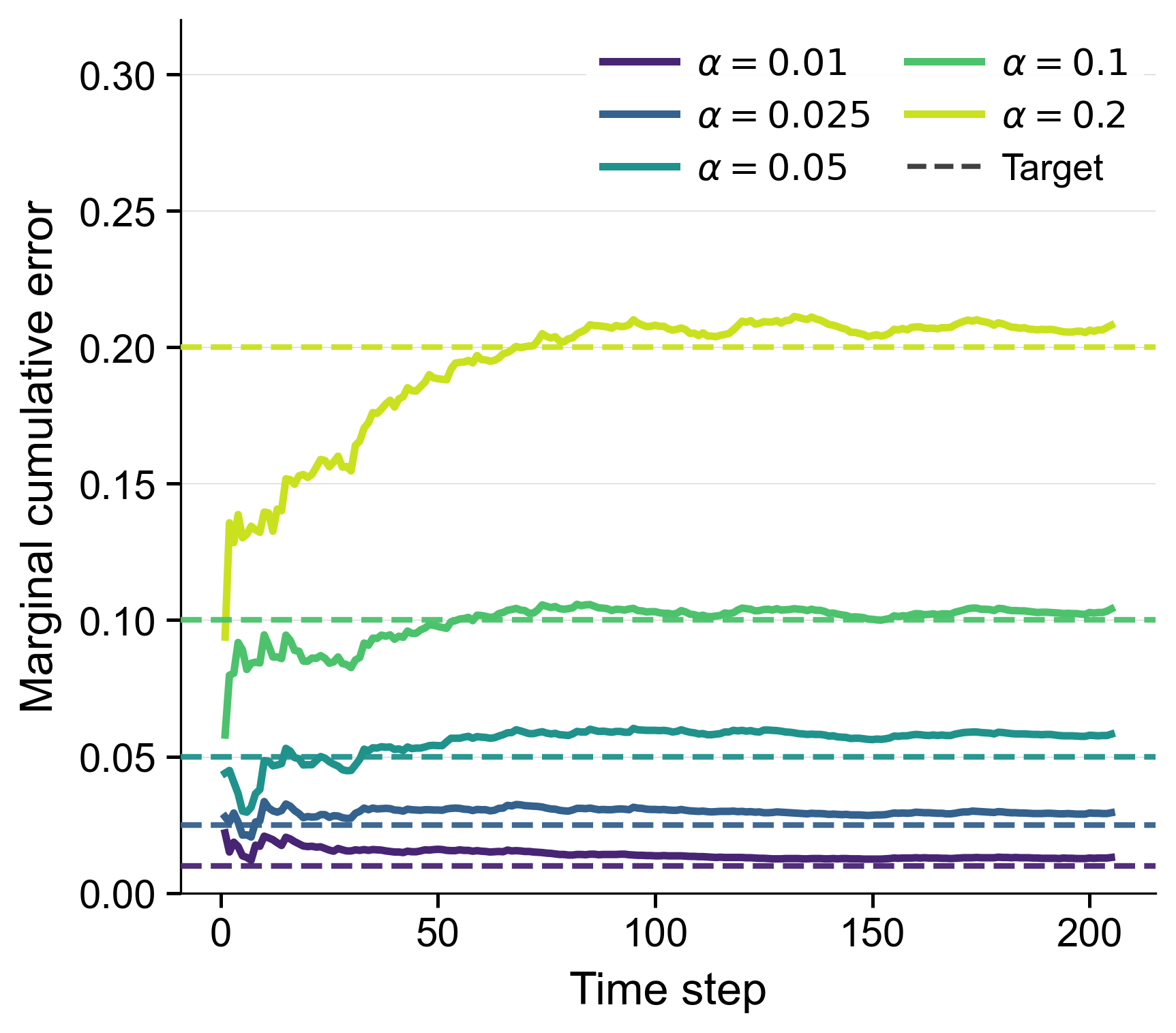}
        \caption{Random, across target levels}
        \label{fig:sensitivity_cumulative_error_alpha}
    \end{subfigure}
    \caption{Sensitivity analysis of marginal cumulative error on
    pancreas tissue with the global NuClass score source. Rows correspond to the
    error-guided, adjacent, and random batch release schemes, while columns
    vary the reviewed fraction $p$, the random seed, and the target error level
    $\alpha$. Dashed horizontal lines indicate the target error level used in
    each setting. In each column, only the parameter named in the column
    heading is varied; the other two retain their default values, $p=0.1$,
    seed $0$, and $\alpha=0.1$, used in
    Sections~\ref{subsec:cumulative_coverage_results} and~\ref{subsec:additional_experiment_results}.}
    \label{fig:sensitivity_cumulative_error}
\end{figure}

We next assess the sensitivity of CARED to the choice of reviewed fraction $p$, random seed, and 
target error level $\alpha$, using the pancreas cohort with the global NuClass
score under the three batch-release
schemes described in Section~\ref{subsec:exp_set}. 
Figure~\ref{fig:sensitivity_cumulative_error} summarizes the cumulative coverage error (which upper bounds the annotation error) trajectories as we vary the review fraction $p\in \{0.01, 0.02, 0.05, 0.1, 0.2\}$, random seeds, and the  target level $\alpha\in \{0.01,0.025,0.05,0.1,0.2\}$. 

Across all settings, the marginal coverage error moves towards the target level. Even with as few as a fraction of $p=0.01$ of reviewed cases, the trajectories can still stabilize and approach the target level. 
Reducing the review rate does slow the convergence: the audit-sampling term in the proof of Proposition~\ref{prop:convergence_rates} scales a $\{(1-p)/(pN)\}^{1/2}$, where $N$ is the total number of cells. The online-update term in Proposition~\ref{prop:convergence_rates} scales as $1/(\gamma_k pN_k)$, so a smaller $p$ shrinks the effective step size and the levels adapt more slowly. The corresponding class-wise results are reported in Appendix~\ref{subsec:additional_sensitivity_analyses} (Figure~\ref{fig:appendix_sensitivity_cumulative_error}). They show the same overall pattern, with larger finite-sample fluctuations only for rare cell types such as endocrine cells, consistent with the class-frequency dependence in Proposition~\ref{prop:convergence_rates}; the marginal trajectories remain stable.

CARED converges as expected under multiple seeds, which determine the initial region and the audit samples. Across seeds (middle column in Figure~\ref{fig:sensitivity_cumulative_error}), trajectories differ in the early stage and approach the target level from below or above depending on the starting region, but end close to $\alpha$ under every scheme. 

Appendix~\ref{sec:discussions} (Figure~\ref{fig:appendix_stepsize_sensitivity}) further examines sensitivity to the scale of the fixed class-wise step sizes. On the pancreas task with the global NuClass score, scaling the default step sizes by factors from $0.1$ to $10$ consistently pulls the cumulative class-wise errors toward the target level $\alpha=0.1$, demonstrating stable calibration over a broad range of step-size scales.

\section*{Discussion}

We proposed CARED, which decides which AI-generated cell-type labels can be accepted without expert review while the annotation workflow drifts. Its validity is guaranteed by the random audit design alone, which yields long-run class-conditional control of miscoverage, and hence of annotation error, over both the full stream and the unreviewed cells, without exchangeability or stationarity conditions. On whole-slide images from pancreas, lung, and ovary tissues and on a real expert annotation stream, it tightly controls the error within cell types while allowing up to 70\% of cells to bypass expert review. It builds prediction sets for difficult cases whose AI annotations are not accurate enough for later resolution. Its performance is robust to the choice of the review fraction and random seeds. 
Overall, CARED offers a perspective on how a small amount of human feedback can calibrate uncertainty quantification and yield effective and safe decisions. 

Several directions remain open. First, our guarantee is asymptotic, and finite-sample confidence bounds for the realized error rate will be useful. Second, our error control is over the entire stream instead of the cases selected for automated annotation; the random audit design may be extended to calibrate the selective errors. Finally, we focus primarily on retrospective evaluation, and a prospective deployment in real-time annotation tasks could further quantify the saved benefits and impact on downstream analysis and decisions. 

\FloatBarrier
\bibliographystyle{apalike}
\bibliography{reference}

\clearpage
\appendix

\section{Details on utility module}\label{sec:utility_details}
\setcounter{equation}{0}
\renewcommand{\theequation}{\thesection.\arabic{equation}}
\setcounter{algorithm}{0}
\renewcommand{\thealgorithm}{\thesection.\arabic{algorithm}}

This appendix gives further details on the optional utility module introduced in Section~\ref{subsec:utility_module}: the online selection algorithm, the structural-coherence metric, and the construction of the score transformation $f_\Theta$.

\FloatBarrier

\subsection{Structural coherence and surrogate loss}\label{subsec:utility_structure_surrogate}

\paragraph{Structural coherence.} The structural-coherence metric captures coarse organization in the label
space. Let $\{\mathcal S_1,\ldots,\mathcal S_M\}$ be a partition of $[K]$
into $M$ subtrees, where labels in the same subtree are structurally close
(for example, by lineage, function, or morphology). Consider a map
$g:[K]\to[M]$ that assigns each class to its subtree. For case $i$ and
subtree $m$, define
\[
  \mathrm{Occ}_{i,m}
  =
  \max_{k:g(k)=m}
  \ind\{k\in \widehat C_{t,i}(\Theta)\},
  \qquad
  \mathrm{Full}_{i,m}
  =
  \min_{k:g(k)=m}
  \ind\{k\in \widehat C_{t,i}(\Theta)\},
\]
and
\[
  \mathrm{Out}_{i,m}
  =
  \max_{m'\neq m}\mathrm{Occ}_{i,m'}.
\]
Here $\mathrm{Occ}_{i,m}$ indicates whether subtree $m$ occurs in the
prediction set, $\mathrm{Full}_{i,m}$ indicates whether the entire subtree is
included, and $\mathrm{Out}_{i,m}$ indicates whether the set also contains a
label outside subtree $m$. We call a prediction set structurally coherent if
\begin{equation}\label{eq:struct_violation}
  V_i(\Theta)
  :=
  \sum_{m=1}^M
  \mathrm{Occ}_{i,m}
  (1-\mathrm{Full}_{i,m})
  \mathrm{Out}_{i,m}
  =
  0.
\end{equation}
This criterion penalizes a set that crosses subtree boundaries while including
only a partial fragment of at least one touched subtree. It permits a small
set contained in one subtree and a set formed by several complete subtrees,
both of which represent uncertainty coherently. The batch-level score in
\eqref{eq:hard_obj} is
\begin{equation}\label{eq:j-struct}
  \mathcal J_{\mathrm{struct}}(\Theta)
  =
  \frac{1}{B_t}
  \sum_{i=1}^{B_t}
  \ind\{V_i(\Theta)=0\}.
\end{equation}

The rule in \eqref{eq:struct_violation} is one admissible structural
preference, not a required component of the conformal framework. Other
domain-specific preferences, such as homogeneity, diversity, or a deeper
hierarchy, can replace the structural component in the hard utility objeAsctive
and its corresponding soft relaxation, while leaving the adaptive conformal
framework in Section~\ref{subsec:online_conf} unchanged.

\paragraph{Surrogate loss.} The hard utility objective in \eqref{eq:hard_obj} contains three operations that are not directly amenable to gradient-based updates: the membership indicators $\ind\{k\in\widehat C_{t,i}(\Theta)\}$, the set-size indicator $\ind\{|\widehat C_{t,i}(\Theta)|\le 1\}$ in $J_{\mathrm{skip}}(\Theta)$, and the structural-coherence indicator $\ind\{V_i(\Theta)=0\}$ in $J_{\mathrm{struct}}(\Theta)$. We construct the surrogate loss $L_{\mathrm{soft}}(\Theta)$ by replacing the operations above in \eqref{eq:hard_obj} as follows.
\begin{itemize}
  \item \textbf{Membership indicators.} Given the class-wise conformal thresholds \(\{q_{k,t}\}_{k=1}^K\) computed under parameter \(\Theta\),
    \(\ind\{k\in\widehat C_{t,i}(\Theta)\}
    =
    \ind\{f_\Theta(s_t(x_{t,i}))_k\ge q_{k,t}\}\)
    is replaced by
    \[z_{i,k}(\Theta)
    =
    \sigma\!\left(
    \frac{f_\Theta(s_t(x_{t,i}))_k-q_{k,t}}
    {T_{\mathrm{temp}}}
    \right),\]
    where \(T_{\mathrm{temp}}\) controls the smoothness of the approximation.
  \item  \textbf{Set-size indicator.}
    We interpret \(z_{i,k}(\Theta)\) as the soft inclusion probability of class \(k\). Then
    \(
\prod_{k=1}^K \bigl(1-z_{i,k}(\Theta)\bigr)
\)
corresponds to including no class, while
\(
z_{i,k}(\Theta)
\prod_{\ell\ne k}
\bigl(1-z_{i,\ell}(\Theta)\bigr)
\)
corresponds to including class \(k\) and no other class. Summing the latter
over \(k\) therefore corresponds to including exactly one class.
    Therefore, the indicator \(\ind\{|\widehat C_{t,i}(\Theta)|\le 1\}\) is replaced by
    \[
    \prod_{k=1}^K\bigl(1-z_{i,k}(\Theta)\bigr)
    +
    \sum_{k=1}^K
    z_{i,k}(\Theta)
    \prod_{\ell\ne k}
    \bigl(1-z_{i,\ell}(\Theta)\bigr).
    \]
  \item \textbf{Structural-coherence indicator.}
In \eqref{eq:struct_violation}, we replace the membership indicators in
\(\operatorname{Occ}_{i,m}\), \(\operatorname{Full}_{i,m}\), and
\(\operatorname{Out}_{i,m}\) by \(z_{i,k}(\Theta)\), while retaining the
max and min operations. Suppose this yields a surrogate $\widehat V_i(\Theta)$, which reflects the extent of structural violation. We therefore replace \(\ind\{V_i(\Theta)=0\}\) by one minus this expression, so that larger values
continue to represent greater structural coherence. The retained max and min operations are differentiable almost everywhere and admit subgradients at ties, so they can be handled directly in the gradient updates. 
\end{itemize}

\subsection{Online selection algorithm}\label{subsec:utility_online_selection}

Algorithm~\ref{alg:online_selection} gives the online selection procedure sketched in Section~\ref{subsec:utility_module}. It selects the score-transformation parameter $\Theta_t$ before the audit set is sampled. 

\begin{algorithm}[htbp]
\caption{Online training of $\Theta_t$ at time step $t$}\label{alg:online_selection}
\begin{algorithmic}[1]
\REQUIRE Batch $D_t$ with raw score vectors $\{s_t(x):x\in D_t\}$; calibration sets
$\{D_{k,\mathrm{cal}}(t)\}_{k=1}^K$ (denoted by
$D_{\mathrm{cal}}(t)$); current online levels
$\{\alpha_{k,t}\}_{k=1}^K$; candidate
number $N$; learning rate $\eta$.
\ENSURE Selected parameter $\Theta_t$ and score transformation $f_{\Theta_t}$.
\STATE $\mathcal J^* \leftarrow \mathcal J(0)$; $\Theta \leftarrow 0$; $\Theta^* \leftarrow 0$.
\FOR{$n=1,\ldots,N$}
  \STATE Compute class-wise conformal quantiles $\{q_{k,t}\}_{k=1}^K$ on $D_{\mathrm{cal}}(t)$ using $\Theta$
  \STATE $\Theta \leftarrow \Theta-\eta\,\nabla_{\Theta}\mathcal L_{\mathrm{soft}}(D_t;\Theta,\{q_{k,t}\}_{k=1}^K)$ \COMMENT{gradient step}
  \STATE Using the updated $\Theta$, recompute $\{q_{k,t}\}_{k=1}^K$ according to Equation~\eqref{eq:quantile}, construct prediction sets on $D_t$ according to Equation~\eqref{eq:set_cons}, and evaluate the utility objective $\mathcal J(\Theta)$
  \IF{$\mathcal J(\Theta)<\mathcal J^*$}
    \STATE $\mathcal J^*\leftarrow \mathcal J(\Theta)$;\quad $\Theta^*\leftarrow \Theta$
  \ENDIF
\ENDFOR
\STATE Set $\Theta_t\leftarrow\Theta^*$.
\end{algorithmic}
\end{algorithm}

\subsection{Score transformation construction}\label{subsec:utility_score_transform}
\begin{algorithm}[H]
\caption{Score transformation $f_\Theta$}\label{alg:score_update}
\begin{algorithmic}[1]
\REQUIRE Score vector $s\in[0,1]^K$; subtree partition $\{\mathcal S_1,\ldots,\mathcal S_M\}$ of $[K]$; parameter matrices $\Theta^{\mathrm{in}},\Theta^{\mathrm{out}}\in\mathbb R^{K\times M}$.
\ENSURE Transformed score vector $\tilde s=f_\Theta(s)\in\mathbb R^K$.
\STATE $s_{\max}\leftarrow \max_k s_k$
\STATE $u_k\leftarrow \operatorname{logit}(s_k/s_{\max})$ for all $k$
\STATE $\mathcal U\leftarrow [K]$ \COMMENT{unprocessed classes}
\FOR{$r=1,\ldots,M$}
  \STATE $\ell_r\leftarrow \arg\max_{k\in\mathcal U}u_k$ \COMMENT{leader class}
  \STATE $m\leftarrow$ subtree index such that $\ell_r\in\mathcal S_m$
  \FOR{$k\in(\mathcal S_m\cap\mathcal U)\setminus\{\ell_r\}$}
    \STATE $u_k\leftarrow u_k+\Theta^{\mathrm{in}}_{k,r}$ \COMMENT{updates scores associated to the leader subtree}
  \ENDFOR
  \FOR{$j\in\mathcal U\setminus\mathcal S_m$}
    \STATE $u_j\leftarrow u_j-\Theta^{\mathrm{out}}_{j,r}$ \COMMENT{updates scores associated to other subtrees}
  \ENDFOR
  \STATE $\mathcal U\leftarrow \mathcal U\setminus\mathcal S_m$
\ENDFOR
\STATE $\tilde s_k\leftarrow s_{\max}\cdot\sigma(u_k)$ for all $k$ \COMMENT{$\tilde s_k\in(0,s_{\max})$}
\STATE \textbf{return} $\tilde s=f_\Theta(s)\in\mathbb R^K$
\end{algorithmic}
\end{algorithm}
Algorithm~\ref{alg:score_update} gives the construction of the score transformation $f_\Theta$ used in Section~\ref{subsec:utility_module}. Given a score vector $s$ and a parameter $\Theta$, it returns the transformed score $\tilde s=f_\Theta(s)$. The construction of $f_\Theta$ is guided by three principles. First, initializing $\Theta^{\mathrm{in}}$ and $\Theta^{\mathrm{out}}$ at zero should yield the identity map $f_{0}(s)=s$, so the unmodified score vector is always available as a baseline candidate. Second, every transformed entry should lie in $(0,s_{\max})\subset(0,1)$, preserving the score range required by the conformal construction. Finally, the transformation should respect the prescribed subtree structure, coordinating score changes within each subtree while allowing different subtrees to change differently.
For numerical stability, in implementation we lower-bound \(s_{\max}\) by \(10^{-12}\) and clip each normalized score \(s_k/s_{\max}\) to \([10^{-4},1-10^{-4}]\) before applying the logit transformation. These numerical safeguards avoid undefined values at the endpoints and are omitted from Algorithm~\ref{alg:score_update} for notational simplicity.


\section{Additional theoretical results}\label{sec:additional_theory}

\subsection{Unbiased randomized rounding}\label{subsec:rounding_scheme}

When the target reviewed fraction is fixed at $p$, the desired
audit budget $pB_t$ may not be an integer. The following randomized rounding
scheme implements this budget while preventing the resulting rounding error
from accumulating systematically over time, and yields the class-wise
asymptotic relations required in the proofs of Lemma~\ref{lem:class_review_fraction}
and Theorem~\ref{thm:long_run_validity};
its proof is deferred to Appendix~\ref{proof:unbiased_rounding}.

\begin{prop}[Unbiased randomized rounding]\label{prop:unbiased_rounding}
Fix $p\in(0,1)$. Let $\mathcal G'_t$ be the time-$t$ sigma-field generated by all
random quantities determined after $D_t$ is realized but before choosing $p_t$.
At time $t$, before $A_t$ is sampled, set
\[
q_t:=pB_t-\lfloor pB_t\rfloor
\]
and draw
$\xi_t\mid\mathcal G'_t\sim\operatorname{Bernoulli}(q_t)$. Now sample the audit set with size
\[
|A_t|:= \lfloor pB_t\rfloor+\xi_t,
\]
Under Assumption~\ref{ass:main_guarantee}(i)--(iii), the realized
reviewed fractions $p_t = |A_t|/B_t$ under this rule satisfy, for every class $k\in[K]$,
\[
\frac{\sum_{t=1}^T(p_t-p)|D_{k,t}|}
{\sum_{t=1}^T |D_{k,t}|}\to 0
\quad\text{a.s.},
\qquad
\frac{\sum_{t=1}^T(p_t-p)
\sum_{x\in D_{k,t}}\mathrm{err}_{k,t}(x)}
{\sum_{t=1}^T |A_{k,t}|}\to 0
\quad\text{a.s.}
\]
\end{prop}

\subsection{A cumulative batch-wise deviation guarantee}\label{subsec:sum_guarantee}

The guarantees in Section~\ref{sec:theory} control cumulative class-wise
errors, the relevant notion for long-run validity. In practice, one
also wants to understand how the procedure behaves locally across individual
batches \citep{gibbs2024conformal}. A natural way to
quantify such local behavior is to aggregate the absolute per-batch deviations
from the target error level. The result below gives an upper bound for this cumulative
local-deviation measure, and discusses its asymptotic properties as the time index $T$ and batch size $|D_{k,t}|$ get large. Note that in this subsection, the incoming batch $D_t$ may be
selected using review feedback observed before time $t$ (as in the error-guided scheme).

We first set up the problem. Fix a class $k$.
Here we consider only nonempty class-$k$ batches because we stipulate that
there is no deviation on empty sets. As a simplification, we assume that
$|D_{k,t}|>0$ for all $t$.

\paragraph{More notation.}
Write the batch-wise class-$k$ error rates as
\[
\mathrm{BatErr}_{k,t}^D
:=
\frac{\sum_{x\in D_{k,t}}\mathrm{err}_{k,t}(x)}{|D_{k,t}|},
\qquad
\mathrm{BatErr}_{k,t}^A
:=
\frac{\sum_{x\in A_{k,t}}\mathrm{err}_{k,t}(x)}{|A_{k,t}|},
\]
whenever $|A_{k,t}|>0$. If $|A_{k,t}|=0$, we define
$\mathrm{BatErr}_{k,t}^A:=\mathrm{BatErr}_{k,t}^D$.
The quantity of interest is the cumulative batch-wise deviation of the
all-cell error rate from the target level:
\[
\sum_{t=1}^T |\mathrm{BatErr}_{k,t}^D - \alpha|.
\]
Define the effective batch step size and the clipped online level by
\[
\eta_{k,t}:=\gamma_k|A_{k,t}|,
\qquad
\bar\alpha_{k,t}:=\operatorname{clip}_{[0,1]}(\alpha_{k,t}).
\]

\begin{assumption}
\label{ass:sum_guarantee_bounds}
Throughout this subsection, assume the following:
\begin{enumerate}[label=(\roman*), topsep=0pt, itemsep=0pt]
\item The lower bound of the batch size $B^-$ satisfies that
\[
B^- > \max\!\left\{\frac{1}{p},\frac{1}{1-p}\right\}.
\]
\item
For every class $k \in [K]$, the effective batch step size satisfies
\[
\eta_{k,t}\leq 1
\quad\text{for all }t.
\]
\item 
For every class $k \in [K]$ and time $t,$ the scores 
\[
\{ \tilde{s}_{k,t}(x) : x \in D_{k,\mathrm{cal}}(t) \cup D_{k,t} \}
\]
are distinct.
\end{enumerate}
\end{assumption}

The discussion on batch-wise deviation is sufficiently informative only when the batch sizes are not excessively small.
Assumption \ref{ass:sum_guarantee_bounds}(i) can be interpreted as "at least one cell per batch is chosen for online review on average", which is a plausible condition for large-scale settings. This condition leads to intermediate consequences
\begin{equation}\tag{B.1}\label{eq:batch-audit-bounds}
B^->2 \qquad \text{and} \qquad
0<p^-\le p_t\le p^+<1
\quad\text{for all }t,
\end{equation}
for some deterministic constants $p^-,p^+$. Specifically, the second consequence holds with
$p^-=p-\frac{1}{B^-}>0$ and $p^+=p+\frac{1}{B^-}<1.$ Assumption~\ref{ass:sum_guarantee_bounds}(ii) can be satisfied when the step size $\gamma_k$
is chosen reasonably small; a sufficient condition is
$\gamma_k<\frac{1}{B^+}$ for every class $k$.

\begin{definition}[Calibration discrepancy]\label{def:calibration_discrepancy}
For each time $t$ with $|D_{k,t}|>0$, define the calibration discrepancy as
\[
\Delta_{k,t}
:=
\sup_{u\in[0,1]}
\left|
\frac{\sum_{x\in D_{k,t}}\ind\{\tilde s_{k,t}(x)\le \hat q_{k,t}(u)\}}{|D_{k,t}|}
-
\frac{\sum_{x\in D_{k,\mathrm{cal}}(t)}\ind\{\tilde s_{k,t}(x)\le \hat q_{k,t}(u)\}}{|D_{k,\mathrm{cal}}(t)|}
\right|
+
\frac{1}{|D_{k,\mathrm{cal}}(t)|}.
\]
\end{definition}

For \(\alpha_{k,t}\in[0,1]\), the clipped \(\bar\alpha_{k,t}=\alpha_{k,t}\).
By the class-wise conformal construction in Section~\ref{subsec:online_conf}, a class-\(k\) error occurs when \(\widetilde s_{k,t}(x)<\widehat q_{k,t}(\bar\alpha_{k,t})\); Assumption~\ref{ass:sum_guarantee_bounds}(iii) ensures that no score in \(D_{k,t}\) equals this threshold, so
\[
\mathrm{BatErr}_{k,t}^{D}
=
\frac{\sum_{x\in D_{k,t}}
\mathbf 1\{\widetilde s_{k,t}(x)\le \widehat q_{k,t}(\bar\alpha_{k,t})\}}
{|D_{k,t}|}.
\]
Here, \(\widehat q_{k,t}(\bar\alpha_{k,t})\) is the smallest threshold for which at least a \(\bar\alpha_{k,t}\) fraction of the calibration scores lie below the threshold. Assumption~\ref{ass:sum_guarantee_bounds}(iii) makes the calibration scores distinct, so this empirical fraction can exceed \(\bar\alpha_{k,t}\) by at most one calibration observation, or \(1/|D_{k,\mathrm{cal}}(t)|\). Therefore,
\begin{equation}\tag{B.2}\label{eq:calib_error}
\left|\bar\alpha_{k,t}-\operatorname{BatErr}_{k,t}^{D}\right|\le \Delta_{k,t}.
\end{equation}
When \(\alpha_{k,t}\notin[0,1]\), the same inequality follows directly from the quantile conventions in Section~\ref{subsec:online_conf}.
The first term in $\Delta_{k,t}$ captures the calibration mismatch caused by
distribution shift between the current class-$k$ batch and the corresponding
calibration set; whereas the second term in $\Delta_{k,t}$ accounts for the
finite-sample correction, which is negligible since $|D_{k,\text{cal}}(t)|$ is large.

Before establishing an informative upper bound on the cumulative absolute
errors, we impose additional assumptions that may not hold in every practical
setting. Nevertheless, they facilitate an interpretable characterization of
the batchwise fluctuations of the errors produced by CARED. We consider two
distinct sets of assumptions, each leading to a different guarantee: one
constrains the range of the class-$k$ batch sizes $|D_{k,t}|$, whereas the
other assumes a nondecreasing sequence formed by the effective step sizes $\eta_{k,t}$.
Subsections~\ref{subsubsec:bounded_class_batch_size_range} and
\ref{subsubsec:nondecreasing_effective_steps} present the corresponding
results, respectively.
\subsubsection{Cumulative deviation under constrained batch size}
\label{subsubsec:bounded_class_batch_size_range}

\begin{assumption}[Bounded batch size range]
\label{ass:bounded_class_batch_size_range}
For any class $k$, there exist deterministic constants
$d_k>0$ and $M_k\ge1$ such that
\[
d_k\le |D_{k,t}|\le M_k d_k
\qquad\text{for every }t.
\]
That is, the number of class-$k$ cells remains controlled across batches.
\end{assumption}

\begin{prop}[Cumulative batch-wise deviation guarantee]
\label{prop:bounded_class_batch_size_range}
Consider any class $k\in[K]$, suppose that Assumptions~\ref{ass:main_guarantee},
\ref{ass:sum_guarantee_bounds}, and
\ref{ass:bounded_class_batch_size_range} hold. Then, for
any $\epsilon\in(0,1)$, define
\[
\begin{aligned}
\mathcal{R}_{k,T}(\epsilon)
&:=\frac{1}{p^-}\Bigg[
\frac{T}{2}\sqrt{\frac{B^-}{B^--1}\,p^+(1-p^-)\,\frac{M_k}{d_k}}
+\frac{5}{2}\sqrt{T\,p^+(1-p^-)\,\frac{M_k}{d_k}\log\frac{2}{\epsilon}}
+\frac{5M_k}{12}\log\frac{2}{\epsilon}
\Bigg].
\end{aligned}
\]
With probability at least $1-\epsilon$, it holds that
\[
\sum_{t=1}^T |\mathrm{BatErr}_{k,t}^D-\alpha|
\le
\left(1+\frac{p^+}{p^-}M_k\right)
\sum_{t=1}^T\Delta_{k,t}
+\mathcal{R}_{k,T}(\epsilon).
\]
\end{prop}

\begin{corollary}[Asymptotic cumulative batch-wise deviation guarantee]
\label{cor:bounded_class_batch_size_range}
Under the setting of Proposition~\ref{prop:bounded_class_batch_size_range},
we consider a sequence of instances for which $T\to\infty$ and the batch size
becomes as large as possible, as in $d_k\to\infty$. Then, for any
$\epsilon \in (0,1)$, with
probability at least $1-\epsilon$,
\[
\sum_{t=1}^T |\mathrm{BatErr}_{k,t}^D-\alpha|
\le
(1+M_k)\sum_{t=1}^T\Delta_{k,t}+\mathcal{O}\Big(\frac{T}{\sqrt{d_k}}\Big)+o_T(T).
\]
\end{corollary}

Thus, in the combined regime where $T\to\infty$ and $d_k\to\infty$, 
all terms other than the calibration mismatch are asymptotically negligible 
relative to $T$. Therefore, the calibration mismatch is the only 
asymptotically non-negligible contribution that can accumulate linearly with 
the horizon, with coefficient $1+M_k$, which approaches $2$ as $M_k\to1$.

\subsubsection{Cumulative deviation under nondecreasing effective step size}
\label{subsubsec:nondecreasing_effective_steps}

\begin{assumption}[Nondecreasing effective step size]
\label{ass:nondecreasing_effective_steps}
Consider any class $k \in [K]$ and any horizon $T$, assume that almost surely
\[
0<\eta_{k,1}\le\eta_{k,2}\le\cdots\le\eta_{k,T}<1.
\]
That is, the effective step sizes across batches form a nondecreasing sequence.
\end{assumption}

\begin{prop}[Cumulative batch-wise deviation guarantee]
\label{prop:nondecreasing_effective_steps}
Fix a class $k\in[K]$. Suppose that Assumptions~\ref{ass:main_guarantee},
\ref{ass:sum_guarantee_bounds}, and
\ref{ass:nondecreasing_effective_steps} hold. Define
\begin{equation}\label{eq:mu and sigma}\tag{B.3}
\mu_{k,t}:=
\frac{1}{\sqrt{2p^-|D_{k,t}|}}
+
\exp\!\left(-\frac{p^-|D_{k,t}|}{12}\right),
\end{equation}
and
\begin{equation}\label{eq:envelope}\tag{B.4}
S_{k,T}(\epsilon):=
\sum_{t=1}^T\mu_{k,t}
+
\sqrt{2T\log\frac{1}{\epsilon}}+\frac{1}{3}\log\frac{1}{\epsilon}.
\end{equation}
Then, for every
$\epsilon\in(0,1)$, with probability at least $1-\epsilon$,
\[
\begin{aligned}
\sum_{t=1}^T |\mathrm{BatErr}_{k,t}^D-\alpha|
&\le
2\sum_{t=1}^T\Delta_{k,t} + S_{k,T}(\epsilon).
\end{aligned}
\]
\end{prop}

Equations \eqref{eq:mu and sigma} and \eqref{eq:envelope} are constructed in this way because $S_{k,T}(\epsilon)$ forms an upper bound for cumulative sampling fluctuations $\sum_{t=1}^T |\mathrm{BatErr}_{k,t}^D - \mathrm{BatErr}_{k,t}^A|$; this is justified in Lemma~\ref{lem:sampling fluctuation}.

\begin{corollary}[Asymptotic cumulative batch-wise deviation guarantee]
\label{cor:nondecreasing_effective_steps}
Under the setting of Proposition~\ref{prop:nondecreasing_effective_steps},
we consider a sequence of instances for which $T\to\infty$ and the batch size becomes asymptotically large, that is, there exists $d_k$ such that $d_k\leq |D_{k,t}|$ for any $1\leq t\leq T$ and $d_k \to \infty.$
Then, for every $\epsilon\in(0,1)$, with probability at least
$1-\epsilon$,
\[
\sum_{t=1}^T |\mathrm{BatErr}_{k,t}^D-\alpha|
\le
2\sum_{t=1}^T\Delta_{k,t}
+
\mathcal{O}\!\left(
\frac{T}{\sqrt{d_k}}
\right)
+o_T(T).
\]
\end{corollary}

Thus, as the horizon $T$ and the class-$k$ batch size grow
asymptotically large, the calibration mismatch is the only nonnegligible term that
accumulates linearly with time, with coefficient $2$. Unlike the coefficient
in Corollary~\ref{cor:bounded_class_batch_size_range}, this value does not
depend on the class-size range.


\section{Additional details and results of experiments}\label{sec:experimental_details}

\subsection{Experimental details}\label{subsec:appendix_experimental_details}

This subsection collects implementation details and auxiliary experimental
settings omitted from Section~\ref{sec:experiments}. The details
include the data processing, the selection rule for the class-wise step sizes $\gamma_k$, the
structural subtree partition used by the utility-aware module, and the
settings for the experiments in Section~\ref{sec:experiments}.

\paragraph{Dataset processing.} In each dataset (cohort), the H\&E stain was applied to the same tissue section after the Xenium run and imaged at 40$\times$ (0.274 $\mu$m/pixel). For all experiments, the images are downsampled to approximately 0.415 $\mu$m/pixel. 
Cell locations were registered to the H\&E image using 10x Genomics Xenium Ranger, a post-processing pipeline for Xenium data, then filtered by 1,000 cells. 

\paragraph{Selection rule for $\gamma_k$.}
In the experiments, the step sizes are selected before the online stream is run
and then kept fixed over time. To satisfy the assumptions in Appendix~\ref{subsec:sum_guarantee},
we choose $\gamma_k$ based on the class frequency in the test cohort, so that
the effective batch step size $\gamma_k|A_{k,t}|$ remains below one with margin
throughout the stream. Although this simulation uses oracle class-frequency
information, in practice, the frequency can be roughly specified from domain
knowledge. The assignment rule for
$\gamma_k$ is reported in Table~\ref{tab:section5_experiment_settings}.

\paragraph{Structural subtree.}
In our experiments, the structural subtree partition is specified by grouping cell types with
closely related cellular characteristics into the same subtree. Specifically,
the pancreas labels are grouped into three subtrees according to cellular
function, while the ovary and lung labels are organized into two subtrees
according to the benign--malignant distinction. The concrete subtree assignments
are reported in Table~\ref{tab:section5_experiment_settings}.

\paragraph{Parameters used in experiments.}
Table~\ref{tab:section5_experiment_settings} lists the experimental details and
parameter values used in the main experiments.

\begin{table}[!htbp]
\centering
\caption{Details and parameters in the main experiments.}
\label{tab:section5_experiment_settings}
\small
\renewcommand{\arraystretch}{1.2}
\begin{tabular}{>{\centering\arraybackslash}m{0.35\linewidth} >{\centering\arraybackslash}m{0.62\linewidth}}
\toprule
Name and notation & Choice used in the experiments \\
\midrule
Held-out tissues & Pancreas, ovary, and lung \\
Label space for pancreas & Acinar, Ductal, Endocrine, Endothelial, Fibroblast-like, T Cell \\
Subtree partition for pancreas & $\mathcal S_1=\{\mathrm{Acinar},\ \mathrm{Ductal},\ \mathrm{Endocrine}\}$; $\mathcal S_2=\{\mathrm{Endothelial},\ \mathrm{Fibroblast\text{-}like}\}$; $\mathcal S_3=\{\mathrm{T\ Cell}\}$ \\
Label space for ovary & Endothelial, Epithelial Malignant, Macrophage \\
Subtree partition for ovary & $\mathcal S_1=\{\mathrm{Endothelial},\ \mathrm{Macrophage}\}$; $\mathcal S_2=\{\mathrm{Epithelial\ Malignant}\}$ \\
Label space for lung & Endothelial, Epithelial Malignant, T Cell, Tumor Associated Fibroblast \\
Subtree partition for lung & $\mathcal S_1=\{\mathrm{Endothelial},\ \mathrm{T\ Cell},\ \mathrm{Tumor\ Associated\ Fibroblast}\}$; $\mathcal S_2=\{\mathrm{Epithelial\ Malignant}\}$ \\
Target error level $\alpha$ & $0.1$ \\
Target reviewed fraction $p$ & $0.1$ \\
Step size of frequent classes ($>25\%$) & $\gamma_k=10^{-3}$ \\
Step size of moderate classes ($5\%$--$25\%$) & $\gamma_k=2\times 10^{-3}$ \\
Step size of scarce classes ($<5\%$) & $\gamma_k=10^{-2}$ \\
Initial labeled block $A_0$ & The first randomly selected batch\\
Batch-size bounds $B^-,B^+$ & $B^-=490$, $B^+=510$ \\
Calibration window $T_w$ & $30$ batches \\
Utility candidate number $N$ & $100$ \\
Utility learning rate $\eta$ & $0.1$ \\
Utility weights $(\lambda_{\mathrm{skip}},\lambda_{\mathrm{size}},\lambda_{\mathrm{struct}})$ & $(20.0, 1.0, 5.0)$ \\
Soft-membership temperature $T_{\mathrm{temp}}$ & $0.1$ \\
Random seed & $0$ \\
\bottomrule
\end{tabular}
\end{table}

\subsection{More experimental results}\label{subsec:more_experimental_results}

\begin{figure}[!htbp]
    \centering
    \captionsetup[subfigure]{font=scriptsize,justification=centering}
    \begin{subfigure}[t]{0.31\linewidth}
        \centering
        \includegraphics[width=\linewidth]{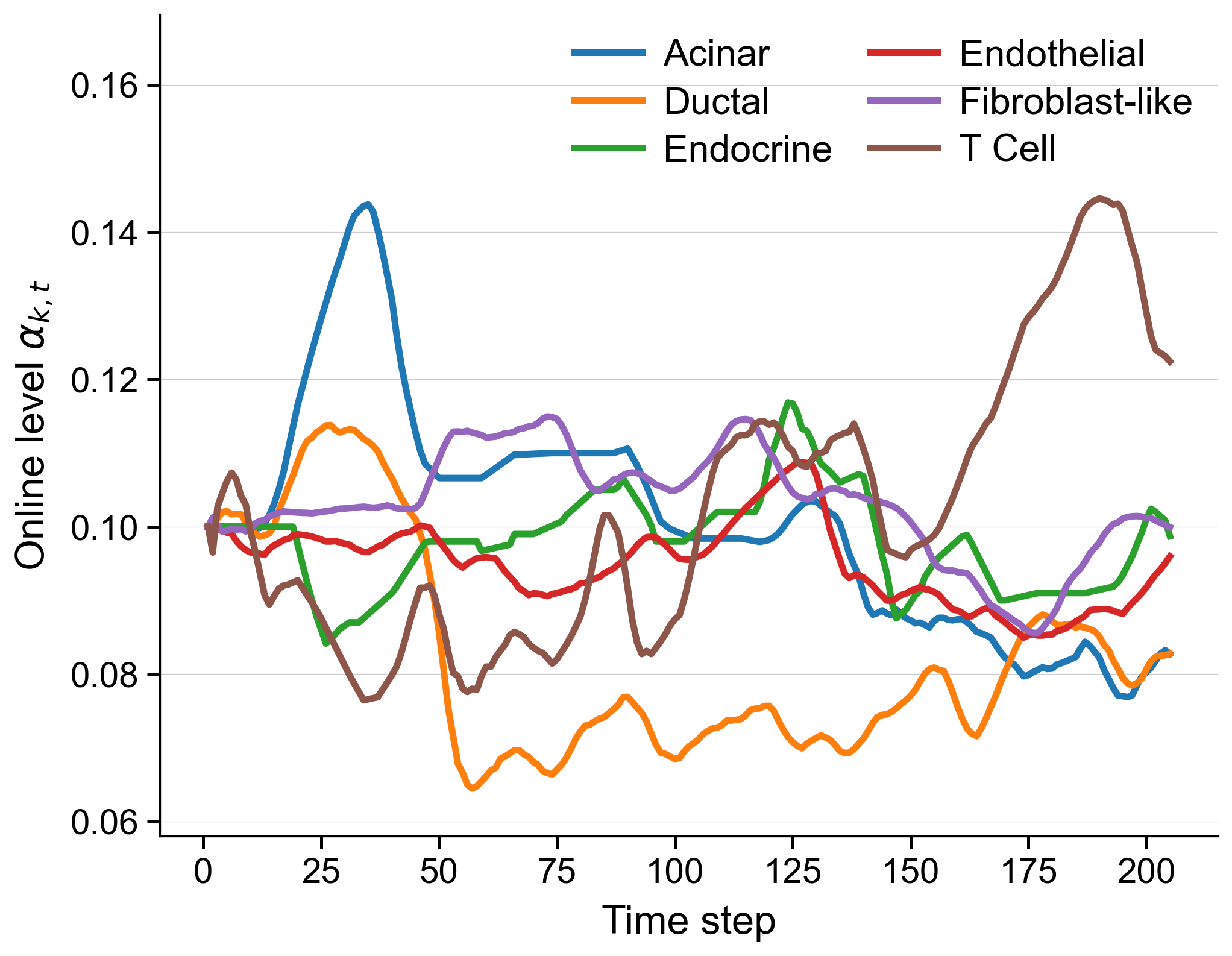}
        \caption{Global, pancreas}
    \end{subfigure}
    \hfill
    \begin{subfigure}[t]{0.31\linewidth}
        \centering
        \includegraphics[width=\linewidth]{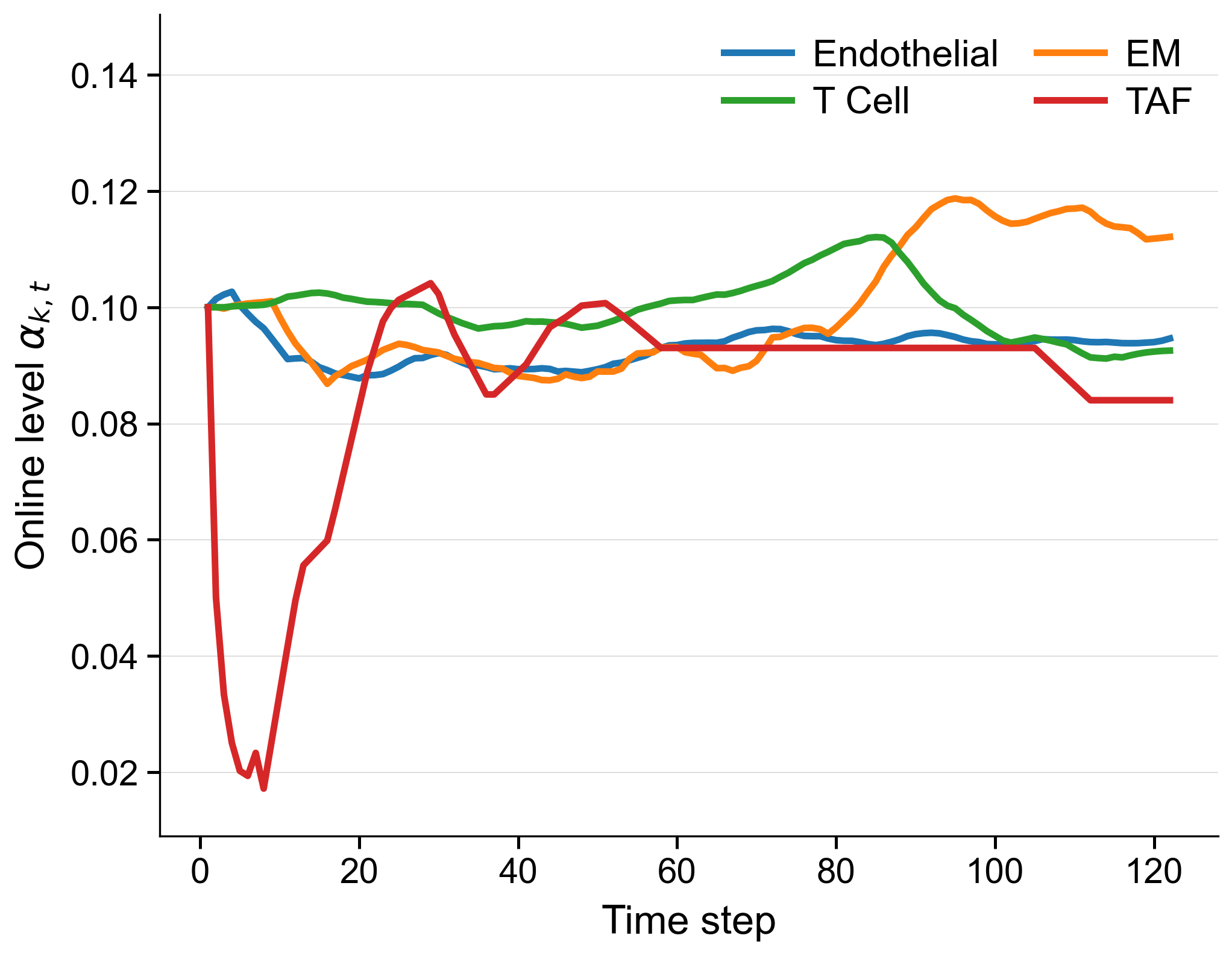}
        \caption{Global, lung}
    \end{subfigure}
    \hfill
    \begin{subfigure}[t]{0.31\linewidth}
        \centering
        \includegraphics[width=\linewidth]{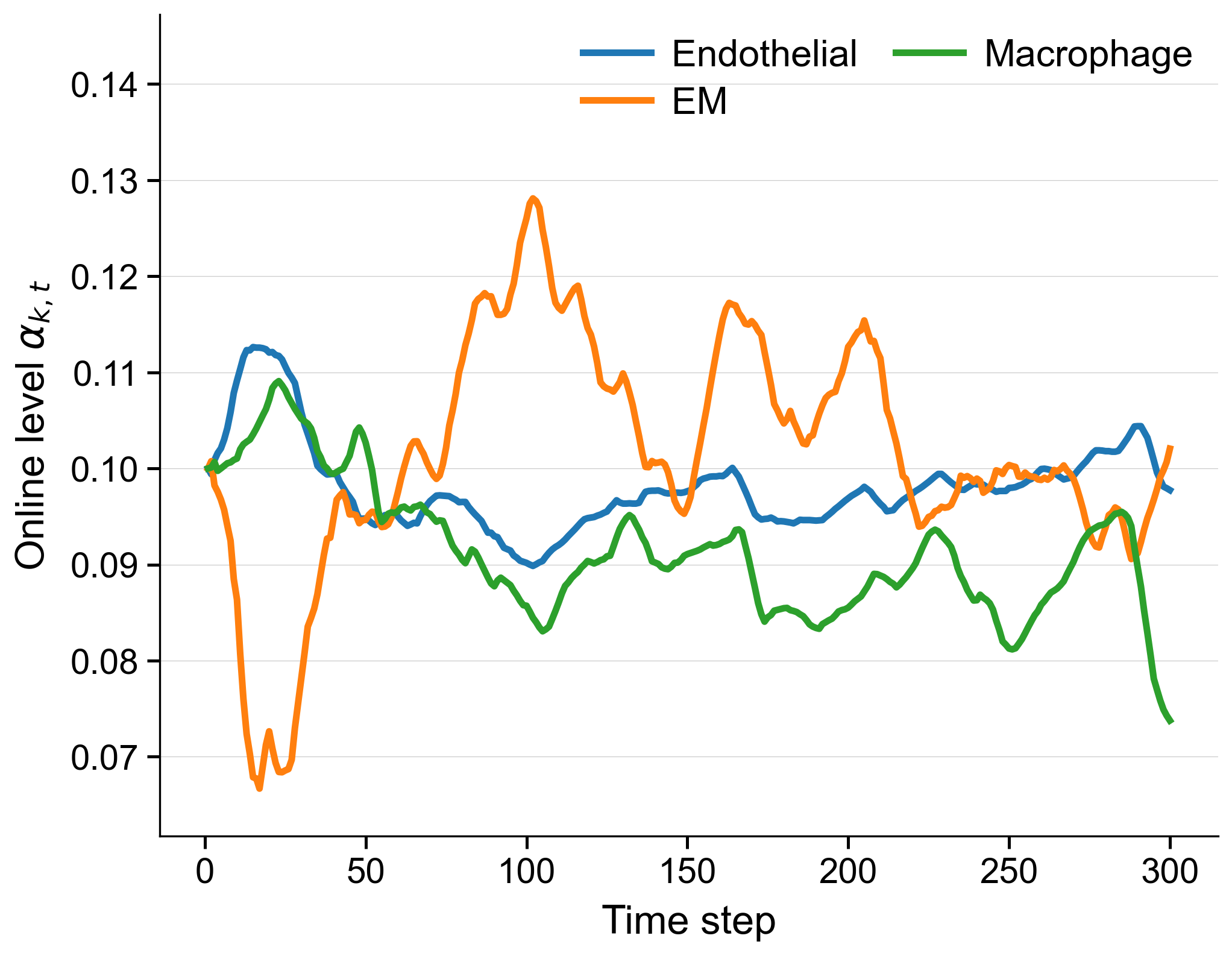}
        \caption{Global, ovary}
    \end{subfigure}
    \caption{Online level trajectories $\alpha_{k,t}$ under the error-guided
    batch-release scheme using the global NuClass score. Panels (a)--(c)
    correspond to pancreas, lung, and ovary, respectively; each curve represents
    a cell type. The curves are smoothed with a window size of 7.}
    \label{fig:appendix_global_alpha_trajectories}
\end{figure}

For reference, we report the online level trajectories $\alpha_{k,t}$ under the \emph{error-guided} batch-release scheme in the experiments reported in Section~\ref{sec:experiments}. These trajectories are shown in
Figure~\ref{fig:appendix_global_alpha_trajectories}.
We then provide two additional sets of experimental results to show CARED's stability. First, we report
the cumulative conformal error trajectories on the three tissues for the remaining
local and gated NuClass score sources, under the error-guided batch-release scheme. Second, we compare the utility metrics obtained from the
utility-aware transformed scores and raw scores
on the lung and ovary cohorts using the global NuClass score.

\begin{figure}[!htbp]
    \centering
    \captionsetup[subfigure]{font=scriptsize,justification=centering}
    \begin{subfigure}[t]{0.22\linewidth}
        \centering
        \includegraphics[width=\linewidth]{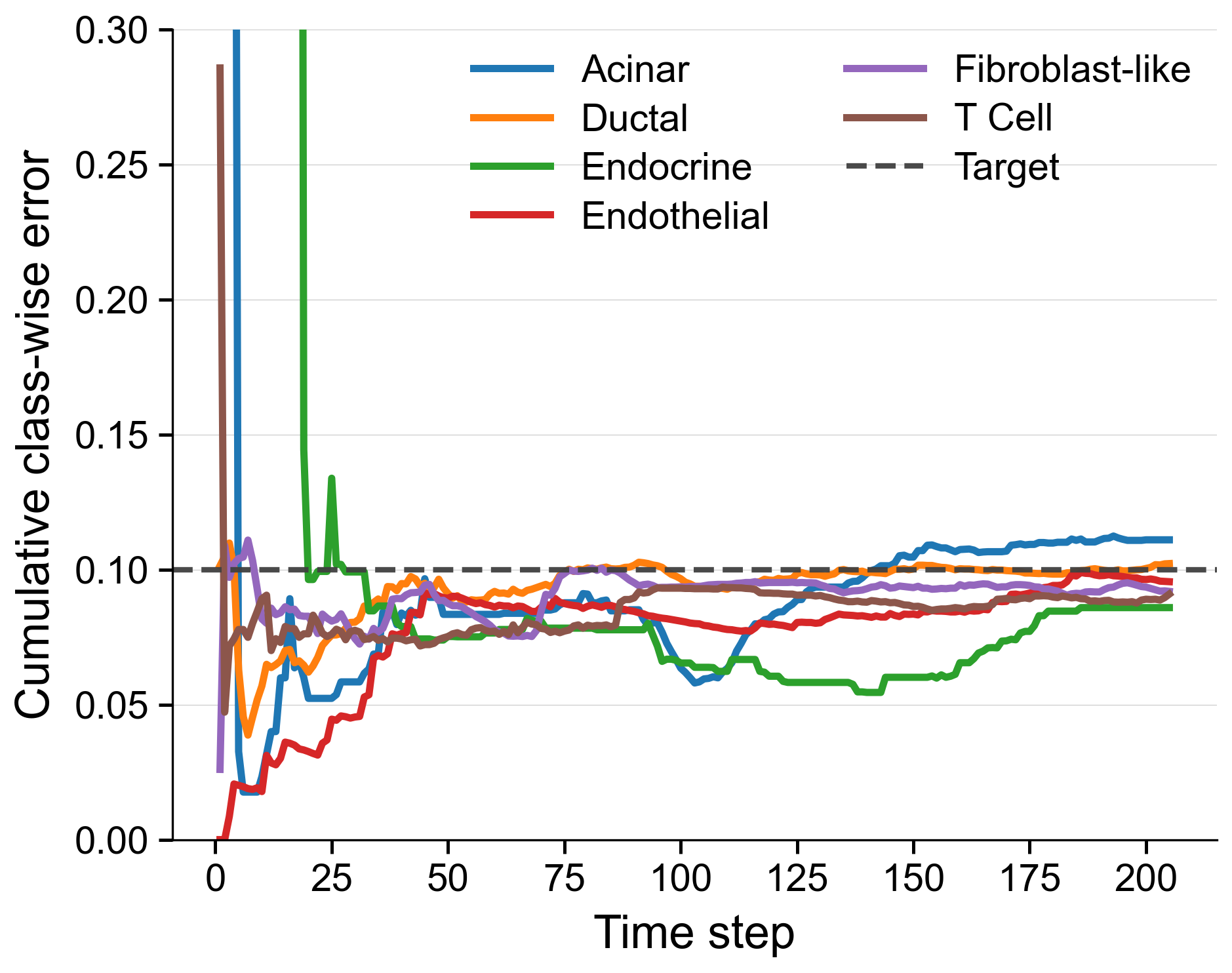}
        \caption{Local, pancreas}
    \end{subfigure}
    \hfill
    \begin{subfigure}[t]{0.22\linewidth}
        \centering
        \includegraphics[width=\linewidth]{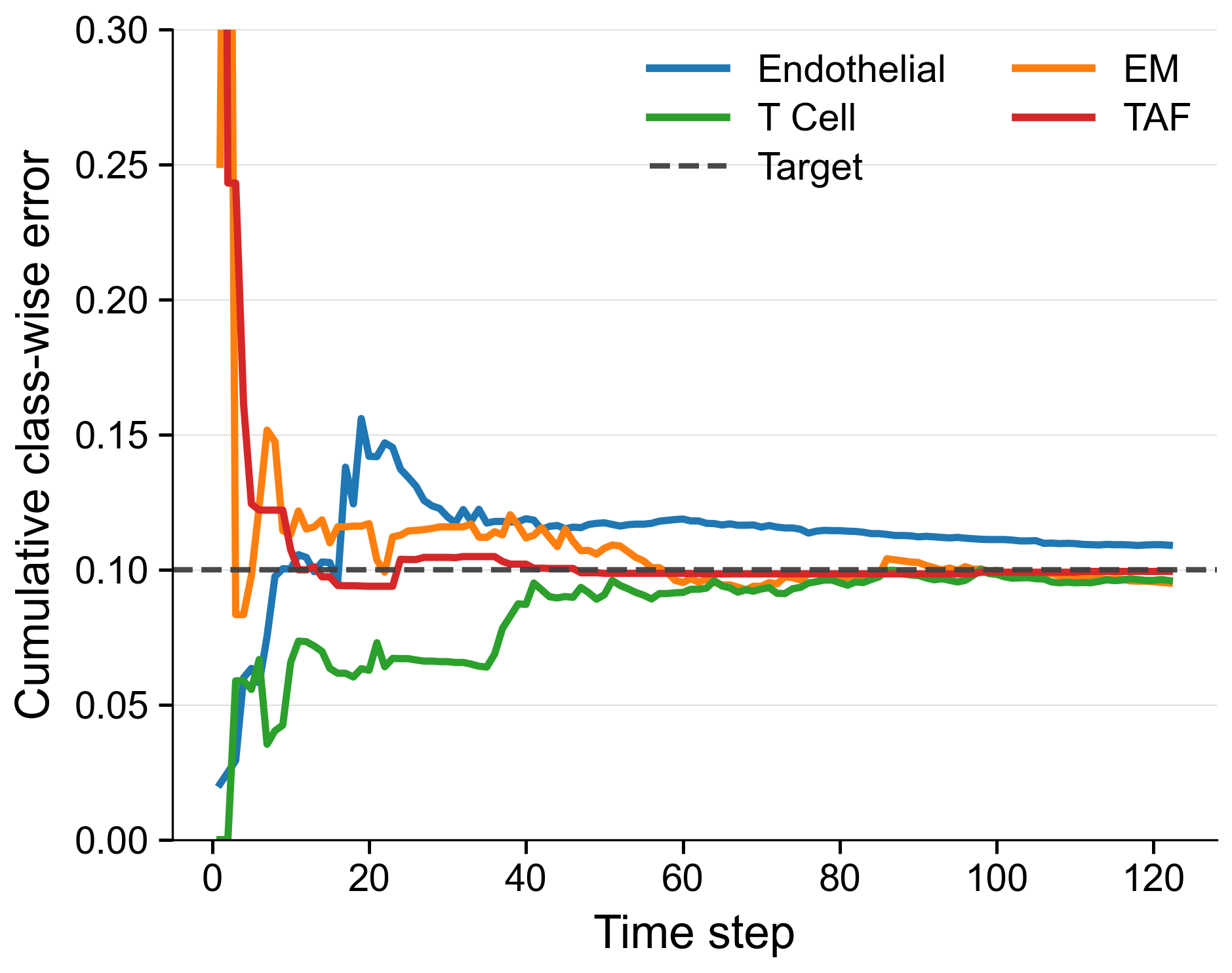}
        \caption{Local, lung}
    \end{subfigure}
    \hfill
    \begin{subfigure}[t]{0.22\linewidth}
        \centering
        \includegraphics[width=\linewidth]{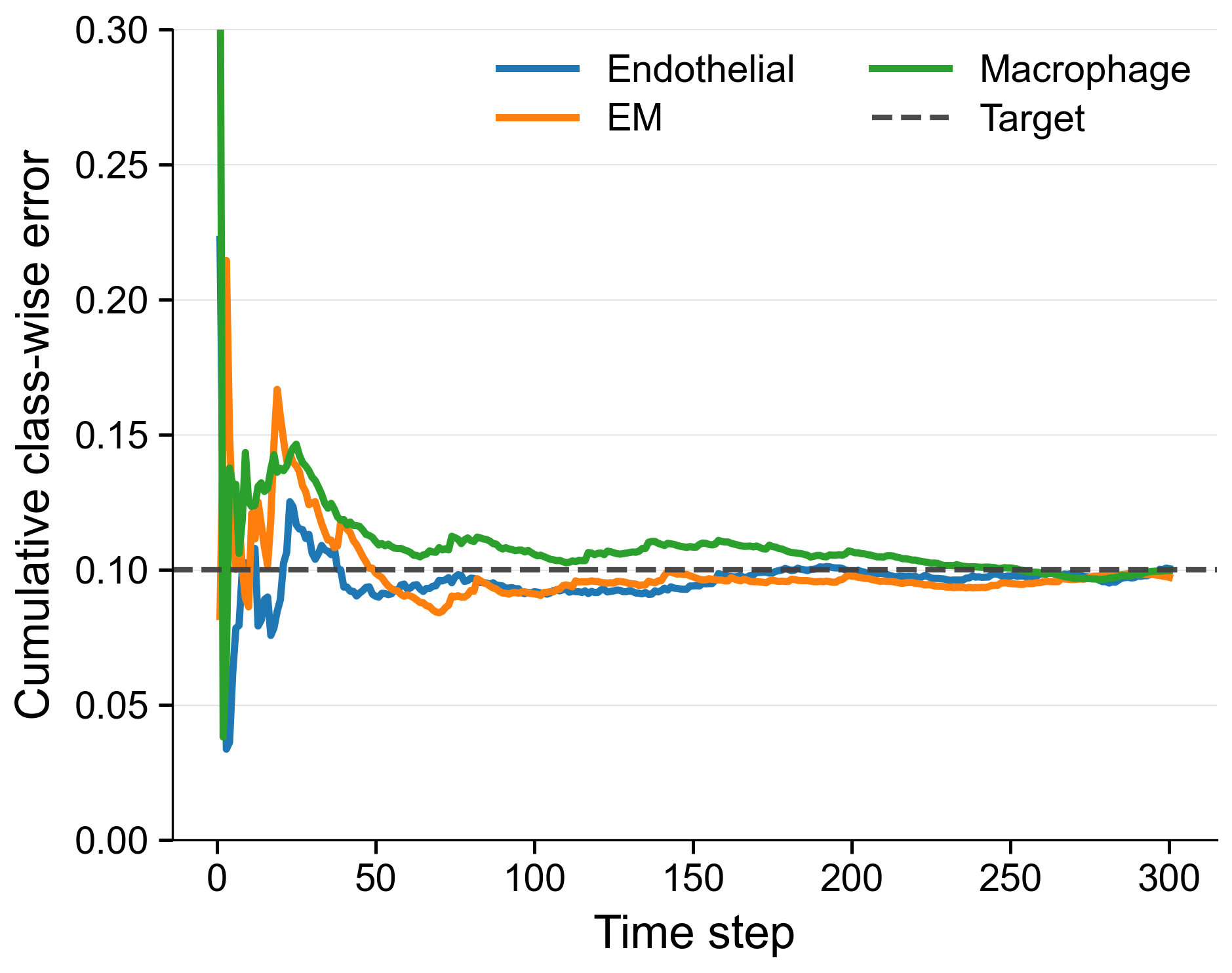}
        \caption{Local, ovary}
    \end{subfigure}
    \hfill
    \begin{subfigure}[t]{0.22\linewidth}
        \centering
        \includegraphics[width=\linewidth]{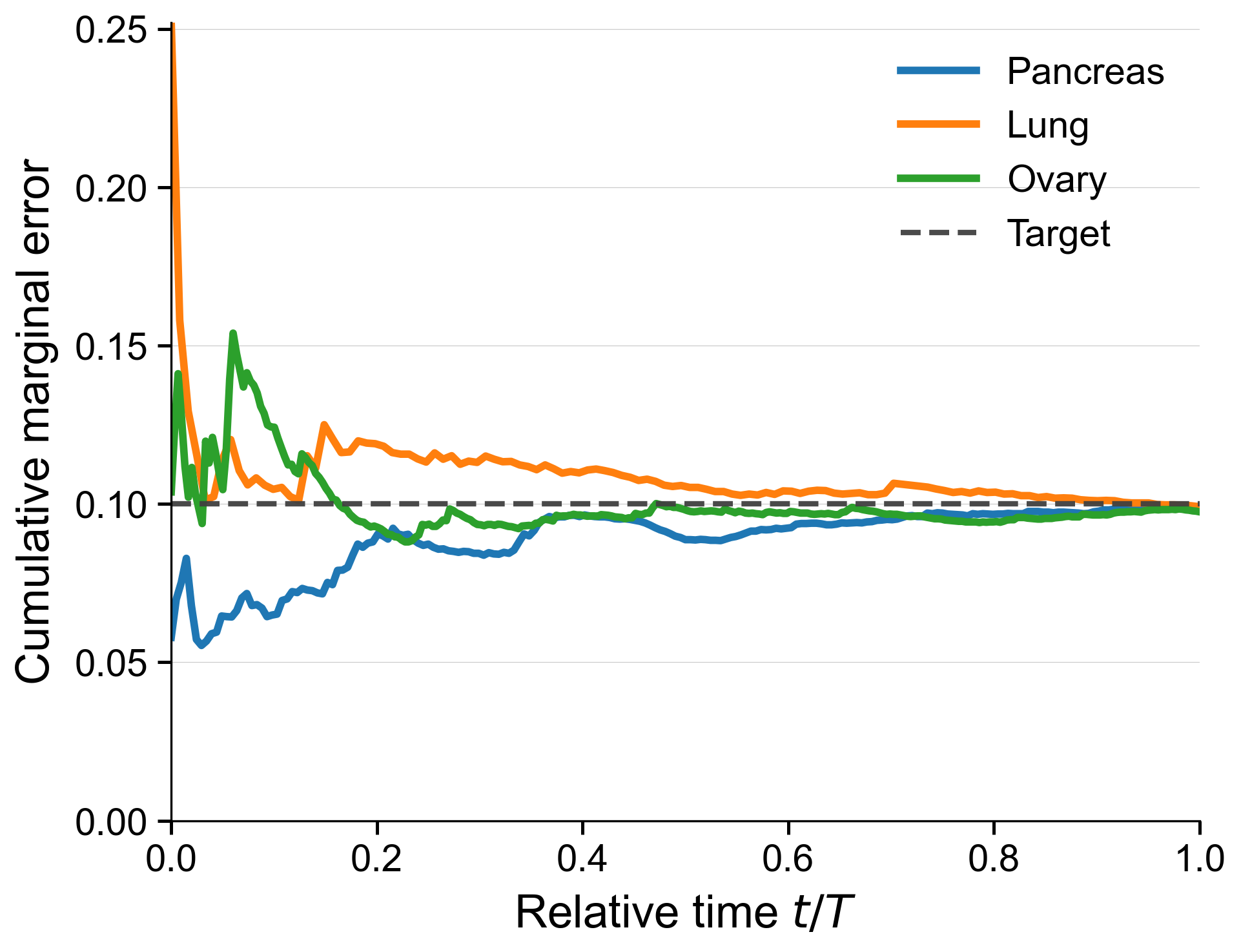}
        \caption{Local, marginal}
    \end{subfigure}

    \smallskip

    \begin{subfigure}[t]{0.22\linewidth}
        \centering
        \includegraphics[width=\linewidth]{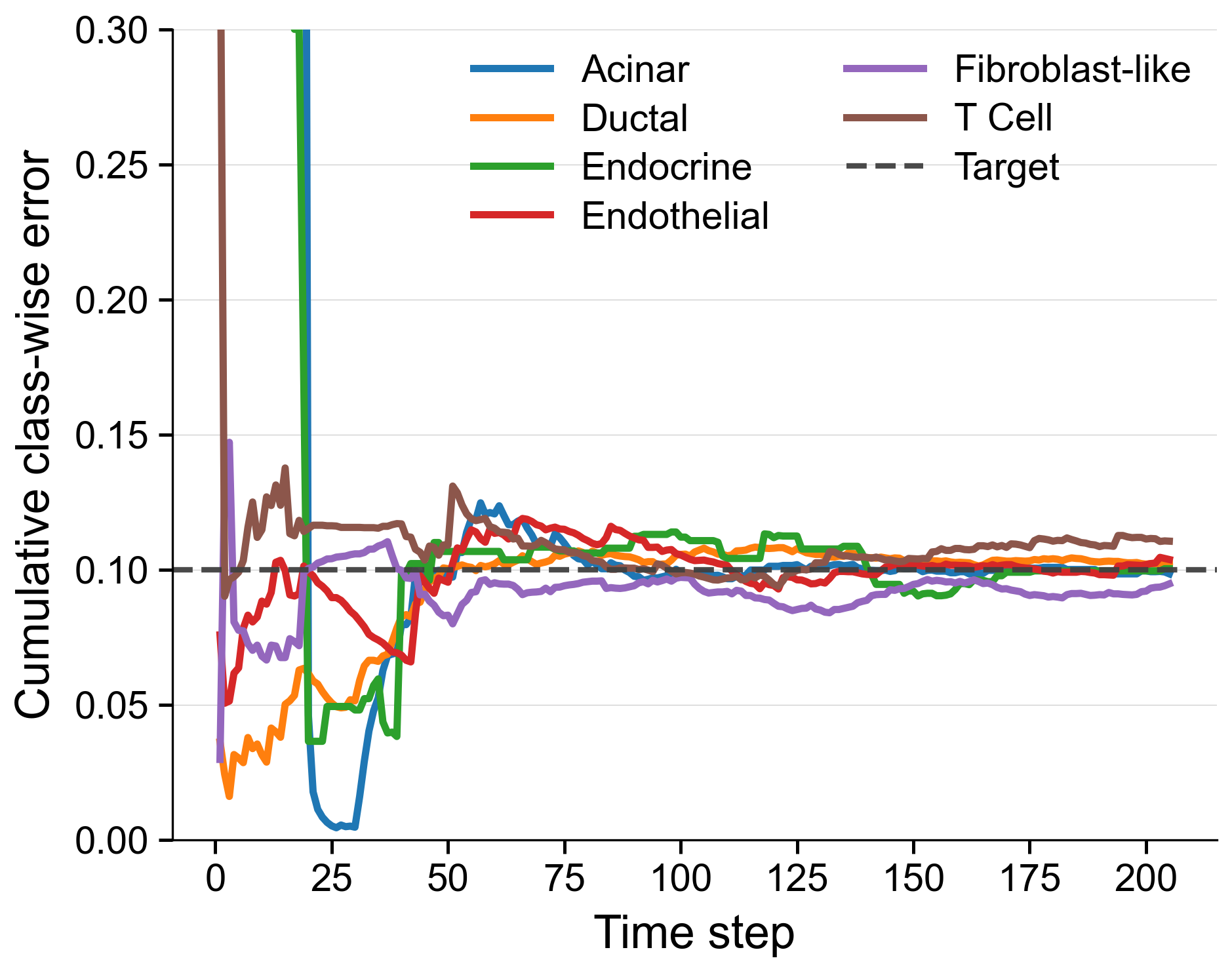}
        \caption{Gated, pancreas}
    \end{subfigure}
    \hfill
    \begin{subfigure}[t]{0.22\linewidth}
        \centering
        \includegraphics[width=\linewidth]{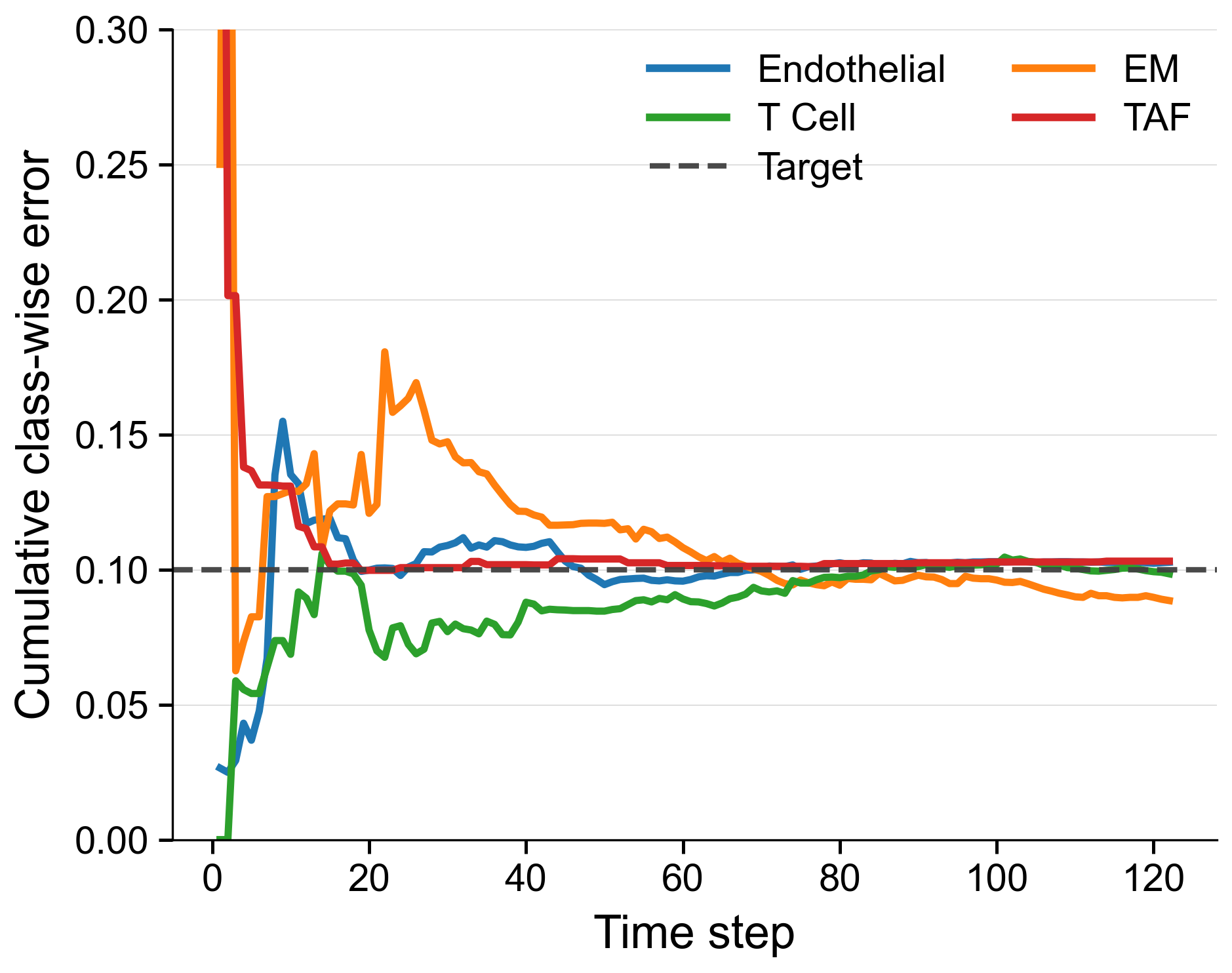}
        \caption{Gated, lung}
    \end{subfigure}
    \hfill
    \begin{subfigure}[t]{0.22\linewidth}
        \centering
        \includegraphics[width=\linewidth]{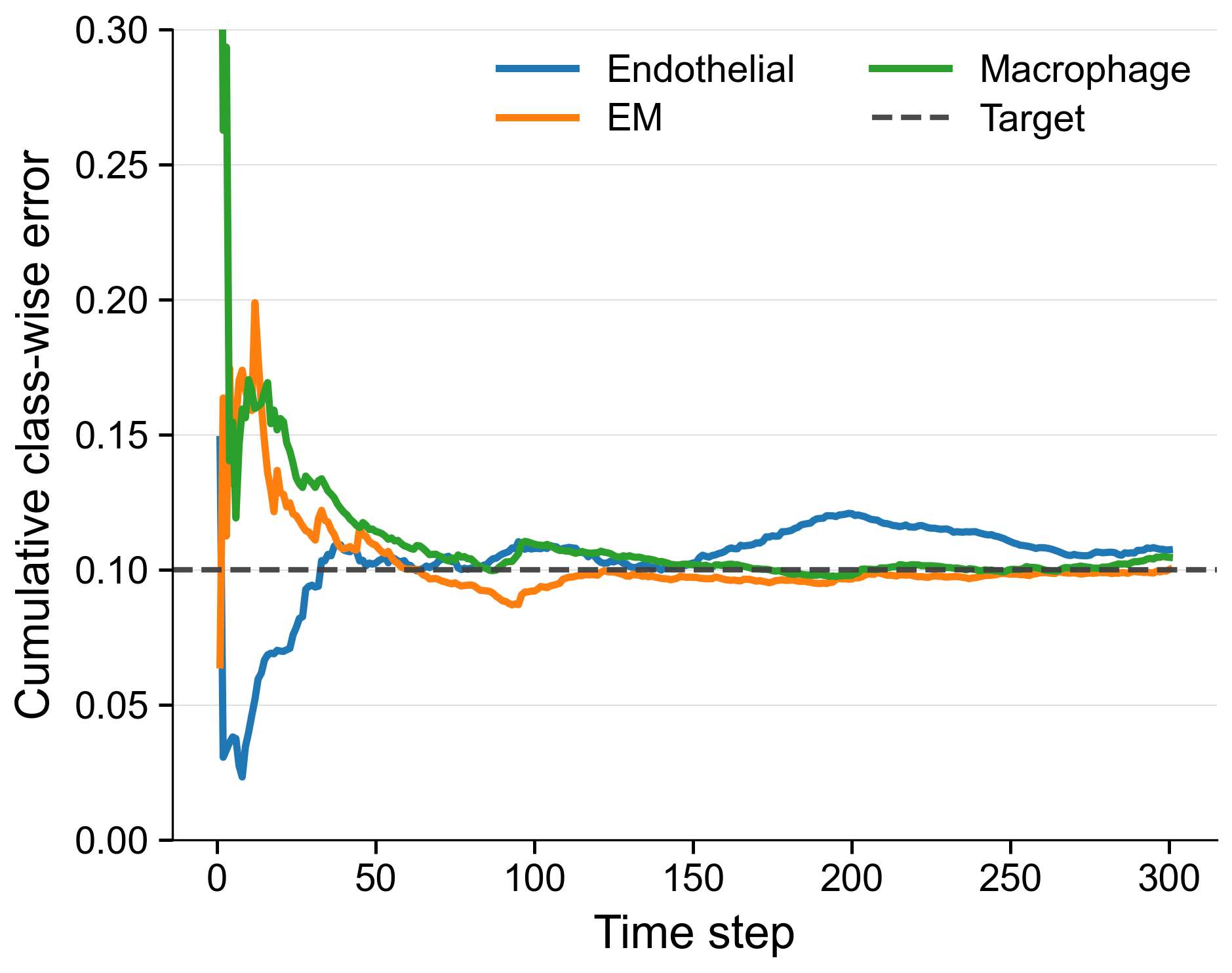}
        \caption{Gated, ovary}
    \end{subfigure}
    \hfill
    \begin{subfigure}[t]{0.22\linewidth}
        \centering
        \includegraphics[width=\linewidth]{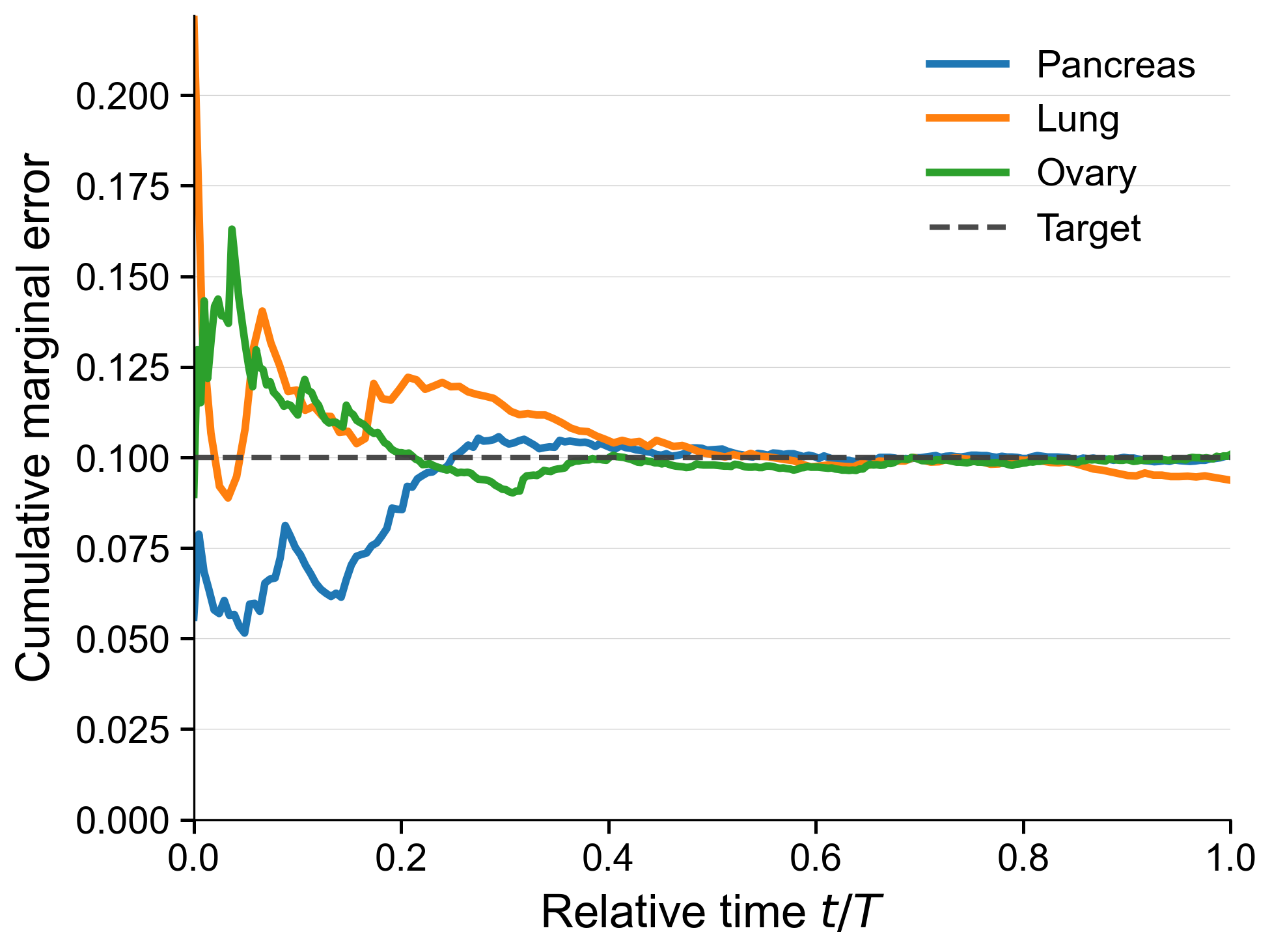}
        \caption{Gated, marginal}
    \end{subfigure}
    \caption{Additional cumulative conformal error trajectories under the
    error-guided batch-release scheme. Rows correspond to the local and gated
    NuClass scores, while columns show pancreas, lung, ovary, and marginal
    results. The first three columns report class-wise errors within each tissue;
    the marginal panels compare the marginal errors of the three tissues against
    the fraction of processed batches. Dashed horizontal lines mark the target
    error level $\alpha=0.1$.}
    \label{fig:appendix_remaining_score_errors}
\end{figure}

\begin{figure}[p]
    \centering
    \captionsetup[subfigure]{font=scriptsize,justification=centering}
    \includegraphics[width=0.44\linewidth]{figures/5/5.3_utility_legend.png}

    \smallskip

    \begin{subfigure}[t]{0.34\linewidth}
        \centering
        \includegraphics[width=\linewidth]{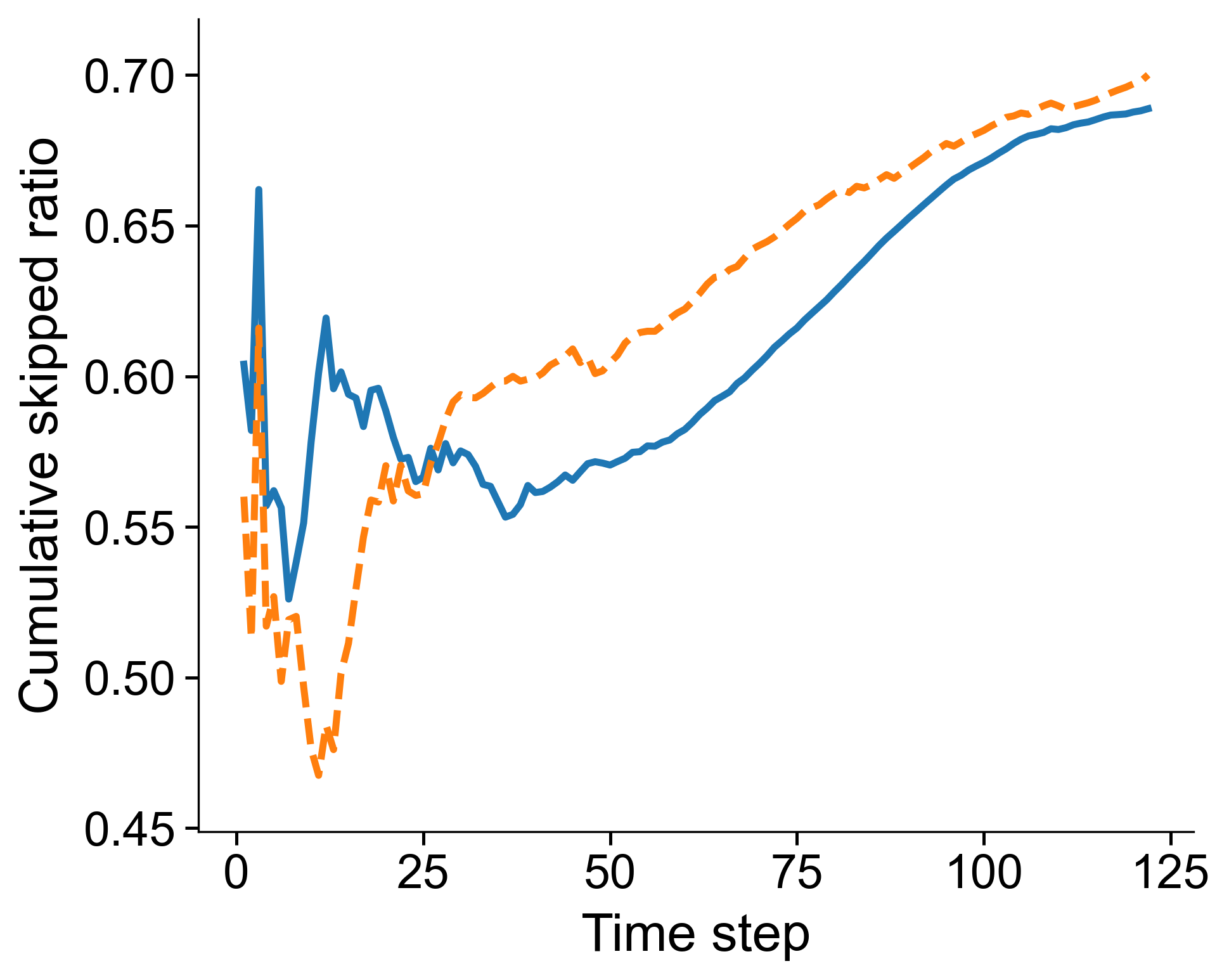}
        \caption{Lung, skipped ratio}
    \end{subfigure}%
    \hspace{0.015\linewidth}%
    \begin{subfigure}[t]{0.34\linewidth}
        \centering
        \includegraphics[width=\linewidth]{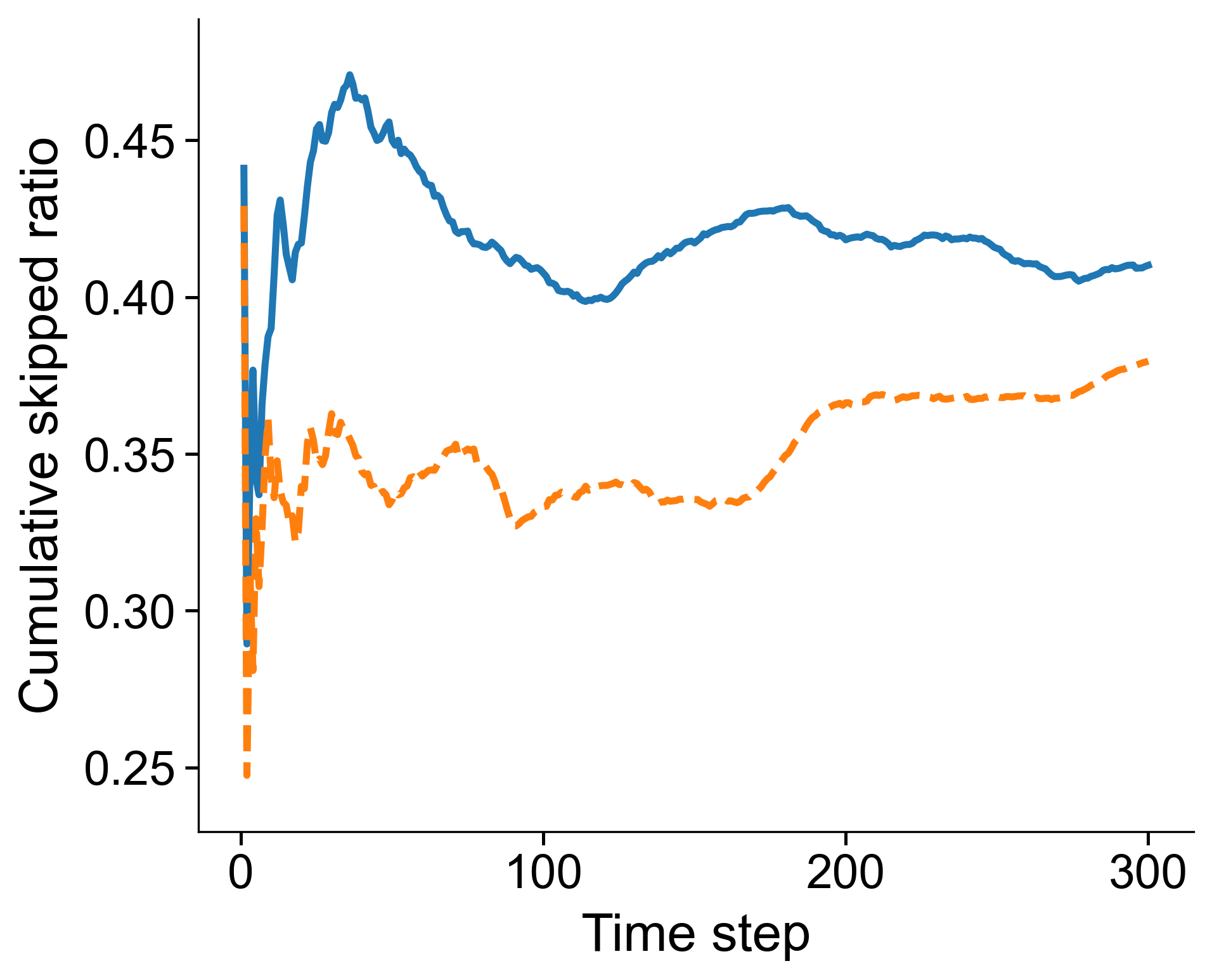}
        \caption{Ovary, skipped ratio}
    \end{subfigure}

    \smallskip

    \begin{subfigure}[t]{0.34\linewidth}
        \centering
        \includegraphics[width=\linewidth]{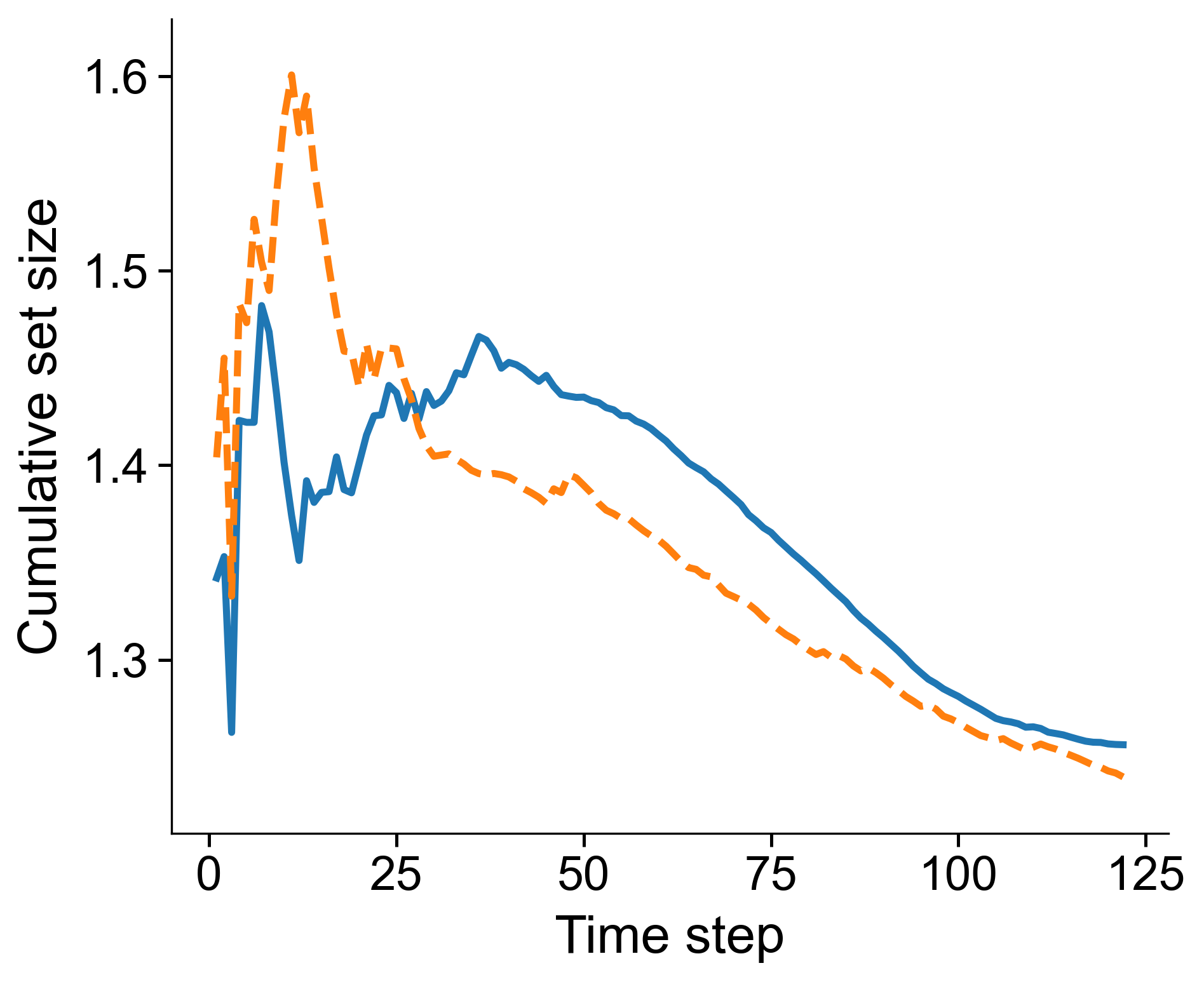}
        \caption{Lung, set size}
    \end{subfigure}%
    \hspace{0.015\linewidth}%
    \begin{subfigure}[t]{0.34\linewidth}
        \centering
        \includegraphics[width=\linewidth]{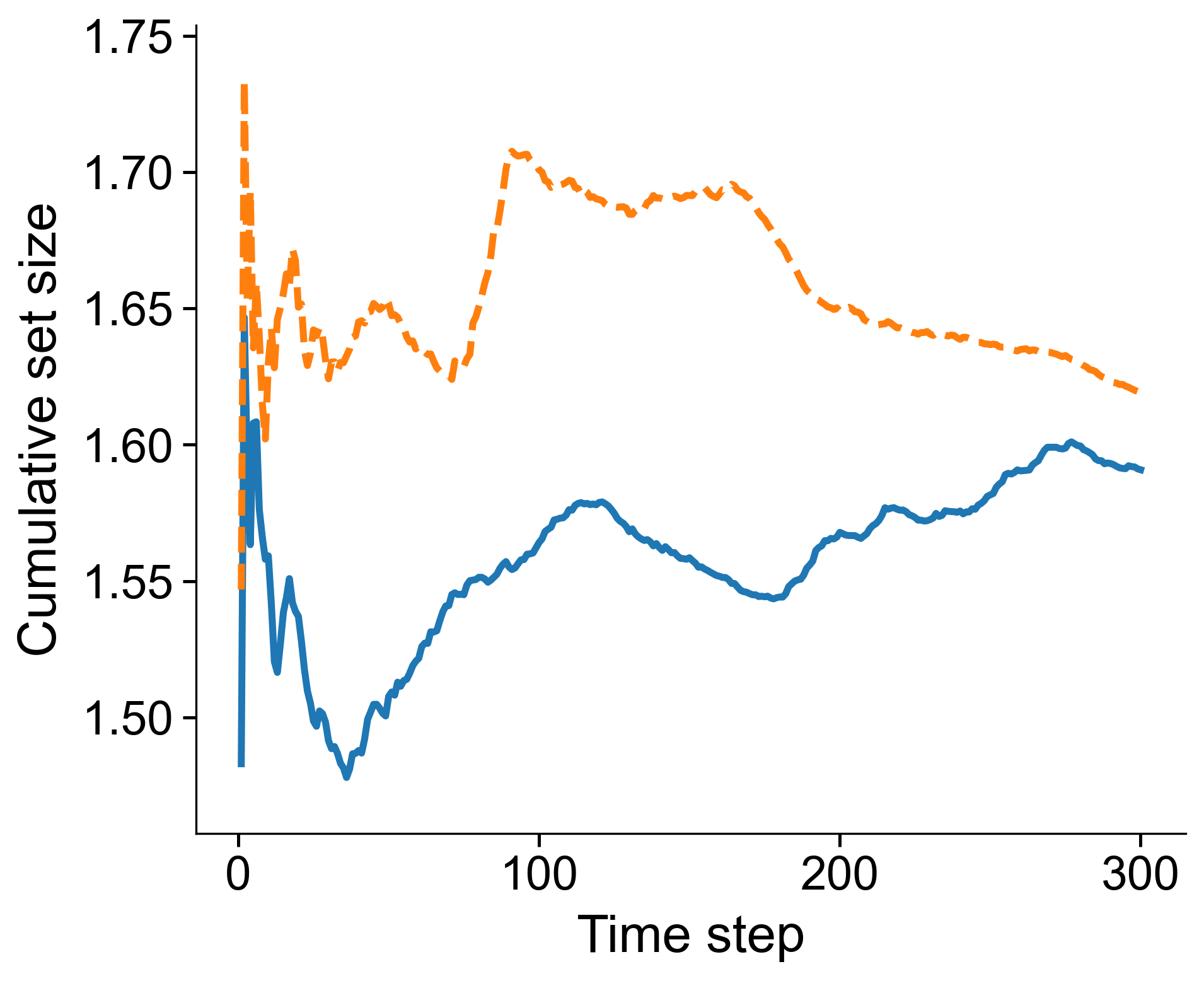}
        \caption{Ovary, set size}
    \end{subfigure}

    \smallskip

    \begin{subfigure}[t]{0.34\linewidth}
        \centering
        \includegraphics[width=\linewidth]{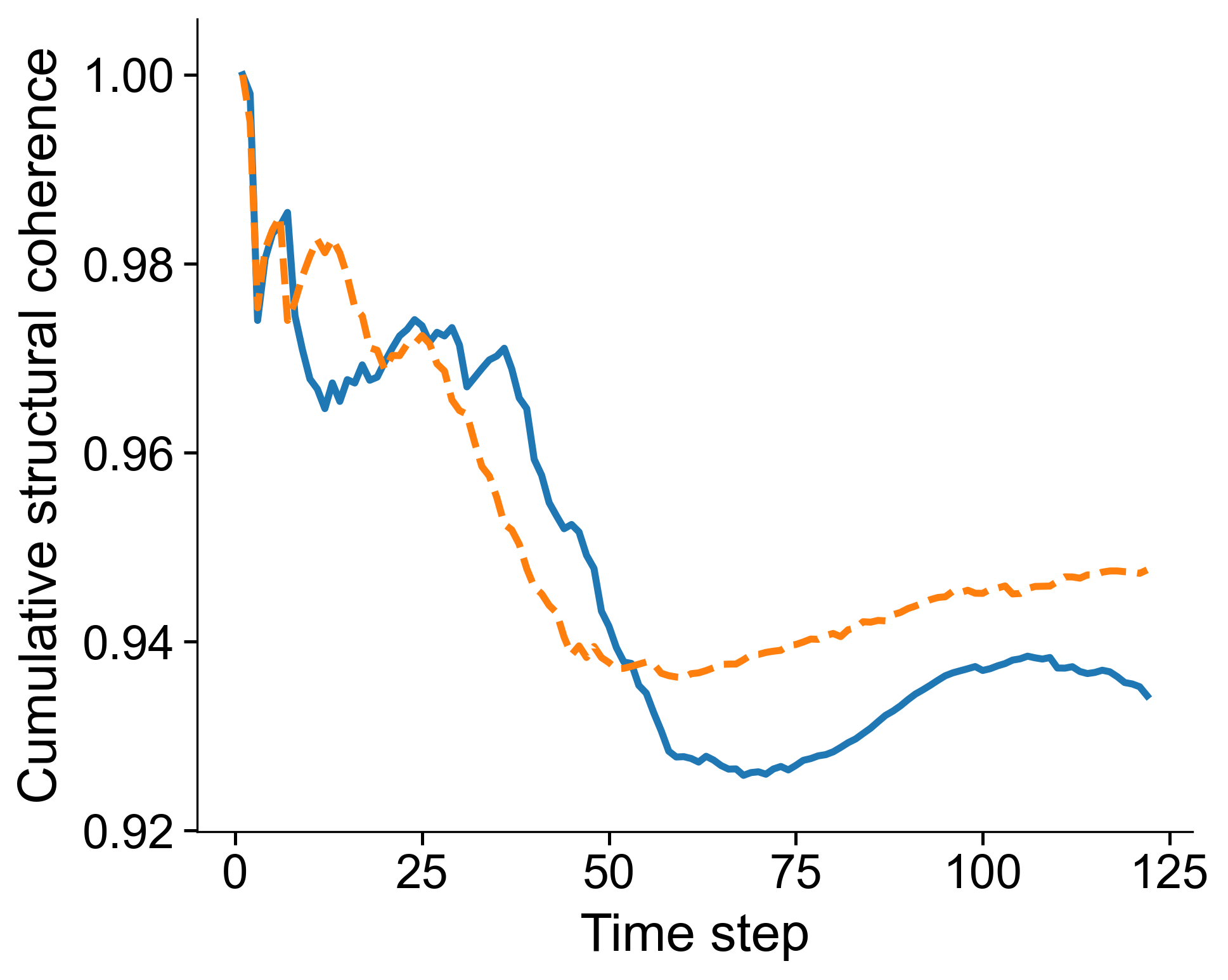}
        \caption{Lung, structural coherence}
    \end{subfigure}%
    \hspace{0.015\linewidth}%
    \begin{subfigure}[t]{0.34\linewidth}
        \centering
        \includegraphics[width=\linewidth]{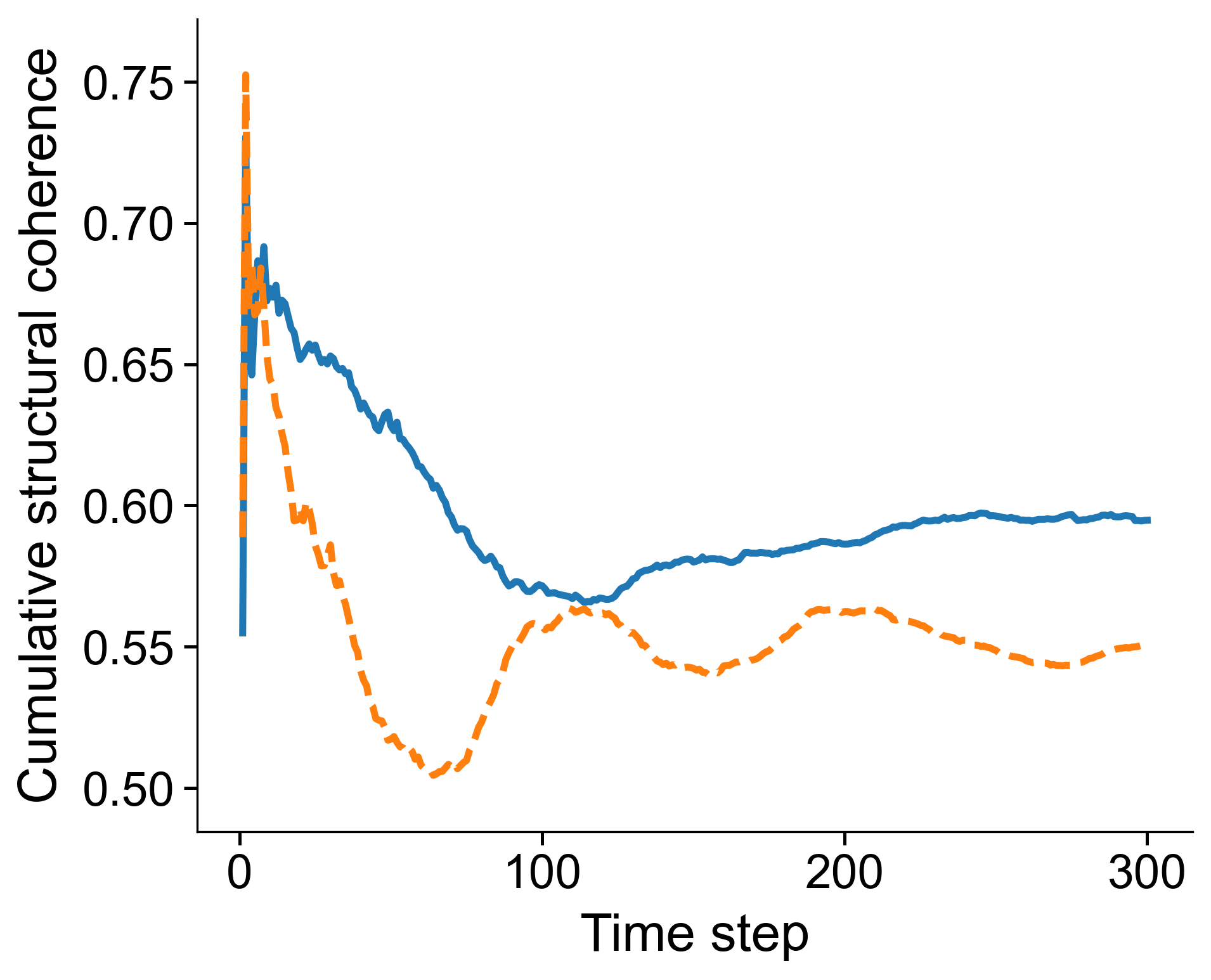}
        \caption{Ovary, structural coherence}
    \end{subfigure}
    \caption{Additional cumulative utility comparison under the error-guided
    batch-release scheme using the global NuClass score. Rows correspond to
    the skipped ratio, average prediction set size, and structural coherence,
    while columns report lung and ovary. Solid blue curves use the utility-aware
    transformed score, while dashed orange curves use the raw score without
    the utility module. Higher skipped ratio and structural coherence and
    smaller prediction sets indicate better utility.}
    \label{fig:appendix_lung_ovary_utility}
\end{figure}

Together with the results under the global score source in
Section~\ref{sec:experiments},
Figure~\ref{fig:appendix_remaining_score_errors} extends the same empirical
validity check to the remaining local and gated NuClass score sources, showing that the method
continues to control class-wise errors across the considered tissues and score
sources.
We next examine the utility results on the lung and ovary tasks in
Figure~\ref{fig:appendix_lung_ovary_utility}. For the lung results, the
utility-aware score transformation shows a minor disadvantage relative to the
raw-score baseline in the cumulative utility metrics. This does not contradict
the intended scope of our framework, since the utility module is guaranteed to
deliver prediction sets with no lower utility for a fixed batch of cells, while
the resulting score transformation may still induce a less favorable trajectory
of selected batches and the online levels $\alpha_{k,t}$ when cumulative utility
is evaluated over the full stream. Moreover, the lung annotation task is not the
primary target setting for the utility module: the skipped ratio is already high
in this task, and the subtree structure has already been satisfied well. The more
important evidence is the six-class pancreas annotation task, where the utility
module shows clearer gains, as shown in
Figure~\ref{fig:utility_pancreas_global}.

\subsection{Additional sensitivity analyses}\label{subsec:additional_sensitivity_analyses}

\subsubsection{Class-wise cumulative error}\label{subsubsec:sensitivity_cumulative_error}

Section~\ref{subsec:sensitivity_cumulative_error} reports sensitivity analyses
for the marginal cumulative error under variations of the reviewed fraction
$p$, random seed, target error level $\alpha$, and batch release scheme. Here,
for each sensitivity setting, we report the final cumulative empirical error
for each cell type after processing the full annotation stream, and compare
these class-wise errors across the three batch release schemes.

\begin{figure}[p]
    \centering
    \setlength{\parskip}{0pt}
    \captionsetup[subfigure]{font=scriptsize,justification=centering}
    \includegraphics[width=0.86\linewidth]{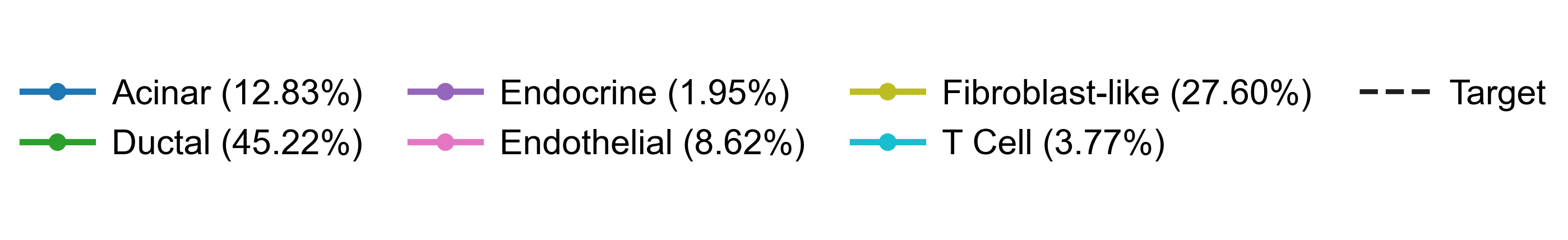}
    \vspace{0.2em}

    \begin{subfigure}[t]{0.28\linewidth}
        \centering
        \includegraphics[width=\linewidth]{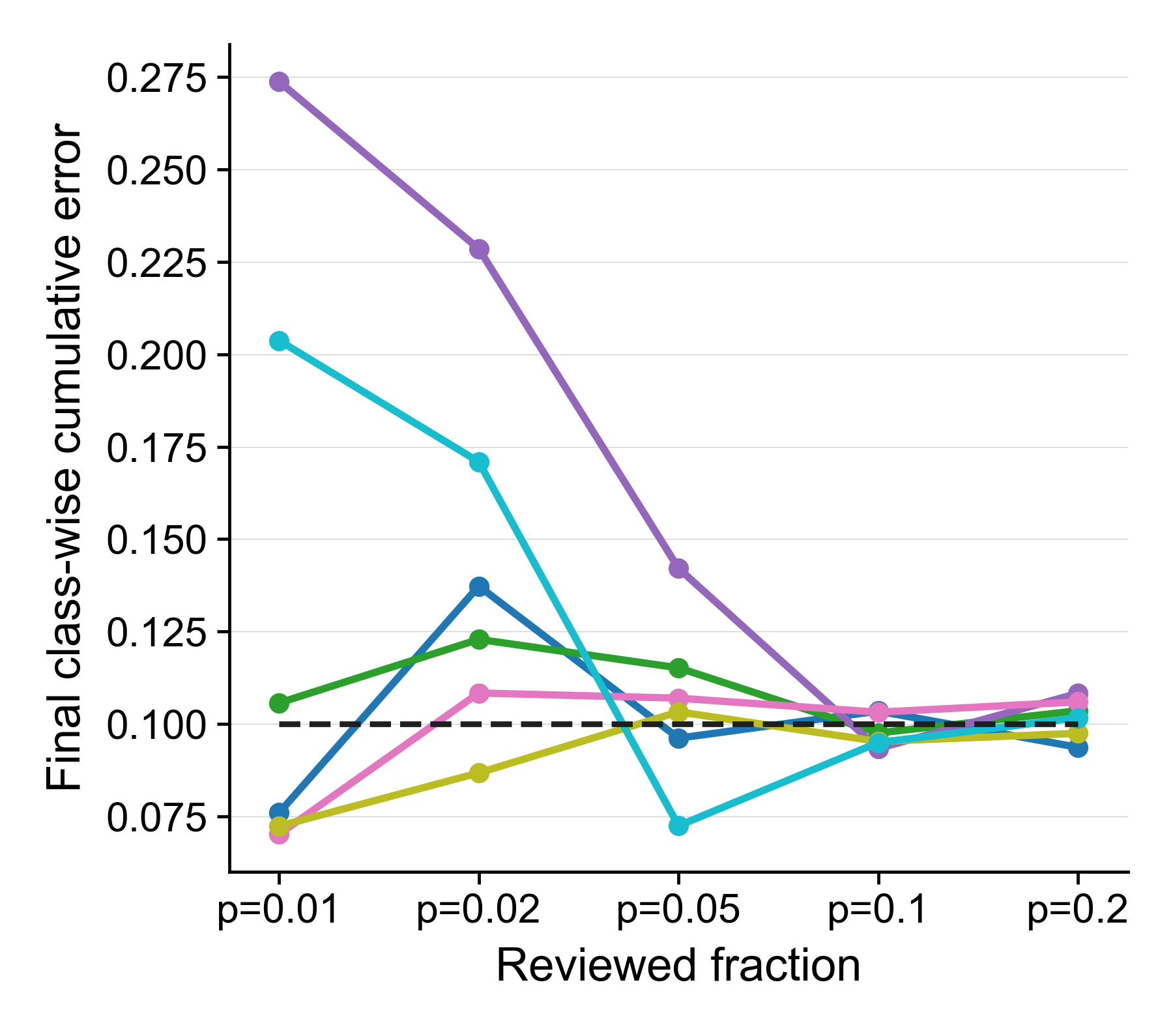}
        \caption{Error-guided, reviewed fraction}
        \label{fig:appendix_sensitivity_cumulative_error_errorg_a}
    \end{subfigure}
    \hfill
    \begin{subfigure}[t]{0.28\linewidth}
        \centering
        \includegraphics[width=\linewidth]{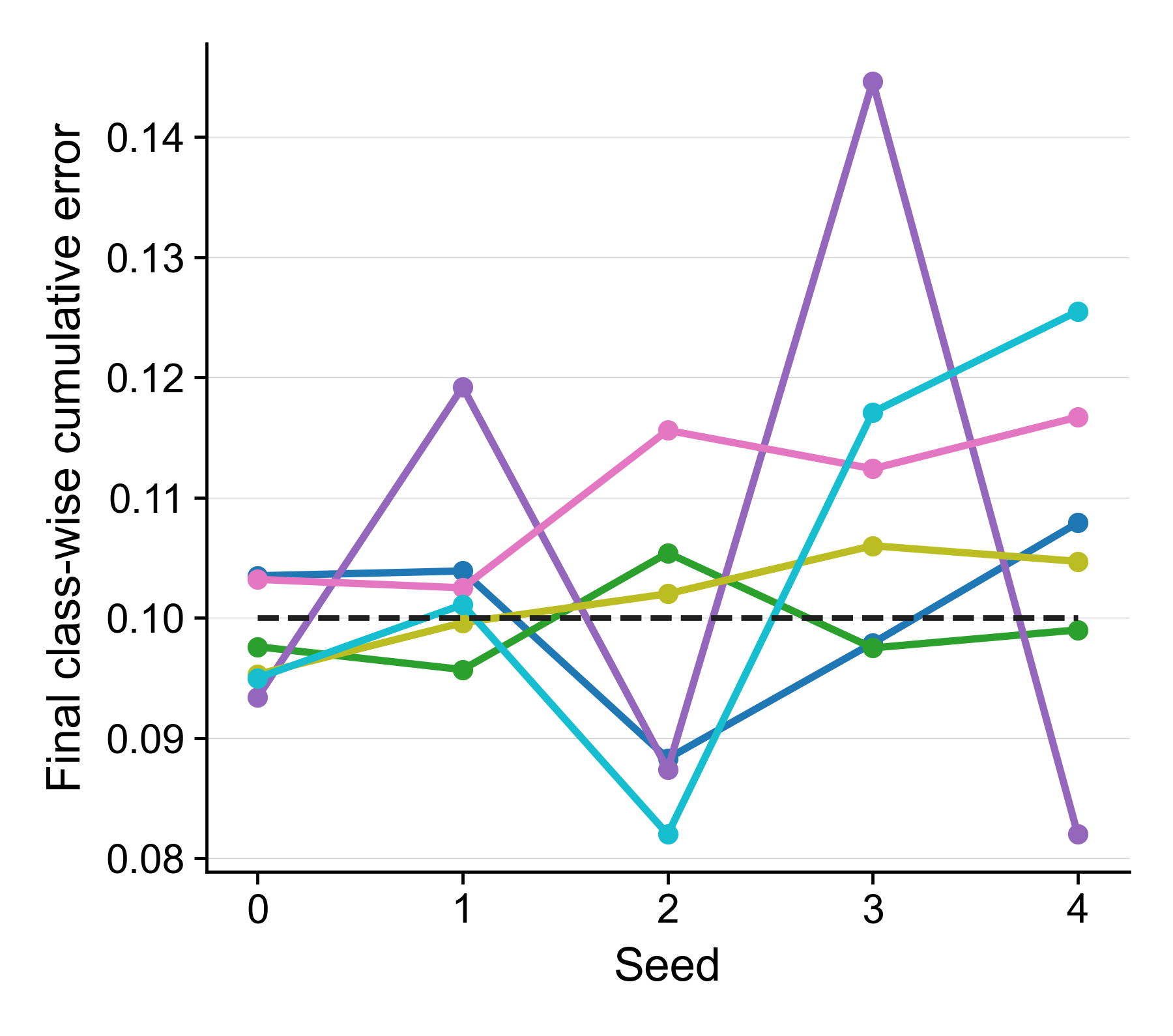}
        \caption{Error-guided, seed}
        \label{fig:appendix_sensitivity_cumulative_error_errorg_b}
    \end{subfigure}
    \hfill
    \begin{subfigure}[t]{0.28\linewidth}
        \centering
        \includegraphics[width=\linewidth]{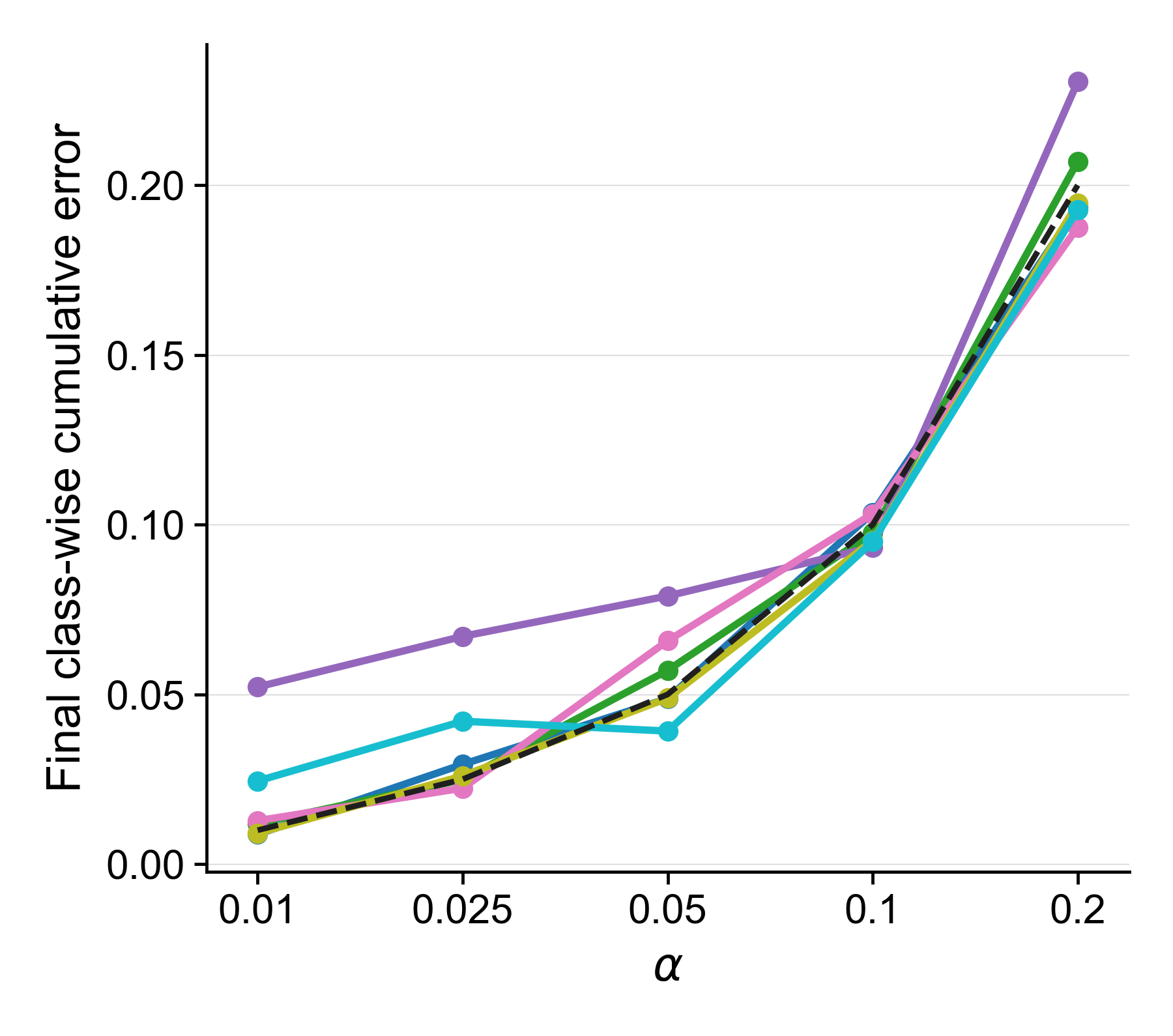}
        \caption{Error-guided, target error}
        \label{fig:appendix_sensitivity_cumulative_error_errorg_c}
    \end{subfigure}

    \vspace{1pt}

    \begin{subfigure}[t]{0.28\linewidth}
        \centering
        \includegraphics[width=\linewidth]{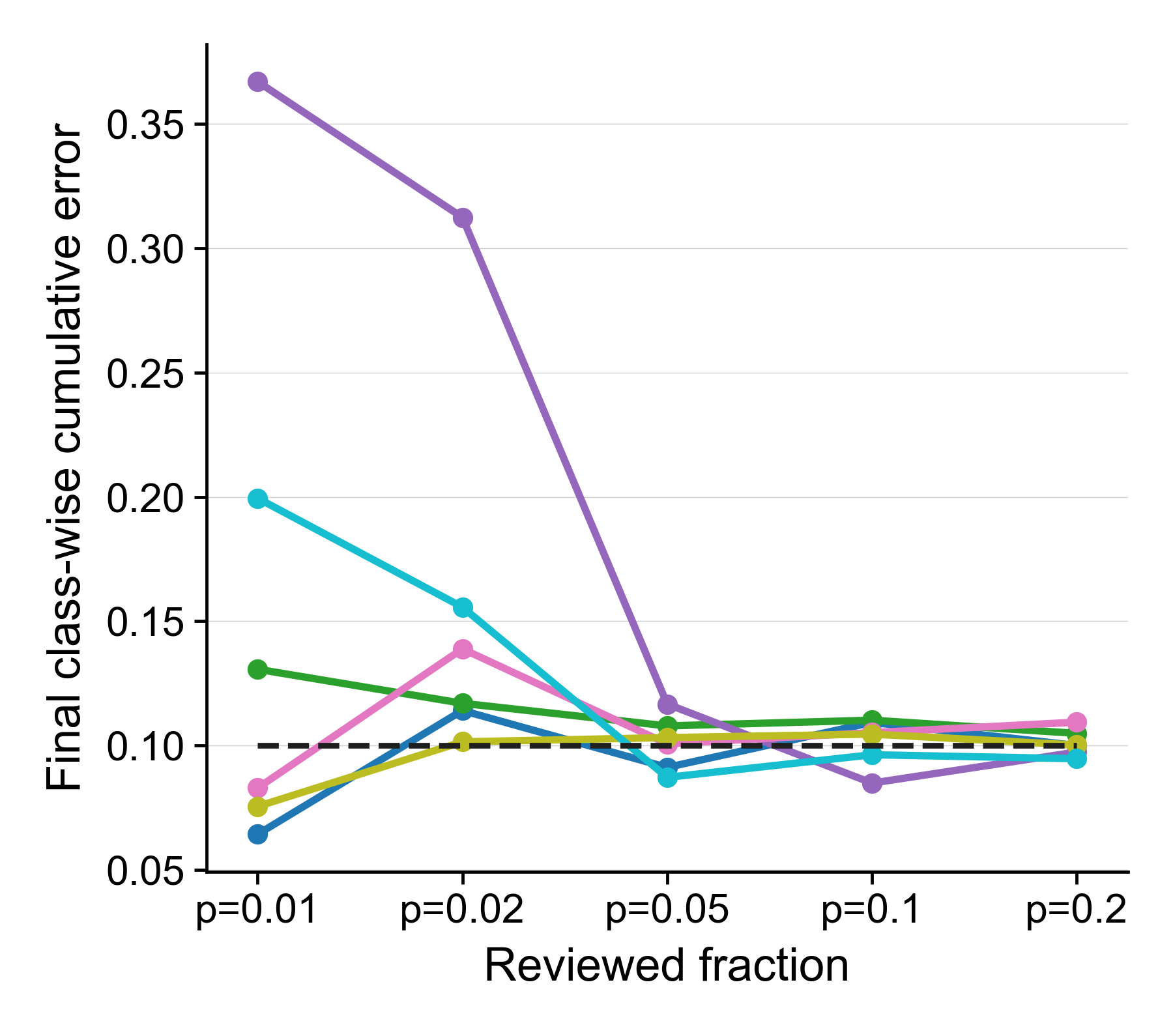}
        \caption{Adjacent, reviewed fraction}
        \label{fig:appendix_sensitivity_cumulative_error_adjacent_a}
    \end{subfigure}
    \hfill
    \begin{subfigure}[t]{0.28\linewidth}
        \centering
        \includegraphics[width=\linewidth]{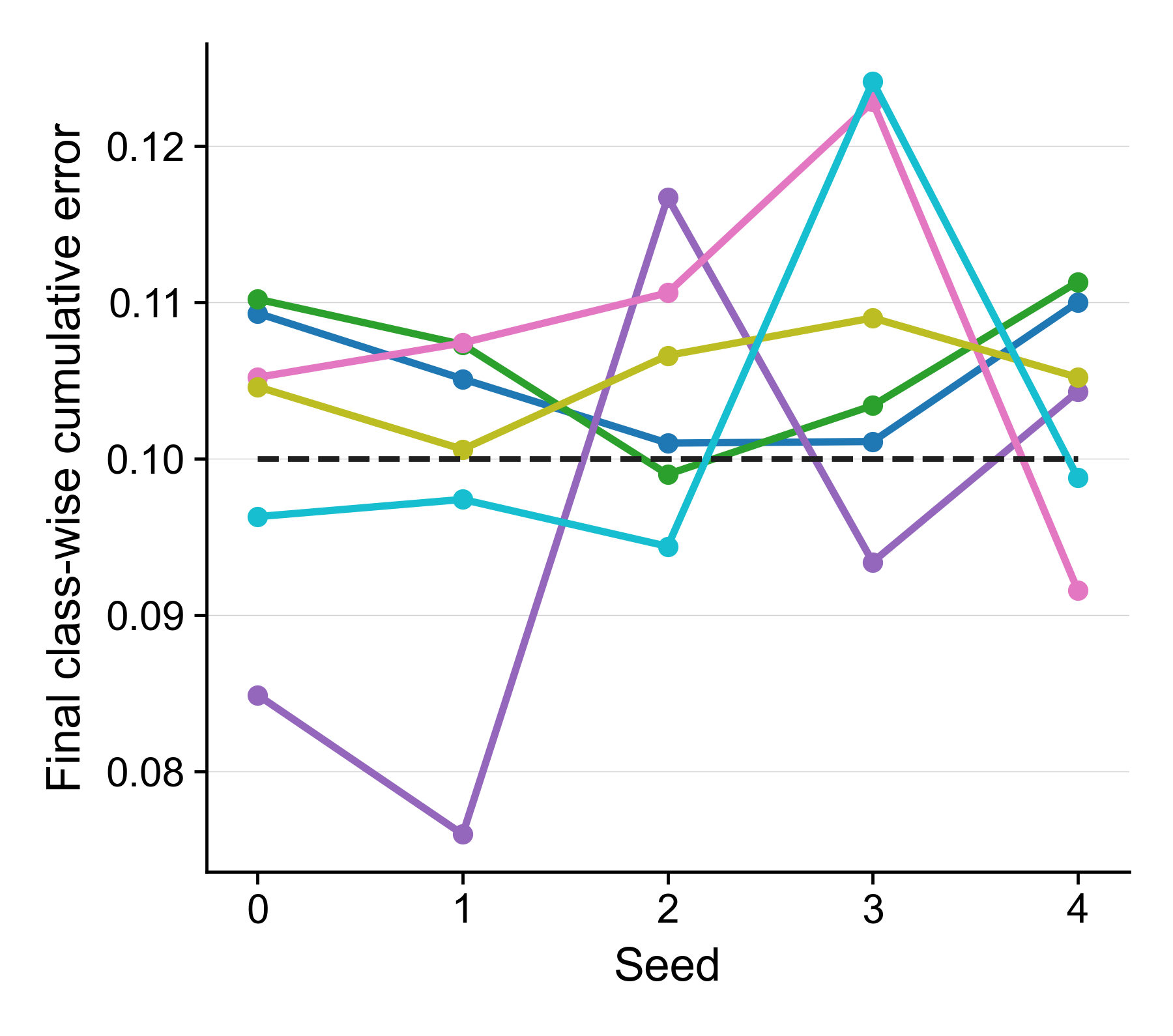}
        \caption{Adjacent, seed}
        \label{fig:appendix_sensitivity_cumulative_error_adjacent_b}
    \end{subfigure}
    \hfill
    \begin{subfigure}[t]{0.28\linewidth}
        \centering
        \includegraphics[width=\linewidth]{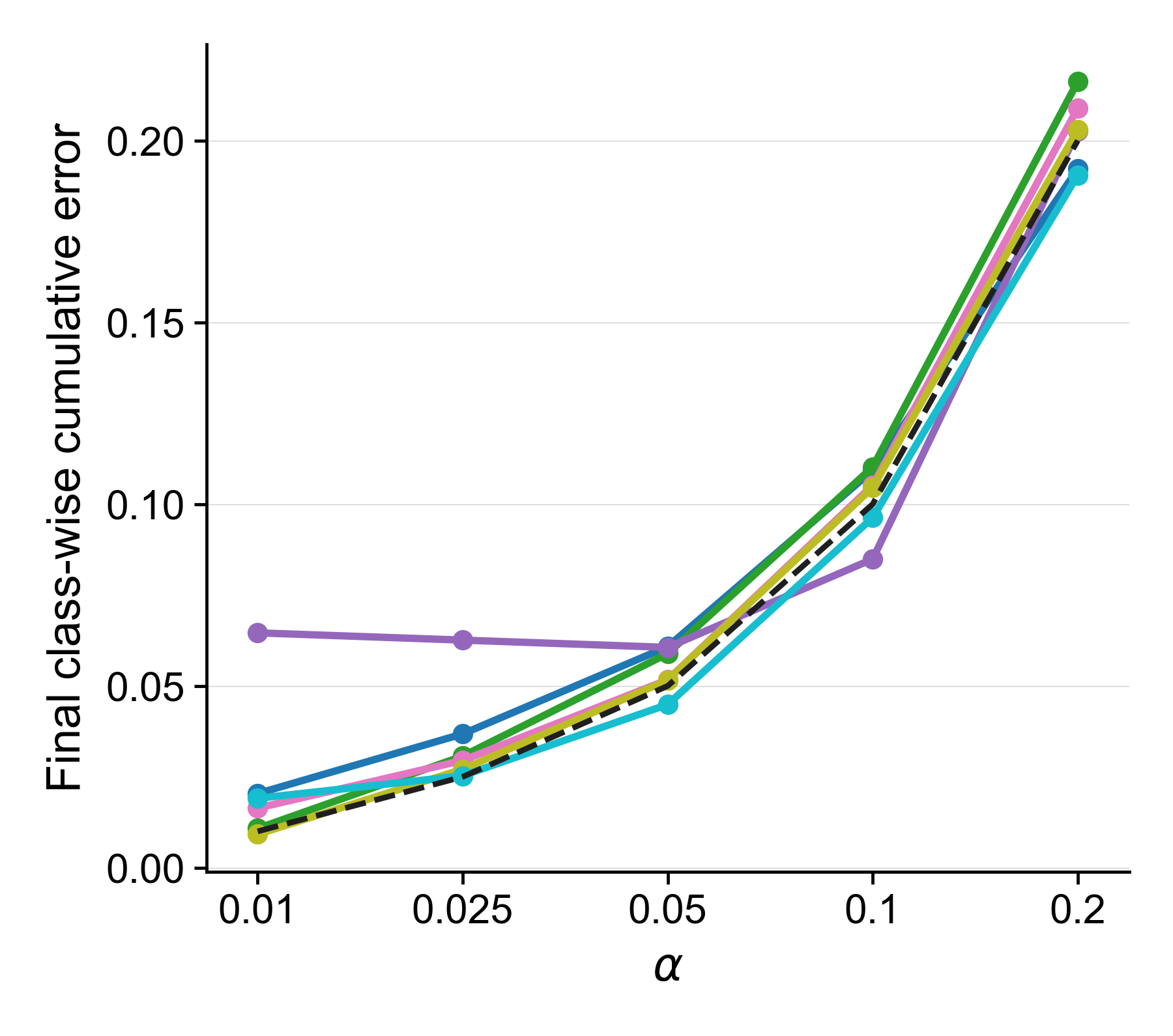}
        \caption{Adjacent, target error}
        \label{fig:appendix_sensitivity_cumulative_error_adjacent_c}
    \end{subfigure}

    \vspace{1pt}

    \begin{subfigure}[t]{0.28\linewidth}
        \centering
        \includegraphics[width=\linewidth]{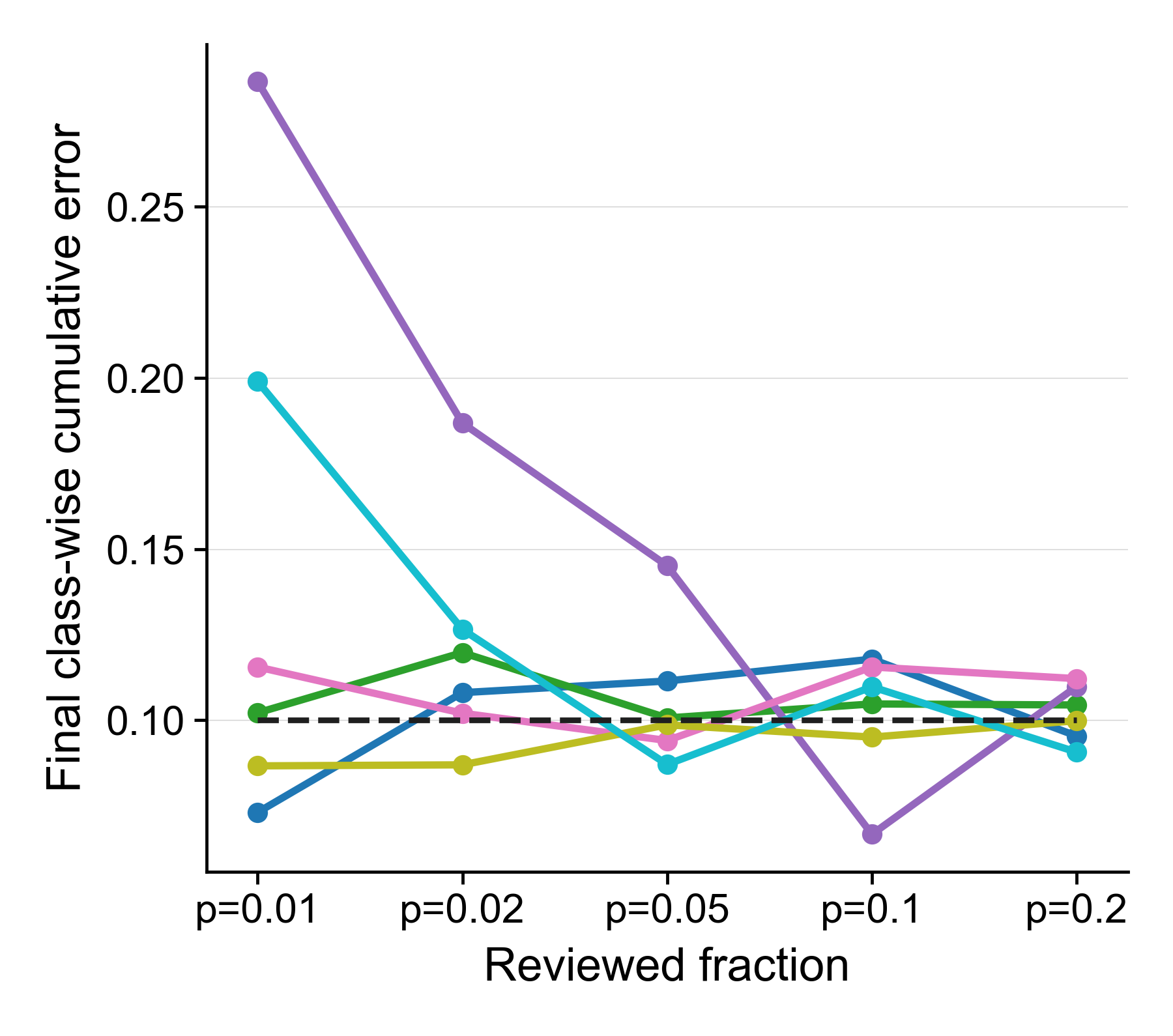}
        \caption{Random, reviewed fraction}
        \label{fig:appendix_sensitivity_cumulative_error_a}
    \end{subfigure}
    \hfill
    \begin{subfigure}[t]{0.28\linewidth}
        \centering
        \includegraphics[width=\linewidth]{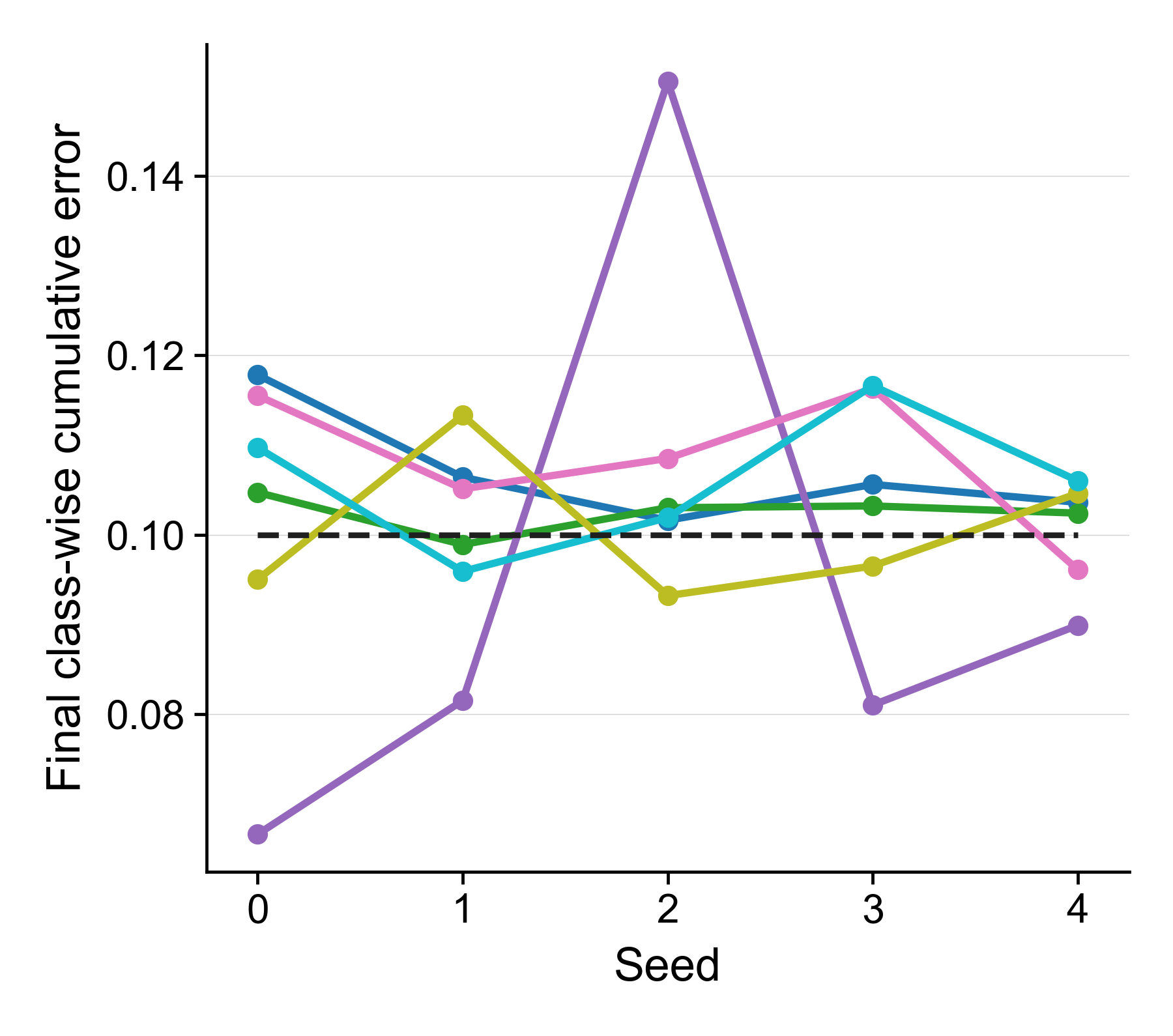}
        \caption{Random, seed}
        \label{fig:appendix_sensitivity_cumulative_error_b}
    \end{subfigure}
    \hfill
    \begin{subfigure}[t]{0.28\linewidth}
        \centering
        \includegraphics[width=\linewidth]{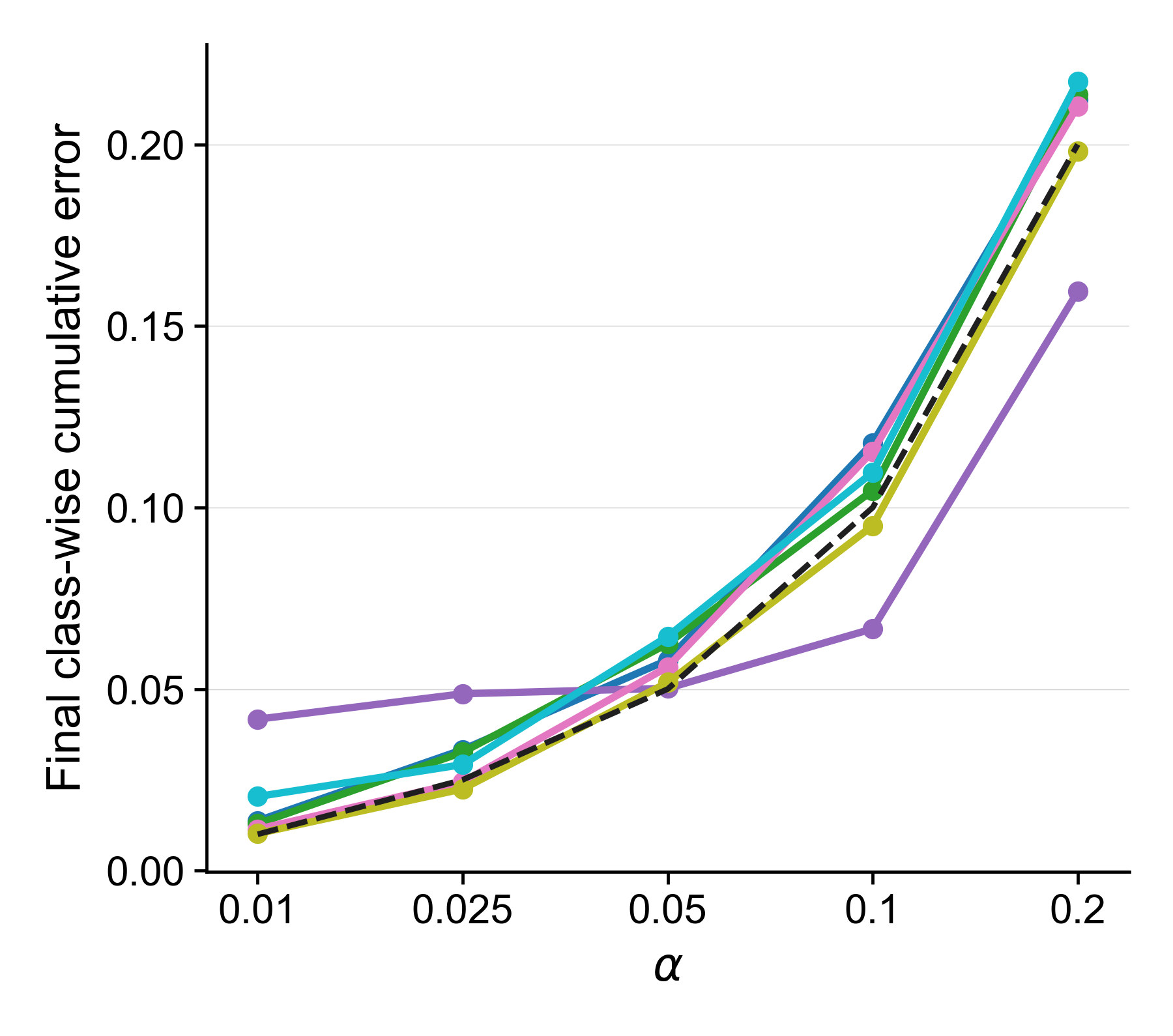}
        \caption{Random, target error}
        \label{fig:appendix_sensitivity_cumulative_error_c}
    \end{subfigure}

    \caption{Class-wise sensitivity analysis of final cumulative empirical error
    on pancreas tissue with the global score source. Rows correspond to the
    error-guided, adjacent, and random batch release schemes, while columns
    vary the reviewed fraction $p$, the random seed, and the target error level
    $\alpha$. Dashed horizontal lines indicate the target error level used in
    each setting, and percentages in the shared legend denote the empirical
    fraction of each cell type in the stream.}
    \label{fig:appendix_sensitivity_cumulative_error}
\end{figure}
\begin{figure}[p]
    \centering
    \setlength{\parskip}{0pt}
    \captionsetup[subfigure]{font=scriptsize,justification=centering}
    \includegraphics[width=0.86\linewidth]{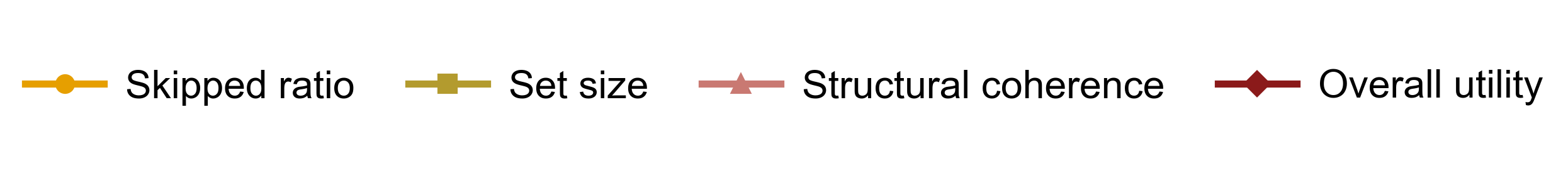}
    \vspace{0.2em}

    \begin{subfigure}[t]{0.28\linewidth}
        \centering
        \includegraphics[width=\linewidth]{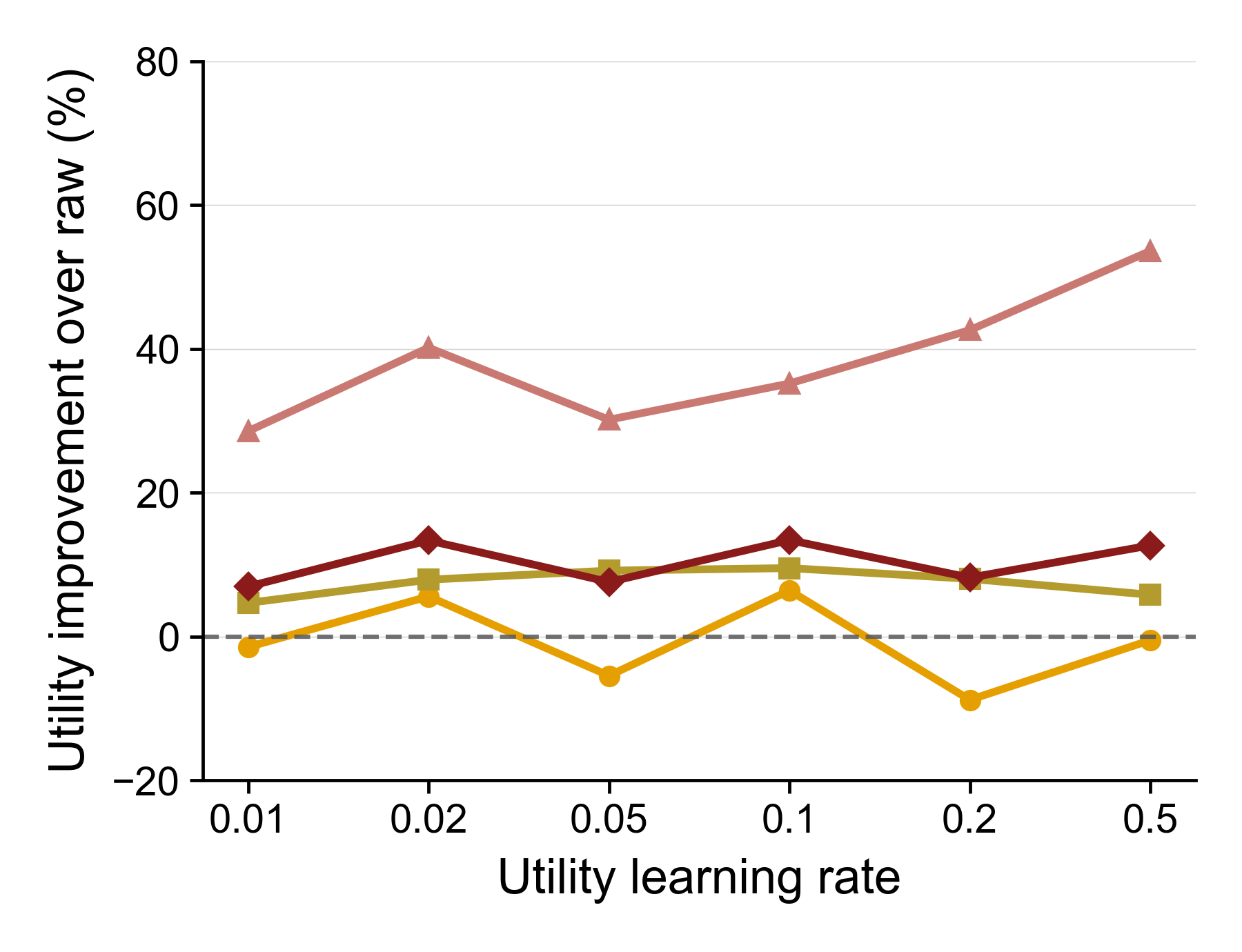}
        \caption{Error-guided, utility lr}
        \label{fig:appendix_sensitivity_utility_metrics_errorg_a}
    \end{subfigure}
    \hfill
    \begin{subfigure}[t]{0.28\linewidth}
        \centering
        \includegraphics[width=\linewidth]{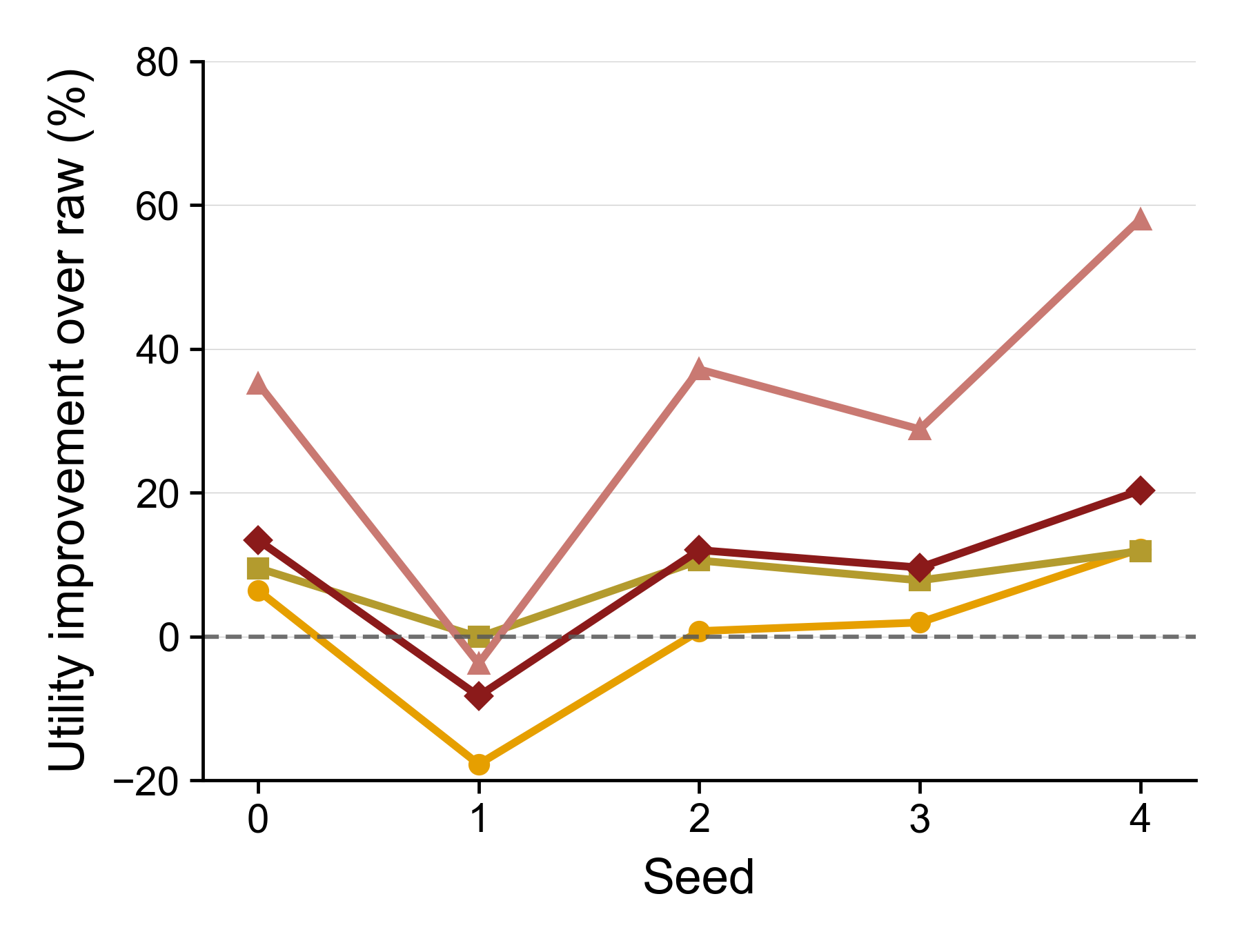}
        \caption{Error-guided, seed}
        \label{fig:appendix_sensitivity_utility_metrics_errorg_b}
    \end{subfigure}
    \hfill
    \begin{subfigure}[t]{0.28\linewidth}
        \centering
        \includegraphics[width=\linewidth]{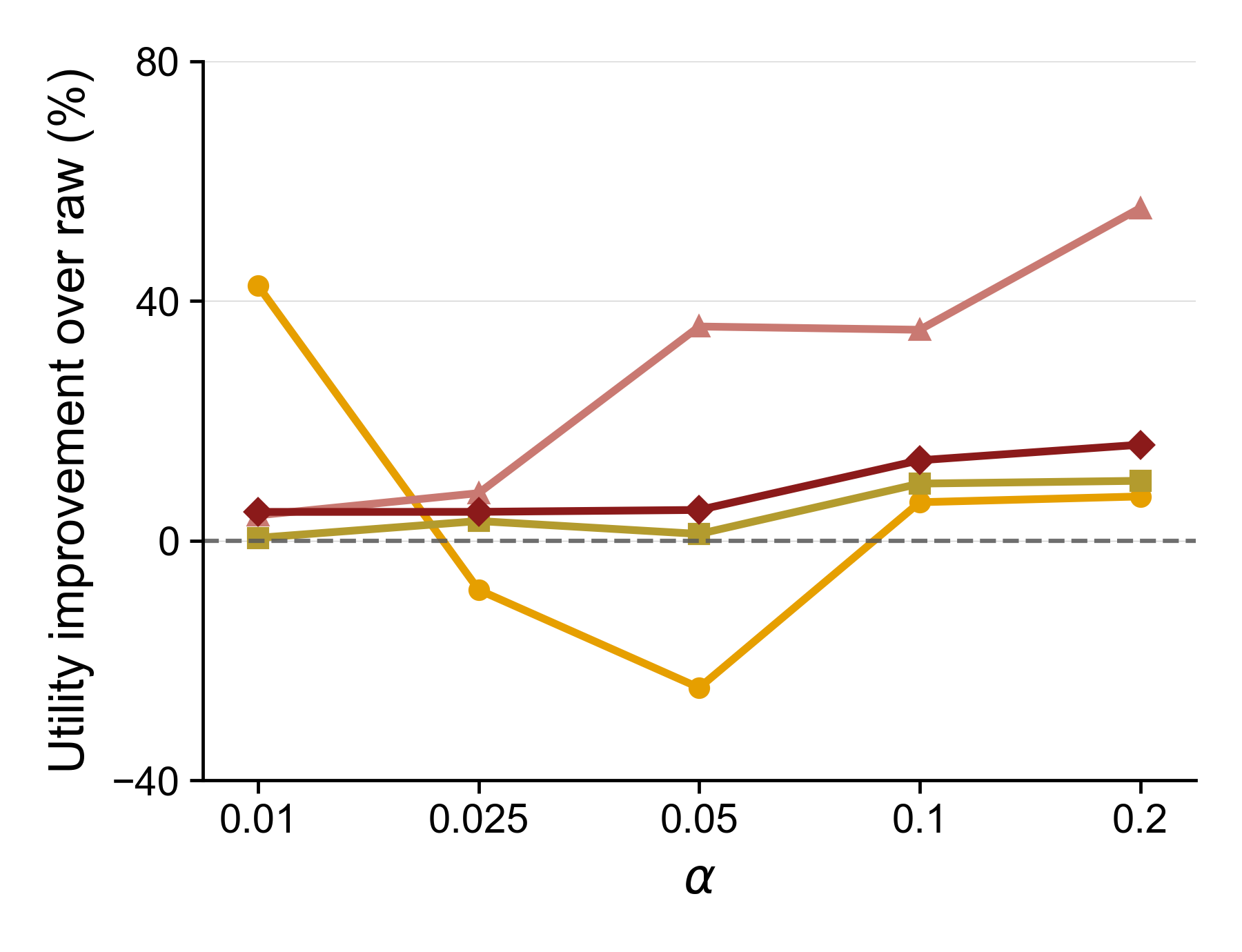}
        \caption{Error-guided, target error}
        \label{fig:appendix_sensitivity_utility_metrics_errorg_c}
    \end{subfigure}

    \vspace{1pt}

    \begin{subfigure}[t]{0.28\linewidth}
        \centering
        \includegraphics[width=\linewidth]{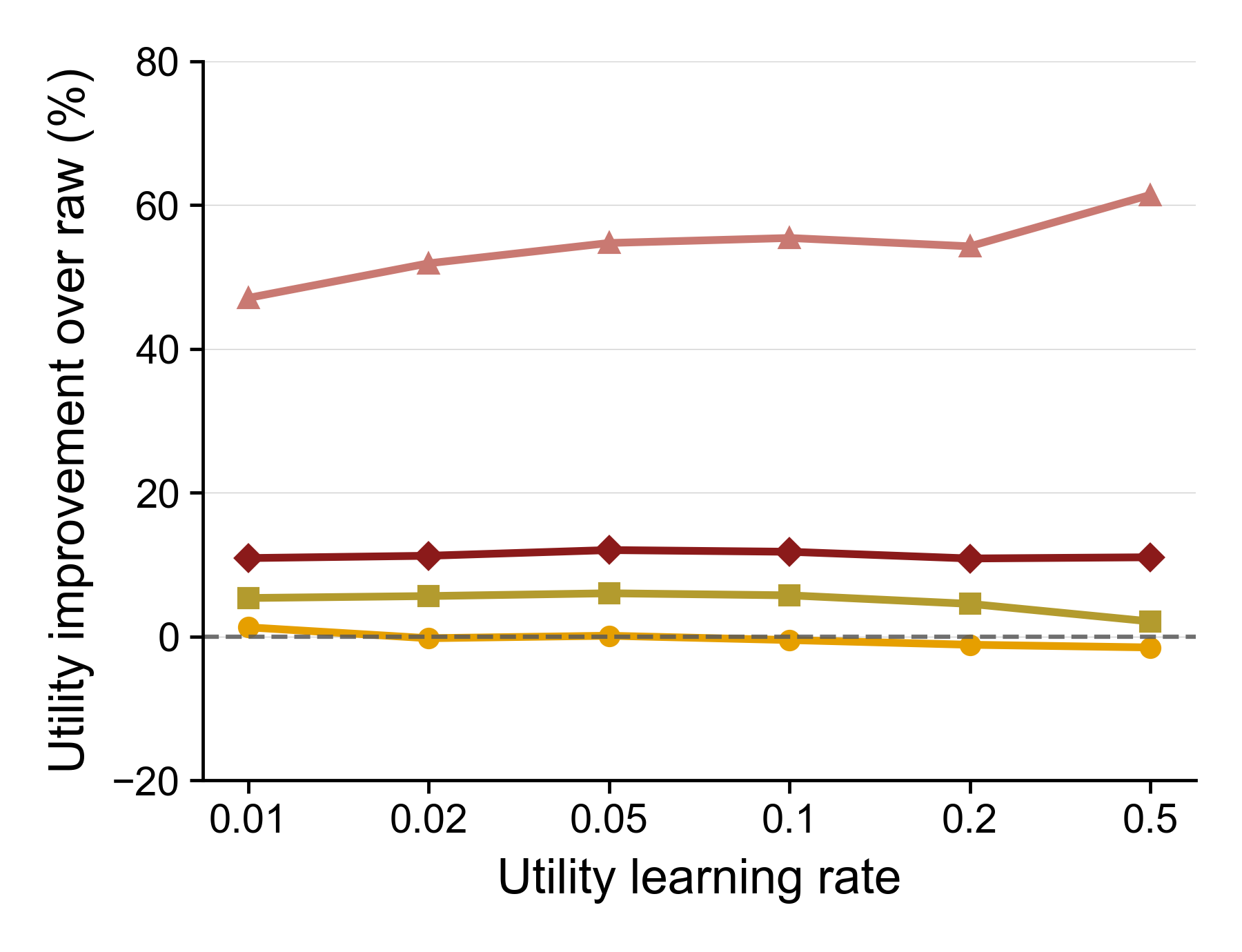}
        \caption{Adjacent, utility lr}
        \label{fig:appendix_sensitivity_utility_metrics_adjacent_a}
    \end{subfigure}
    \hfill
    \begin{subfigure}[t]{0.28\linewidth}
        \centering
        \includegraphics[width=\linewidth]{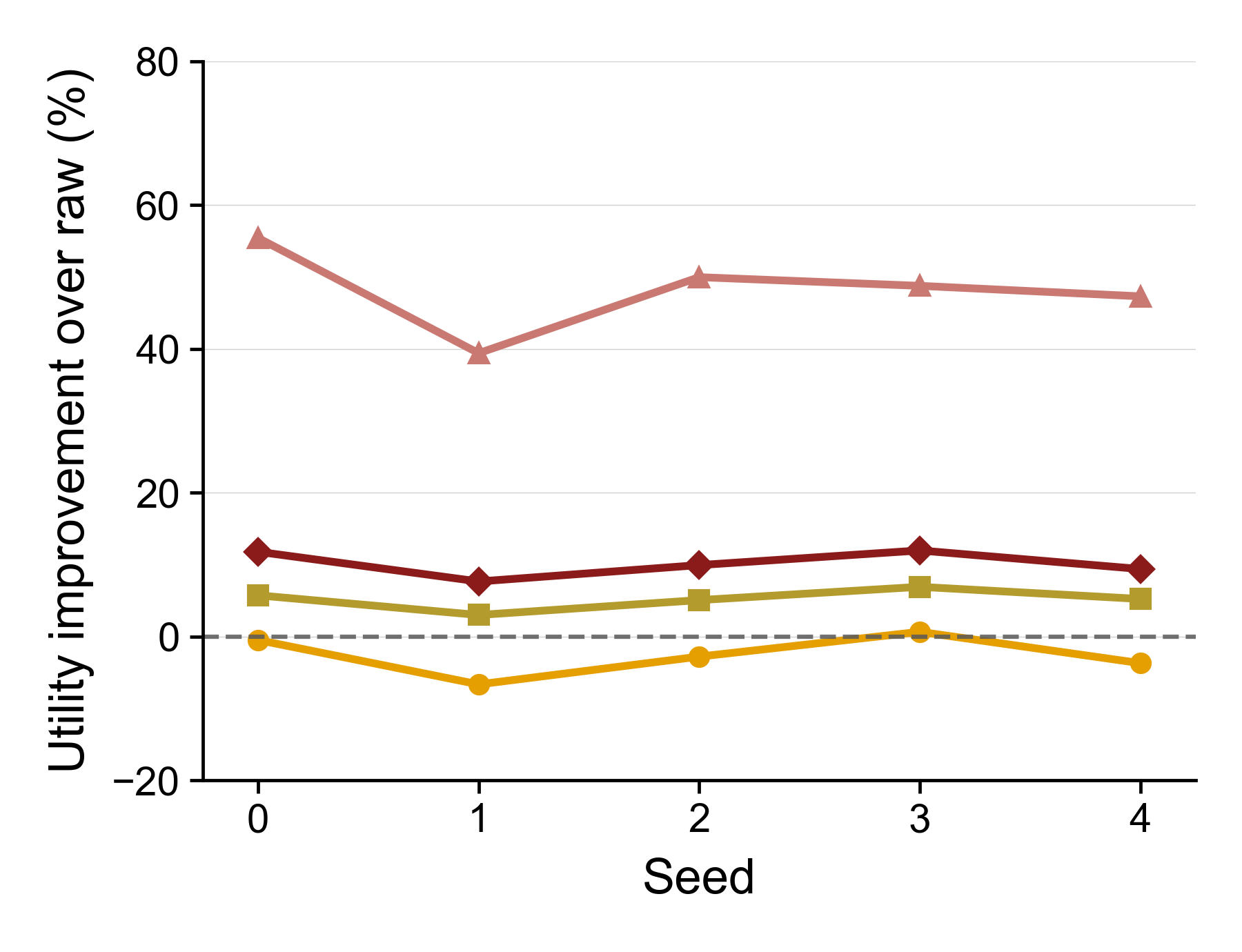}
        \caption{Adjacent, seed}
        \label{fig:appendix_sensitivity_utility_metrics_adjacent_b}
    \end{subfigure}
    \hfill
    \begin{subfigure}[t]{0.28\linewidth}
        \centering
        \includegraphics[width=\linewidth]{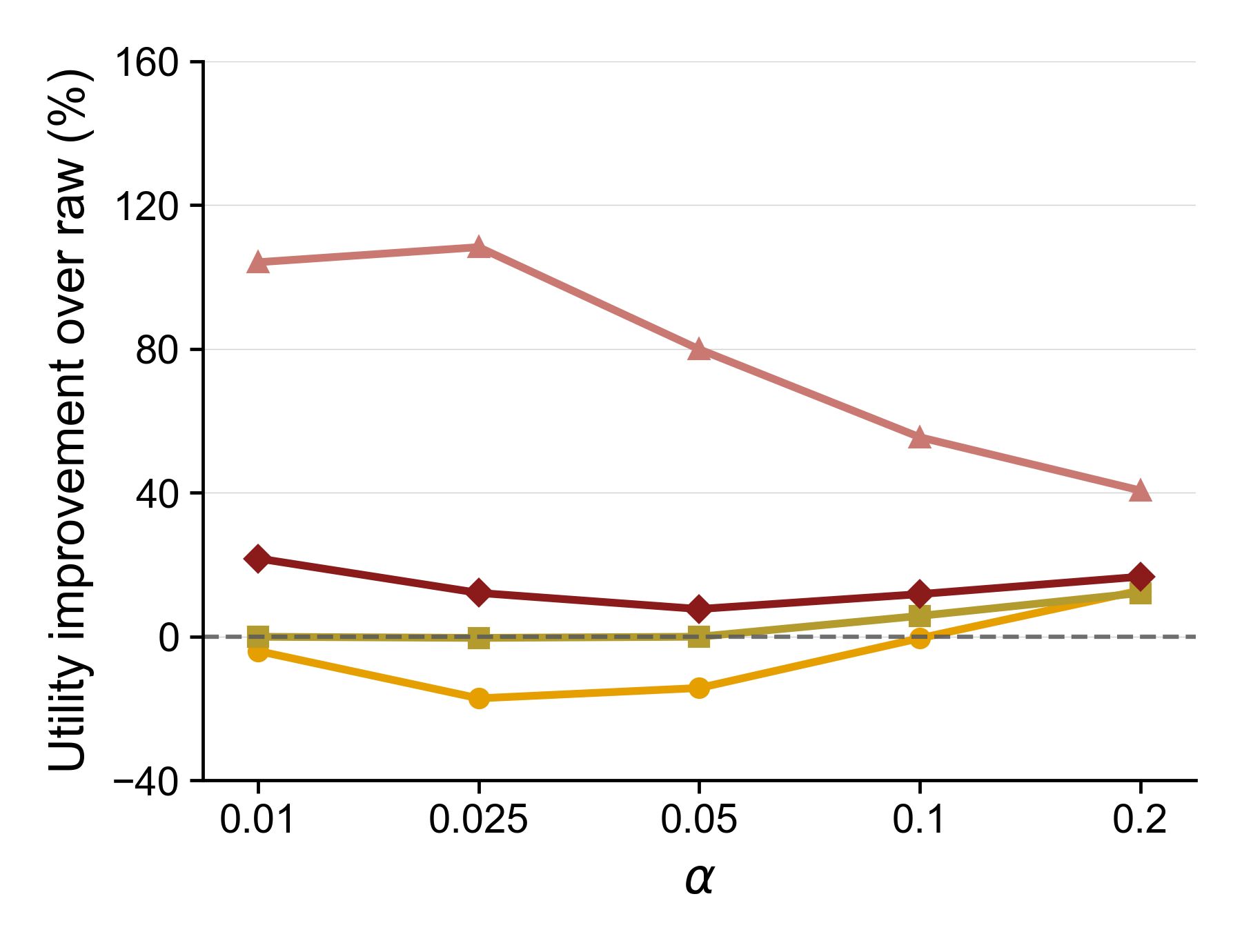}
        \caption{Adjacent, target error}
        \label{fig:appendix_sensitivity_utility_metrics_adjacent_c}
    \end{subfigure}

    \vspace{1pt}

    \begin{subfigure}[t]{0.28\linewidth}
        \centering
        \includegraphics[width=\linewidth]{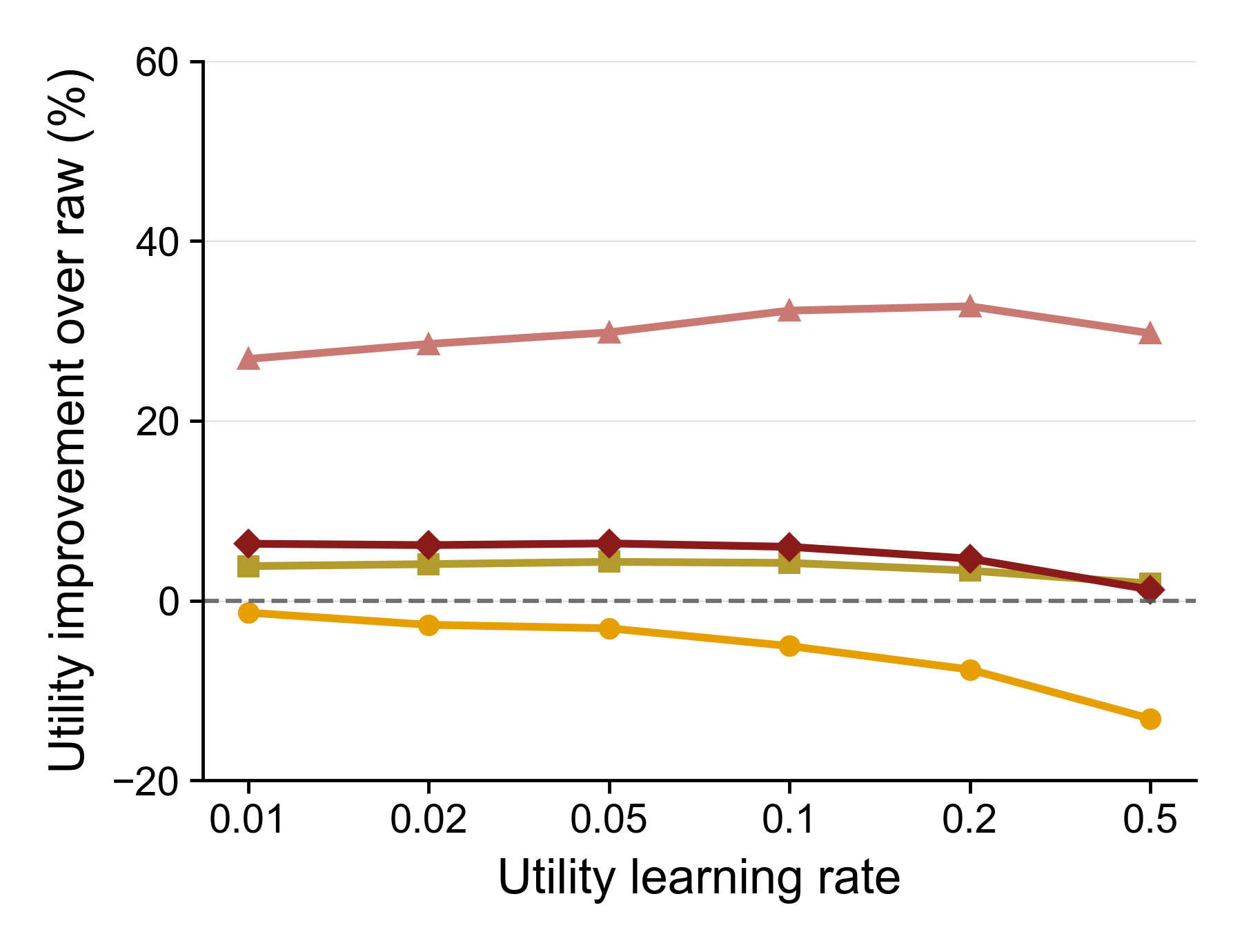}
        \caption{Random, utility lr}
        \label{fig:appendix_sensitivity_utility_metrics_a}
    \end{subfigure}
    \hfill
    \begin{subfigure}[t]{0.28\linewidth}
        \centering
        \includegraphics[width=\linewidth]{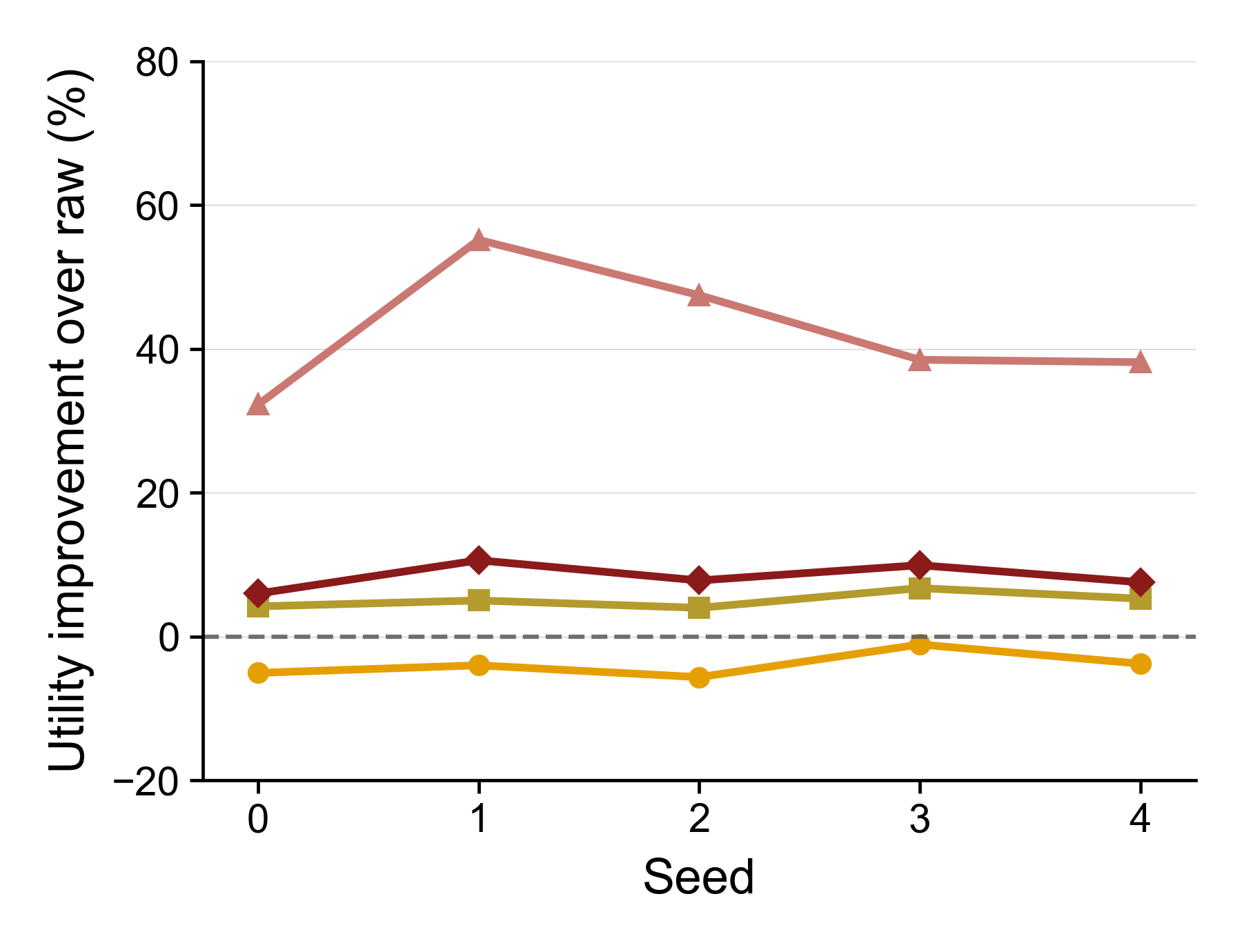}
        \caption{Random, seed}
        \label{fig:appendix_sensitivity_utility_metrics_b}
    \end{subfigure}
    \hfill
    \begin{subfigure}[t]{0.28\linewidth}
        \centering
        \includegraphics[width=\linewidth]{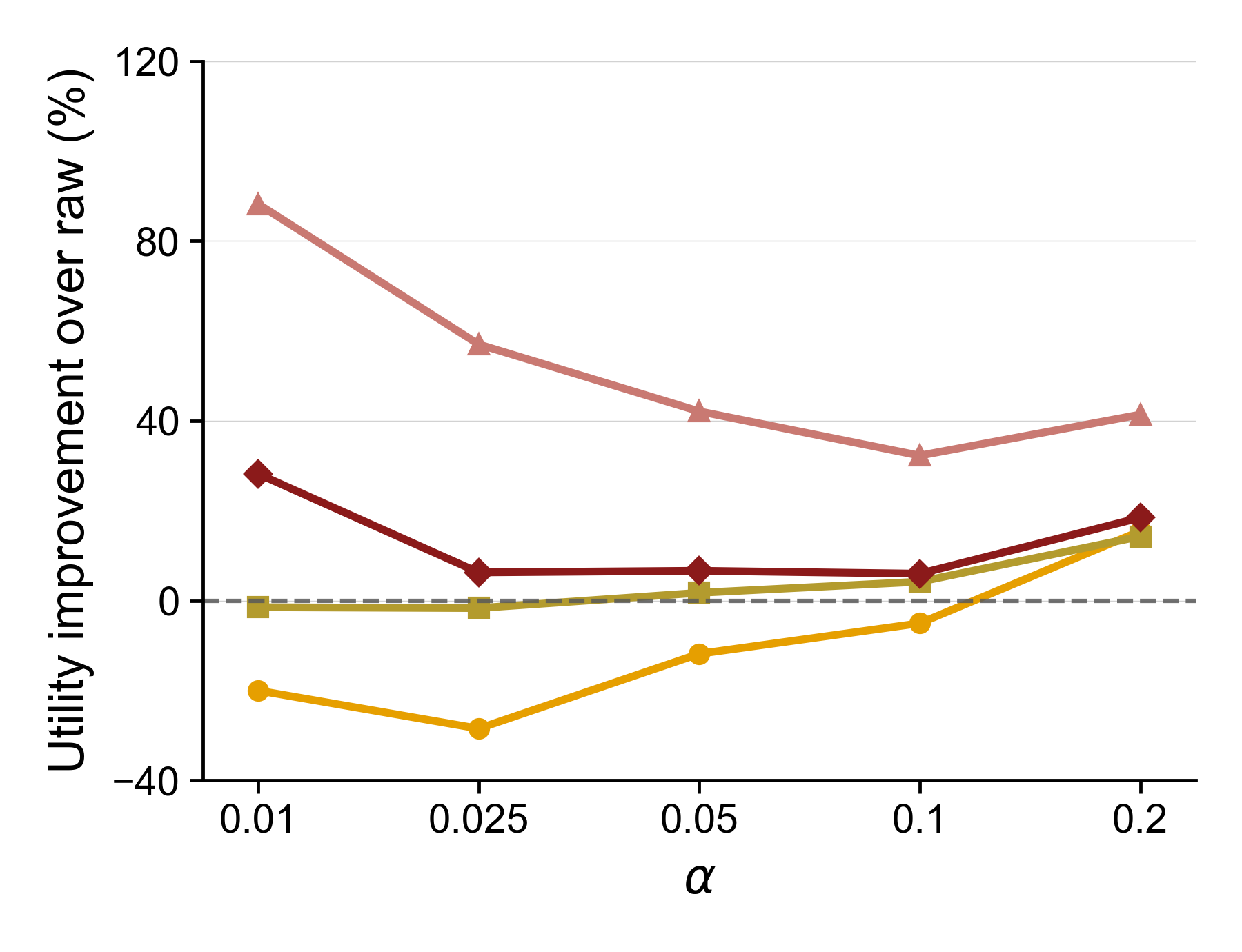}
        \caption{Random, target error}
        \label{fig:appendix_sensitivity_utility_metrics_c}
    \end{subfigure}

    \caption{Sensitivity analysis of utility metrics on the six-class pancreas
    annotation task with the global score source. Rows correspond to the error-guided,
    adjacent, and random batch release schemes, while columns vary the
    utility learning rate $\eta$, the random seed, and the target error level
    $\alpha$. Each panel compares the utility metrics obtained from the
    utility-aware transformed score and the raw score. The shared legend
    identifies the four utility metrics.}
    \label{fig:appendix_sensitivity_utility_metrics}
\end{figure}
\afterpage{\clearpage}

Figure~\ref{fig:appendix_sensitivity_cumulative_error} can be further
interpreted through Proposition~\ref{prop:convergence_rates}, which implies
that the deviation of the class-wise cumulative empirical error from the target
level decreases at the inverse-square-root rate in the total number of cells
from that class. Therefore, for a fixed finite stream, rarer cell types are
expected to show greater sensitivity. This accounts for the relatively large
fluctuations for endocrine cells, which constitute only $1.95\%$ of the stream,
and also for the milder instability of T cells and endothelial cells, which are two relatively
scarce classes in this experiment.  Note that the marginal
cumulative-error trajectories in
Section~\ref{subsec:sensitivity_cumulative_error} remain stable, and the
remaining class-wise curves in
Figure~\ref{fig:appendix_sensitivity_cumulative_error} do not show comparable
fluctuations. Thus, the class-wise fluctuations reflect finite-sample
variability rather than a failure of the empirical error-control pattern.

\subsubsection{Utility metrics}\label{subsubsec:sensitivity_utility_metrics}

We further examine the stability of the utility improvements on the six-class
pancreas annotation task, as summarized in
Figure~\ref{fig:appendix_sensitivity_utility_metrics}.

For each batch release scheme and parameter setting, we compare the prediction
sets produced by the utility-aware transformed score with those obtained from
the raw scores. The three columns vary the utility learning rate $\eta$, the
random seed, and the target error level $\alpha$, and the three rows correspond
to the error-guided, adjacent, and random batch release schemes. The plotted
value is the percentage improvement over the raw-score baseline, computed from
the cumulative stream-level utility metrics after processing the full
annotation stream. Overall utility is the weighted aggregate of skipped ratio,
set size, and structural coherence, using
$(\lambda_{\text{skip}}, \lambda_{\text{size}}, \lambda_{\text{struct}})
= (20, 1, 5)$ as in Appendix~\ref{subsec:appendix_experimental_details}; its
percentage improvement is normalized by the coordinate-wise worst-case six-class reference
$(0,6,0)$, corresponding to zero skipped ratio, full set size, and zero
structural coherence.

Figure~\ref{fig:appendix_sensitivity_utility_metrics} shows that the utility
module consistently improves overall utility over the raw-score baseline in
almost every considered case. The component-wise gains are more uneven: structural
coherence improves substantially in most settings, whereas skipped ratio and
set size can change more modestly or vary with the sensitivity parameter. This
is expected because the aggregate utility balances three competing criteria,
and the relative emphasis can be adjusted through the utility weights.


\section{Discussions on stepsize selection}\label{sec:discussions}


The main experiments use a constant class-wise step size $\gamma_k$ in the
online update \eqref{eq:alpha-update}. This choice is aligned with the
distribution-free guarantee established in Section~\ref{sec:theory}: for each
class $k$, the online level is updated as
\[
    \alpha_{k,t+1}
    =
    \alpha_{k,t}
    +
    \gamma_k
    \sum_{x\in A_{k,t}}
    \bigl(\alpha-\operatorname{err}_{k,t}(x)\bigr),
\]
where $\gamma_k$ is fixed over time and selected according to the class
frequency rule reported in Appendix~\ref{subsec:appendix_experimental_details}.

Decaying stepsize schedules provide a natural alternative. For the standard
adaptive conformal prediction setting, \citet{angelopoulos2024online}
show that step sizes of order $\gamma_t\propto t^{-1/2-\varepsilon}$, with
$0<\varepsilon<1/2$, can retain long-run convergence of cumulative error while improving the
stability of local error fluctuations. However, their local convergence theory
is developed for the i.i.d.\ setting, and therefore should not be read as a
general local-error guarantee under distribution shift. A complementary
analysis of adaptive conformal prediction under distribution
drift highlights the same trade-off from the
non-i.i.d.\ perspective: smaller or decaying step sizes can stabilize the
calibration-quantile trajectory within relatively stationary segments, but may
adapt more slowly after distributional changes \citep{liang2026optimal}. Thus,
decaying step sizes are a
reasonable stability-oriented variant, but not a uniformly preferable
replacement for the constant step sizes used in our main experiments.

\begin{figure}[!htbp]
    \centering
    \includegraphics[width=0.95\linewidth]{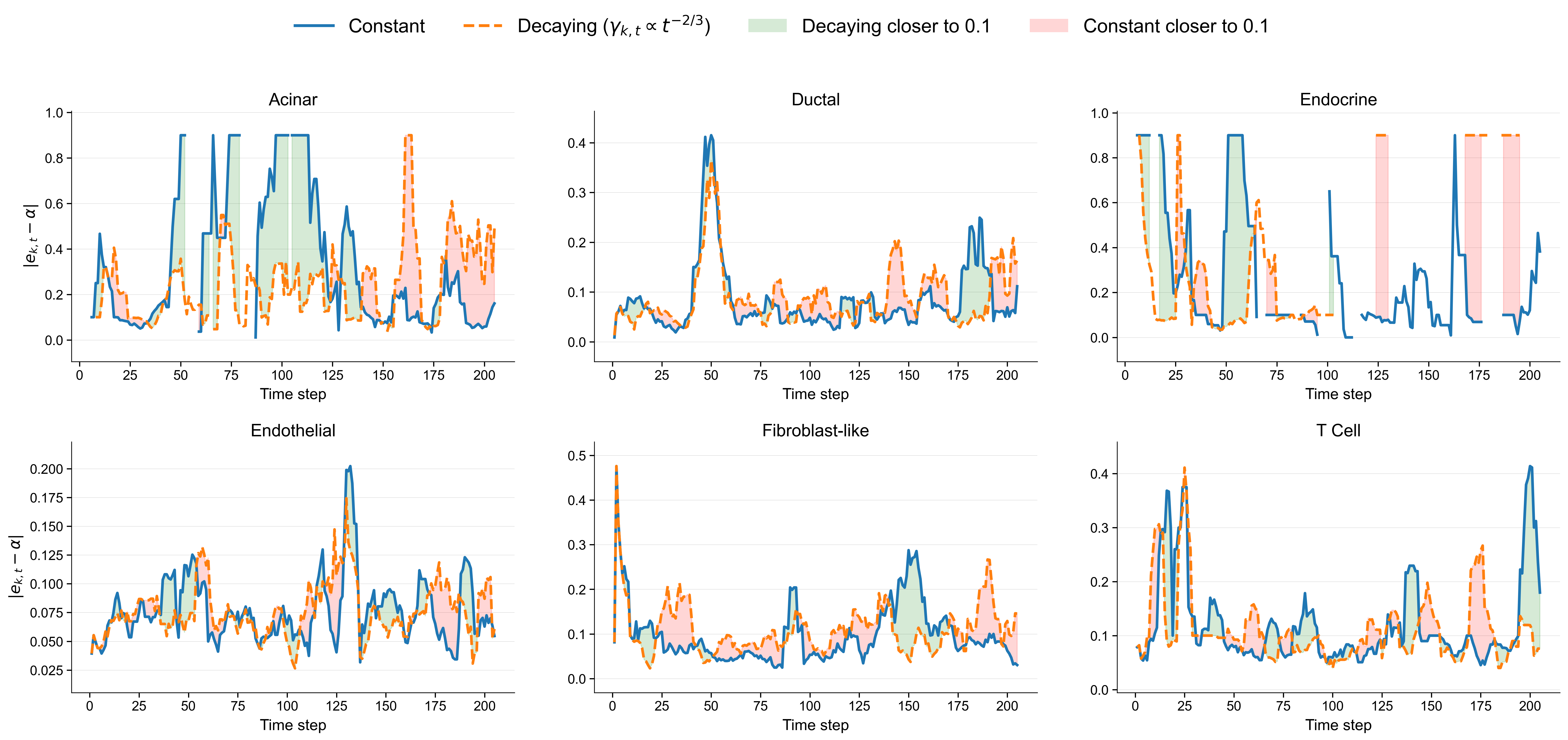}
    \caption{Comparison between constant and decaying class-wise step sizes on
    the pancreas experiment with the global NuClass score. The curves report
    per-batch all-cell deviations $|\mathrm{BatErr}_{k,t}^D-\alpha|$ for each class under
    the constant schedule used in the main experiments and the decaying
    schedule $\gamma_{k,t}=\gamma_k t^{-2/3}$. Shaded intervals indicate which
    schedule is closer to the target error level $\alpha=0.1$ at each time.}
    \label{fig:appendix_decaying_stepsize}
\end{figure}

To examine this issue, we rerun the pancreas experiment with the global NuClass
score using the same initial class-wise scales $\gamma_k$ as in the main
experiment, but replace the constant update by
\[
    \gamma_{k,t}=\gamma_k t^{-2/3}.
\]
Figure~\ref{fig:appendix_decaying_stepsize} compares the resulting per-batch
all-cell deviations $|\mathrm{BatErr}_{k,t}^D-\alpha|$ with those obtained from the
constant stepsize update. The comparison shows a mixed pattern: the decaying
schedule is closer to the target in some intervals and classes, while the
constant schedule is closer in others. This is consistent with the
stability--adaptation trade-off above, and explains why we keep the constant
schedule as the default while reporting the decaying schedule as an empirical
variant.

\begin{figure}[!htbp]
    \centering
    \includegraphics[width=0.95\linewidth]{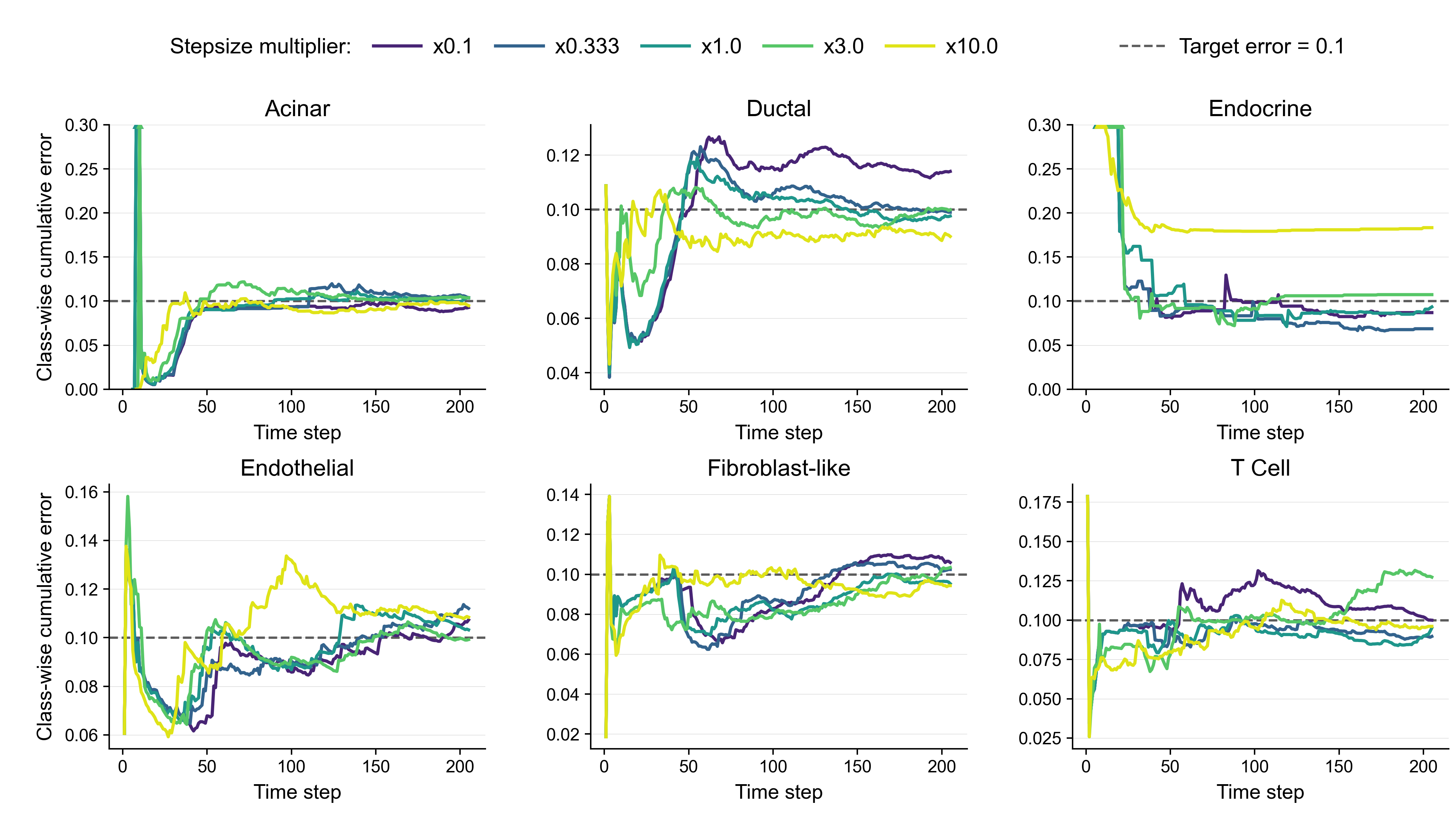}
    \caption{Sensitivity of cumulative class-wise error trajectories to the
    scale of the constant step sizes on the pancreas experiment with the global
    NuClass score. Each curve uses the same class-wise stepsize rule as in the
    main experiments, multiplied by the factor shown in the legend. The dashed
    line marks the target error level $\alpha=0.1$.}
    \label{fig:appendix_stepsize_sensitivity}
\end{figure}

In practice, the numerical choice of $\gamma_k$ is usually guided by empirical
considerations rather than by an exact calculation. Although
Theorem~\ref{thm:long_run_validity} shows that the cumulative class-wise error
converges asymptotically for any fixed positive step size, the finite-stream
behavior still depends on the scale of $\gamma_k$. To illustrate this
sensitivity, Figure~\ref{fig:appendix_stepsize_sensitivity} compares the
constant stepsize update after multiplying the default class-wise step sizes
by factors ranging from $0.1$ to $10$. Across this range,
class-wise errors are generally pulled toward the target level $\alpha=0.1$,
although the finite-sample trajectories differ by class. The main exception is
the endocrine class under large stepsize multipliers, where the finite-stream
trajectory remains visibly above the target for longer. This behavior is
consistent with the scarce-class instability discussed in
Appendix~\ref{subsubsec:sensitivity_cumulative_error}: endocrine cells account
for only a small fraction of the stream, so their class-wise cumulative error is
more sensitive in finite samples. This supports using the reported $\gamma_k$
values as a pragmatic default rather than a deliberately optimal tuning.

\FloatBarrier

\section{Technical proofs}\label{sec:technical_proof}

\subsection{Proof of Proposition~\ref{prop:reviewed_error}}\label{proof:reviewed_error}

Telescoping the update rule~\eqref{eq:alpha-update} from the initial value
$\alpha_{k,1}=\alpha$,
\[
\alpha_{k,T+1}
=
\alpha
+
\gamma_k\sum_{t=1}^T\sum_{x\in A_{k,t}}\bigl(\alpha-\mathrm{err}_{k,t}(x)\bigr).
\]
Rearranging, $\sum_{t=1}^T\sum_{x\in A_{k,t}}\mathrm{err}_{k,t}(x)
-\alpha\sum_{t=1}^T|A_{k,t}|
=(\alpha-\alpha_{k,T+1})/\gamma_k$.
Dividing both sides by $\sum_{t=1}^T|A_{k,t}|$ and taking absolute values yields
the exact identity
\begin{equation}\label{eq:D1}\tag{E.1}
\bigl|{\text{CovErr}_{k,T}^{A}}-\alpha\bigr|
=
\frac{|\alpha_{k,T+1}-\alpha|}{\gamma_k\sum_{t=1}^T|A_{k,t}|}.
\end{equation}
It remains to bound $|\alpha_{k,T+1}-\alpha|$.
Set ${\bar M_k:=\sup_{t\geq1}|A_{k,t}|}$.
Since $\mathrm{err}_{k,t}(x)\in\{0,1\}$, the maximum single-step change satisfies
\begin{equation}\label{eq:D2}\tag{E.2}
|\alpha_{k,t+1}-\alpha_{k,t}|
=
\gamma_k\Bigl|\sum_{x\in A_{k,t}}\bigl(\alpha-\mathrm{err}_{k,t}(x)\bigr)\Bigr|
\le
\gamma_k|A_{k,t}|\max\{\alpha,1-\alpha\}
\le
\gamma_k\bar M_k\max\{\alpha,1-\alpha\}.
\end{equation}
We claim that for all $t\ge1$,
\begin{equation}\label{eq:D3}\tag{E.3}
\alpha_{k,t}\in\bigl[-\gamma_k\bar M_k(1-\alpha),\;1+\gamma_k\bar M_k\alpha\bigr].
\end{equation}
We establish the lower bound; the upper bound is symmetric. Fix $t\ge1$.
If $\alpha_{k,t}\ge0$, the lower bound is immediate. Otherwise, let $t'<t$
be the last time with $\alpha_{k,t'}\ge0$, which exists since
$\alpha_{k,1}=\alpha\ge0$. By~\eqref{eq:D2},
\[
\alpha_{k,t'+1}
\ge
\alpha_{k,t'}-\gamma_k\bar M_k(1-\alpha)
\ge
-\gamma_k\bar M_k(1-\alpha).
\]
If $\alpha_{k,s}<0$ for some $s$ with $t'<s\le t$, the quantile
convention~\eqref{eq:quantile} sets $\hat q_{k,s}(\alpha_{k,s})=-\infty$, so
class $k$ belongs to every prediction set and $\mathrm{err}_{k,s}(x)=0$ for
all $x\in A_{k,s}$. The update then gives
\[
\alpha_{k,s+1}-\alpha_{k,s}
=
\gamma_k|A_{k,s}|\alpha
\ge
0,
\]
so $\alpha_{k,s+1}\ge\alpha_{k,s}$; this holds regardless of whether
$|A_{k,s}|=0$. By the choice of $t'$, $\alpha_{k,s}<0$ for every $s$ with
$t'<s\le t$, so applying this monotonicity for $s=t'+1,\ldots,t-1$ yields
\[
\alpha_{k,t}\ge\alpha_{k,t'+1}\ge-\gamma_k\bar M_k(1-\alpha).
\]
Hence~\eqref{eq:D3} holds, and consequently
\begin{equation}\label{eq:D4}\tag{E.4}
|\alpha_{k,T+1}-\alpha|
\le
\max\{\alpha,1-\alpha\}+\gamma_k\bar M_k\max\{\alpha,1-\alpha\}.
\end{equation}
Substituting~\eqref{eq:D4} into~\eqref{eq:D1}:
\[
|{\text{CovErr}_{k,T}^{A}}-\alpha|
\le
\frac{\max\{\alpha,1-\alpha\}}{\gamma_k\sum_{t=1}^T|A_{k,t}|}
+
\frac{\bar M_k\max\{\alpha,1-\alpha\}}{\sum_{t=1}^T|A_{k,t}|}.
\]
The first term matches the stated bound.
For the second term, Lemma~\ref{lem:class_review_fraction} gives
${\sum_{t=1}^T|A_{k,t}|}/{\sum_{t=1}^T|D_{k,t}|}\to p$ a.s.
By the assumed frequency lower bound and $B_t\ge B^-$,
almost surely there exists some $T_0<\infty$
such that for all $T\ge T_0$,
\[
\sum_{t=1}^T|A_{k,t}|\ge\frac{p\,q_k\,T\,B^-}{2}.
\]
Since $|A_{k,t}|\le|A_t|\le p_t B^+$, we have $\bar M_k\le (\sup_t p_t) B^+$. Using unbiased randomized rounding, it follows
\[
\sup_t p_t
\leq
\sup_t \frac{\lceil p B_t \rceil}{B_t}
\leq
p+\frac{1}{B^-},
\qquad
\frac{\sup_t p_t}{p}
\le
1+\frac{1}{pB^-}.
\]
Therefore, on the same probability-one event,
for all $T\ge T_0$,
\[
\frac{\bar M_k\max\{\alpha,1-\alpha\}}{\sum_{t=1}^T|A_{k,t}|}
\le
\frac{2(\sup_t p_t)B^+}{p\,q_k\,T\,B^-}
\le
\frac{1}{T} \cdot 
\frac{2B^+}{B^- q_k}\big(1+\frac{1}{pB^-} \big) \quad \text{a.s.},
\]
Since both terms vanish as $T\to\infty$, we further have
$$|{\text{CovErr}_{k,T}^{A}}-\alpha|\to 0 \quad \text{a.s.}$$
This proves the proposition.
\hfill$\square$

\subsection{Proof of Theorem~\ref{thm:long_run_validity}}\label{proof:long_run_validity}

Fix a class $k\in[K]$. We first prove the convergence of accumulated error on all cells.
{Combining~\eqref{eq:D1} and~\eqref{eq:D4}, and noting that
$\bar M_k\leq B^+$, gives
\[
\left|\text{CovErr}_{k,T}^{A}-\alpha\right|
\leq
\frac{\max\{\alpha,1-\alpha\}(1+\gamma_k B^+)}
{\gamma_k\sum_{t=1}^T|A_{k,t}|}
\longrightarrow 0
\quad\text{a.s.}
\]
By Assumption~\ref{ass:main_guarantee}(ii) and Lemma~\ref{lem:class_review_fraction},
\[
\sum_{t=1}^T |D_{k,t}|\to\infty,
\qquad
\frac{\sum_{t=1}^T|A_{k,t}|}{\sum_{t=1}^T|D_{k,t}|}\to p>0
\quad\text{a.s.},
\]
and hence $\sum_{t=1}^T|A_{k,t}|\to\infty$ a.s., which justifies the above
convergence.}
Thus, by the triangle inequality, it suffices to show
\begin{equation}\label{eq:D5}\tag{E.5}
\bigl|{\text{CovErr}_{k,T}^{D}}-{\text{CovErr}_{k,T}^{A}}\bigr|\to0
\quad \text{a.s.}
\end{equation}
For convenience, write
$$Z_{k,t}:=\sum_{x\in A_{k,t}}\mathrm{err}_{k,t}(x),$$
and the centered sampling residual as
$$W_{k,t}:=Z_{k,t}-p_t{\sum_{x\in D_{k,t}}\mathrm{err}_{k,t}(x)}.$$
By the definitions above,
\[
{\text{CovErr}_{k,T}^{A}}-{\text{CovErr}_{k,T}^{D}}
=
\frac{\sum_{t=1}^T W_{k,t}
+\sum_{t=1}^T p_t{\sum_{x\in D_{k,t}}\mathrm{err}_{k,t}(x)}}
{\sum_{t=1}^T|A_{k,t}|}
-
\frac{\sum_{t=1}^T{\sum_{x\in D_{k,t}}\mathrm{err}_{k,t}(x)}}
{\sum_{t=1}^T|D_{k,t}|}.
\]
We then use
$\sum_{t=1}^T p_t{\sum_{x\in D_{k,t}}\mathrm{err}_{k,t}(x)}
=p\sum_{t=1}^T{\sum_{x\in D_{k,t}}\mathrm{err}_{k,t}(x)}
+\sum_{t=1}^T(p_t-p){\sum_{x\in D_{k,t}}\mathrm{err}_{k,t}(x)}$.
Factoring the full-batch average from the resulting middle two terms gives
\begin{equation}\label{eq:D6}\tag{E.6}
{\text{CovErr}_{k,T}^{A}}-{\text{CovErr}_{k,T}^{D}}
=
\underbrace{
\frac{\sum_{t=1}^T W_{k,t}}{\sum_{t=1}^T|A_{k,t}|}
}_{\mathrm{I}}
+
\underbrace{
{\text{CovErr}_{k,T}^{D}}
\left(
\frac{p\sum_{t=1}^T|D_{k,t}|}{\sum_{t=1}^T|A_{k,t}|}-1
\right)
}_{\mathrm{II}}
+
\underbrace{
\frac{\sum_{t=1}^T(p_t-p)
{\sum_{x\in D_{k,t}}\mathrm{err}_{k,t}(x)}}
{\sum_{t=1}^T|A_{k,t}|}
}_{\mathrm{III}}.
\end{equation}
We now treat the three terms in \eqref{eq:D6} separately.

\textit{Term~I.}
We first verify the martingale conditions for $W_{k,t}$.
Using the audit indicators $E_{t,i}=\mathbf{1}\{x_{t,i}\in A_t\}$,
we can write
\[
W_{k,t}
=
\sum_{i:\,y_{t,i}=k}
\mathrm{err}_{k,t}(x_{t,i})(E_{t,i}-p_t).
\]
Conditional on $\mathcal G_t$, the error indicators and $p_t$
are fixed. Uniform sampling without replacement gives
\[
\mathbb E[E_{t,i}\mid\mathcal G_t]=p_t,
\qquad
\operatorname{Var}(E_{t,i}\mid\mathcal G_t)=p_t(1-p_t).
\]
Consequently,
\begin{equation}\label{eq:D7}\tag{E.7}
\mathbb E[W_{k,t}\mid\mathcal G_t]=0.
\end{equation}
Moreover, when $B_t\ge2$, for distinct $i,j$,
\[
\operatorname{Cov}(E_{t,i},E_{t,j}\mid\mathcal G_t)
=
-\frac{p_t(1-p_t)}{B_t-1}
\le0.
\]
Since the error indicators are nonnegative and binary, for $B_t\ge2$
the nonpositive covariance terms yield
\begin{equation}\label{eq:D8}\tag{E.8}
\begin{aligned}
\mathbb E[W_{k,t}^2\mid\mathcal G_t]
&=
\operatorname{Var}(W_{k,t}\mid\mathcal G_t)\\
&\le
p_t(1-p_t)
\sum_{x\in D_{k,t}}\mathrm{err}_{k,t}(x)^2\\
&=
p_t(1-p_t)
\sum_{x\in D_{k,t}}\mathrm{err}_{k,t}(x)\\
&\le
p_t(1-p_t)|D_{k,t}|
\le \frac14|D_{k,t}|.
\end{aligned}
\end{equation}
When $B_t=1$, the audit set is deterministic conditional on
$\mathcal G_t$, so $W_{k,t}=0$ and \eqref{eq:D8} holds as well.
It also holds when $|D_{k,t}|=0$, since the defining sum is empty. Thus,
\eqref{eq:D7} and \eqref{eq:D8} hold for all $t$.
Equations~\eqref{eq:D7} and \eqref{eq:D8} verify the hypotheses of
Lemma~\ref{lem:weighted_mds_slln} with $b_t=|D_{k,t}|$ and $C=1/4$.
This sequence is nonnegative, $\mathcal G_t$-measurable, and satisfies
$\sum_{t=1}^T b_t\to\infty$ a.s.\ by Assumption~\ref{ass:main_guarantee}(ii).
Therefore,
${\sum_{t=1}^T W_{k,t}}/{\sum_{t=1}^T|D_{k,t}|}\to0$ a.s.
Lemma~\ref{lem:class_review_fraction} gives
${\sum_{t=1}^T|D_{k,t}|}/{\sum_{t=1}^T|A_{k,t}|}\to1/p$ a.s., so
\[
\textrm{Term I}
=
\frac{\sum_{t=1}^T W_{k,t}}{\sum_{t=1}^T|D_{k,t}|}
\cdot
\frac{\sum_{t=1}^T|D_{k,t}|}{\sum_{t=1}^T|A_{k,t}|}
\to
0\cdot\frac{1}{p}=0
\quad\text{a.s.}
\]

\textit{Term~II.}
By Lemma~\ref{lem:class_review_fraction},
$\sum_{t=1}^T|A_{k,t}|/\sum_{t=1}^T|D_{k,t}|\to p$ a.s.,
so
$$p\sum_{t=1}^T|D_{k,t}|/\sum_{t=1}^T|A_{k,t}|-1\to0 \quad \text{a.s.}$$
Since $0\le{\text{CovErr}_{k,T}^{D}}\le1$,
Term~II converges to zero almost surely.

\textit{Term~III.}
By Proposition~\ref{prop:unbiased_rounding},
\[
\frac{\sum_{t=1}^T (p_t-p)\sum_{x\in D_{k,t}}\mathrm{err}_{k,t}(x)}
{\sum_{t=1}^T|A_{k,t}|}
\to0
\quad\text{a.s.},
\]
which is precisely this term.

Hence all three terms in~\eqref{eq:D6} vanish a.s., proving~\eqref{eq:D5} and therefore
$$|{\text{CovErr}_{k,T}^{D}}-\alpha|\to0\quad \text{a.s.}$$

For the unreviewed cells, since $U_{k,t}=D_{k,t}\setminus A_{k,t}$,
writing the total error on $D_{k,t}$ as the sum of those on $A_{k,t}$ and
$U_{k,t}$ and rearranging:
\begin{equation}\label{eq:D9}\tag{E.9}
{\text{CovErr}_{k,T}^{U}}-{\text{CovErr}_{k,T}^{D}}
=
\frac{\sum_{t=1}^T|A_{k,t}|}{\sum_{t=1}^T|U_{k,t}|}
\bigl({\text{CovErr}_{k,T}^{D}}-{\text{CovErr}_{k,T}^{A}}\bigr).
\end{equation}
By Lemma~\ref{lem:class_review_fraction},
$\sum_{t=1}^T|A_{k,t}|/\sum_{t=1}^T|U_{k,t}|\to p/(1-p)$, which is finite.
Since~\eqref{eq:D5} holds,
identity~\eqref{eq:D9} yields
$$|{\text{CovErr}_{k,T}^{U}}-{\text{CovErr}_{k,T}^{D}}|\to0
\quad \text{a.s.},$$
and the triangle inequality completes the proof.
\hfill$\square$

\subsection{Proof of Corollary~\ref{cor:annotation_error_control}}
\label{proof:annotation_error_control}

\textit{Proof of Corollary~\ref{cor:annotation_error_control}.}
Fix a class $k\in[K]$. For any $x_{t,i}\in D_{k,t}$ with $I_{t,i}=1$,
Section~\ref{subsec:online_conf} gives
$\widehat C_t^+(x_{t,i})=\{\hat y_{t,i}\}$. Since
$\widehat C_t(x_{t,i})\subseteq\widehat C_t^+(x_{t,i})$, an incorrect
automated annotation implies that the original prediction set does not contain
the true class:
\[
\ind\{\hat y_{t,i}\neq k\}
\le
\ind\{k\notin\widehat C_t(x_{t,i})\}.
\]
Summing this inequality over the automatically annotated class-$k$ cells and
then over all batches gives
\[
\text{AnnoErr}_{k,T}^{D}
\le
\text{CovErr}_{k,T}^{D},
\qquad
\text{AnnoErr}_{k,T}^{U}
\le
\text{CovErr}_{k,T}^{U}.
\]
Theorem~\ref{thm:long_run_validity} shows that both quantities on the
right-hand side converge almost surely to $\alpha$. Therefore,
\[
\limsup_{T\to\infty}\text{AnnoErr}_{k,T}^{D}
\le\alpha,
\qquad
\limsup_{T\to\infty}\text{AnnoErr}_{k,T}^{U}
\le\alpha
\quad\text{a.s.}
\]
\hfill$\square$

\subsection{Proof of Proposition~\ref{prop:convergence_rates}}\label{proof:convergence_rates}
Fix a class $k\in[K]$. We retain the notation from the proof of Theorem~\ref{thm:long_run_validity}. In particular, \eqref{eq:D6} decomposes $\text{CovErr}_{k,T}^{A}-\text{CovErr}_{k,T}^{D}$ into the three underbraced terms denoted there by $\mathrm{I}$, $\mathrm{II}$, and $\mathrm{III}$. The proof below gives rate bounds for these three terms. By the triangle inequality,
\[
\left|\text{CovErr}_{k,T}^{D}-\alpha\right|
\le
\left|\text{CovErr}_{k,T}^{D}-\text{CovErr}_{k,T}^{A}\right|
+
\left|\text{CovErr}_{k,T}^{A}-\alpha\right|.
\]
Proposition~\ref{prop:reviewed_error} gives
\[
\left|\text{CovErr}_{k,T}^{A}-\alpha\right|
=
\mathcal{O}\!\left(\left(\sum_{t=1}^{T}|A_{k,t}|\right)^{-1}\right)
=
\mathcal{O}(T^{-1})
\quad\text{a.s.}
\]
Moreover, the class-frequency condition in Proposition~\ref{prop:reviewed_error} together with Assumption~\ref{ass:main_guarantee}(i) implies
\[
\sum_{t=1}^{T}|D_{k,t}|\asymp T
\qquad \text{a.s.},
\]
while Lemma~\ref{lem:class_review_fraction} gives
\[
\sum_{t=1}^{T}|A_{k,t}|\sim p\sum_{t=1}^{T}|D_{k,t}|
\quad\text{a.s.}
\]
It is therefore enough to prove that each of $\mathrm{I}$, $\mathrm{II}$, and $\mathrm{III}$ in \eqref{eq:D6} is $\mathcal{O}_p(T^{-1/2})$.

\textit{Term~I in \eqref{eq:D6}.} By~\eqref{eq:D7}, $(W_{k,t})$ is a martingale difference sequence, and~\eqref{eq:D8} gives $\mathbb{E}[W_{k,t}^2\mid\mathcal{G}_t]\le \frac14|D_{k,t}|$. Hence, by the orthogonality of martingale differences and $|D_{k,t}|\le B^+$,
\[
\mathbb{E}\left[\left(\sum_{t=1}^{T}W_{k,t}\right)^2\right]
=
\mathbb{E}\left[\sum_{t=1}^{T}\mathbb{E}[W_{k,t}^2\mid\mathcal{G}_t]\right]
\le
\frac14 B^+T.
\]
Chebyshev's inequality therefore gives $\sum_{t=1}^{T}W_{k,t}=\mathcal{O}_p(\sqrt{T})$. Using $\sum_{t=1}^{T}|A_{k,t}|\asymp T$ a.s.,
\[
\text{Term I}
=
\frac{\sum_{t=1}^{T}W_{k,t}}{\sum_{t=1}^{T}|A_{k,t}|}
=
\mathcal{O}_p(T^{-1/2}).
\]

For the two rounding terms in the remaining parts of the decomposition, let $\mathcal{G}_t'$ denote the time-$t$ sigma-field generated by all random quantities determined after $D_t$ is realized but before the randomized rounding variable that determines $p_t$ is drawn. Thus $\mathcal{G}_t'$ precedes the pre-review sigma-field $\mathcal{G}_t$, in which $p_t$ is already realized and the annotation subset $A_t$ has not yet been sampled. Under the unbiased randomized rounding rule,
\[
p_t-p=\frac{\xi_t-q_t}{B_t},
\qquad
q_t=pB_t-\lfloor pB_t\rfloor,
\qquad
\xi_t\mid\mathcal{G}_t'\sim\operatorname{Bernoulli}(q_t),
\]
and therefore
\begin{equation}\label{eq:D11}\tag{E.11}
\mathbb{E}[p_t-p\mid\mathcal{G}_t']=0,
\qquad
\mathbb{E}[(p_t-p)^2\mid\mathcal{G}_t']
=
\frac{q_t(1-q_t)}{B_t^2}
\le
\frac{1}{4B_t^2}.
\end{equation}
Since $|D_{k,t}|$ and $\sum_{x\in D_{k,t}}\operatorname{err}_{k,t}(x)$ are $\mathcal{G}_t'$-measurable and $0\le\sum_{x\in D_{k,t}}\operatorname{err}_{k,t}(x)\le |D_{k,t}|\le B_t$, \eqref{eq:D11} implies
\begin{equation}\label{eq:D12}\tag{E.12}
\begin{aligned}
\mathbb{E}[(p_t-p)|D_{k,t}|\mid\mathcal{G}_t']&=0,\\
\mathbb{E}\left[((p_t-p)|D_{k,t}|)^2\mid\mathcal{G}_t'\right]&\le\frac14\ind\{|D_{k,t}|>0\},
\end{aligned}
\end{equation}
and
\begin{equation}\label{eq:D13}\tag{E.13}
\begin{aligned}
\mathbb{E}\left[(p_t-p)\sum_{x\in D_{k,t}}\operatorname{err}_{k,t}(x)\,\middle|\,\mathcal{G}_t'\right]&=0,\\
\mathbb{E}\left[\left((p_t-p)\sum_{x\in D_{k,t}}\operatorname{err}_{k,t}(x)\right)^2\middle|\mathcal{G}_t'\right]&\le\frac14\ind\{|D_{k,t}|>0\}.
\end{aligned}
\end{equation}

\textit{Term~II in \eqref{eq:D6}.} Write the numerator of the bracketed factor as
\[
p\sum_{t=1}^{T}|D_{k,t}|-\sum_{t=1}^{T}|A_{k,t}|
=
-\sum_{t=1}^{T}Y_{k,t}-\sum_{t=1}^{T}(p_t-p)|D_{k,t}|,
\]
where $Y_{k,t}:=|A_{k,t}|-p_t|D_{k,t}|$. The proof of Lemma~\ref{lem:class_review_fraction} establishes \eqref{eq:lemma-e2-cond-mean} and \eqref{eq:lemma-e2-cond-var}; hence $(Y_{k,t})$ is a martingale difference sequence and $\mathbb{E}[Y_{k,t}^2\mid\mathcal{G}_t]\le C|D_{k,t}|$. The same second-moment calculation as in Term~I gives $\sum_{t=1}^{T}Y_{k,t}=\mathcal{O}_p(\sqrt{T})$. By~\eqref{eq:D12}, $((p_t-p)|D_{k,t}|)$ is also a martingale difference sequence with respect to the rounding filtration, and
\[
\mathbb{E}\left[\left(\sum_{t=1}^{T}(p_t-p)|D_{k,t}|\right)^2\right]
\le
\frac{T}{4},
\]
so $\sum_{t=1}^{T}(p_t-p)|D_{k,t}|=\mathcal{O}_p(\sqrt{T})$. Therefore, the bracketed factor in Term~II is $\mathcal{O}_p(T^{-1/2})$ after dividing by $\sum_{t=1}^{T}|A_{k,t}|\asymp T$, and since $0\le\text{CovErr}_{k,T}^{D}\le1$, we conclude
\[
\text{Term II}=\mathcal{O}_p(T^{-1/2}).
\]

\textit{Term~III in \eqref{eq:D6}.} By~\eqref{eq:D13}, $\big((p_t-p)\sum_{x\in D_{k,t}}\operatorname{err}_{k,t}(x)\big)$ is also a martingale difference sequence with respect to the rounding filtration. Its conditional second moments are bounded by $1/4$, so the same fixed-$T$ second-moment argument gives
\[
\sum_{t=1}^{T}(p_t-p)\sum_{x\in D_{k,t}}\operatorname{err}_{k,t}(x)
=
\mathcal{O}_p(\sqrt{T}).
\]
Dividing by $\sum_{t=1}^{T}|A_{k,t}|\asymp T$ therefore yields $\text{Term III}=\mathcal{O}_p(T^{-1/2})$.
Combining the three terms via~\eqref{eq:D6},
\[
\left|\text{CovErr}_{k,T}^{D}-\text{CovErr}_{k,T}^{A}\right|
=
\mathcal{O}_p(T^{-1/2}).
\]
Adding the $\mathcal{O}(T^{-1})$ contribution from Proposition~\ref{prop:reviewed_error} and using $\sum_{t=1}^{T}|D_{k,t}|\asymp T$, the triangle inequality gives
\[
\left|\text{CovErr}_{k,T}^{D}-\alpha\right|
=
\mathcal{O}_p\left(\left(\sum_{t=1}^{T}|D_{k,t}|\right)^{-1/2}\right)
=
\mathcal{O}_p(T^{-1/2}).
\]

For the unreviewed rate, identity~\eqref{eq:D9} and Lemma~\ref{lem:class_review_fraction} give
\[
\left|\text{CovErr}_{k,T}^{U}-\text{CovErr}_{k,T}^{D}\right|
=
\frac{\sum_{t=1}^{T}|A_{k,t}|}{\sum_{t=1}^{T}|U_{k,t}|}
\left|\text{CovErr}_{k,T}^{D}-\text{CovErr}_{k,T}^{A}\right|
=
\mathcal{O}_p(1)\cdot\mathcal{O}_p(T^{-1/2}),
\]
and the triangle inequality gives
\[
\left|\text{CovErr}_{k,T}^{U}-\alpha\right|
=
\mathcal{O}_p\left(\left(\sum_{t=1}^{T}|D_{k,t}|\right)^{-1/2}\right)
=
\mathcal{O}_p(T^{-1/2}).
\]
\hfill$\square$

\subsection{Proof of Proposition~\ref{prop:unbiased_rounding}}\label{proof:unbiased_rounding}

Fix a class $k\in[K]$. The rounding moment bounds in~\eqref{eq:D12} and
\eqref{eq:D13} show that $((p_t-p)|D_{k,t}|)$ and
$\bigl((p_t-p){\sum_{x\in D_{k,t}}\mathrm{err}_{k,t}(x)}\bigr)$
are martingale difference sequences with conditional second moments bounded by
a constant multiple of $|D_{k,t}|$. Applying
Lemma~\ref{lem:weighted_mds_slln} to these two sequences gives
\[
\frac{\sum_{t=1}^T(p_t-p)|D_{k,t}|}{\sum_{t=1}^T|D_{k,t}|}\to0,
\qquad
\frac{\sum_{t=1}^T(p_t-p)
{\sum_{x\in D_{k,t}}\mathrm{err}_{k,t}(x)}}
{\sum_{t=1}^T|D_{k,t}|}\to0
\quad\text{a.s.}
\]

It remains only to replace the denominator in the second convergence by
$\sum_{t=1}^T|A_{k,t}|$. Conditional on $\mathcal G_t$, where $p_t$ has already
been realized and the {audit set} $A_t$ has not yet been sampled,
$|A_{k,t}|{=\sum_{i:y_{t,i}=k}E_{t,i}}$ is hypergeometric with mean
$p_t|D_{k,t}|$ and variance bounded by $C|D_{k,t}|$ for a deterministic
constant $C<\infty$. Therefore Lemma~\ref{lem:weighted_mds_slln}, together with
the first convergence above, gives
\[
\frac{\sum_{t=1}^T|A_{k,t}|}{\sum_{t=1}^T|D_{k,t}|}
=
\frac{\sum_{t=1}^T\{|A_{k,t}|-p_t|D_{k,t}|\}}
{\sum_{t=1}^T|D_{k,t}|}
+
\frac{\sum_{t=1}^T(p_t-p)|D_{k,t}|}{\sum_{t=1}^T|D_{k,t}|}
+
p
\to p
\quad\text{a.s.}
\]
Since $p>0$, it follows that
\[
\frac{\sum_{t=1}^T(p_t-p)
{\sum_{x\in D_{k,t}}\mathrm{err}_{k,t}(x)}}
{\sum_{t=1}^T|A_{k,t}|}
=
\frac{
\sum_{t=1}^T(p_t-p)
{\sum_{x\in D_{k,t}}\mathrm{err}_{k,t}(x)}
/\sum_{t=1}^T|D_{k,t}|
}{
\sum_{t=1}^T|A_{k,t}|/\sum_{t=1}^T|D_{k,t}|
}
\to0
\quad\text{a.s.}
\]
Together with the first convergence, this establishes the two asymptotic
relations claimed in the proposition. Since $K$ is finite, the conclusions
hold simultaneously for all classes after intersecting the corresponding
probability-one events.
\hfill$\square$

\subsection{Proofs for Proposition~\ref{prop:bounded_class_batch_size_range} and Corollary~\ref{cor:bounded_class_batch_size_range}}
\label{proof:bounded_class_batch_size_range}

\paragraph{Proof of Proposition~\ref{prop:bounded_class_batch_size_range}.}

Fix a class $k\in[K]$ and write
\[
x_{k,t}:=\alpha_{k,t}-\alpha.
\]
Recall that $\bar\alpha_{k,t}$ denotes $\alpha_{k,t}$ clipped to the interval
$[0,1]$. Hence, $x_{k,t}$ and $\bar\alpha_{k,t}-\alpha$ have the same sign and
$|\bar\alpha_{k,t}-\alpha|\le |x_{k,t}|$. Since
$0\le\eta_{k,t}\le1$, we have
\[
|x_{k,t}-\eta_{k,t}(\bar\alpha_{k,t}-\alpha)|
=|x_{k,t}|-\eta_{k,t}|\bar\alpha_{k,t}-\alpha|.
\]
The update rule and the triangle inequality therefore give
\begin{equation}\label{eq:D14-new}\tag{E.14}
\eta_{k,t}|\bar\alpha_{k,t}-\alpha|
\le |x_{k,t}|-|x_{k,t+1}|
+\eta_{k,t}\left(
|\bar\alpha_{k,t}-\mathrm{BatErr}_{k,t}^D|+|\mathrm{BatErr}_{k,t}^A-\mathrm{BatErr}_{k,t}^D|
\right).
\end{equation}
Define
\[
Z_{k,t}:=|A_{k,t}|\,|\mathrm{BatErr}_{k,t}^A-\mathrm{BatErr}_{k,t}^D|,
\qquad
Y_{k,t}:=\bigl(|A_{k,t}|-p_t|D_{k,t}|\bigr)
\left(
|\bar\alpha_{k,t}-\mathrm{BatErr}_{k,t}^D|
-|\bar\alpha_{k,t}-\alpha|
\right).
\]
Summing~\eqref{eq:D14-new}, using
$\eta_{k,t}=\gamma_k|A_{k,t}|$, and dividing by $\gamma_k>0$ gives
\[
\sum_{t=1}^T |A_{k,t}|\,|\bar\alpha_{k,t}-\alpha|
\le
\frac{|x_{k,1}|-|x_{k,T+1}|}{\gamma_k}
+\sum_{t=1}^T |A_{k,t}|\,|\bar\alpha_{k,t}-\mathrm{BatErr}_{k,t}^D|
+\sum_{t=1}^T Z_{k,t}.
\]
The initialization $\alpha_{k,1}=\alpha$ implies $x_{k,1}=0$, so the
telescoping term is nonpositive and can be discarded. Moreover,
\[
\begin{aligned}
|A_{k,t}|\left(
|\bar\alpha_{k,t}-\alpha|-|\bar\alpha_{k,t}-\mathrm{BatErr}_{k,t}^D|
\right)
&=p_t|D_{k,t}|\left(
|\bar\alpha_{k,t}-\alpha|-|\bar\alpha_{k,t}-\mathrm{BatErr}_{k,t}^D|
\right)
-Y_{k,t}.
\end{aligned}
\]
Thus, adding and subtracting $p_t|D_{k,t}|$ in the preceding sum gives
\[
\begin{aligned}
\sum_{t=1}^T p_t|D_{k,t}|\,|\bar\alpha_{k,t}-\alpha|
&\le
\sum_{t=1}^T p_t|D_{k,t}|\,|\bar\alpha_{k,t}-\mathrm{BatErr}_{k,t}^D| + \sum_{t=1}^T Z_{k,t}+\sum_{t=1}^T Y_{k,t}.
\end{aligned}
\]
Using $p_t|D_{k,t}|\ge p^-d_k$ then yields
\begin{equation}\label{eq:D15-new}\tag{E.15}
\begin{aligned}
p^-d_k\sum_{t=1}^T|\bar\alpha_{k,t}-\alpha|
&\le
\sum_{t=1}^T p_t|D_{k,t}|\,|\bar\alpha_{k,t}-\mathrm{BatErr}_{k,t}^D|
+\sum_{t=1}^TZ_{k,t}+\sum_{t=1}^TY_{k,t}.
\end{aligned}
\end{equation}
Indeed, by~\eqref{eq:calib_error},
$|\bar\alpha_{k,t}-\mathrm{BatErr}_{k,t}^D|\le\Delta_{k,t}$, while
$p_t|D_{k,t}|\le p^+M_kd_k$. Hence,
\[
\sum_{t=1}^T p_t|D_{k,t}|\,|\bar\alpha_{k,t}-\mathrm{BatErr}_{k,t}^D|
\le
p^+M_kd_k\sum_{t=1}^T\Delta_{k,t}.
\]

We next control the remaining two sums. Conditional on $\mathcal G_t$,
the reviewed cells are sampled uniformly without replacement from $D_t$.
The term $Z_{k,t}$ is the absolute centered sample sum of the class-$k$
errors, and its conditional second moment is
\begin{equation}\label{eq:D16-new}\tag{E.16}
\begin{aligned}
\mathbb E\!\left[
Z_{k,t}^2
\,\middle|\,\mathcal G_t
\right]
&=
p_t(1-p_t)\frac{{B_t}}{{B_t}-1}\,
|D_{k,t}|\mathrm{BatErr}_{k,t}^D(1-\mathrm{BatErr}_{k,t}^D)\\
&\le
\frac14\frac{B^-}{B^--1}\,p^+(1-p^-)M_kd_k.
\end{aligned}
\end{equation}
Consequently, the conditional Cauchy--Schwarz inequality gives
\[
\mathbb E[Z_{k,t}\mid\mathcal G_t]
\le
\frac12\sqrt{
\frac{B^-}{B^--1}\,p^+(1-p^-)M_kd_k}.
\]
Moreover, $0\le Z_{k,t}\le |D_{k,t}|/4\le M_kd_k/4$. Since $B^-\ge2$,
the conditional variance of
$Z_{k,t}-\mathbb E[Z_{k,t}\mid\mathcal G_t]$ is at most
$p^+(1-p^-)M_kd_k/2$.

For $Y_{k,t}$, uniform sampling gives
$\mathbb E[|A_{k,t}|\mid\mathcal G_t]=p_t|D_{k,t}|$. Since the second
factor in the definition of $Y_{k,t}$ is determined before review,
$(Y_{k,t})_{t=1}^T$ is a martingale difference sequence. This factor has
absolute value at most one, so $|Y_{k,t}|\le M_kd_k$, and its conditional
variance is bounded by
\[
\operatorname{Var}(|A_{k,t}|\mid\mathcal G_t)
=p_t(1-p_t)\frac{{B_t}-|D_{k,t}|}{{B_t}-1}|D_{k,t}|
\le p^+(1-p^-)M_kd_k.
\]
Applying the rescaled form of Lemma~\ref{lem:freedman_application} to
$(Z_{k,t}-\mathbb E[Z_{k,t}\mid\mathcal G_t])_{t=1}^T$ and
$(Y_{k,t})_{t=1}^T$, and taking a union bound, shows that with probability at
least $1-\epsilon$,
\begin{equation}\label{eq:D17-new}\tag{E.17}
\begin{aligned}
\sum_{t=1}^TZ_{k,t}
&\le
\frac{T}{2}\sqrt{
\frac{B^-}{B^--1}\,p^+(1-p^-)M_kd_k}
+\sqrt{p^+(1-p^-)M_kd_kT\log\frac{2}{\epsilon}}
+\frac{M_kd_k}{12}\log\frac{2}{\epsilon},\\
\sum_{t=1}^TY_{k,t}
&\le
\sqrt{2p^+(1-p^-)M_kd_kT\log\frac{2}{\epsilon}}
+\frac{M_kd_k}{3}\log\frac{2}{\epsilon}.
\end{aligned}
\end{equation}
The two square-root deviation terms in~\eqref{eq:D17-new} have total
coefficient $1+\sqrt{2}<5/2$, and the two logarithmic terms sum to
$(5M_kd_k/12)\log(2/\epsilon)$. Substituting~\eqref{eq:D17-new} and the
calibration bound into~\eqref{eq:D15-new}, and then dividing by $p^-d_k$,
gives
\[
\sum_{t=1}^T|\bar\alpha_{k,t}-\alpha|
\le
\frac{p^+}{p^-}M_k\sum_{t=1}^T\Delta_{k,t}
+\mathcal R_{k,T}(\epsilon).
\]
Finally, the triangle inequality and~\eqref{eq:calib_error} imply
\[
|\mathrm{BatErr}_{k,t}^D-\alpha|
\le
|\mathrm{BatErr}_{k,t}^D-\bar\alpha_{k,t}|
+|\bar\alpha_{k,t}-\alpha|
\le
\Delta_{k,t}+|\bar\alpha_{k,t}-\alpha|.
\]
Summing this inequality and applying the preceding bound yields
\[
\sum_{t=1}^T|\mathrm{BatErr}_{k,t}^D-\alpha|
\le
\left(1+\frac{p^+}{p^-}M_k\right)
\sum_{t=1}^T\Delta_{k,t}
+\mathcal R_{k,T}(\epsilon),
\]
which proves the proposition.
\hfill$\square$

\paragraph{Proof of Corollary~\ref{cor:bounded_class_batch_size_range}.}

Under randomized rounding, take
$B^-:=\min_{1\le t\le T}{B_t}$. The bounded
class-size condition implies $B^-\ge d_k$, so $B^-\to\infty$ as
$d_k\to\infty$. Hence $p^-=p-1/B^-$ and $p^+=p+1/B^-$ satisfy
\[
\frac{p^+}{p^-}=1+\mathcal O\!\left(\frac{1}{d_k}\right),
\qquad
\frac{1}{p^-}=\mathcal O(1).
\]
Also, $\Delta_{k,t}\le2$, and therefore
\[
\left(\frac{p^+}{p^-}-1\right)M_k
\sum_{t=1}^T\Delta_{k,t}
=\mathcal O\!\left(\frac{T}{d_k}\right).
\]
For fixed $\epsilon$, the explicit expression for
$\mathcal R_{k,T}(\epsilon)$ gives
\[
\mathcal R_{k,T}(\epsilon)
=
\mathcal O\!\left(\frac{T}{\sqrt{d_k}}\right)+o_T(T).
\]
Substituting these two bounds into
Proposition~\ref{prop:bounded_class_batch_size_range}, and absorbing
$\mathcal O(T/d_k)$ into $\mathcal O(T/\sqrt{d_k})$, proves the corollary.
\hfill$\square$

\subsection{Proofs for Proposition~\ref{prop:nondecreasing_effective_steps} and Corollary~\ref{cor:nondecreasing_effective_steps}}
\label{proof:nondecreasing_effective_steps}

\paragraph{Proof of Proposition~\ref{prop:nondecreasing_effective_steps}.}

Write $x_{k,t}:=\alpha_{k,t}-\alpha$. Dividing~\eqref{eq:D14-new} by
$\eta_{k,t}>0$, and then using~\eqref{eq:calib_error} and the triangle
inequality, gives
\begin{equation}\label{eq:D18-new}\tag{E.18}
|\mathrm{BatErr}_{k,t}^D-\alpha|
\le
2\Delta_{k,t}+|\mathrm{BatErr}_{k,t}^A-\mathrm{BatErr}_{k,t}^D|
+\frac{|x_{k,t}|-|x_{k,t+1}|}{\eta_{k,t}}.
\end{equation}
By summation by parts, $x_{k,1}=0$, and the nondecreasing effective step
sizes,
\begin{equation}\label{eq:D19-new}\tag{E.19}
\sum_{t=1}^T\frac{|x_{k,t}|-|x_{k,t+1}|}{\eta_{k,t}}
\le0.
\end{equation}
Summing~\eqref{eq:D18-new} and applying~\eqref{eq:D19-new} yields
\[
\sum_{t=1}^T|\mathrm{BatErr}_{k,t}^D-\alpha|
\le
2\sum_{t=1}^T\Delta_{k,t}
+\sum_{t=1}^T|\mathrm{BatErr}_{k,t}^A-\mathrm{BatErr}_{k,t}^D|.
\]
By Lemma~\ref{lem:sampling fluctuation}, the second term on the right-hand
side is at most $S_{k,T}(\epsilon)$ with probability at least $1-\epsilon$,
which proves the proposition.
\hfill$\square$

\paragraph{Proof of Corollary~\ref{cor:nondecreasing_effective_steps}.}

By assumption, $|D_{k,t}|\ge d_k$ for every $t\le T$. For fixed
$\epsilon\in(0,1)$, the definition of $S_{k,T}(\epsilon)$ therefore gives
\[
\begin{aligned}
S_{k,T}(\epsilon)
&\le
T\left\{
\frac{1}{\sqrt{2p^-d_k}}
+\exp\!\left(-\frac{p^-d_k}{12}\right)
\right\}
+\sqrt{2T\log\frac{1}{\epsilon}}
+\frac{1}{3}\log\frac{1}{\epsilon}\\
&=
\mathcal O\!\left(\frac{T}{\sqrt{d_k}}\right)+o_T(T),
\end{aligned}
\]
where $p^-$ is fixed and positive. Combining this bound with
Proposition~\ref{prop:nondecreasing_effective_steps} proves the corollary.
\hfill$\square$


\section{Auxiliary lemmas}\label{sec:auxiliary_lemmas}

\begin{lemma}[Weighted strong law for martingale differences]
\label{lem:weighted_mds_slln}
Let $(\mathcal G_t)_{t\ge 1}$ be a filtration, and let $Y_t$ be
$\mathcal G_{t+1}$-measurable with
\[
\mathbb E[Y_t\mid\mathcal G_t]=0.
\]
Suppose there exists a nonnegative $\mathcal G_t$-measurable sequence
$(b_t)_{t\ge 1}$ and a constant $C<\infty$ such that
\[
\mathbb E[Y_t^2\mid\mathcal G_t]\le Cb_t
\quad\text{a.s.}
\]
If $\sum_{t=1}^T b_t\to\infty$ almost surely, then
\[
\frac{\sum_{t=1}^T Y_t}{\sum_{t=1}^T b_t}\to 0
\quad\text{a.s.}
\]
\end{lemma}

\begin{proof}
Set $S_0=1$ and $S_t=1+\sum_{i=1}^t b_i$.
Then $S_t$ is positive, nondecreasing, and $\mathcal G_t$-measurable. For every
$t\ge1$,
\[
\frac{b_t}{S_t^2}
\le
\frac{b_t}{S_{t-1}S_t}
=
\frac{1}{S_{t-1}}-\frac{1}{S_t},
\]
and consequently,
\[
\sum_{t=1}^\infty\frac{b_t}{S_t^2}\le1.
\]
Hence,
\[
\sum_{t=1}^{\infty}
\frac{\mathbb E[Y_t^2\mid\mathcal G_t]}{S_t^2}
\le
C\sum_{t=1}^{\infty}\frac{b_t}{S_t^2}
\le C<\infty
\quad\text{a.s.}
\]
Since $S_T\to\infty$ almost surely, the case $p=2$ of
Theorem~2.18 in \cite{hall1980martingale} yields
\[
\frac{\sum_{t=1}^T Y_t}{S_T}\to0
\quad\text{a.s.},
\]
Finally, $S_T/\sum_{t=1}^T b_t\to1$ almost surely, which gives the stated
conclusion.
\end{proof}

\begin{lemma}[Class-wise reviewed fraction]
\label{lem:class_review_fraction}
Under Assumption~\ref{ass:main_guarantee}, for every class $k\in[K]$, the fraction of reviewed cells converges to the target fraction $p$ with probability 1, i.e.,
\[
\frac{\sum_{t=1}^T |A_{k,t}|}{\sum_{t=1}^T |D_{k,t}|}
\to p
\quad\text{a.s.}
\]
\end{lemma}

\begin{proof}
Fix a class $k\in[K]$ and define
\[
Y_{k,t}:=|A_{k,t}|-p_t|D_{k,t}|.
\]
It suffices to show that
\[
\frac{\sum_{t=1}^{T}Y_{k,t}}{\sum_{t=1}^{T}|D_{k,t}|}\to0
\quad\text{a.s.}
\]
Indeed, by Proposition~\ref{prop:unbiased_rounding},
\[
\frac{\sum_{t=1}^{T}(p_t-p)|D_{k,t}|}
{\sum_{t=1}^{T}|D_{k,t}|}\to0
\quad\text{a.s.},
\]
and
\[
\frac{\sum_{t=1}^{T}|A_{k,t}|}{\sum_{t=1}^{T}|D_{k,t}|}
=
\frac{\sum_{t=1}^{T}Y_{k,t}}{\sum_{t=1}^{T}|D_{k,t}|}
+
\frac{\sum_{t=1}^{T}(p_t-p)|D_{k,t}|}
{\sum_{t=1}^{T}|D_{k,t}|}
+p .
\]
Now we verify the conditions of Lemma~\ref{lem:weighted_mds_slln}. Recall that $\mathcal G_t$ denotes the pre-review sigma-field after the
prediction sets for $D_t$ have been formed and before the {audit set} $A_t$ is sampled. Conditional on
$\mathcal G_t$, the incoming batch $D_t$, the class count $|D_{k,t}|$, and the
reviewed fraction $p_t$ are fixed. By Assumption~\ref{ass:main_guarantee}(iii),
$|A_{k,t}|$ follows a hypergeometric distribution with
population size ${B_t}$, number of class-$k$ cells $|D_{k,t}|$, and draw size
$p_t{B_t}$. Therefore
\[
\mathbb E(|A_{k,t}|\mid\mathcal G_t)
=
\frac{|A_t|}{{B_t}}|D_{k,t}|
=
p_t|D_{k,t}|,
\]
and consequently
\begin{equation}\tag{F.1}\label{eq:lemma-e2-cond-mean}
\mathbb E(Y_{k,t}\mid\mathcal G_t)=0 .
\end{equation}
If $B_t=1$, the conditional distribution of $|A_{k,t}|$ is degenerate and its
variance is zero. If $B_t\ge2$, the hypergeometric variance formula gives
\[
\operatorname{Var}(|A_{k,t}|\mid\mathcal G_t)
=
|A_t|\frac{|D_{k,t}|}{{B_t}}
\left(1-\frac{|D_{k,t}|}{{B_t}}\right)
\frac{{B_t}-|A_t|}{{B_t}-1}.
\]
For $B_t\ge2$, the factors after $|A_t|$ are bounded by one and
$|A_t|\le {B_t}$, whereas for $B_t=1$ the variance is zero. Thus there
exists a constant $C<\infty$ such that
\begin{equation}\tag{F.2}\label{eq:lemma-e2-cond-var}
\mathbb E(Y_{k,t}^2\mid\mathcal G_t)
=
\operatorname{Var}(|A_{k,t}|\mid\mathcal G_t)
\le
C|D_{k,t}|.
\end{equation}
By \eqref{eq:lemma-e2-cond-mean} and \eqref{eq:lemma-e2-cond-var}, taking
$b_t=|D_{k,t}|$, which is nonnegative and $\mathcal G_t$-measurable, in
Lemma~\ref{lem:weighted_mds_slln} yields
\[
\frac{\sum_{t=1}^{T}Y_{k,t}}{\sum_{t=1}^{T}|D_{k,t}|}\to0
\quad\text{a.s.},
\]
where $\sum_{t=1}^{T}|D_{k,t}|\to\infty$ follows directly from
Assumption~\ref{ass:main_guarantee}(ii).
Combining this with the preceding decomposition proves the result.
\end{proof}

\begin{lemma}[Cumulative sampling fluctuation]
    \label{lem:sampling fluctuation}
Fix a class $k\in[K]$. Suppose that Assumptions~\ref{ass:main_guarantee} and
\ref{ass:sum_guarantee_bounds} hold. Then, for any horizon $T\ge1$ and any
$\zeta\in(0,1)$, with probability at least $1-\zeta$,
\begin{equation}\nonumber\label{eq:lemma-e3-cumulative}
\sum_{t=1}^T |\mathrm{BatErr}_{k,t}^A-\mathrm{BatErr}_{k,t}^D|
\le
\sum_{t=1}^T\mu_{k,t}
+
\sqrt{2T\log\frac{1}{\zeta}}+\frac{1}{3}\log\frac{1}{\zeta}
=S_{k,T}(\zeta),
\end{equation}
where $\mu_{k,t}$ is defined in~\eqref{eq:mu and sigma}.
\end{lemma}

\begin{proof}
Fix $t$ and define
\[
\mathcal H_t:=\mathcal G_t\vee\sigma(|A_{k,t}|),
\]
where $\mathcal G_t$ is the previously defined pre-review sigma-field 
and $|A_{k,t}|$ denotes the number of reviewed
class-$k$ cells.
Conditional on $\mathcal H_t$ and on the event $|A_{k,t}|=m>0$, the number
of erroneous reviewed class-$k$ cells
$
\sum_{x\in A_{k,t}}\mathrm{err}_{k,t}(x)
$
has a hypergeometric distribution. The population size is $|D_{k,t}|$, the
success count is $\sum_{x\in D_{k,t}}\mathrm{err}_{k,t}(x)$, and the draw
size is $m$. Since
\[
\mathrm{BatErr}_{k,t}^A
=
\frac{1}{m}\sum_{x\in A_{k,t}}\mathrm{err}_{k,t}(x),
\]
the conditional mean is $\mathrm{BatErr}_{k,t}^D$. If $|D_{k,t}|=1$, then the conditional
distribution is degenerate and the variance is zero. If $|D_{k,t}|\ge2$, the
hypergeometric variance formula gives
\[
\begin{aligned}
\operatorname{Var}\!\left(
\mathrm{BatErr}_{k,t}^A
\,\middle|\,\mathcal H_t,\ |A_{k,t}|=m
\right)
&=
\frac{1}{m^2}\,
m \mathrm{BatErr}_{k,t}^D(1-\mathrm{BatErr}_{k,t}^D)
\frac{|D_{k,t}|-m}{|D_{k,t}|-1}\\
&\le
\frac{\mathrm{BatErr}_{k,t}^D(1-\mathrm{BatErr}_{k,t}^D)}{m}
\le
\frac{1}{4m},
\end{aligned}
\]
where the last inequality uses $0\le \mathrm{BatErr}_{k,t}^D\le1$. Thus the same upper bound
holds for all $|D_{k,t}|\ge1$. Together with the convention
$\mathrm{BatErr}_{k,t}^A=\mathrm{BatErr}_{k,t}^D$ when $|A_{k,t}|=0$, this gives
\begin{equation}\tag{F.3}\label{eq:lemma-e3-conditional-square}
\mathbb E\!\left[
(\mathrm{BatErr}_{k,t}^A-\mathrm{BatErr}_{k,t}^D)^2
\,\middle|\,\mathcal H_t
\right]
\le
\frac{{\ind}\{|A_{k,t}|>0\}}{4|A_{k,t}|}.
\end{equation}
Since $\mathcal G_t\subseteq\mathcal H_t$, by the tower property and conditional
Jensen's inequality, \eqref{eq:lemma-e3-conditional-square} gives
\begin{equation}\tag{F.4}\label{eq:lemma-e3-jensen}
\begin{aligned}
\mathbb E\!\left[
|\mathrm{BatErr}_{k,t}^A-\mathrm{BatErr}_{k,t}^D|
\,\middle|\,\mathcal G_t
\right]
&=
\mathbb E\!\left[
\mathbb E\!\left[
|\mathrm{BatErr}_{k,t}^A-\mathrm{BatErr}_{k,t}^D|
\,\middle|\,\mathcal H_t
\right]
\,\middle|\,\mathcal G_t
\right]\\
&\le
\mathbb E\!\left[
\left\{
\mathbb E\!\left[
(\mathrm{BatErr}_{k,t}^A-\mathrm{BatErr}_{k,t}^D)^2
\,\middle|\,\mathcal H_t
\right]
\right\}^{1/2}
\,\middle|\,\mathcal G_t
\right]\\
&\le
\mathbb E\!\left[
\frac{{\ind}\{|A_{k,t}|>0\}}{2\sqrt{|A_{k,t}|}}
\,\middle|\,\mathcal G_t
\right].
\end{aligned}
\end{equation}

Define the high-count event
\[
{\mathcal E_{k,t}^{\mathrm{high}}}
=\{|A_{k,t}|\ge p^-|D_{k,t}|/2\}.
\]
Conditional on $\mathcal G_t$, the count $|A_{k,t}|$ follows a hypergeometric
distribution with mean $p_t|D_{k,t}|$. Since $p_t\ge p^-$, the
multiplicative Chernoff bound for sampling without replacement yields
\begin{equation}\tag{F.5}\label{eq:lemma-e3-hyper-tail}
\begin{aligned}
\mathbb P(({\mathcal E_{k,t}^{\mathrm{high}}})^c\mid\mathcal G_t)
&=
\mathbb P\!\left(
|A_{k,t}|<\frac12 p^-|D_{k,t}|
\,\middle|\,\mathcal G_t
\right)\\
&\le
\mathbb P\!\left(
|A_{k,t}|<\frac12 p_t|D_{k,t}|
\,\middle|\,\mathcal G_t
\right)\\
&\le
2\exp\!\left(-\frac{p_t|D_{k,t}|}{12}\right)
\le
2\exp\!\left(-\frac{p^-|D_{k,t}|}{12}\right).
\end{aligned}
\end{equation}

Now define the centered fluctuation
\[
V_{k,t}:=|\mathrm{BatErr}_{k,t}^A-\mathrm{BatErr}_{k,t}^D|-
\mathbb E[|\mathrm{BatErr}_{k,t}^A-\mathrm{BatErr}_{k,t}^D|\mid\mathcal G_t].
\]
Then $V_{k,t}\in\mathcal G_{t+1}$,
$\mathbb E(V_{k,t}\mid\mathcal G_t)=0$, and $|V_{k,t}|\le1$.
It remains to bound the predictable part of $|\mathrm{BatErr}_{k,t}^A-\mathrm{BatErr}_{k,t}^D|$ and the
conditional second moment of $V_{k,t}$. On the high-count event
${\mathcal E_{k,t}^{\mathrm{high}}}$, the two factors
${\ind}\{|A_{k,t}|>0\}/(2\sqrt{|A_{k,t}|})$ and
${\ind}\{|A_{k,t}|>0\}/(4|A_{k,t}|)$ are bounded by
$(2p^-|D_{k,t}|)^{-1/2}$ and $(2p^-|D_{k,t}|)^{-1}$, respectively. On
$({\mathcal E_{k,t}^{\mathrm{high}}})^c$, they are bounded by
$1/2$ and $1/4$. Combining these
bounds with \eqref{eq:lemma-e3-jensen} and \eqref{eq:lemma-e3-hyper-tail}
gives
\begin{equation}\tag{F.6}\label{eq:lemma-e3-drift-bound}
\begin{aligned}
\mathbb E\!\left[
|\mathrm{BatErr}_{k,t}^A-\mathrm{BatErr}_{k,t}^D|
\,\middle|\,\mathcal G_t
\right]
&\le
\frac{1}{\sqrt{2p^-|D_{k,t}|}}
+
\frac12\,\mathbb P(({\mathcal E_{k,t}^{\mathrm{high}}})^c
\mid\mathcal G_t)\\
&\le
\frac{1}{\sqrt{2p^-|D_{k,t}|}}
+
\exp\!\left(-\frac{p^-|D_{k,t}|}{12}\right)
=\mu_{k,t}
\end{aligned}
\end{equation}
For the centered fluctuation, since $|V_{k,t}|\le1$, the predictable quadratic variation
satisfies
\begin{equation}\tag{F.8}\label{eq:lemma-e3-quadratic-variation}
\sum_{t=1}^T\mathbb E[V_{k,t}^2\mid\mathcal G_t]
\le T
\quad\text{a.s.}
\end{equation}
Applying Lemma~\ref{lem:freedman_application} to the martingale
difference sequence $(V_{k,t})_{t=1}^T$ therefore gives, with probability at
least $1-\zeta$,
\begin{equation}\tag{F.9}\label{eq:lemma-e3-freedman}
\sum_{t=1}^T V_{k,t}
\le
\sqrt{2T\log\frac{1}{\zeta}}+\frac{1}{3}\log\frac{1}{\zeta}.
\end{equation}
By the definition of $V_{k,t}$ and~\eqref{eq:lemma-e3-drift-bound},
\[
\begin{aligned}
\sum_{t=1}^T |\mathrm{BatErr}_{k,t}^A-\mathrm{BatErr}_{k,t}^D|
&=
\sum_{t=1}^T
\mathbb E[|\mathrm{BatErr}_{k,t}^A-\mathrm{BatErr}_{k,t}^D|\mid\mathcal G_t]
+
\sum_{t=1}^T V_{k,t}\\
&\le
\sum_{t=1}^T\mu_{k,t}
+
\sqrt{2T\log\frac{1}{\zeta}}+\frac{1}{3}\log\frac{1}{\zeta}
=S_{k,T}(\zeta),
\end{aligned}
\]
which proves the lemma.
\end{proof}

\begin{lemma}[Corollary of Freedman's inequality]
\label{lem:freedman_application}
Let $(\mathcal G_t)_{t\geq 0}$ be a filtration. Suppose $V_t$ is
$\mathcal G_{t+1}$-measurable with
$
\mathbb E[V_t\mid \mathcal G_t]=0$ and $|V_t|\le 1 \text{ a.s.}
$
Assume further that 
\[
\sum_{t=1}^T\mathbb E(V_t^2\mid\mathcal G_t)\le B\quad\text{a.s.}
\]
for some constant $B>0$. Then for any $\zeta\in(0,1)$, with probability at
least $1-\zeta$,
\[
\sum_{t=1}^T V_t
\le
\sqrt{2B\log\frac{1}{\zeta}}
+
\frac{1}{3}\log\frac{1}{\zeta}.
\]
\end{lemma}

\begin{proof}
Theorem~4.1 of \citet{freedman1975tail} states: for a martingale difference
sequence with increments bounded above by one, if $S_n=\sum_{t=1}^n V_t$
denotes its partial sum and $W_n=\sum_{t=1}^n\mathbb E(V_t^2\mid\mathcal G_t)$
its predictable quadratic variation, then for all $a,b>0$,
\[
\mathbb P\!\left(
\exists n:\ S_n\ge a\ \text{and}\ W_n\le b
\right)
\le
\left(\frac{b}{a+b}\right)^{a+b}e^a.
\]
Since $W_T\le B$ a.s.\ by assumption, the event $\{S_T\ge a\}$ implies the
event in the display (at $n=T$). We therefore apply the theorem with $b=B$ and
\[
a=\sqrt{2B\log\frac{1}{\zeta}}+\frac{1}{3}\log\frac{1}{\zeta}.
\]
Set $L=\log(1/\zeta)$. Since $b=B$, the choice of $a$ gives
$a=\sqrt{2bL}+L/3$. We compute
\[
\begin{aligned}
\left(\frac{b}{a+b}\right)^{a+b}e^a
&=
\exp\!\left\{
-(a+b)\log\!\left(1+\frac{a}{b}\right)+a
\right\}\\
&=
\exp\!\left\{
-b\left[
\left(1+\frac{a}{b}\right)
\log\!\left(1+\frac{a}{b}\right)
-\frac{a}{b}
\right]
\right\}.
\end{aligned}
\]
Setting $r=L/b$, we have $a/b=\sqrt{2r}+r/3$.
We first establish
\begin{equation}\tag{F.7}
\label{eq:bennett-inversion}
\left(1+\sqrt{2r}+\frac{r}{3}\right)
\log\!\left(1+\sqrt{2r}+\frac{r}{3}\right)
-\sqrt{2r}-\frac{r}{3}
\ge r,
\qquad r\ge0,
\end{equation}
To prove \eqref{eq:bennett-inversion}, define
\[
F(x)=
\left(1+x+\frac{x^2}{6}\right)
\log\!\left(1+x+\frac{x^2}{6}\right)
-x-\frac{2x^2}{3},
\qquad x\ge0.
\]
Direct differentiation gives $F(0)=F'(0)=0$ and
\[
F''(x)=\frac{1}{3}\left[
\log\!\left(1+x+\frac{x^2}{6}\right)
-1+\frac{1}{1+x+x^2/6}
\right]\ge0,
\]
where the inequality follows from $\log z\ge1-z^{-1}$ for $z\ge1$.
Thus $F(x)\ge0$ for all $x\ge0$. Taking $x=\sqrt{2r}$ proves
\eqref{eq:bennett-inversion}.
Applying \eqref{eq:bennett-inversion} with $r=L/b$ yields
\[
b\left[
\left(1+\frac{a}{b}\right)
\log\!\left(1+\frac{a}{b}\right)-\frac{a}{b}
\right]
\ge br=L.
\]
Consequently,
\[
\left(\frac{b}{a+b}\right)^{a+b}e^a
\le e^{-L}=\zeta.
\]
Therefore $\mathbb P(S_T\ge a)\le\zeta$, which proves the stated bound.
\end{proof}

\end{document}